\documentclass[12pt]{iopart}
\usepackage{graphicx}
\usepackage{dcolumn}
\usepackage{epsfig}
\usepackage{bm}
\usepackage{subfig}
\def\rp{$R_p\hspace{-1em}/\ \ $}

\def\dbd{$0\nu\beta\beta$}

\newcommand{\beq}{\begin{equation}}
\newcommand{\eeq}{\end{equation}}

\newcommand{\barr}{\begin{eqnarray}}
\newcommand{\earr}{\end{eqnarray}} 
\begin{document}   
\topmargin -0.50in

\title[Neutrinoless double beta decay]{Theory of neutrinoless double beta decay}

\author{J D Vergados$^{1,2}$, H  Ejiri$^{3,4}$ and F \v Simkovic$^{5,6}$}
\address{$^1$Theoretical Physics Division, University of Ioannina, 
GR--451 10, Ioannina, Greece.}
\address{$^2$CERN, Theory Division, Geneva, Switzerland.\ead{vergados@uoi.gr}}
\address{$^3$RCNP, Osaka University, Osaka, 567-0047, Japan.}
 \address{$^4$ Nuclear Science, Czech Technical University, Brehova, Prague, Czech Republic.
        \ead{ejiri@rcnp.osaka-u.ac.jp}}
\address{$^5$Laboratory of Theoretical Physics, JINR,
141980 Dubna, Moscow region, Russia.}
\address{$^6$Department of Nuclear Physics and Biophysics, 
Comenius University, Mlynsk\'a dolina F1, SK--842 15
Bratislava, Slovakia.\ead{Fedor.Simkovic@fmph.uniba.sk
}}

\date{\today}

		\begin{abstract}
Neutrinoless double beta decay, which is a very old and yet elusive process, is reviewed. Its observation will signal that lepton number is not conserved and the neutrinos are Majorana particles. More importantly it is our best hope for determining the absolute neutrino mass scale at the level of a  few tens of meV. 
 To achieve the last goal certain hurdles have to be overcome involving particle, nuclear and experimental physics.\\
Nuclear	physics is important for extracting the useful information from the data. One must accurately  evaluate the relevant nuclear matrix elements, a formidable task.
To this end, we review the sophisticated nuclear structure approaches recently been developed, which give confidence that the needed nuclear matrix 
elements can be reliably calculated employing different methods: 
a) the various versions of the Quasiparticle Random Phase Approximations, 
b) the interacting boson model, c) the energy density functional method 
and d) the large basis Interacting Shell Model.
 It is encouraging that, for the 
light neutrino mass term at least, these  vastly different approaches now give comparable results.\\ 
From an experimental point of view it is challenging, since the life times are long and one has to 
fight against formidable backgrounds. One needs large isotopically enriched sources and detectors 
with high energy resolution, low thresholds and very low background.\\
If a signal is found, it will be a tremendous accomplishment. Then, of course, the real task 
is going to be the extraction  of the neutrino mass from the observations. This is not trivial, 
since current particle models  predict the presence of many  mechanisms other than 
the neutrino mass, which may contribute or even dominate this process. We will, in particular, 
consider the following processes:
\begin{enumerate} 
\item The neutrino induced, but neutrino mass independent contribution. 
\item Heavy left and/or right handed neutrino mass contributions. 
\item Intermediate  scalars (doubly charged etc).
\item  Supersymmetric (SUSY) contributions. 
\end{enumerate}
We will show that it is  possible to disentangle the various mechanisms and unambiguously 
extract the important  neutrino mass scale, if all the signatures of the reaction are 
searched in a sufficient number of nuclear isotopes. 
\end{abstract}
\submitto{Reports on Progress in Physics}
\maketitle

\tableofcontents
\pagestyle{plain}
\newpage 

\section[A brief history of double beta decay]{A brief history of double beta decay}

A brief history of the double-beta decay is presented.

\subsection{The early period}

Double beta decay (DBD), namely the two-neutrino double-beta decay 
($2\nu\beta\beta$-decay)
\begin{equation}
(A, Z) \to (A, Z + 2) + e^- + e^- + {\overline\nu}_e + {\overline\nu}_e,
\label{eq:a2}   
\end{equation}
was first considered in publication \cite{GOEPMAY}
of Maria Goeppert-Mayer in 1935. It was Eugene Wigner, who
suggested this problem  to the author of \cite{GOEPMAY} about one year after 
the Fermi weak interaction theory appeared. In the work of Maria Goeppert-Mayer
\cite{GOEPMAY} an expression for the $2\nu\beta\beta$-decay rate was 
derived and a half-life of  $10^{17}$ years was estimated by assuming a Q-value 
of about 10 MeV. 

Two years later (1937) Ettore  Majorana formulated a new theory of 
neutrinos, whereby the neutrino $\nu$ and the antineutrino $\overline\nu$ are indistinguishable, 
and suggested antineutrino induced $\beta^-$-decay for experimental
verification of this hypothesis \cite{emajorana}. 
Giulio Racah was the first, who proposed testing 
Majorana's theory with real neutrinos by chain of reactions
\begin{eqnarray}
(A,Z) \rightarrow (A,Z+1) + e^- + \nu, ~~~\nu + (A',Z') \rightarrow (A',Z'+1) + e^-,
\end{eqnarray}
which is allowed in the case of the Majorana neutrino and forbidden in the case of Dirac
neutrino \cite{racah}. In 1939, Wolfgang Furry considered for the first time
neutrinoless double beta decay
($0\nu\beta\beta$-decay),
\begin{eqnarray}
(A,Z) \rightarrow (A,Z+2) + e^- + e^-,
\label{eq:a}   
\end{eqnarray}
a Racah chain of reactions with virtual neutrinos ($(A,Z+1) \equiv (A',Z')$) \cite{Fur39}. Here $A,A'$ are the nuclear mass numbers and $Z,Z'$ the charges of the nuclei involved. The available energy $\Delta$ is equal to the $Q$-value of the reaction, i.e. the mass difference of the ground states of the two atoms involved.\\
In 1952 Henry Primakoff  \cite{Pr52}
calculated the electron-electron angular correlations and electron energy spectra 
for both the $2\nu\beta\beta$-decay and the $0\nu\beta\beta$-decay, producing a
useful tool for distinguishing between the two processes.

At that time nothing was known about the chirality suppression of the 
$0\nu\beta\beta$-decay. It was believed that, due to a considerable phase-space
advantage, the $0\nu\beta\beta$-decay mode dominates the double beta decay rate. 
Starting 1950 this phenomenon was exploited in early geochemical, radiochemical 
and counter experiments\footnote{ For more detailed
historical review of the experimental activities in the field of double-beta
decay we recommend reader a recent review on this subject by
A. Barabash \cite{barabhis}.}. It was found that the measured lower limit on the 
$\beta\beta$-decay half-life far exceeds the values expected for this process,
$T^{0\nu}_{1/2} \sim 10^{12}-10^{15}$ years. In 1955 the Raymond Davis experiment \cite{davisno},
which searched for the antineutrinos from reactor via nuclear reaction
${\overline{\nu}}_e + {^{37}Cl} \rightarrow {^{37}Ar} + e^-$, produced a zero result.
The above experiments were interpreted as proof that the neutrino 
was not a Majorana particle, but a Dirac particle. This prompted the
introduction of the lepton number to distinguish the neutrino from its
antiparticle. The assumption of lepton number conservation allows the 
$2\nu\beta\beta$-decay but forbids the $0\nu\beta\beta$-decay, in which lepton
number is changed by two units. 

In 1949 Fireman reported the first observation of the $\beta\beta$-decay of $^{124}Sn$ 
in a laboratory experiment \cite{fire1}, but he disclaimed it later 
\cite{fire2}. The first geochemical observation of the $\beta\beta$-decay,
with an estimated half-life $T_{1/2} (^{130}Te) = 1.4 \times 10^{21}$
years, was announced by Ingram and Reynolds in 1950 \cite{ing}. 
Extensive studies have been made by Gentner and Kirsten \cite{kir67,kir67b} and others \cite{tak66,sir72} on such rare-gass isotopes as $^{82}$Kr, 
$^{128}$Xe, and $^{130}$Xe, which are $\beta\beta$-decay products of $^{82}$Se $^{128}$Te, and $^{130}$Te, 
respectively,  obtaining half lives around 10$^{21}$y for $^{130}$Te. 

\subsection{The period of scepticism}

Shortly after its theoretical formulation by Lee and Yang, parity violation in the 
weak interaction was established by two epochal experiments. 
In 1957 Wu et al. discovered the asymmetry in the angular distribution 
of the $\beta$-particles emitted relative to the spin orientation of 
the parent nucleus $^{60}Co$. A year later Goldhaber et al \cite{G58} discovered  that the neutrinos are polarized and left handed 
by measuring the polarization of a  photon, moving back to back with the neutrino, 
produced  by the de-excitation of a $^{152}Eu^*$ nucleus 
after K-capture. In 1958 the seemingly confused situation was simplified in the
form of the vector-axial vector (V -A) theory of weak interactions
describing maximal parity violation in agreement with available data. 
In order to account for the chiral symmetry breaking of the weak interaction 
only left handed fermions participate and the mediating particles 
must be vectors of spin 1, which are left handed in the sense that they couple only to left handed fermions. 

The maximal parity violation is easily realized in the lepton sector 
by using the two-component theory of a massless neutrino, proposed 
in 1957 by L. Landau, T.D. Lee, C.N. Yang and A. Salam 
(This idea was first developed by H. Weyl in 1929, but it was 
rejected by Pauli in 1933 on the grounds that it
violates parity.). In this theory, neutrinos are left handed and antineutrinos
are right-handed, leading automatically to the V - A couplings. 

With the discovery of parity violation, it became apparent that the 
Majorana/Dirac character of the electron neutrino was still in question. 
The particles that participate in the $0\nu\beta\beta$-decay reaction 
at nucleon level are right-handed antineutrino  ${\overline \nu}_e$
and left handed neutrino $\nu_e$: 
\begin{eqnarray}
n \rightarrow p + e^- + {\overline \nu}^{R}_e, ~~~
{\nu}^{L}_e + n \rightarrow p + e^-.  
\end{eqnarray}
Thus even if the neutrino is a (massless) Majorana particle, since the first neutrino has 
the wrong helicity for absorption by a neutron, the
absence of the $0\nu\beta\beta$-decay  implies neither a Dirac 
electron neutrino nor a conserved lepton number. 

The requirement that both lepton number conservation and the $\gamma_5$ 
invariance of the weak current had to be violated, in order the 
$0\nu\beta\beta$-decay to occur, discouraged experimental searches.

\subsection{The period of Grand Unified Theories}

The maximal violation of parity (and of charge-conjugation)
symmetry is accommodated in the Standard Model (SM),
which describes jointly weak and electromagnetic interactions. 
This model was developed largely upon the empirical observations 
of nuclear beta decay during the latter half of the past century. 
Despite the phenomenological success of the SM, the 
fundamental origin of parity violation has not been understood. 
In spite of the fact that the SM represents the simplest and 
the most economical theory, it has  not been considered as the ultimate
theory of nature. It was assumed that, most likely, it describes  a low energy approximation to a more fundamental theory. 

With the development of modern gauge theories during the last quarter  of the previous century, perceptions began to change.
In the SM it became apparent that the assumption of lepton number
conservation led to the neutrino being strictly massless, thus preserving
the $\gamma_5$-invariance of the weak current. With the
development of Grand Unified Theories (GUT's) of the electroweak and strong 
interactions, the prejudice has grown that lepton number conservation 
was the result of a global symmetry not of a gauge symmetry and had to be broken at some level. 
In other words modern GUT's and supersymmetric (SUSY) extensions of the SM 
suppose that such conservation laws of the SM may be violated to 
some small degree. The lepton number may only appear to be conserved
at low energies because of the large grand unified mass scale $\Lambda_{GUT}$
governing its breaking. Within the proposed see-saw mechanism one
expects the neutrino to acquire a small Majorana mass of a size
$\sim ~(\mbox{light~mass})^2/\Lambda_{GUT}$, where ``light mass'' is typically
that of a quark or charged lepton. The considerations of a sensitivity
of the $0\nu\beta\beta$-decay experiments to a neutrino mass 
$m_\nu \sim 1~eV$ became the genesis of a new interest to double beta
decay. Thus the  interest in $0\nu\beta\beta$-decay  was resurrected through the pioneering work of Kotani and his group
		\cite {DTNOT}, which brought it again to the attention of the nuclear physics 
		community.

Neutrino masses require either the existence of right-handed neutrinos
or require violation of the lepton number (LN)  so that Majorana masses are
possible. So, one is forced to go beyond the minimal models again, 
whereby LF and/or LN violation can be allowed in the theory. 
A good candidate for such a theory is  the left-right symmetric 
model of Grand Unification inaugurated by Salam, Pati, Mohapatra 
and Senjanovi\'c \cite{pasa74,mopa75,mose75} and especially models based on SO(10), which 
have first been proposed by Fritzsch and Minkowski \cite{fri}, 
with their supersymmetric versions \cite{kumo93,aula98,dumo99}. 
The left-right symmetric models,  
representing generalization of the $SU(2)_L \otimes U(1)$ SM, 
predict not only  that the neutrino is a Majorana particle, that means 
it is up to a phase identical with its antiparticle, but automatically 
predict the neutrino has a mass and a weak right-handed interaction.

In the left-right symmetric models the LN conservation is
broken by the presence of the Majorana neutrino mass.
The LN violation is also inherent in those SUSY theories 
whereby R-parity, defined as 
$R_p = (-1)^{3B+L+2S}$, with $S$, $B$, and $L$ being the spin, baryon and lepton
number, respectively is not a conserved quantity anymore.

The $0\nu\beta\beta$-decay, which involves the emission of 
two electrons and no neutrinos, has been found as 
a powerful tool to study the LN conservation. Schechter and Valle
proved that, if the $0\nu\beta\beta$-decay 
takes place, regardless of the mechanism causing it, the neutrinos are Majorana particles 
with non-zero mass \cite{SVa82,klko96}. It was recognized that 
the GUT's and R-parity violating
SUSY models offer a plethora of the $0\nu\beta\beta$-decay 
mechanisms triggered by exchange of neutrinos, neutralinos, 
gluinos, leptoquarks, etc. \cite{FKSS97,FKS98,WKS99}.

The experimental effort concentrated on high $Q_{\beta\beta}$ isotopes,
in particular on $^{48}$Ca, $^{76}$Ge, $^{82}$Se, $^{96}$Zr, $^{100}$Mo,
$^{116}$Cd, $^{130}$Te, $^{136}$Xe and $^{150}$Nd \cite{zdes02,eji05,AEE08}.  
In 1987 the first actual
laboratory observation of the two neutrino double beta decay ($2\nu\beta\beta$-decay) was done 
for $^{82}$Se by M. Moe and collaborators \cite{ell87}, who used a time
projection chamber. Within the next few years, experiments employing 
counters were able to detect $2\nu\beta\beta$-decay of many nuclei. 
In addition, the experiments searching for the signal of the $0\nu\beta\beta$-decay
pushed by many orders of magnitude the experimental 
lower limits for the $0\nu\beta\beta$-decay half-life of different nuclei.

\subsection{The period of massive neutrinos - the current period} 

Various early measurements of neutrinos produced in the sun, in the 
atmosphere, and by accelerators suggested that neutrinos might 
oscillate from one "flavor" (electron, muon, and tau) to 
another,  expected as a consequence of non-zero neutrino mass. Non-zero neutrino mass can be accommodated by fairly straightforward 
extensions of the SM of particle physics. Starting 1998 we have a convincing evidence about the existence of neutrino masses 
due to SuperKamiokande \cite{SUPERKAMIOKANDE}, SNO, \cite{SOLAROSC} KamLAND \cite{KAMLAND} and other experiments. 
 
 Thus neutrino oscillations 
have supplied  additional information in constructing  Grand Unified Theories
of physics. It also has provided additional input for cosmologists
and opened new perspectives for observation of the $0\nu\beta\beta$-decay.

So far the $2\nu\beta\beta$-decay has been recorded for eleven nuclei 
($^{48}$Ca, $^{76}$Ge, $^{82}$Se, $^{96}$Zr, $^{100}$Mo, $^{116}$Cd, 
$^{128}$Te, $^{130}$Te, $^{150}$Nd, $^{136}$Xe, $^{238}$U) \cite{zdes02,eji05,AEE08}.  
In addition, the $2\nu\beta\beta$-decay 
of $^{100}$Mo and $^{150}$Nd to $0^+$ excited state of the daughter 
nucleus has been observed  and the two-neutrino double electron capture
process in $^{130}Ba$ has been recorded. 
Experiments studying $2\nu\beta\beta$-decay  are presently approaching 
a qualitatively new level, when high-precision measurements 
are performed not only for half-lives but also for all other observables 
of the process. 
As a result, a trend is emerging towards thorough investigation of all aspects of 
$2\nu\beta\beta$-decay, and this will furnish very important information about the 
values of nuclear matrix elements, the parameters of various theoretical models, and so on. 
In this connection, one may expect advances in the calculation of nuclear matrix elements and 
in the understanding of the nuclear-physics aspects of double beta decay.

Neutrinoless double beta decay has not yet been confirmed. 
The strongest limits on the half-life of the $0\nu\beta\beta$-decay were set 
in Heidelberg-Moscow \cite{bau99},
NEMO3 \cite{nemoiii05,tre11}, CUORICINO \cite{te130} and KamLAND-Zen \cite{kamlandzen} experiments: 
\begin{eqnarray}
T_{1/2}^{0\nu}({}^{76}\mbox{Ge}) &\geq& 1.9 \times 10^{25} ~\mathrm{y}, 
~~~~~~~
T_{1/2}^{0\nu} ({}^{100}\mbox{Mo}) \geq 1.0 \times 10^{24} ~\mathrm{y},  
\nonumber\\
T_{1/2}^{0\nu}({}^{130} \mbox{{Te})} &\geq&  3.0 \times 10^{24} ~\mathrm{y},
~~~~~~~
T_{1/2}^{0\nu} ({}^{136}\mbox{Xe}) \geq 5.7 \times 10^{24} ~\mathrm{y}.
\end{eqnarray}
There is also a  claim for an observation of the $0\nu\beta\beta$-decay of $^{76}$Ge with 
half-life $T^{0\nu}_{1/2} = 2.23^{+0.44}_{-0.31}\times 10^{25}$ years \cite{evidence1,evidence2}.
One of the the goals of the upcoming GERDA experiment \cite{gerda} is to put this claim to a test by improving the 
sensitivity limit of the detection by more than an order of magnitude. 
The next generation experiments, which will be using several other candidate 
nuclei, will eventually be able to achieve this goal as well~\cite{AEE08}.

\subsection{The period of Majorana neutrinos?}

There is a hope that {\it the period of Majorana neutrinos} is not far. This period
should start by a direct and undoubtable observation of the $0\nu\beta\beta$-decay.   
It would establish that neutrinos are Majorana particles, and a measurement of the
decay rate, when combined with neutrino oscillation data and a reliable calculation
of nuclear matrix elements, would yield insight into all three neutrino mass 
eigenstates.

\section[An overview]{An overview}
\label{overview}
The question of neutrino masses and mixing is one of the most important issues 
of modern particle physics. It has already been discussed in a number of excellent 
reviews \cite{Ver86,BilPet87,KING04,MohSmyr06,MohEtal07} and its relevance to 
the $0\nu\beta\beta$-decay will also  be briefly discussed in this report. 
		
%
Today, seventy five years later, $0\nu\beta\beta$-decay (\ref{eq:a}),  
		 continues to be one of the most interesting processes.  The experimental status and prospects regarding this process will be reviewed in section  \ref{sect:exp}. 
               The corresponding non exotic $2\nu\beta\beta$ (\ref{eq:a2})   
		  has been observed in many systems, see section \ref{sect:exp}.
\begin{figure}[!t]
\begin{center}
\includegraphics [width=0.67\textwidth]{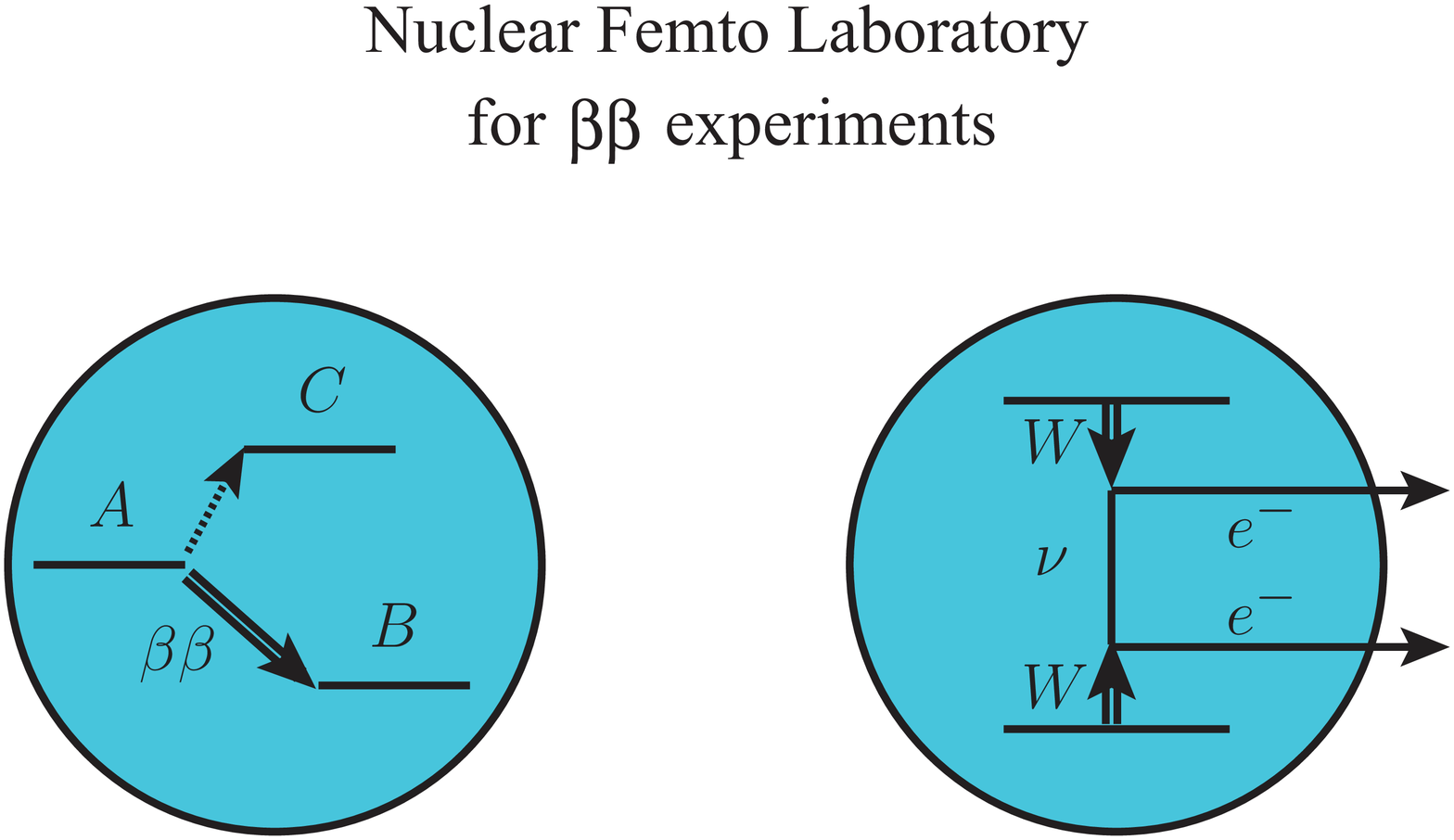}
\caption{Schematic diagrams of $\beta \beta $ decays in nuclear femto(10$^{-15}$m) laboratories, where single $\beta
$-decay is not allowed and neutrinoless DBD is much enhanced. Left hand side: DBD decay scheme. Right hand side: Neutrinoless DBD with the Majorana $\nu $ exchange between two nucleons \cite{eji10}.}
\label{fig:6.1}
\end{center}
\end{figure} 

		  If the neutrinos
                 are Majorana particles other related processes in which the charge of the nucleus
                 is decreased by two units may also occur, if they happen to be allowed by
                 energy and angular momentum conservation laws, e.g.
                 \beq
                 (A,Z)\rightarrow~(A,Z-2)+e^++e^+~~~~(0\nu\mbox{ positron emission}).
                \label{eq:b}   
		\eeq
	Here the available energy is	$\Delta=Q-4 m_e c^2$, i.e.,  everything else being equal, it is somewhat kinematically disfavored compared to the usual two electron emission for which the available energy $\Delta$ is equal to the $Q$ value.
	
	Electron positron conversion:
		 \beq
                 (A,Z)+e_b^-~\rightarrow~(A,Z-2)+e^+~~(0\nu~\mbox{ electron positron
                 conversion}).
                \label{eq:c}   
		\eeq
$\Delta=Q-2 m_e c^2-\varepsilon_b$, where $\varepsilon_b$ is a binding energy of the absorbed atomic electron.

The resonant neutrinoless double electron capture ($0\nu$ECEC),
\begin{equation}
(A,Z) + e^-_b + e^-_b \rightarrow (A,Z-2)^{**},\mbox{ (resonant }0\nu\mbox{ double electron capture }),
\end{equation}
was first considered very long time ago by Winter \cite{WINTER}. It is always allowed, whenever (\ref{eq:c}) is. This reaction was expounded  in more detail later 
\cite{Ver83,BeRuJar83} as a two step process: In the first step the
two neutral atoms, (A,Z) and the excited (A,Z-2), get admixed via the lepton number violating interaction.
In the second step the (A,Z-2) atom and, possibly, the nucleus de-excite. The available energy is $\Delta=Q-B_{2h}$, $B_{2h}$ being the energy of two electron holes in the atomic shells of daughter nucleus.\\
Decays to excited states are in some cases possible. Then in addition to x rays one has various decay modes with emission of a single $\gamma$, a pair of $\gamma$'s, internal electron-positron pair formation and emission of electron by 
internal conversion \cite{DK93}, i.e. 
		 \beq
                 (A,Z)+e_b^-+e_b^-\rightarrow~(A,Z-2)+X,\quad X= \gamma, 2\gamma,~e^+e^-,~e^-_{int}
                \label{eq:d}   
		\eeq
		
   The life time expected was very long, since 
the above mixing amplitude was tiny compared to the energy difference of the two atoms involved. 
It  has recently, however, been gaining in importance \cite{SIMKO11,DEC11} after ion Penning traps 
\cite{BLAUM06} made it possible to accurately  determine the $Q$ values, which gave rise to the 
the presence of resonances. This, in turn, could lead to an increase of the width by many orders 
of magnitude, see section \ref{sec:twoecap} for details.

                Another lepton number violating process, not hindered by energy conservation in any nuclear system, involves the
                 neutrinoless bound muon capture \cite{PR81,DKFS04}, 
		 \beq
                 (A,Z)+\mu_b^-~\rightarrow~(A,Z-2)+e^+~~(0\nu\mbox{ muon-positron 
                 conversion}).
                \label{eq:e}   
		\eeq
                The best experimental limit on the muon to positron conversion branching ratio
                has been established at PSI \cite{PSI} for the $^{48}Ti$ nuclear target.
                The muonic analogue of neutrinoless double beta decay \cite{MMM94,SFKS02}, 
		 \beq
                 (A,Z)+\mu_b^-~\rightarrow~(A,Z-2)+\mu^+~~(0\nu\mbox{ muon-muon
                 conversion}),
                \label{eq:mu}   
		\eeq
                 has never been searched for experimentally. For bound muon in atom 
                 energy conservation of the process is very restrictive. It is satisfied only for three isobars,
                 namely $^{44}$Ti, $^{72}$Se and $^{82}$Sr.

                The above processes are expected to occur whenever one has
		lepton number violating interactions. Lepton number, being a global quantity,
		is not sacred, but it is expected to be broken at some level. In short, these
		processes pop up almost everywhere, in every theory. On the other hand since,
                if there exist lepton number violating interactions, the neutrinos have to be 
                Majorana particles, all the above processes can, in principle, decide whether
                or not the
		neutrino is a Majorana particle, i.e. it coincides with its own antiparticle. 
                This is true even if these processes are induced not by intermediate neutrinos
                but by other mechanisms as we will see below.

                Neutrinoless double beta decay (Eq. (\ref {eq:a})) seems to be the most likely
                to yield the  information \cite{Ver86,HS84,DTK85,Tom91,SC98,FS98,Ver02,RODEJ11} we are after. For this
                reason we will focus our discussion on this reaction, but we will pay some
                attention to resonant neutrinoless double electron capture, which has recently been revived 
                \cite{Ver83,BeRuJar83,SIMKO11,DEC11,SujWy04,SimKriv09,BELLI11}, since its observation seems 
                to be a realistic possibility \cite{SIMKO11,DEC11}.
                We will only peripherally discuss the other less interesting processes \cite{Ver83}.  

		From a nuclear physics \cite {SC98,FS98,SDSJ97,RCN95,CNPR96,SSDV92,CPZ90}
		point of view, calculating the relevant nuclear matrix elements is indeed
		a challenge. First almost all nuclei, which can undergo double beta decay,
		are far from closed shells and some of them are even deformed. One thus faces
		a formidable task. Second the nuclear matrix elements are  small compared to
		 a canonical value, like the one associated with the matrix element to the (energy non
		allowed) double Gamow-Teller resonance or a small fraction of some appropriate sum
		rule. Thus, effects which are normally negligible, become important here.
		Third in many models the dominant mechanism for $0\nu\beta\beta$-decay
		does not involve intermediate light neutrinos, but very heavy particles.
		Thus one must be able to cope with the short distance behavior of the
		relevant operators and wave functions (see section \ref{sec:0nuME} for details).

		From the experimental point of view it is also very challenging to measure
		the slowest perhaps process accessible to observation. Especially, if one realizes
		 that even, if one obtains only lower bounds on the life
		time for this $0\nu\beta\beta$-decay, the extracted limits  on the theoretical model
		parameters may be comparable, if not better, and complementary to those
		extracted from the most ambitious accelerator experiments.

		The recent discovery of neutrino oscillations \cite{SUPKAM,AHARMIN,CHOOZ,ARAKI}  have given the first evidence of physics
		beyond the Standard Model (SM)  and in particular they indicate that the
		neutrinos are massive
		particles. They were able to show that the neutrinos are admixed, determined two of the mixing angles and set a stringent limit on the third (for a global analysis see, e.g.,  \cite{SCHWETZ}). Furthermore they determined one  square mass difference and the absolute value of the other. Neutrino oscillations, however, cannot  determine:
		\begin{itemize}
		\item Whether the neutrinos are Majorana or Dirac particles. \\
		 It is obviously important to proceed further and decide on this important issue.
		 Neutrinoless double beta decay can achieve this, even if, as we have mentioned,  there might
		be processes  that dominate over  the conventional intermediate neutrino mechanism of $0\nu\beta\beta$-decay. It has been known that
		whatever the lepton number violating process is, which gives rise to $0\nu\beta\beta$-decay,
		it can be used to generate a Majorana mass for the neutrino \cite{SVa82}. This mechanism, however,  may not be the dominant  mechanism for  generating the neutrino mass \cite{DuLindZub11b}.
		\item The scale of the neutrino masses.\\
		 These experiments  can measure only mass squared differences.\\
		 This task can be accomplished by astrophysical observations or  via other experiments involving  low energy weak decays, like triton decay or electron capture, or the $0\nu\beta\beta$-decay.  It seems that for a  neutrino mass in meV, ($10^{-3}$ eV), region, the best process to achieve this is the $0\nu\beta\beta$-decay. The extraction of neutrino masses from such observations will be discussed in detail and compared with each other later (see section \ref{sec:extrnumass}).
		 \item The neutrino hierarchy\\
		 They cannot at present decide which scenario is realized in nature, namely the degenerate, the normal hierarchy or the inverted hierarchy. They may be able to distinguish between the two hierarchies  in the future by observing the wrong sign muons in neutrino factories \cite{CDGG-CHMR10,Agar11}.
		 \end{itemize}
		 For details on such issues see a recent review \cite{G-GM08}.
		 
		The study of the $0\nu\beta\beta$-decay is further stimulated by the 
		development of GUT's and supersymmetric models (SUSY) 
		representing extensions of the $SU(2)_L \otimes U(1)$ standard model. 
		The GUT's and SUSY offer a variety of mechanisms
		which allow the $0\nu\beta\beta$-decay to occur \cite{Moh98}.

		 The best known mechanism leading to $0\nu\beta\beta$-decay 
		is via the exchange of a Majorana neutrino between the two
		decaying neutrons \cite{Ver86,HS84,DTK85,Tom91,SC98,FS98,PSV96}. 
		 Nuclear physics allows us to study the 
		light ($m_\nu \ll m_e$) and heavy ($m_\nu \gg m_p)$ neutrino components separately. In the presence of only
		left handed currents and for the light intermediate neutrino components,
                the obtained amplitude
		is proportional to a suitable average neutrino mass, which vanishes in the limit 
                in which the neutrinos become Dirac particles. On the other hand
		in the case of heavy Majorana neutrino components the amplitude is proportional
		to the average of the inverse of the neutrino mass, i.e. it is again suppressed.
                In the
		 presence of right handed currents  one can have a contribution similar
		 to the one above for heavy neutrinos but involving a different (larger)
		average inverse  mass with some additional suppression due to the
                the fact the right handed gauge boson, if it exists, is  heavier than the
                usual left handed one. 

		In the presence of right handed currents it is also possible to have
		interference between the leptonic left and right currents, $j_{L}-j_{R}$ 
		interference. In this case the amplitude in momentum space becomes proportional
		to the 4-momentum of the neutrino and, as a result, only the light neutrino
		components become important. One now has two possibilities. First the two
		hadronic currents have a chirality structure of the same kind, i.e. $J_L-J_R$.
		Then one can extract from the data a
		dimensionless parameter $\lambda$, which is proportional to the square of
		the ratio of the masses of the L and R gauge bosons, $\kappa=(m_{L}/m_{R})^2$.
		Second the two hadronic currents are left handed, which can happen via the
		mixing $\epsilon$ of the two bosons. The relevant lepton number violating parameter
                $\eta$ is now proportional to this mixing $\epsilon$. Both of these parameters,
                however, also involve
		the neutrino mixing. They are, in a way, proportional to the mixing between the 
		light and heavy neutrinos.

		 In gauge theories one has, of course, many more possibilities. Exotic 
		intermediate scalars may mediate $0\nu\beta\beta$-decay \cite{Ver86}. These are
		not favored in current gauge theories and are not going to be further
		discussed. In superstring inspired models one may have singlet fermions in
		addition to the usual right handed neutrinos. Not much progress has been made
		on the phenomenological side of these models and they are not going to be
		discussed further.

		 In recent years supersymmetric models are taken seriously and semirealistic
		calculations are taking place. In standard calculations one invokes
		universality at the GUT's scale, employing  a set of five
                independent   parameters,
                and uses the
		renormalization group equation to obtain all parameters (couplings and
		particle masses) at low energies.  Hence, since such parameters are in
		principle calculable in terms of the five input parameters, one can use
                experimental data to constrain the input parameters.  One, then, can use the
                $0\nu\beta\beta$-decay experiments to constrain
		the R-parity violating couplings, which cannot be specified by the theory 
		\cite{FKSS97,Moh86,Ver87,HKK95,PAES99,WKS97,FKS98a,FKS98b,HIRSCH01}. 
		Recent review articles \cite {AEE08,SC98,FS98,RODEJ11}
                also give  a detailed account of
                some of the latest developments in this field. 

		 From the above discussion it is clear that one has to consider the case of
		heavy intermediate particles. One thus has to tackle problems related to
                the very short ranged
		operators in the presence of the nuclear repulsive core. If the interacting
		nucleons are point-like one gets negligible contributions. We know, however
		that the nucleons are not point like, but that they have structure described by a
                quark bag with a size that can be determined experimentally. It can also be
                accounted for by
		a form factor, which can be calculated in the quark model or parametrized
		in a dipole shape with a parameter determined by experiment. This approach, 
                 first considered by Vergados \cite{Ver81}, has now been 
		adopted by almost everybody. The resulting effective operator has
		a range somewhat less than the inverse of the proton mass (see sect. 4 below).
		 
		 Another approach in handling this problem consists of considering particles
                 other than the nucleons present
		in the nuclear soup. For $0^+\rightarrow 0^+$ the most important such particles
		are the pions. One thus may consider the double beta decay of pions in flight
		between nucleons, like
		\beq
				   \pi^- \longrightarrow \pi^+ + e^- + e^-~~~~~  ,~~~~~~
				   n \longrightarrow p +  \pi^+ + e^- + e^-.
		\label{eq:1}   
		\eeq
		 Recognition of such contribution first appeared as a remark
                 by the genius of Pontecorvo \cite{ponte68} in the famous
                 paper in which he suggested that the ratio of the lifetimes
                 of the $^{128}$Te and $^{130}$Te isotopes, which merely
                 differ by two neutrons, is essentially independent of nuclear
                 physics. He did not perform any estimates of such a
                 contribution. Such estimates and calculations were first 
                 performed by Vergados \cite{Ver82} in the case
                 of heavy intermediate neutrinos, .i.e. vector and axial vector
                 currents. It was found that it
		 yields results of the same order as the nucleon mode with the above recipe
		for treating the short range behavior. It was revived by the Tuebingen group 
		\cite{FKSS97,FKS98,WKS99,FKS98a} in the context of R-parity violating interactions, i.e. 
                scalar, pseudoscalar and tensor currents arising out of neutralino and gluino
                exchange, and it was found to dominate.

		 In a yet another approach one may estimate the presence of six quark clusters
                 in the nucleus. Then, since the change of charge takes place in the same
                 hadron  there is
		no suppression due to the short nature of the operator, even if it is a 
                $\delta$-function. One only needs a reliable method for estimating
                the probability of finding these clusters in a nucleus \cite{Ver85}. 

		 All the above approaches seem reasonable and lead to quite similar results.
                The matrix elements obtained are not severely suppressed.
		This gives us a great degree of confidence that the resulting matrix elements
		are sufficiently reliable, allowing double beta decay to probe very important
		physics.

		 The other recent development is the better description of nucleon current
		by including momentum dependent terms, such as the modification of the axial
		current due to PCAC and the inclusion of the weak magnetism terms. These
		contributions have been considered previously \cite{PSV96,tom85}, but only
		in connection with
		the extraction of the $\eta$ parameter mentioned above. Indeed these terms
		were very important in this case since they compete with the p-wave lepton wave
		function, which, with the usual currents, provides the lowest non vanishing
		contribution. Since in the mass term only s-wave lepton wave functions are
                relevant such terms have hitherto  been neglected.

		 It was recently found \cite {SPVF}, however, that for light
		neutrinos the inclusion of these momentum dependent terms reduces the nuclear
		matrix element by about $25\%$, independently of the nuclear model employed.
		On the other hand for heavy neutrinos the effect can be larger and it depends
                on the nuclear wave functions.
		 The reason for expecting them to be relevant is that the average momentum
		$\langle q\rangle$ of the exchanged neutrino is expected to be large \cite{SEIL92}. In the
		case of a light intermediate neutrino the  mean nucleon-nucleon separation is
		about 2 fm which implies that the average momentum $\langle q\rangle$ is about 100 MeV/c.  
		In the case of a heavy neutrino
		exchange the mean inter nucleon distance is considerably smaller 
		and the average momentum $\langle q\rangle$ is supposed to be considerably larger. 

		 Since $0\nu\beta\beta$ decay is a two step process, in
                principle, one needs to
		construct and sum over all the intermediate nuclear states, a formidable job
		indeed in the case of the shell model calculations. Since, however, the
		average neutrino momentum is much larger compared to the nuclear excitations,
		one can invoke closure using some average excitation energy (this does not
		apply in the case of $2\nu\beta\beta$ decays). Thus one need construct only
		the initial and final $0^+$ nuclear states. This is not useful in Quasiparticle Random Phase
		Approximation  (QRPA), since one must construct the intermediate states anyway. In any
		case, it was explicitly shown, taking advantage of the momentum space formalism
		developed by Vergados \cite {Ver90}, that this approximation is very good
		 \cite {PV90,FKPV91}. The same conclusion was reached independently by others
		 \cite{SKF90} via a more complicated technique relying on coordinate space.

		Granted that one takes into account all the above ingredients in order to 
		obtain quantitative answers for
		the lepton number violating parameters from the results of
		$0\nu\beta\beta$-decay experiments, it is necessary to evaluate the relevant
		nuclear matrix elements with high reliability. The most extensively used
		methods  are the large basis Interacting Shell Model (ISM) calculations,  (for a
                recent review see \cite{SC98}) and QRPA( for a recent 
		review see \cite {FS98,SC98}). The ISM is forced to use few single particle
		orbitals, while this restriction does not apply in the case of QRPA. The latter suffers,
		of course, from the approximations inherent in the RPA method. So a direct
		comparison between them is not possible.

		 The shell model calculations have a long history 
		\cite{CPZ90,SSDV92,Ver76,HSS82,SV83,ZBR90,ZB93} in double beta decay
		calculations. In recent years it has lead to large matrix calculations  in
		traditional as well as Monte Carlo types of formulations
		\cite{SDSJ97,RCN95,CNPR96,Retal96,NSM96,KDL97}. For a more complete set of
		references as well as a discussion of the appropriate
                effective interactions see Ref. \cite{SC98}).

		 There have been a number of QRPA calculations covering almost all nuclear
		targets \cite{VZ86,CAT87,MBK88,EVJP91,RFSK91,GV92,SC93,CS94,SSVP97,SPF98,cheoun,MUT97}.
                 These involve a number of collaborations, but the most
                 extensive and complete calculations in one way or another
                 include the Tuebingen group.
		 We also have seen some refinements of QRPA, like proton neutron pairing and
		inclusion of 
		renormalization effects due to Pauli principle corrections
                \cite{TS95,SSF96}. Other less conventional approaches, like
                operator expansion techniques have also been employed
                \cite{FS98}. 

                Recently, calculations based on the Projected 
                Hartree-Fock-Bogoliubov (PHFB) method \cite{phfb}, the Interacting Boson Model (IBM)  \cite{IBM09}
                and the Energy Density Functional (EDF) method  \cite{edf} 
                entered the field of such calculations. The above schemes, in 
                conjunction with the other
		improvements mentioned above, offer some optimism in our efforts for obtaining
		nuclear matrix
		elements  accurate enough to allow us to extract reliable values of the lepton
		violating parameters from the data. 
		
	       As we have mentioned neutrinoless double beta decays (DBD) are concerned with 
               fundamental properties of neutrinos. These properties arise out of interactions involving  
               high energy scales, which are of great interests from view points of particle physics 
               and cosmology. On the other hand, DBD processes are nuclear rare-decays in the low energy 
               scale, which are studied experimentally by low-energy and low-background nuclear 
               spectroscopy, as given in review articles \cite{eji05,AEE08}.

               Double beta decays are low-energy second-order weak processes with 
               $Q_{\beta \beta }\approx $2 - 3 MeV. Decay rates of 2$\nu\beta\beta$-decay within 
               the SM are of the order of 10$^{-20}$/y, and the rates of 0$\nu\beta\beta$-decay 
               beyond the SM are even many orders of magnitudes smaller than  
               2$\nu\beta\beta$-decay rates, depending on the $Q_{\beta \beta}$ value and 
               the effective Majorana neutrino mass $\langle m_\nu\rangle$ (for definition see
               Eq. (\ref{eq:1.5a})) in case of the light neutrino  mass process. Then the 0$\nu\beta\beta$-decay 
               half lives are of the orders of $T^{0\nu}_{1/2} \approx $ 10$^{27}$ y and 10$^{29}$ y in cases of 
               the IH (inverted hierarchy) mass of $\langle m_{\nu }\rangle \approx$ 30 meV and the NH (normal hierarchy) 
               mass of $\langle m_{\nu } \rangle \approx$ 3 meV, respectively.  
   
               For experimental studies of such rare decays, large detectors with ton-scale 
               DBD isotopes are needed  to get 0$\nu\beta\beta$-decay signals in case of the IH $\nu$ mass. 
               Here the signals are very rare and are as low as  $E_{\beta \beta }\approx $2 - 3 MeV. 
               Background (BG) signal rates, however, are huge in the energy region of $E_B \le $ 3 MeV. 
               Thus it is crucial to build ultra low BG detectors to find the rare and small 
               0$\nu\beta\beta$-decay signals among huge BGs in the low energy region. We are going 
               to review this (see section \ref{sect:exp}) in the case
	       of most of the nuclear targets of experimental
		interest \cite{eji05,AEE08,Ejiri00} ($^{76}$Ge, $^{82}$Se, $^{96}$Zr, $^{100}$Mo,
		$^{116}$Cd, $^{128}$Te, $^{130}$Te, $^{136}$Xe,
		$^{150}$Nd).

		\section[The  neutrino mass matrix]{The  neutrino mass matrix in various models}
		\label{sec:numass}
		Within the SM of elementary particles, with the particle content of 
                the gauge bosons $A_{\mu},Z_{\mu}$ and $W_{\mu}^{\pm,0}$, the Higgs scalar isodoublet 
                $\Phi=(\phi^0,\phi^-)$ (and its conjugate $\Phi^{*}$) and the fermion fields arranged in:
		\begin{itemize}
		\item  Isodoublets: $(u_{\alpha L},d_{\alpha L})$ and $(\nu_{\alpha L},e_{\alpha L})$ 
                for quarks and leptons respectively and
		\item Isosinglets: $u_{\alpha R},d_{\alpha R}$ and $e_{\alpha R}$ 
		\end{itemize}
		where $\alpha$ is a family index taking three values,  the neutrinos are massless. They can not obtain mass after the symmetry breaking, like the quarks and the charged leptons, since the right handed neutrino is absent.
\subsection{Neutrino masses at tree level}		
		The minimal extension of the SM that would yield mass for the neutrino is to introduce an isosinglet right handed neutrino. Then one can have a Dirac mass term  arising via coupling of the leptons and Higgs as follows:
\begin{eqnarray}
		&&h_{\alpha,\beta} (\bar{\nu}_{\alpha_L}\bar{e}_{\alpha_L})
		\left(
\begin{array}{c}\phi^0\\
                    \phi^-                
\end{array}                      
\right ) \nu_{\beta R} \rightarrow 
\nonumber\\
&&
h_{\alpha,\beta} (\bar{\nu}_{\alpha_L}\bar{e}_{\alpha_L})
		\left(
\begin{array}{c}v/\sqrt{2}\\
                    0                  
\end{array}                      
\right ) \nu_{\beta R}~~~ \mbox{ or }~~~m^D_{\alpha,\beta}	 = h_{\alpha,\beta}\frac{ v}{\sqrt{2}}
\label{Eq:mDirac}
\end{eqnarray}
	Thus one can have:
	\beq
	{\mathcal M}=
		\left(\bar{\nu}_L,\bar{\nu}_L^c\right)\left(\begin{array}{cc}0&\quad m^D
		\\
		(m^D)^T& 0
		\end{array}\right ) \left(\begin{array}{c}\nu_R\\ \nu_R^c  \end{array} \right)
	\eeq
In the above expression, as well as in analogous expressions below, we have explicitly indicated not only the $6\times6$ matrix, but the states on which this matrix acts and $m^D$ is the above $3\times3$ matrix. One then obtains 6 Majorana eigenvectors which pair wise can be associated  to opposite eigenvalues. Their sum and their difference may equally well  be selected as physical states which correspond to three Dirac neutrinos and their charged conjugate (antineutrinos). This is fine within the above minimal extension. In grand unified theories, however, one is faced with the problem that these neutrinos are going to be very heavy with a mass similar to that of up quarks, which is clearly unacceptable. So such a model is inadequate\footnote{There may exist light Dirac neutrinos in theories formulated in extra dimensions, see e.g. the recent review by Smirnov \cite{SMIRNOV04}. If these neutrinos do not couple to the usual leptons they are of little interest to us. If they do and it so happens that  the standard neutrinos are Majorana, they also become Majorana, except in the case of very fine tuning.}. Besides, such neutrinos viewed as Majorana with opposite CP eigenvalues or as Dirac particles cannot contribute to $0\nu\beta \beta$ decay.

 The next extension is to introduce a Majorana type mass involving the isosinglet neutrinos and an additional isosinglet Higgs field, which can acquire a large vacuum expectation value, an idea essentially put forward by Weinberg \cite{WEINBERG79} long time ago. Thus the neutrino mass matrix becomes:
	\beq
	{\mathcal M}=
		\left(\bar{\nu}_L,\bar{\nu}_L^c\right)\left(\begin{array}{cc}0&\quad m^D
		\\
		(m^D)^T& m_R
		\end{array}\right ) \left(\begin{array}{c}\nu_R\\ \nu_R^c  \end{array} \right)
		\label{Eq:see-saw}
	\eeq
Thus, provided that the Majorana mass matrix has only very large eigenvalues, one obtains an effective Majorana $3\times 3$ matrix:
\beq
{\mathcal M}_{\nu}=-\bar {\nu}_L(m^D)^T M_R^{-1}m_D\nu^c_R,
\eeq
which can provide small neutrino masses provided that the eigenvalues of the matrix $M_R$ are sufficiently large. $M_R$ can be arbitrarily large, since the the new scale, associated with the  vacuum expectation of the isosinglet, does not affect the low energy scale arising from the vacuum expectation value of the standard Higgs particles. This is the celebrated see-saw mechanism. More precisely the type I see-saw mechanism, since, as we will see below, there exist other see-saw types (for a summary see, e.g., Abada {\it et al}\cite{SeeSaw07}). \\
Thus with the above mechanism the neutrino flavors get admixed, the resulting eigenstates are Majorana particles and lepton number violating interactions, like $0\nu\beta \beta$ decay, become possible.

Other extensions of the SM are possible, which do not require the addition of the right handed neutrinos but additional exotic scalars  or fermions \cite{MohEtal07,MohSmyr06}, e.g.
\begin{itemize}
\item An isotriplet $\Delta$ of Higgs scalars whose charge decomposition is $\delta^{--},\delta^-,\delta^0$. \\Then this leads to the coupling:
$$
		(h_T)_{\alpha,\beta} (\bar{\nu}_{\alpha_L}\bar{e}_{\alpha_L})
		\left(
\begin{array}{lr}\delta^-&-\delta^0\\
                    \delta^{--}& \delta^-\\                                  
\end{array}                      
\right )\left(
\begin{array}{lr}e^c_{\beta R}\\
                    -\nu^c_{\beta R}\\                  
\end{array}                      
\right )$$
This after the isotriplet acquires a vacuum expectation value becomes
$$	(h_T)_{\alpha,\beta} (\bar{\nu}_{\alpha_L}\bar{e}_{\alpha_L})
		\left(
\begin{array}{lr}0&-v_{\Delta}/\sqrt{2}\\
                    0&0                                  
\end{array}                      
\right )
\left (
\begin{array}{c}e^c_{\beta R}\\
                    -\nu^c_{\beta R}                                 
\end{array}                      
\right )$$
yielding the neutrino Majorana mass matrix
\beq
m^{M}_{\alpha,\beta}=(h_T)_{\alpha,\beta}\frac{v_{\Delta}}{\sqrt{2}}\bar{\nu}_{\alpha_L} \nu^c_{\beta R}
\label{Eq:tripletmass}
\eeq
If, for some reason, the introduction of such an isotriplet, acquiring a small vacuum expectation value, is not preferred, the Majorana mass matrix can be obtained assuming that the isotriplet $\Delta$ possesses a  cubic coupling $\mu_{\Delta} $ with  two standard Higgs doublets \cite{Ver86,MagWet80,LazShafWet81,MohSen81} (see Fig. \ref{Fig:tripletDelta}). 
 \begin{figure}[!t]
\begin{center}
\includegraphics[scale=0.4]{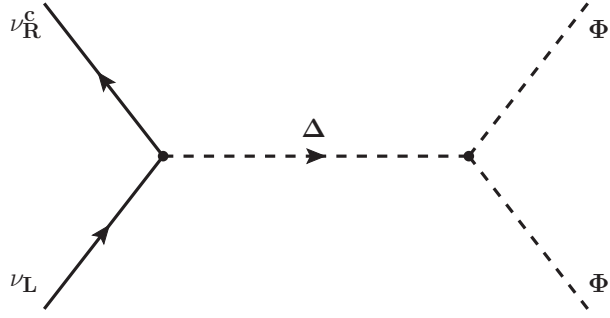}
 \caption{ The tree level contribution to the neutrino mass mediated by  an isotriplet scalar.}
 \label{Fig:tripletDelta}
\end{center}
\end{figure}
Then one finds an effective Majorana neutrino mass from Eq. (\ref{Eq:tripletmass}) via the substitution
 $$\frac{v_{\Delta}}{\sqrt{2}}\rightarrow \frac{v^2}{2}\frac{\mu_{\Delta}}{m^2_{\Delta}},$$
 where $v/\sqrt{2}$ is the vacuum expectation value of the standard Higgs doublet (see Eq. (\ref{Eq:mDirac}) ), $m_{\Delta}$ is the mass of $\delta^0$. This is sometimes refer to as see-saw mechanism II \cite{MohSmyr06,MohEtal07}.\\
 As we will see later this mechanism  may lead to a new contribution to neutrinoless double beta decay via the direct decay of $\delta^{--}$ into two electrons.
 \item An isotriplet of fermions with hypercharge zero $(\Sigma^+,\Sigma^0,\Sigma^-)$.\\
 In this case the leptons couple to the isotriplet via the higgs doublet( see \cite{Abada08,SeeSawIII08} and references therein):
 $$
		(h_{\Sigma})_{\alpha} (\bar{\nu}_{\alpha_L}\bar{e}_{\alpha_L})
		\left(
\begin{array}{lr}\Sigma ^0_R/\sqrt{2}&\Sigma ^+_R\\
                    \Sigma ^-_R&-\Sigma ^0_R/\sqrt{2}                                
\end{array}                      
\right )\left(
\begin{array}{c}\phi^0\\
                    \phi^-                 
\end{array}                      
\right )\rightarrow$$
$$
		(h_{\Sigma})_{\alpha} (\bar{\nu}_{\alpha_L}\bar{e}_{\alpha_L})
		\left(
\begin{array}{lr}\Sigma ^0_R/\sqrt{2}&\Sigma ^+_R\\
                    \Sigma ^-_R&-\Sigma ^0_R/\sqrt{2}                                
\end{array}                      
\right )\left(
\begin{array}{c}v/\sqrt2\\
                    0                  
\end{array}                      
\right )=$$
$$(h_{\Sigma})_{\alpha} v\left( \frac{1}{2}\bar {\nu}_{\alpha L} \Sigma^0_R +\frac{1}{\sqrt{2}}\bar{e}_{\alpha L} \Sigma^-_R \right )
$$
Then a subsequent coupling of the isotriplet to the leptons yields an effective Majorana coupling of the form:
\beq
m^M_{\alpha,\beta}=-(h_{\Sigma})_{\alpha}(h_{\Sigma})_{\beta}\frac{v^2}{2}\frac{1}{m_{\Sigma}}
\eeq
where $ m_{\Sigma}$ is the mass of the neutral component of the isotriplet (see Fig. \ref{Fig:tripletSigma} and ref. \cite{Abada08}). 
 \begin{figure}[!t]
\begin{center}
\includegraphics[scale=0.4]{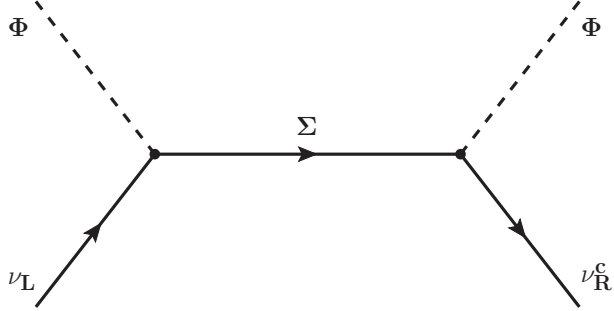}
 \caption{ The tree level contribution to the neutrino mass mediated by  a fermion isotriplet}
 \label{Fig:tripletSigma}
\end{center}
\end{figure}
 It is sometimes referred as mechanism see-saw III. 
 This mechanism, however, by itself cannot constitute a viable neutrino mass generator  since it leads to two eigenstates with zero mass. This can be circumvented \cite{Abada08,SeeSawIII08},
 but then the model becomes more complicated.
\end{itemize}

\subsection{Neutrino masses at the loop level}
 There many ways to obtain neutrino masses at the 1-loop level \cite{Pascoli08}, which have nicely been summarized by Smirnov \cite{SMIRNOV04}. We will only discuss one such case, which arises in the presence of R-parity violating supersymmetry, which leads to a viable neutrino mass spectrum \cite{HVFC00}, in the sense that it can yield three massive neutrinos, if one includes not only the tree level contribution arising from the bilinear terms \cite{BFK98}, but both type the 1-loop contributions \cite{GKS04,GKSA06} shown in Fig. \ref{Fig:RPS_nu}. This is interesting, since, in such models, as we will see below, one can have particles other than neutrinos contributing to $0\nu\beta\beta$-decay.
 \begin{figure}[!t]
\begin{center}
\includegraphics[scale=0.45]{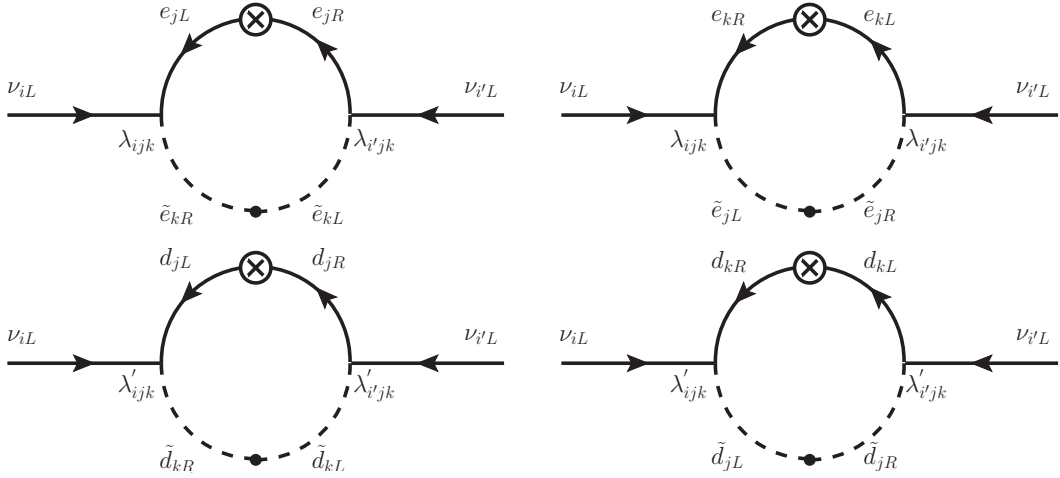}
\caption{ The squark-quark (lower panel) and slepton-lepton (upper panel)
1-loop diagrams contributing to the neutrino mass in the presence of R-parity violating supersymmetric theories.}
 \label{Fig:RPS_nu}
\end{center}
\end{figure}
%
The above Majorana matrices are symmetric, in general complex, matrices.

\subsection{SUSY, GUT's and Family symmetries}
In many models, like the standard see-saw, the smallness of neutrino mass requires the existence of a heavy mass scale. The coexistence of two mass scales can naturally be accommodated in supersymmetry. In minimal supersymmetric extensions of the SM \cite{Pascoli06} one construct the see-saw matrix of Eq. (\ref{Eq:see-saw}), see e.g. the review \cite{KING04}. Furthermore this can be extended to larger symmetries, e.g. two commuting symmetries, a Grand Unified Symmetry $G_{\mbox{\tiny{GUT}}}$ and a family symmetry $G_f$. The family symmetry could be continuous, like $SU(3),\,SU(2)$ or $U(1)$ or discreet  $Z_N$, $S_3$ or $S_3\times S_3$ etc.  We are not going to further elaborate on such situations, which have been summarized in  recent reviews \cite{KING04,AltFer10,IKOOST10} to which we direct the interested reader. We should also mention that $U(1)$ flavor symmetries arise naturally in superstring inspired models \cite{AntLeoRiz90,LeoRiz99,FarKounRiz90}, in particular for the heterotic string and the 4-d fermionic constructions. D-brane models have also paved the way for completely new structures \cite{RamRobRoss93,IbanRich09,AnasKirLion09,LEONTARIS09} and in particular a very interesting formulation D-brane inspired mass textures \cite{LeonVlach10}.

All the models considered in this section lead to a light effective neutrino mass acting jointly or separately. Almost all of them involve parameters, which can be adjusted to fit phenomenology. It is not clear which one, if any, is going to ultimately be the theoretically preferred one.
\subsection[Neutrino mixing]{Neutrino mixing}
\label{sec:numixing}
We have seen above that in general the neutrino mass matrix is a complex symmetric matrix. It can, however, be diagonalized by separate left and write unitary transformations, which can take the form \cite{Ver86}:
 \barr
	S_L\leftrightarrow
\left(\nu^0_L,\nu^{0c}_L\right )&=&\left (\begin{array}{lr}
	S^{(11)}& S^{(12)}\\
                    S^{(21)}& S^{(11)}                  
\end{array}                      
\right )
\left (\begin{array}{c}
	                                  \nu'_L\\
	                                  N'_L               
\end{array}                      
\right ),\nonumber\\
	S_R\leftrightarrow
\left(\nu^{0c}_R,\nu^0_{R} \right )&=&\left (\begin{array}{lr}
	\left(S^{(11)} \right )^{*}& \left(S^{(12)} \right )^{*}\\
                    \left(S^{(21)} \right )^{*}& \left(S^{(22)} \right )^{*}                 
\end{array}                      
\right )
\left (\begin{array}{c}
	                                  \nu'_R\\
	                                  N'_R                
\end{array}                      
\right )
\label{Eq:mixing}
\earr
where we have added the superscript 0 to stress that they are the states entering the weak interactions. $S^{(ij)}$ are $3 \times 3$ matrices with $S^{(11)}$ and $S^{(22)}$ approximately unitary, while $S^{(12)}$ and $S^{(22)}$ are very small. 
$(\nu'_L,N'_L )$ and $(\nu'_R,N'_R)$ are the eigenvectors from the left and the right  respectively. 
Thus the neutrino mass in the new basis takes the form:
\beq
{\mathcal M}_{\nu}=\sum_{j=1}^3 \left ( m_j \bar{\nu}'_{jL} \nu'_{jR}+M_j \bar{N}'_{jL} N'_{jR} \right )+H.C.
\eeq
This matrix can be brought into standard form by writing:
$$
m_j=|m_i|e^{-i \alpha_j},\quad M_j=|M_i|e^{-i \Phi_j}
$$
and defining:
$$
\nu_j=\nu'_{jL}+e^{-i \alpha_j}\nu'_{jR}\quad N_j=\nu'_{jL}+e^{-i \Phi_j}N'_{jR}
$$
Then
\beq
{\mathcal M}_{\nu}=\sum_{j=1}^3 \left (| m_j| \bar{\nu}_{j} \nu_{j}+|M_j| \bar{N}_{j} N_{j} \right )
\eeq
Note, however, that:
\barr
\nu^c&=&\nu'_{jR}+e^{i \alpha_j}\nu'_{jL}=e^{i \alpha_j}\left (\nu'_{jL}+e^{-i \alpha_j}\nu'_{jR} \right )=e^{i \alpha_j}\nu_j\nonumber\\
N^c&=&N'_{jR}+e^{i \Phi_j}N'_{jr}=e^{i \Phi_j}\left (\nu'_{jR}+e^{-i \Phi_j}N'_{jR} \right )=e^{i \Phi_j}N_j
\label{Eq:Majphase}
\earr
i.e. they are Majorana neutrinos with the given Majorana phases. Furthermore
$$
\nu_{iL}=\nu'_{iL},\quad \nu_{iR}=e^{-i \alpha_j}\nu'_{iR},\quad N_{iL}=N'_{iL},\quad N_{iR}=e^{-i \Phi_j}\nu'_{iR}
$$
The second of Eqs. (\ref{Eq:mixing}) can now be written as 
\beq
S_R\leftrightarrow
\left(\nu^{0c}_R,\nu^0_R \right )=\left (\begin{array}{lr}
	\left(S^{(11)} \right )^{*}& \left(S^{(12)} \right )^{*}\\
                    \left(S^{(21)}\right )^{*}& \left(S^{(22)} \right )^{*}                  
\end{array}                      
\right )
\left (\begin{array}{c}
	                                  e^{i \alpha}\nu_R\\
	                                  e^{i \Phi}N_R                
\end{array}                      
\right )
\label{Eq:mixing2}
\eeq
where $e^{i \alpha}$ and  $e^{i \Phi}$ are diagonal matrices containing the above Majorana phases.
 The neutrinos interact with the charged leptons via the charged current (see below). So the effective coupling of the neutrinos to the charged leptons involves the mixing of the electrons $S^e$. Thus  the standard mixing matrix appearing in the absence of right-handed neutrinos is:
 \beq
 U_{PMNS}=U=U^{(11)}=\left (S^{(e)} \right )^{+} S^{(11)}
 \eeq
  The other entries are defined analogously:
  \beq
 U^{(ij)}=\left (S^{(e)} \right )^{+} S^{(ij)},\quad (ij)=(12),(21),(22)
 \eeq
 In particular the usual electronic neutrino is written as:
 \beq
 \nu_{eL}=\sum_j\left (U^{(11)}_{ej}\nu_j +U^{(12)}_{ej}N_j \right )
 \label{Eq:leftnu}
 \eeq
  \beq
 \nu_{eR}=\sum_j\left (U^{(21)}_{ej}\nu_j +U^{(22)}_{ej}N_j \right )
  \label{Eq:rightnu}
 \eeq
 In other words the left handed neutrino may have a small heavy  component, while the  right handed neutrino may have a small light component. Note also that the neutrinos appearing in weak interactions can be Majorana particles in the special case that all Majorana phases are the same.
 
The Pontecorvo-Maki-Nakagawa-Sakata neutrino mixing matrix $U_{PMNS}$ is parametrized 
by
\beq
 U_{PMNS} = R_{23} {\tilde R}_{13} R_{12},
 \eeq
where the matrices $R_{ij}$ are rotations in $ij$ space, i.e.,
\begin{eqnarray}
 R_{23} &=& \left(
\begin{array}{lll}
  1 &   0     & 0 \\
  0 &  c_{12} & s_{12} \\
  0 & -s_{12} & c_{12} 
\end{array}
\right),~
 R_{13} = \left(
\begin{array}{lll}
 c_{13} &  0  & s_{13} e^{-i\delta} \\
   0   &  1  &   0    \\
-s_{13}e^{i\delta} &  0  & c_{13} 
\end{array}
\right), \nonumber\\
 R_{12} &=& \left(
\begin{array}{lll}
 c_{12} & s_{12} & 0 \\
-s_{12} & c_{12} & 0 \\
  0    &   0    & 1 
\end{array}
\right),
 \end{eqnarray}
 where
 \beq
 c_{ij}\equiv \cos{(\theta_{ij})}, \quad s_{ij}\equiv \sin{(\theta_{ij})}.
 \eeq
$\theta_{12}$, $\theta_{13}$ and $\theta_{23}$
and three mixing angles and $\delta$ is the 
CP-violating phase. Then, we get
 \beq
 U_{PMNS} = \left(
\begin{array}{lll}
 c_{12} c_{13} & c_{13} s_{12} & e^{-i \delta } s_{13} \\
 -c_{23} s_{12}-e^{i \delta } c_{12} s_{13} s_{23} & c_{12}
   c_{23}-e^{i \delta } s_{12} s_{13} s_{23} & c_{13} s_{23} \\
 s_{12} s_{23}-e^{i \delta } c_{12} c_{23} s_{13} & -e^{i \delta }
   c_{23} s_{12} s_{13}-c_{12} s_{23} & c_{13} c_{23}
\end{array}
\right),
\label{pmns}
 \eeq
If neutrinos are Majorana particles $U_{PMNS}$ in Eq. (\ref{pmns}) is 
multiplied by a diagonal phase matrix P, which contains two additional
CP-violating Majorana phases $\alpha_1$ and $\alpha_2$:
\begin{equation}
P = diag(e^{i \alpha_1}, e^{i \alpha_2}, e^{i \delta}).
\end{equation}
	\section[The absolute scale of the neutrino mass]{The elusive absolute scale of the neutrino mass}
		
		\label{sec:extrnumass}
	With the discovery of neutrino oscillations quite a lot of information regarding the neutrino 
    sector has become available (e.g., for  recent reviews see \cite{G-GM08,MohPas05}). More specifically we know: 
	\begin{itemize}
	\item The mixing angles $\theta_{12}$ and $\theta_{23}$ and we have both a lower and an upper  bound on the small angle $\theta_{13}$
	\item we know the mass squared differences:
$$	\Delta^2_{\mbox{\tiny{SUN}}}=\Delta^2_{12}=m_2^2-m_1^2,\quad \mbox{ and } \Delta^2_{\mbox{\tiny{ATM}}}=|\Delta^2_{23}|=|m_3^2-m_2^2|$$
entering the solar and atmospheric neutrino oscillation experiments. Note that we do not know the absolute scale of the neutrino mass and the sign of $\Delta^2_{23} $
	\end{itemize}
For  determination of an absolute scale of the  neutrino mass the relevant neutrino oscillation parameters are
the MINOS value  $\Delta m^2_{\mbox{\tiny{ATM}}}=(2.43 \pm 0.13) \times 10^{-3}~\mathrm{eV}^{2}$ \cite{Minos},
the global fit value $\Delta m^2_{\mbox{\tiny{SUN}}}= (7.65^{+0.13}_{-0.20}) \times 10^{-5}~\mathrm{eV}^{2}$ \cite{SCHWETZ},
the solar-KamLAND value $\tan^2{\theta_{12}}=0.452_{-0.033}^{+0.035}$ \cite{Gando11} 
and the recent Daya Bay observation $\rm{\sin}^2 2 \theta_{13} = 0.092 \pm 0.016~(stat)~\pm 0.005~(syst)$ 
with a significance of 5.2 standard deviations \cite{dayabay}.
We note that non-zero value of mixing angle $\theta_{13}$ was already observed also
by the T2K ($0.04 < \rm{\sin}^2 2\theta_{13} < 0.34$) \cite{T2K}, the
DOUBLE CHOOZ ($\rm{\sin}^2 2 \theta_{13} = 0.085 \pm 0.029~(stat)~\pm 0.042~(syst)$ 
(68\% CL)) \cite{dct13}  and the RENO 
($\rm{\sin}^2 2 \theta_{13} = 0.103 \pm 0.013~(stat)~\pm 0.011~(syst)$) \cite{reno12}
collaborations. 

	Based on the above we have the following scenarios:
\begin{itemize}
\item { Normal Spectrum (NS)}, $m_{1} < m_{2} < m_{3}$:
$${\Delta m^2_{\mbox{\tiny{SUN}}}=m^2_2-m^2_1}~,~{ \Delta m^2_{\mbox{\tiny{ATM}}}=m^2_3-m^2_1}$$
$$m_0={ m_1},~m_2=\sqrt{{ \Delta m^2_{\mbox{\tiny{SUN}}}}+{ m_0^2}}~,~m_3=\sqrt{{\Delta m^2_{\mbox{\tiny{ATM}}}}+{ m_0^2}}$$
\item { Inverted Spectrum (IS)}, $m_{3} < m_{1} < m_{2}$:
$${ \Delta m^2_{\mbox{\tiny{SUN}}}=m^2_2-m^2_1}~,~{ \Delta m^2_{\mbox{\tiny{ATM}}}=m^2_2-m^2_3}$$
$$ m_0=m_3,~m_2 = \sqrt{ \Delta m^2_{\mbox{\tiny{ATM}}}+{ m_0^2}},$$
$$~~~m_1=\sqrt{{ \Delta m^2_{\mbox{\tiny{ATM}}}-\Delta m^2_{\mbox{\tiny{SUN}}}}+{ m_0^2}}$$
\end{itemize}
 The absolute scale $m_0$ of neutrino mass can in principle be determined by the following observations:
 \begin{itemize}
 \item Neutrinoless double beta decay.\\
  As we shall see later (section \ref{sec:numec}) the 
	effective light neutrino mass $\langle m_\nu\rangle$ extracted in such experiments  is given as follows \cite {Ver86,Pascoli05}:
	\begin{eqnarray}
	\langle m_\nu \rangle  &=&  \sum^{3}_k~ (U^{(11)}_{ek})^2 ~ m_k \nonumber\\
                &=& c^2_{12}c^2_{13} e^{2i\alpha_1} m_1 +  c^2_{13} s^2_{12}e^{2i\alpha_2} m_2 + s^2_{13} m_3.            
	\label{eq:1.5a}   
	\end{eqnarray}
		\item The  neutrino mass extracted from ordinary beta decay, e.g. from triton decay \cite{Katrin,otten,Mare}.\\
	\begin{eqnarray}
	\langle m_\nu\rangle_{\mbox{\tiny{decay}}} &=& \sqrt{ \sum^{3}_k~ |U^{(11)}_{ek}|^2  m^2_k}\nonumber\\
                &=& \sqrt{c^2_{12}c^2_{13} m^2_1 +  c^2_{13} s^2_{12} m^2_2 + s^2_{13} m^2_3}.            
	\label{eq:1.5b}   
	\end{eqnarray}
assuming, of course, that the three neutrino states cannot be resolved. 
\item From astrophysical and cosmological observations (see, e.g., the recent summary\cite{Abarajan11}).\\
	\beq
m_{\nu} ~ =  \sum^{3}_k   m_k\leq m_{\mbox{\tiny astro}} 
	\label{eq:1.5c}   
	\eeq
	The current limit  $m_{\mbox{\tiny astro}}$ depends on the type of observation\cite{Abarajan11}. 
        Thus CMB primordial  gives  1.3 eV,  CMB+distance 0.58 eV, galaxy distribution and and lensing 
        of galaxies  0.6 eV. On the other hand  the largest photometric red shift survey yields
   0.28 eV \cite{ThAbdaLah10}. For  purposes of illustration we will take a world average of  $m_{\mbox{\tiny astro}}=0.71$ eV.
 \end{itemize}
The above mass combinations  can be written as follows:
\begin{enumerate}
\item Normal Hierarchy (NH), $m_1\ll m_2 \ll m_3$.\\
In this case we have 
$${ \Delta m^2_{\mbox{\tiny{SUN}}}=m^2_2-m^2_1}~,~{ \Delta m^2_{\mbox{\tiny{ATM}}}=m^2_3-m^2_1}$$
Thus:
\begin{itemize}
\item Triton decay.
\begin{eqnarray}
&&\langle m_\nu\rangle_{\mbox{\tiny{decay}}} =\nonumber\\
&&\sqrt{
{c_{12}^2} {c^2_{13}}{ m^2_1}+{s_{12}^2} {c^2_{13}} \left (\Delta m^2_{\mbox{\tiny{SUN}}}
+{ m_1^2}\right )+s^2_{13}(\Delta m^2_{\mbox{\tiny{ATM}}}+{m_1^2})}
\end{eqnarray}
\item Cosmological bound:
\beq
m_{\nu}=m_1+\sqrt{ \Delta m^2_{\mbox{\tiny{SUN}}}+m_1^2}+\sqrt{(\Delta m^2_{\mbox{\tiny{ATM}}}+{m_1^2})}
\eeq
\item $0\nu\beta\beta$ decay:
\begin{eqnarray}
\langle m_\nu\rangle &=&
{c_{12}^2} {c^2_{13}}{ m_1}e^{2i {\alpha}_1}+{s_{12}^2} {c^2_{13}}e^{2i {\alpha}_2}
\sqrt{ \Delta m^2_{\mbox{\tiny{SUN}}}
+{ m_1^2}} \nonumber\\
&&~~~~~~~~~~~~~~~~~+ s^2_{13}\sqrt{(\Delta m^2_{\mbox{\tiny{ATM}}}+{m_1^2})},
\end{eqnarray}
where $\alpha_1$ and $\alpha_2$ ($\alpha_3=0$) are Majorana CP violating phases 
of the elements $U_{e1}$ and $U_{e2}$  with $U_{ej} = |U_{ej}| e^{i \alpha_j}$
($i=1,2$). 

By assuming NH, i.e., that $m_1$ is negligibly small 
($m_1 \ll \sqrt{\Delta m^2_{\mbox{\tiny{SUN}}}}$, $m_2 \simeq \sqrt{\Delta m^2_{\mbox{\tiny{SUN}}}}$, 
and $m_3 \simeq \sqrt{\Delta m^2_{\mbox{\tiny{ATM}}}}$),  we obtain
\begin{eqnarray}\label{normal1}
|\langle m_\nu \rangle| &\simeq&
|c^2_{13} s^2_{12} \sqrt{\Delta m^{2}_{\mbox{\tiny{SUN}}}}
+ s^2_{13} \sqrt{\Delta m^{2}_{\mbox{\tiny{ATM}}}}e^{-2i\alpha_{2}}| \nonumber  \\
&\leq & 4\cdot 10^{-3}~\mathrm{eV}.
\end{eqnarray}
\end{itemize}
\item  Inverted Hierarchy (IH), $m_3 \ll m_1 < m_2$:
In this case we have
$${\Delta m^2_{\mbox{\tiny{SUN}}}=m^2_2-m^2_1}~,~{\Delta m^2_{\mbox{\tiny{ATM}}}=m^2_2-m^2_3}$$
\begin{itemize}
\item Triton decay.
\barr
\langle m_\nu\rangle_{\mbox{\tiny{decay}}} &=&\left( s^2_{13}{ m^2_3} + { s_{12}^2}{ c^2_{13}}( \Delta m^2_{\mbox{\tiny{ATM}}}+{ m_3^2})\right .\nonumber\\
&&\left .+c^2_{12}{ c^2_{13}}( \Delta m^2_{\mbox{\tiny{ATM}}}-\Delta m^2_{\mbox{\tiny{SUN}}}+{ m_3^2})\right )^{1/2}
\earr
\item Cosmological bound:
\beq
m_{\nu}=m_3+\sqrt{( \Delta m^2_{\mbox{\tiny{ATM}}}+{ m_3^2})}+\sqrt{( \Delta m^2_{\mbox{\tiny{ATM}}}-\Delta m^2_{\mbox{\tiny{SUN}}}+{ m_3^2})}
\eeq
\item $0\nu\beta\beta$ decay:
\begin{eqnarray}
\langle m_\nu\rangle &=& { s^2_{13}}{ m_3} + { s_{12}^2}{ c^2_{13}}e^{2i {\alpha}_2}\sqrt{( \Delta m^2_{\mbox{\tiny{ATM}}}+{ m_3^2})}
\nonumber\\
&&~~~~~~~~
+{ c^2_{12}}{ c^2_{13}}e^{2i {\alpha}_1}\sqrt{( \Delta m^2_{\mbox{\tiny{ATM}}}-\Delta m^2_{\mbox{\tiny{SUN}}}+{ m_3^2})}.
\end{eqnarray}

Since in IH scenario  $m_3$ is negligibly small 
($m_1\simeq m_2 \simeq \sqrt{\Delta m^2_{\mbox{\tiny{Atm}}}}$ and 
$m_3 \ll \sqrt{\Delta m^2_{\mbox{\tiny{ATM}}}}$),  we find
\begin{equation}
|\langle m_\nu\rangle|\simeq \sqrt{ \Delta m^{2}_{\mbox{\tiny{Atm}}}} c^2_{13}
(1- \sin^{2} 2\,\theta_{12}\,\sin^{2}\alpha_{12})^{\frac{1}{2}},
\label{invhierarA}
\end{equation}
where $\alpha_{12}=\alpha_{2}-\alpha_{1}$. 
The phase difference $\alpha_{12}$ is the only unknown parameter in the expression
for $|\langle m_\nu\rangle|$. From (\ref{invhierarA}) we obtain the following inequality \cite{bilpot11}
\begin{equation}\label{invhierar1}
1.5\cdot10^{-2} ~\mathrm{eV}   \leq |\langle m_\nu\rangle|\leq 5.0\cdot 10^{-2}~\mathrm{eV}.
\end{equation}
\end{itemize}
\item { Quasi-degenerate (QD) spectrum}, $m_0 = m_{1}\simeq m_{2}\simeq m_{3}$. Then
\begin{itemize}
\item Triton decay and Cosmology:
\beq
\langle m_\nu\rangle_{\mbox{\tiny{decay}}}= m_0,\quad m_{\nu}= 3 m_0
\eeq
\item $0\nu\beta\beta$ decay:\\
The effective Majorana mass is relatively large in this case  and for both 
types of the neutrino mass spectrum is given by the expression
\begin{equation}\label{degenerate1}
m_{0} |1- 2 c^2_{13} c^2_{12}| \le    |\langle m_\nu\rangle| \le m_{0}.
\end{equation}
\end{itemize}
\end{enumerate}

\begin{figure}[!t]
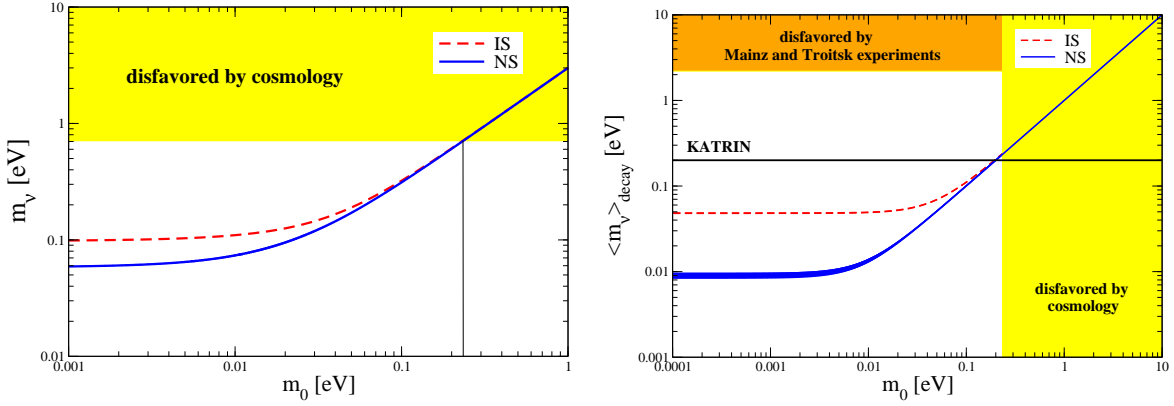

\centering
\begin{tabular}{cc}
\includegraphics[width=0.48\textwidth]{DBDReview_fig5a.eps} &
\includegraphics[width=0.48\textwidth]{DBDReview_fig5b.eps}
\end{tabular}
 \caption{(Color online) The neutrino mass limits in eV as a function of mass of the lowest eigenstate $m_0$ 
also in eV, extracted from cosmology (left panel) triton decay (right panel). From the current upper limit 
of 2.2 eV of the Mainz and Troitsk experiments we deduce a lowest neutrino mass of 2.2 eV both 
for the NS and IS. From the astrophysical limit value of 0.71 eV the corresponding neutrino mass 
extracted is about 0.23 eV for the NS and IS. It is assumed:
$\Delta m^2_{\mbox{\tiny{ATM}}} = (2.43 \pm 0.13) \times 10^{-3}~\mathrm{eV}^{2}$ \cite{Minos},
$\Delta m^2_{\mbox{\tiny{SUN}}}= (7.65^{+0.13}_{-0.20}) \times 10^{-5}~\mathrm{eV}^{2}$ \cite{SCHWETZ},
$\rm{\tan}^2{\theta_{12}}=0.452_{-0.033}^{+0.035}$ \cite{Gando11} and
$\rm{\sin}^2 2 \theta_{13} = 0.092 \pm 0.016$ \cite{dayabay}.
}        
\label{fig:limits}
\end{figure}

\begin{figure}[!t]
\begin{center}
\includegraphics[scale=0.5]{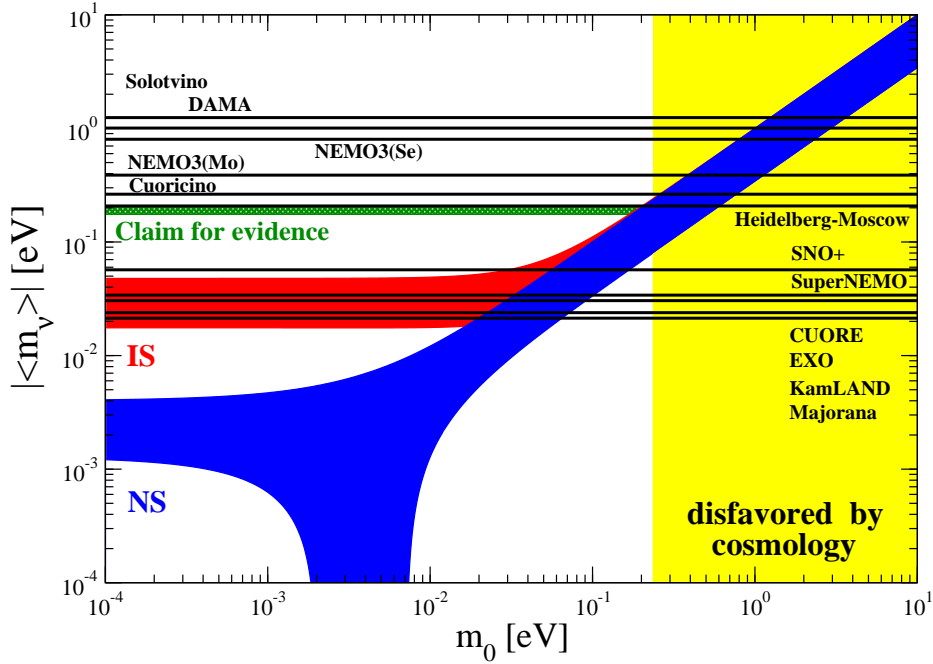}
 \caption{(Color online) We show the allowed range of values for $|\langle m_\nu\rangle|$ as a 
function of the lowest mass eigenstate $m_0$ using the three standard neutrinos 
for the cases of normal (NS, $m_0=m_1$) and inverted
(IS, $m_0=m_3$) spectrum of neutrino masses.
Also shown are the current experimental limits and the expected future 
results \cite{ablimit} (QRPA NMEs with CD-Bonn short-range correlations
and $g^{eff}_A=1.25$ are assumed \cite{src09,fang10}). 
Note that in the inverted hierarchy there is a lower bound, 
which means that in such a scenario the $0\nu\beta\beta$-decay 
should definitely be observed, if the experiments reach the required level.
The same set of neutrino oscillation parameters as in Fig. \ref{fig:limits}
is considered.
}
 \label{Fig:bbnumass}
\end{center}
\end{figure}

The above results are exhibited in Fig. \ref{fig:limits} for the tritium $\beta$-decay and cosmological 
limits as a function of the lowest neutrino mass and 
in Fig. \ref{Fig:bbnumass} for the case of the $0\nu\beta\beta$-decay both for  the NS and the IS scenarios. 
The allowed range values of $|\langle m_\nu\rangle|$ as a function of the lowest mass eigenstate $m_0$ is exhibited. 
For the values of the parameter $\sin^{2}2\theta_{13}$ new Double Chooz data are
used \cite{dct13}. The IH allowed region for $|\langle m_\nu\rangle|$ is presented by the region 
between two parallel lines in the upper part of Fig. \ref{Fig:bbnumass}. The NH
allowed region for $|\langle m_\nu\rangle| \approx$ few meV is compatible with $m_0$ smaller
than 10 meV.  The quasi-degenerate spectrum can be determined, if $m_{0}$ is known 
from future $\beta$-decay experiments 
KATRIN \cite{Katrin,otten} and MARE \cite{Mare} or from cosmological observations.
The lowest value for the sum of the neutrino masses, which can be
reached in future cosmological measurements \cite{Thomas,Serpico,Abazajian},
is about (0.05-0.1) eV. The corresponding values of $m_0$ are in the region,
where the IS and the NS predictions for $|\langle m_{\nu}\rangle|$ differ significantly 
from each other.

From the most precise  experiments on the search for  $0\nu\beta\beta$-decay 
\cite{bau99,te130,kamlandzen} by using of nuclear matrix elements of
Ref. \cite{src09} the following stringent bounds were inferred  (see Table \ref{tab:ej1})
\begin{eqnarray}\label{bounds}
|\langle m_\nu\rangle| & < & (0.20-0.32) ~eV~~ (^{76}\mathrm{Ge}), \nonumber\\
               & < & (0.33-0.46) ~eV~~ (^{130}\mathrm{Te}), \nonumber\\
               & < & (0.17-0.30) ~eV~~ (^{136}\mathrm{Xe}).
\end{eqnarray}
These bounds we obtained using the $0\nu\beta\beta$-decay NMEs of \cite{src09} calculated with
Brueckner two-nucleon short-range correlations. There exist, however, a claim of 
the observation of the $0\nu\beta\beta$-decay of $^{76}\rm{Ge}$ made by some participants 
of the Heidelberg-Moscow collaboration \cite{evidence2}. Their estimated value 
of the effective Majorana mass (assuming a specific value for the NME) 
is $|\langle m_\nu\rangle|\simeq 0.4$ eV. This result will be checked
by an independent experiment relatively soon.  In the new germanium
experiment GERDA \cite{gerda,ger08}, the Heidelberg-Moscow sensitivity  will be reached
in about one year of measuring time.

In future experiments, CUORE\cite{allesa11}, EXO\cite{ack11,Exo}, MAJORANA\cite{ell09},
SuperNEMO \cite{deppisch11}, SNO+ \cite{snoplus},  KamLAND-Zen
and others \cite{barabhis,AEE08,ASB07}, a sensitivity
\begin{equation}\label{sensitiv}
|\langle m_\nu\rangle|\simeq \mathrm{a~few}~10^{-2}~\mathrm{eV}
\end{equation}
is planned to be reached, what is the region of the IH of neutrino masses.
In the case of the normal mass hierarchy $|\langle m_\nu\rangle|$
is too small in order to be probed in the $0\nu\beta\beta$-decay 
experiments of the next generation.

Recently, however, for the explanation of the Reactor Antneutrino Anomaly \cite{RNA11}, 
a light sterile neutrino has been introduced with mass squared difference:
\beq
 \Delta m^2_{24}=|m_2^2-m_4^2|\approx \Delta m^2_{14}=|m_1^2-m_4^2|\ge1.5\mbox{(eV)}^2.
 \eeq
which couples to the electron neutrino with a mixing angle: 
\beq
\sin^2{2 \theta_{14}}=0.14\pm 0.08~~ (95\% ~C.L.).
\eeq
On the other hand in a recent global analysis more than one sterile neutrino are needed 
\cite{KMS11}, with somewhat smaller mass squared differences, but similar couplings. 
Due to such a mixing, even if their couplings are of the usual Dirac type, 
the resulting mass eigenstates are of the Majorana type due to their coupling to the usual neutrino.

The $U(4\times 4)$ neutrino mixing matrix  in the presence of one sterile neutrino 
with a small mixing becomes \cite{BRZ11}:
\begin{equation}
U = R_{34} {\tilde R}_{24} {\tilde R}_{14} R_{23} {\tilde R}_{13} R_{12} P.  
\end{equation}
It depends on 6 mixing angles ($\theta_{14}$, $\theta_{24}$, $\theta_{34}$, $\theta_{12}$, $\theta_{13}$, $\theta_{23}$,
3 Dirac ($\delta_{24}$, $\delta_{14}$, $\delta_{13}$) and 3 Majorana ($\alpha_1$, $\alpha_2$, $\alpha_3$)
CP-violating phases entering the diagonal
P matrix :
\begin{equation}
P = diag\left(e^{i\alpha_1}, e^{i\alpha_2}, e^{i(\alpha_3 +\delta_{13})}, e^{i\delta_{14}}\right).
\end{equation}
Similarly, one can parametrized the $U(5\times 5)$ mixing matrix for two sterile neutrinos as
(10 mixing angles and 5+4 CP violating phases) 
\begin{equation}
U = R_{45} 
   {\tilde R}_{35} {R}_{34} 
   {\tilde R}_{25} {\tilde R}_{24} R_{23} 
   {\tilde R}_{15} {\tilde R}_{14} {\tilde R}_{13} R_{12} P,  
\end{equation}
where $P = diag\left(e^{i\alpha_1}, e^{i\alpha_2}, e^{i(\alpha_3 +\delta_{13})}, e^{i(\alpha_4+\delta_{14})}, e^{i\delta_{15}}\right)$.

 \begin{figure}[!t]
\begin{center}
\includegraphics[scale=0.45]{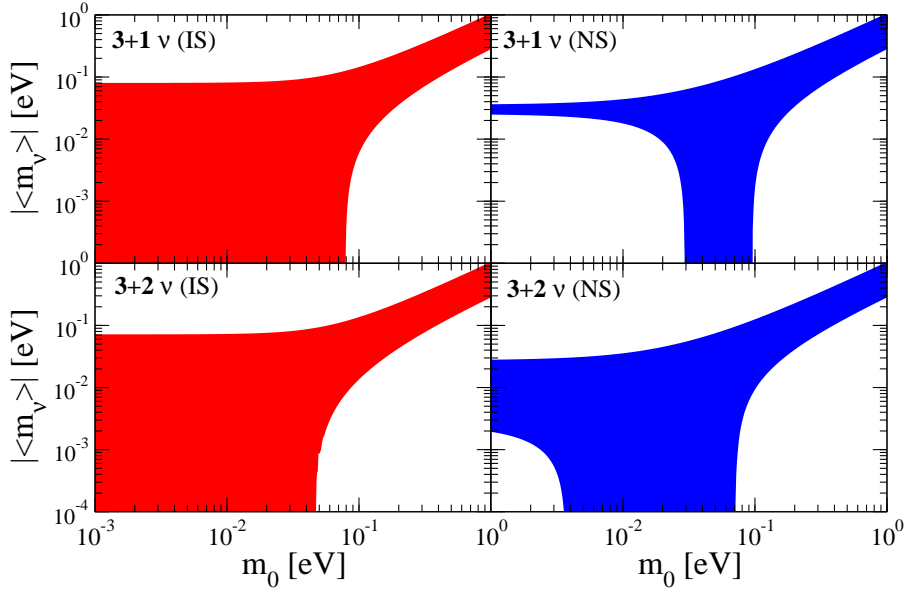}
 \caption{(Color online) The same as in Fig. \ref{Fig:bbnumass}, if one considers one (version 3+1) 
or two (version 3+2)
sterile neutrinos, which are heavier than the standard neutrinos. 
Best fit points for the 3+1 ($\Delta  m^2_{41} = 1.78~eV^2$, $U_{e4}=0.151$)
and 3+2 ($\Delta  m^2_{41} = 0.46~eV^2$, $U_{e4}=0.108$
and $\Delta  m^2_{51} = 0.89~eV^2$, $U_{e5}=0.124$)
scenarios from reactor antineutrino data are taken into account \cite{KMS11}.  
In addition, best fit values 
$\Delta m^2_{\mbox{\tiny{ATM}}} = 2.43\times 10^{-3}~\mathrm{eV}^{2}$ \cite{Minos},
$\Delta m^2_{\mbox{\tiny{SUN}}}= 7.65\times 10^{-5}~\mathrm{eV}^{2}$ \cite{SCHWETZ},
$\rm{\tan}^2{\theta_{12}}=0.452$ \cite{Gando11} and
$\rm{\sin}^2 2 \theta_{13} = 0.092 \pm 0.016$ \cite{dayabay}
are assumed. 
}
\label{Fig:sternumass}
\end{center}
\end{figure}

If the presence of one sterile neutrino, the effective neutrino mass in the $0\nu\beta\beta$-decay 
is given by \cite{BRZ11}
\begin{eqnarray} \label{ostern}
|\langle m_\nu\rangle_{3+1}| &=& | c^2_{12}c^2_{13}c^2_{14}e^{2i\alpha_1} m_1 +  
c^2_{13}c^2_{14}s^2_{12}e^{2i\alpha_2} m_2 \nonumber\\
&&+ c^2_{14}s^2_{13}e^{2i\alpha_3} m_3 +    
s^2_{14}m_4 |.
\end{eqnarray}
We note that from three additional angles $\langle m_\nu\rangle|$ depends only one of them, namely
$\theta_{14}$. If there are two sterile neutrinos we end up with
\begin{eqnarray} \label{tstern}
|\langle m_\nu\rangle_{3+2}| &=& | c^2_{12}c^2_{13}c^2_{14}c^2_{15}e^{2i\alpha_1} m_1 +  
c^2_{13}c^2_{14}c^2_{15}s^2_{12}e^{2i\alpha_2} m_2 \nonumber\\
&&+ c^2_{14}c^2_{15}s^2_{13}e^{2i\alpha_3} m_3 +    
c^2_{15}s^2_{14}e^{2i\alpha_4} m_4 + s^2_{15} m_5|.
\end{eqnarray}
Due to the extra terms in Eqs. (\ref{ostern}) and (\ref{tstern}) and their couplings, 
depending on the extra majorana phases, the sterile could dominate, 
increase or deplete  $\langle m_\nu\rangle$. By assuming best fit values for $|U_{e4}|$,  $m^2_{41}$
and $|U_{e5}|$,  $m^2_{51}$ from reactor antineutrino data 
\cite{KMS11} the allowed range values of $\langle m_\nu\rangle$ 
as a function as a function of the lowest mass eigenstate $m_0$
in the presence of one and two sterile neutrinos is shown in Fig. \ref{Fig:sternumass}.


		\section[The  Majorana neutrino mechanism]{The intermediate Majorana neutrino mechanism in $0\nu\beta\beta$ decay}
    \label{sec:numec}
    To proceed further in our study of the neutrino mediated
		 $0\nu\beta\beta$-decay process it is necessary to study the structure 
		 the  effective weak beta decay Hamiltonian. In general it has both left handed and right handed components. Within the $SU(2)_L\times SU(2)_R\times U(1)$
at low energies it  takes  the current-current form:
\begin{eqnarray}
{\mathcal{H}}^\beta &=& \frac{G_{{F}}}{\sqrt{2}} 
2~\left[\left(\bar{e}_L \gamma_\mu \nu^{0}_{{e L}} \right)
   \left(J^{\mu \dagger}_L + \epsilon J^{\mu \dagger}_R\right) \right.\nonumber\\
&&~~~~~\left.
    +  \left(\bar{e}_R \gamma_\mu \nu^{0}_{{e R}} \right)
   \left(\epsilon J^{\mu \dagger}_L + \kappa J^{\mu \dagger}_R\right)
+ {h.c.}\right],
\label{eq:1.1}   
\end{eqnarray}
Here, $\epsilon$ is mixing of $W_L$ and $W_R$ gauge bosons
\beq
W_L=\cos{\epsilon} W_1 - \sin{\epsilon} W_2,\quad 
W_R=\sin{\epsilon} W_1 + \cos{\epsilon} W_2
\eeq 
where $W_1$ and $W_2$ are the mass eigenstates of the gauge bosons
with masses $M_{W_1}$ and $M_{W_2}$, respectively. 
The mixing is assumed to be small, 
$\sin{\epsilon}\approx \epsilon$,  $\cos{\epsilon}\approx1$, 
and $m_{W_1}\approx m_{W_L}$, $m_{W_2}\approx m_{W_R}$.
$\kappa$ is the mass squared ratio  
\beq
\kappa=\frac{m_{W_1}^2}{m_{W_2}^2}.
\eeq 
$J^{\mu} _L$ is the standard hadronic current in V-A theory:
\begin{eqnarray}
J^{\mu \dagger} _L
=  \sum_i\bar{u}_p(i)\left[ g_V \gamma^\mu
+ i g_M \frac{\sigma^{\mu \nu}}{2 m_p} q_\nu 
 - g_A \gamma^\mu \gamma_5 - g_P q^\mu \gamma_5 \right] u_n(i),
\nonumber\\
\end{eqnarray}
where $u_p(i)$  and $u_n(i)$ are the spinors describing the proton and neutron  $i$. 
$m_p$ is the nucleon mass and $q^\mu$ is the momentum transfer.
$g_V\equiv g_V({q}^{2})$, $g_M\equiv g_M({q}^{2})$, $g_A\equiv g_A({q}^{2})$ and 
$g_P\equiv g_P({q}^{2})$ 
are respectively the vector, weak-magnetism, axial-vector and induced pseudoscalar
form-factors. 

We will see below that it is necessary to also consider the right handed current $J^{\mu \dagger} _R$ 
of the form V+A, 
\begin{eqnarray}
J^{\mu \dagger} _R
&=&  \sum_i\bar{u}_p(i)\left[ g_V \gamma^\mu
+ i g_M \frac{\sigma^{\mu \nu}}{2 m_p} q_\nu 
 + g_A \gamma^\mu \gamma_5 + g_P q^\mu \gamma_5 \right] u_n(i),
\nonumber\\
\end{eqnarray}
even though normally one  expects its contribution to ordinary beta decay to be 
suppressed by a factor $\kappa$ or $\epsilon$. Some relations among form factors
in $J^{\mu \dagger} _L$ and $J^{\mu \dagger} _R$ are considered because the strong and
electromagnetic interactions conserve parity.

$e_L(e_R)$ and $\nu^{0}_{{e L}}(\nu^{0}_{{e R}})$ are field operators
		representing the left (right) handed electrons and  electron
		neutrinos in a  weak interaction basis in which the charged leptons are 
               diagonal.
               We have seen above the the weak neutrino eigenstates can be expressed in terms of the propagating mass eigenstates \cite {Ver86}  (see Eqs. (\ref{Eq:leftnu}) and (\ref{Eq:rightnu})).
The mass eigenstates
		$\nu_k,N_k$ satisfy the Majorana condition: 
		$\nu_k \xi_k = C ~{\overline{\nu}}_k^T$, 
		$N_k \Xi_k = C ~{\overline{N}}_k^T$,
		where C denotes the charge conjugation and $\xi$, $ \Xi$ 
		are phase factors, which guarantee that the eigenmasses are positive ($\xi_k=e^{i \alpha_k},\,\Xi_k=e^{i\Phi_k}$, see Eq. (\ref{Eq:Majphase})). 
		
Before proceeding further we should mention that in the context of the above interaction neutrinoless double beta decay is a two step process. The neutrino is produced via the lepton current in one vertex and propagates in the other vertex. If the two current helicities are  the same one picks out of the neutrino propagator the term:
\beq
\frac{m_j}{q^2-m_j^2}\rightarrow \left\{\begin{array}{c}m_j/q^2,\quad m^2_j \ll q^2\\-{1}/{m_j},\quad m^2_j \gg q^2\end{array} \right .
\eeq
where $q$ is the momentum transferred by the neutrino. In other words the amplitude for light neutrino becomes proportional to its mass, but for a heavy neutrino inversely proportional to its mass.

If the leptonic currents have opposite chirality the one picks out of the neutrino propagator the term:
\beq
\frac{	 \displaystyle{\not} q }{q^2-m_j^2}\rightarrow \frac{	 \displaystyle{\not} q }{q^2},\quad m^2_j \ll q^2
\eeq
i.e. in the interesting case of light neutrino the amplitude does not explicitly depend on the neutrino mass. The kinematics becomes different than that for the mass term.
		\subsection{The Majorana neutrino mass mechanism}

\begin{figure}[!t]
\begin{center}
\subfloat[]
{
\includegraphics[scale=0.4]{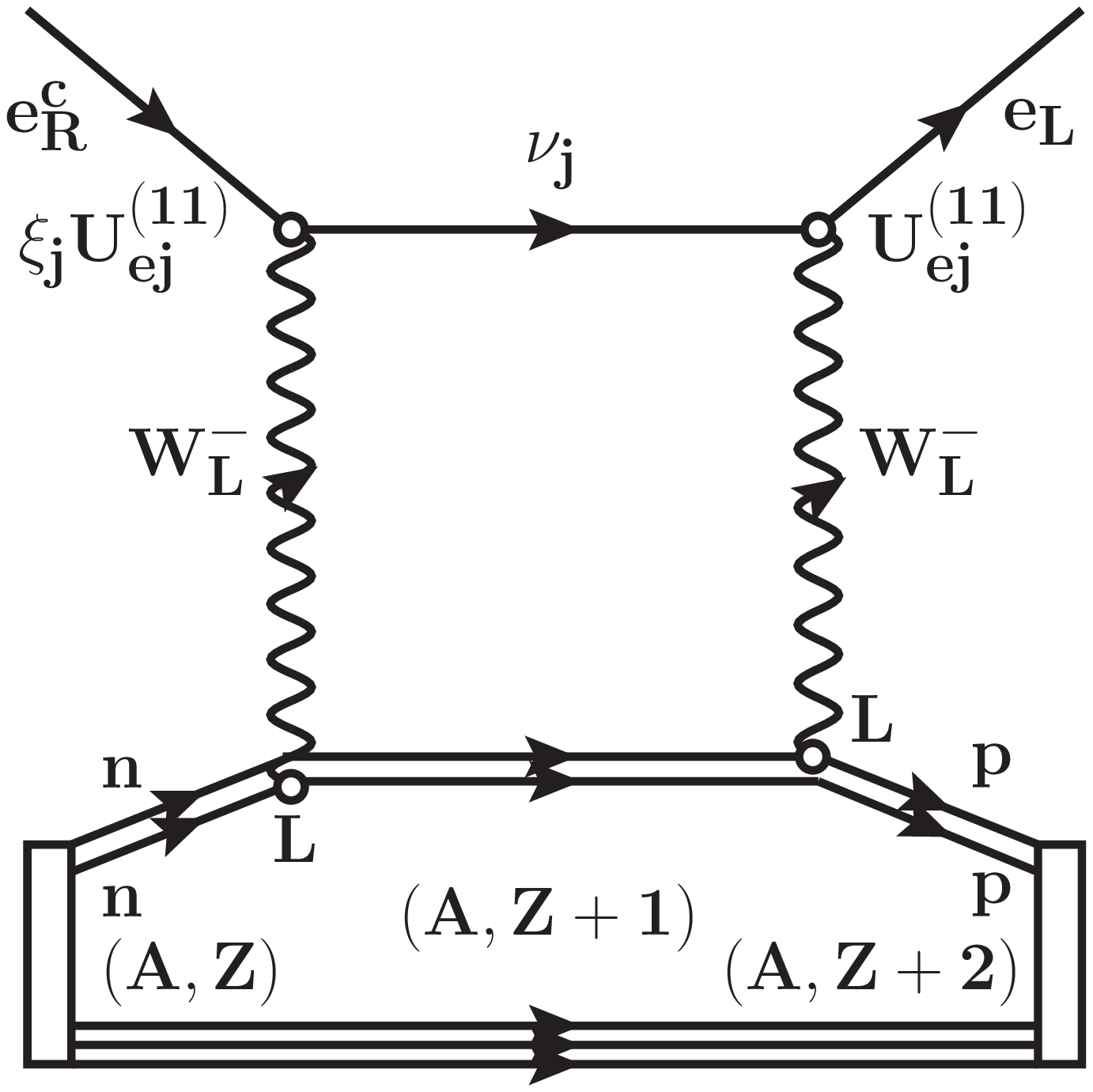}
}
\subfloat[]
{
\includegraphics[scale=0.4]{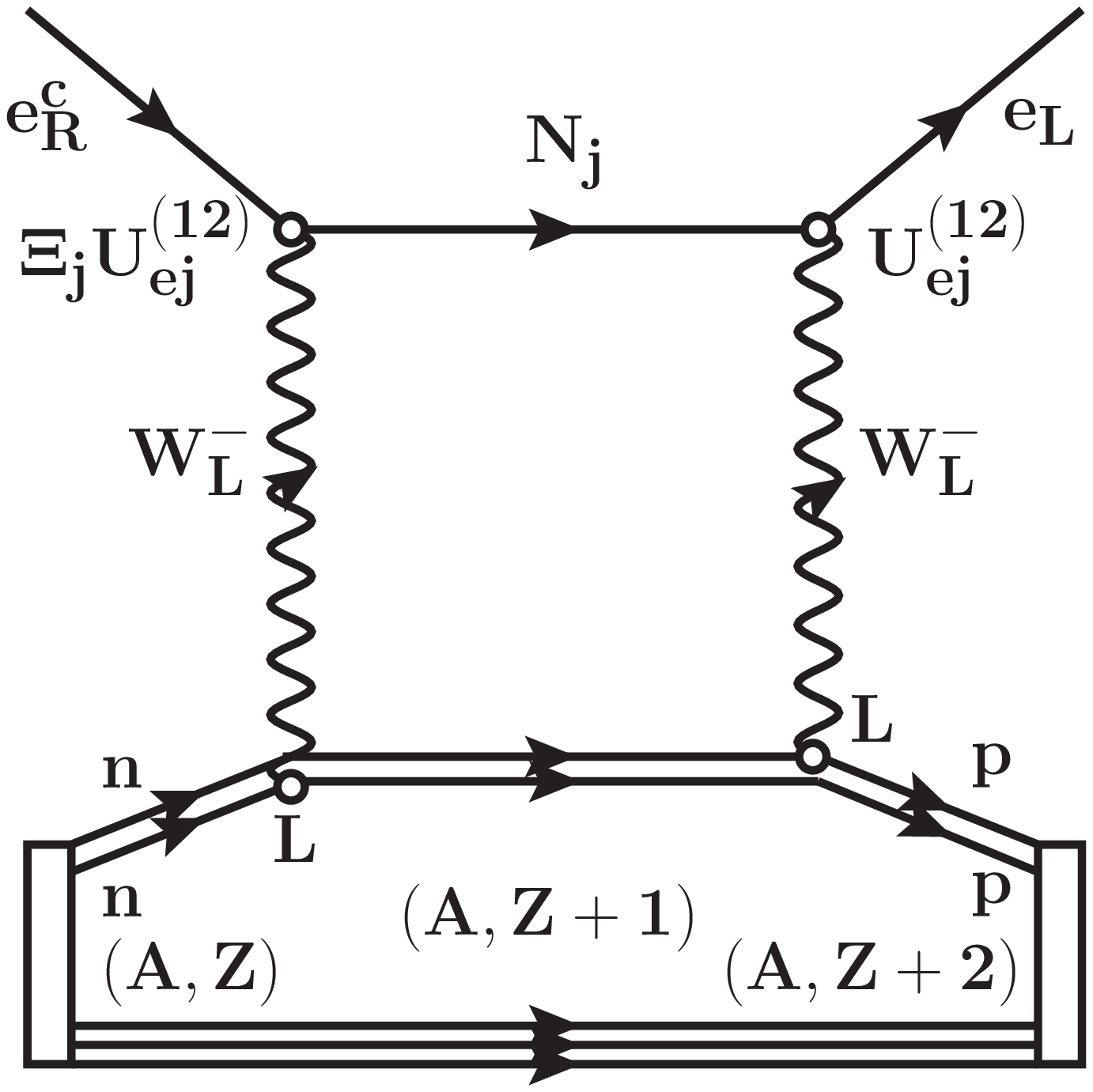}
}\\
 \caption{ The neutrino mass contribution at the nuclear level in the presence of left handed currents for light intermediate neutrino (a) and heavy neutrino (b).}
 \label{Fig:nuLcontr}
\end{center}
\end{figure}

		This mechanism is the most popular and most commonly discussed in the 
		literature (see Fig. \ref{Fig:nuLcontr}).  From this figure we can see read off the couplings and the  phases. We have also seen that for currents of the same chirality  one picks out the mass of the propagating neutrino. Thus the  lepton violating parameter is defined as $\langle m_\nu \rangle/m_e$ with $\langle m_\nu \rangle$ given by Eq. (\ref{eq:1.5a}). 	

\begin{figure}[ht]
\begin{center}
\subfloat[]
{
\includegraphics[scale=0.36]{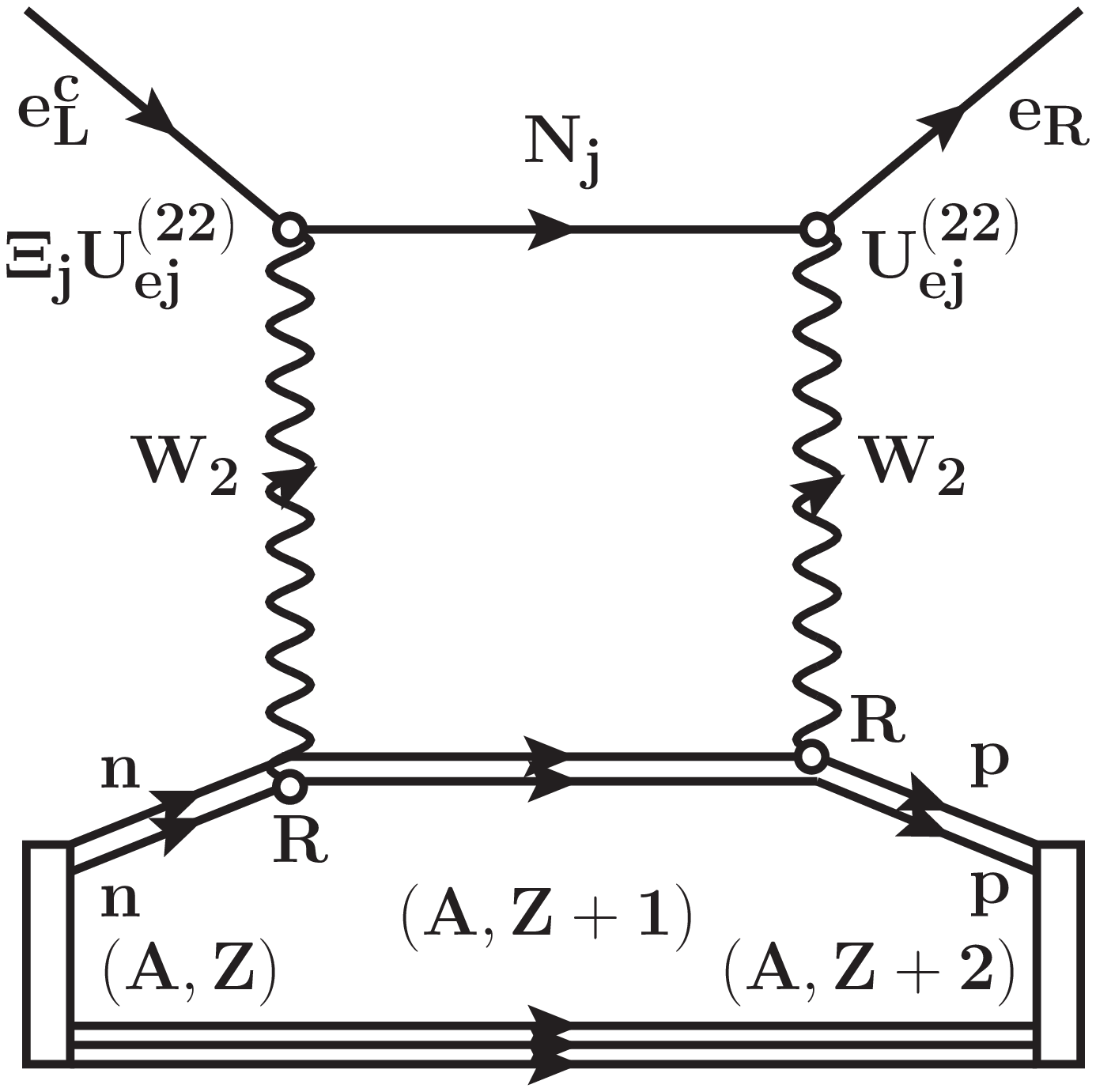}
}
\subfloat[]
{
\includegraphics[scale=0.36]{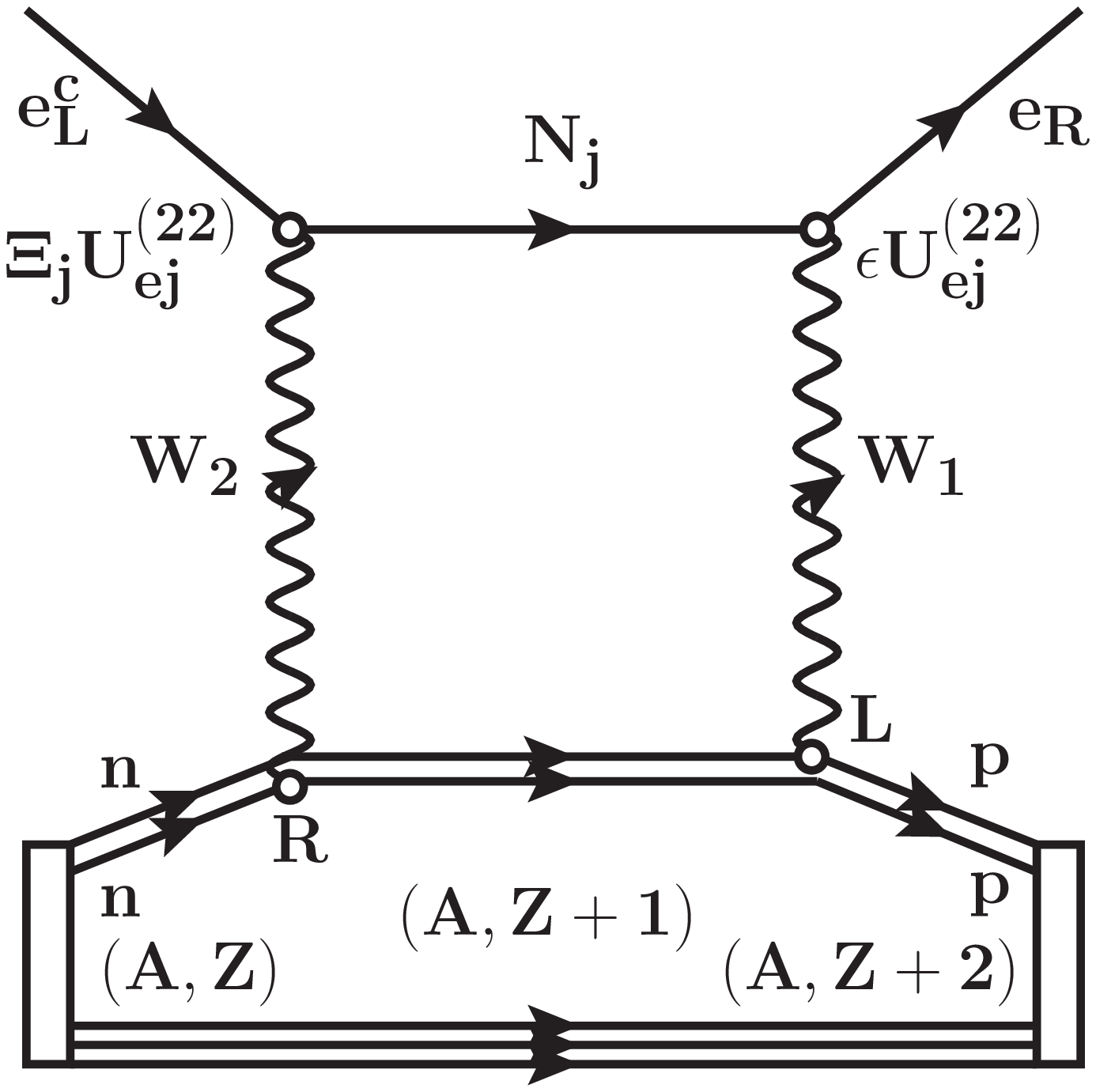}
}\\
\subfloat[]
{
\includegraphics[scale=0.36]{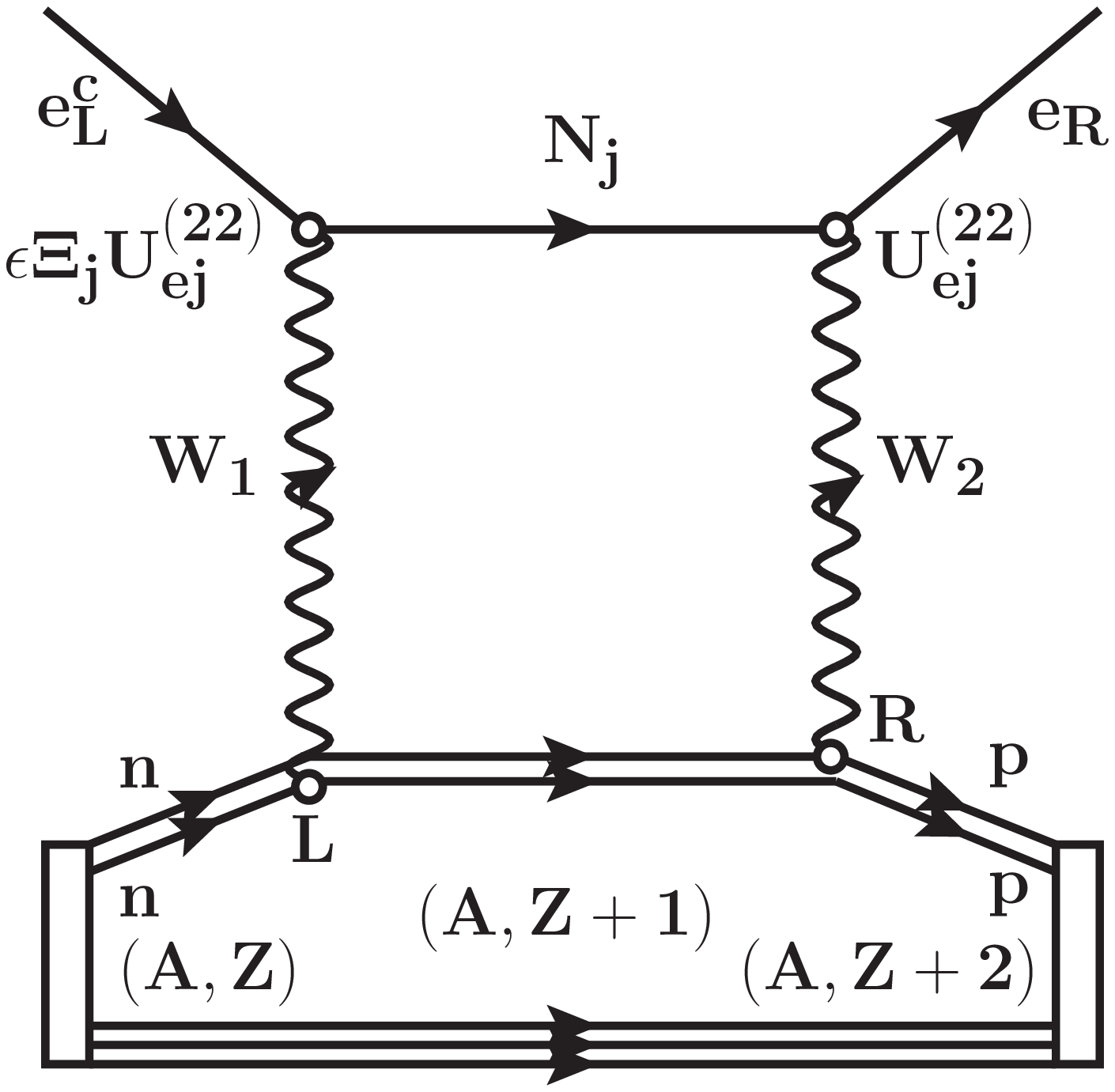}
}
\subfloat[]
{
\includegraphics[scale=0.36]{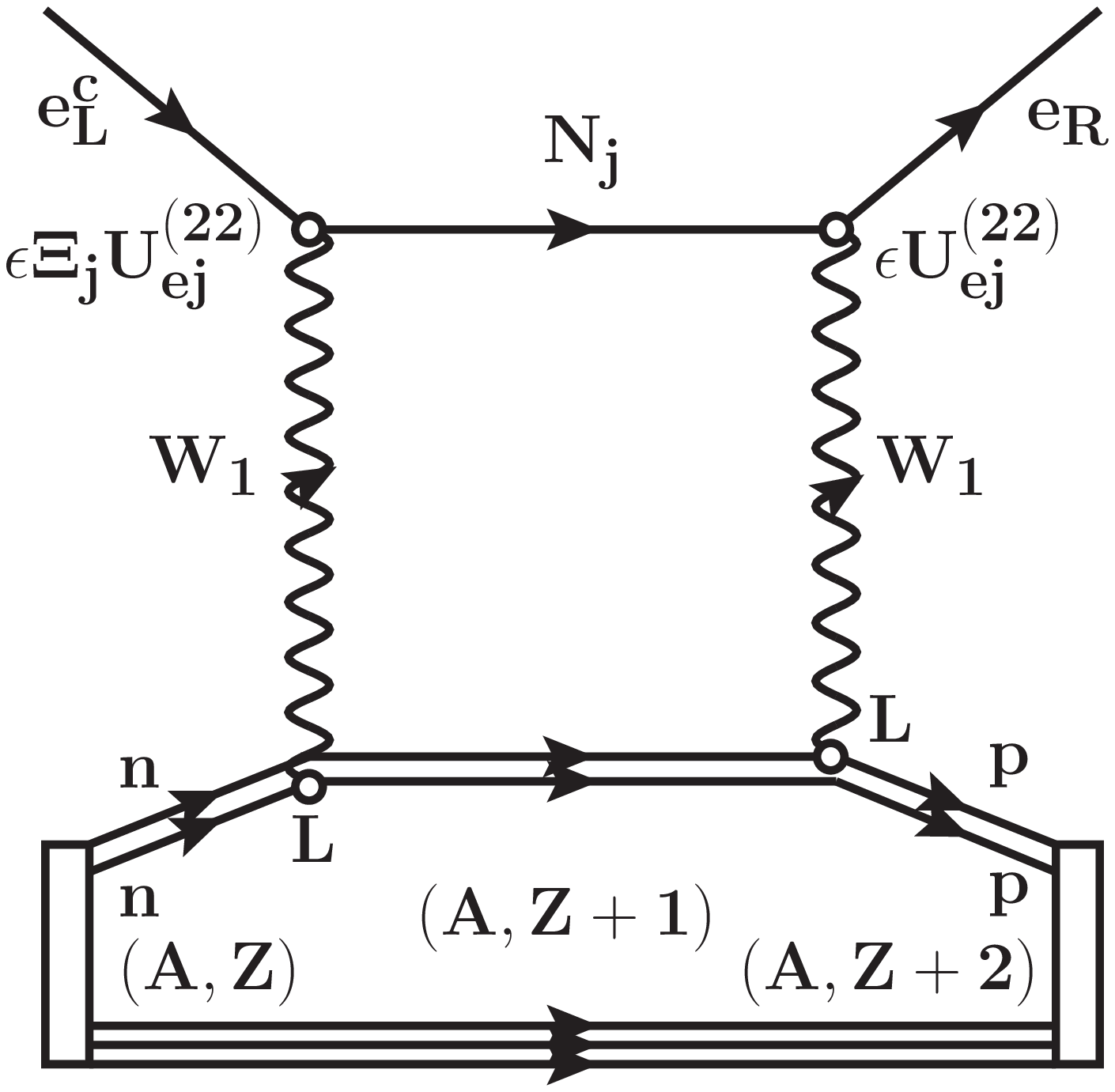}
}\\
 \caption{ The neutrino mass contribution at the nuclear level in the presence of right handed currents proceeding via the right handed boson (a) ($\kappa^2$ suppression factor), the combination of the right handed boson and the mixing between the two bosons (b) and (c) ($\kappa\times \epsilon$ suppression factor) and via the light gauge boson via its mixing with its heavy partner (d) ($\epsilon^2$ suppression factor).}
 \label{Fig:nuRRcontr}
\end{center}
\end{figure}
	 We will consider only $0^+_i\rightarrow 0^+_f$ transitions. 
	Then both outgoing electrons are in the $s_{1/2}$ state. 
	Thus for the ground state transition, restricting ourselves to the  usual light  left handed neutrino
	mass mechanism, we obtain  the following expression for the $0\nu\beta\beta$-decay inverse half-life:
	\beq
	[T_{1/2}^{0\nu}]^{-1} = G_{01}\left|\frac{\langle m_\nu \rangle}{m_e}\right|^2 \left|M^{0\nu}_\nu \right|^2
	\label{Eq:T0nulight}
	\eeq 
	Extraction of $\langle m_\nu \rangle$ from the above life time will have a wide range of implications in physics, it can, e.g., constrain the baryon asymmetry in the universe \cite{Pascoli06b} $|Y_B|$ etc.
	
 Another less popular possibility is the mass contribution arising from DBD in the presence of right handed currents (see Fig. \ref{Fig:nuRRcontr}) or heavy neutrinos in general\cite{Pascoli09}.
			The above expression  can be generalized to include many mass mechanisms 
\cite{Ver86,HS84,DTK85,Tom91,SC98,FS98,PSV96} as follows:
	\begin{equation}
	[T_{1/2}^{0\nu}]^{-1} = G_{01}
        [|X_L|^2+|X_R|^2-\tilde{C}_1^{'}X_LX_R+...].
	\label{eq:1.4}   
	\end{equation}
        The coefficient $\tilde{C}_1^{'}$ is negligible, because these terms do not interfere to leading order due to the different helicity of the 2 electrons. Here  ... indicate other non
        traditional modes (SUSY etc.).
 The nuclear matrix elements entering the above expression are
given in units of $M^{}_{GT}$ 
and are denoted \cite{DTK85} by $\chi$. 
\begin{equation}
X_{L}^{} = \frac{\langle m_\nu\rangle} {m_{e}} M^{0\nu}_\nu + \eta_N^L M^{0\nu}_N,\quad X_{R} =\eta^R_N M^{0\nu}_N,
\label{eq:34}   
\end{equation}
where the nuclear matrix elements $M^{0\nu}_\nu$ and $M^{0\nu}_N$ will be discussed later (see section \ref{sec:TranOper}).
The subscripts $L(R)$ indicate left (right) handed  currents respectively.
	The lepton-number non-conserving parameters, i.e. the
	effective neutrino mass $\langle m_\nu\rangle$ given by Eq. (\ref{eq:1.5a}) and 
	$\eta^{L}_{_N}$ ,$\eta^{R}_{_N}$  are given as follows \cite {Ver86}:
	\beq
	\eta^L_{_N} ~ = ~ \sum^{3}_k~ (U^{(12)}_{ek})^2 ~ 
	~\Xi_k ~ \frac{m_p}{M_k},
	\label{eq:1.5d}   
	\eeq
	\beq
	\eta^R_{_N} ~ =~(\kappa ^2 + \epsilon ^2+2\epsilon \kappa ) \sum^{3}_k~ (U^{22}_{ek})^2 ~ 
	~\Xi_k ~ \frac{m_p}{M_k},
	\label{eq:1.6}   
	\eeq
	with $m_p$ ($m_e$) being the proton (electron) mass.
        $G_{01}$ is the integrated kinematical factor \cite{DK93,Ver83,DTK85,PSV96}.
	The nuclear matrix elements associated with the
	exchange of light and heavy neutrino
	must be computed in a suitable nuclear model. The ellipses ... mean that
	Eq. (\ref{eq:1.4}) can be generalized to the mass term resulting from
       any other intermediate fermion. 

	 At this point we should stress that the main suppression in the case of light
        neutrinos
	comes from the smallness of neutrino masses. In the case of heavy neutrino
	not only from the large values of neutrino masses but the small couplings,
	$U^{(12)}$ for the left handed neutrinos and $\kappa$ and $\epsilon$ for the
	right-handed ones.

	\subsection{The neutrino mass independent mechanism (leptonic left-right interference, $\lambda$ and
	$\eta$ terms).}
\label{sec:etalam}
\begin{figure}[!t]
\begin{center}
\subfloat[]
{
\includegraphics[scale=0.36]{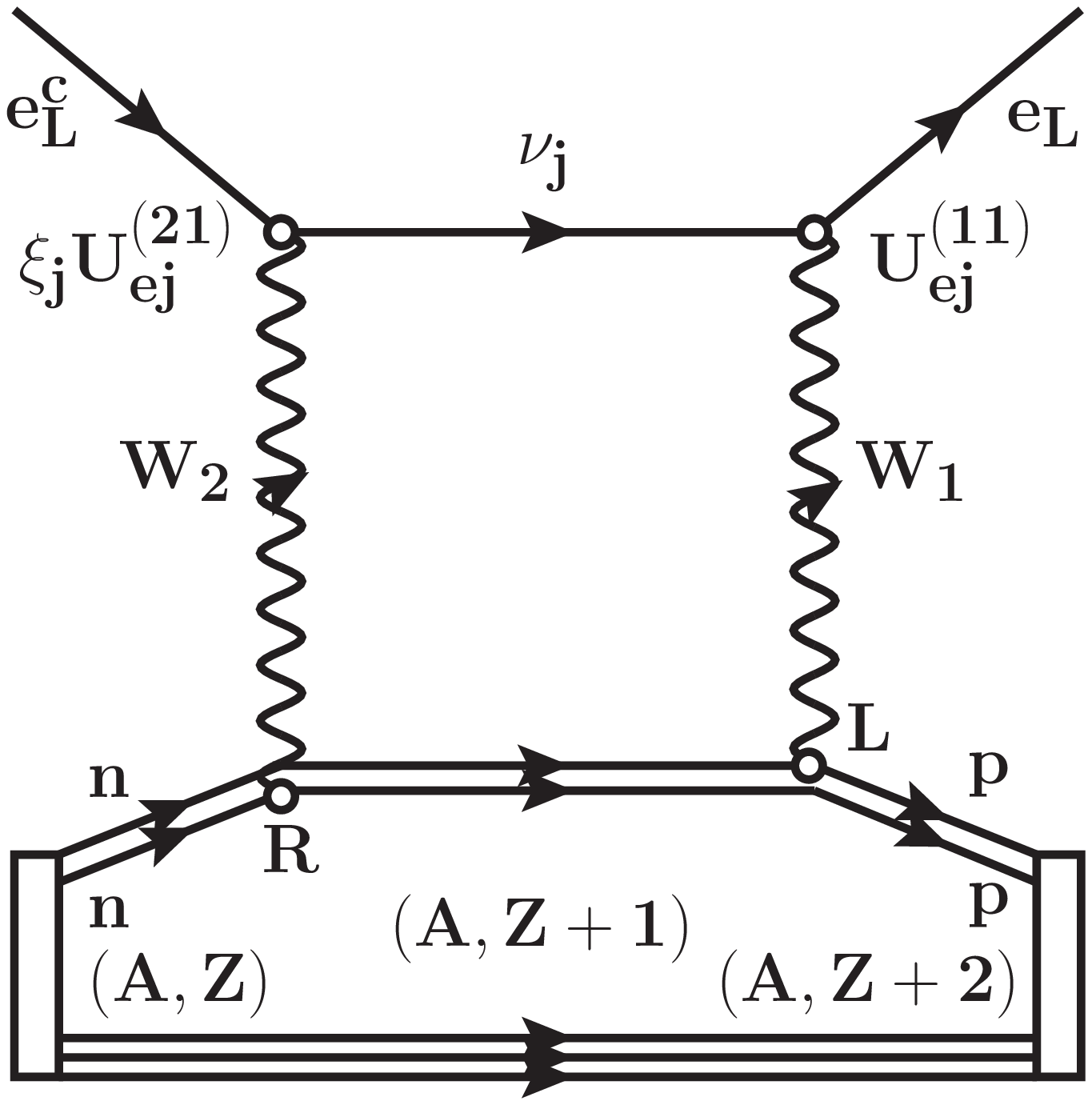}
}
\subfloat[]
{
\includegraphics[scale=0.36]{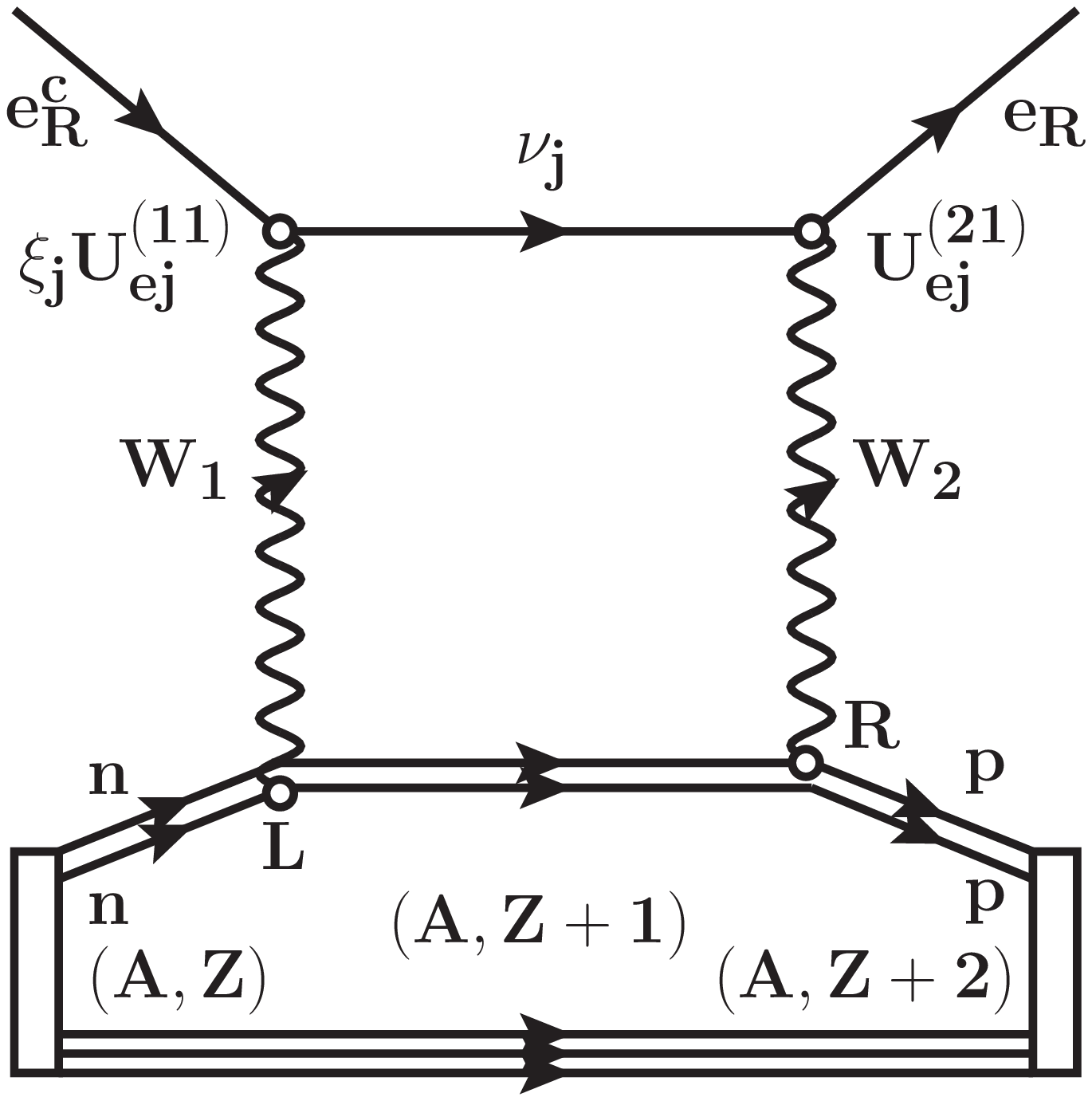}
}\\
\subfloat[]
{
\includegraphics[scale=0.36]{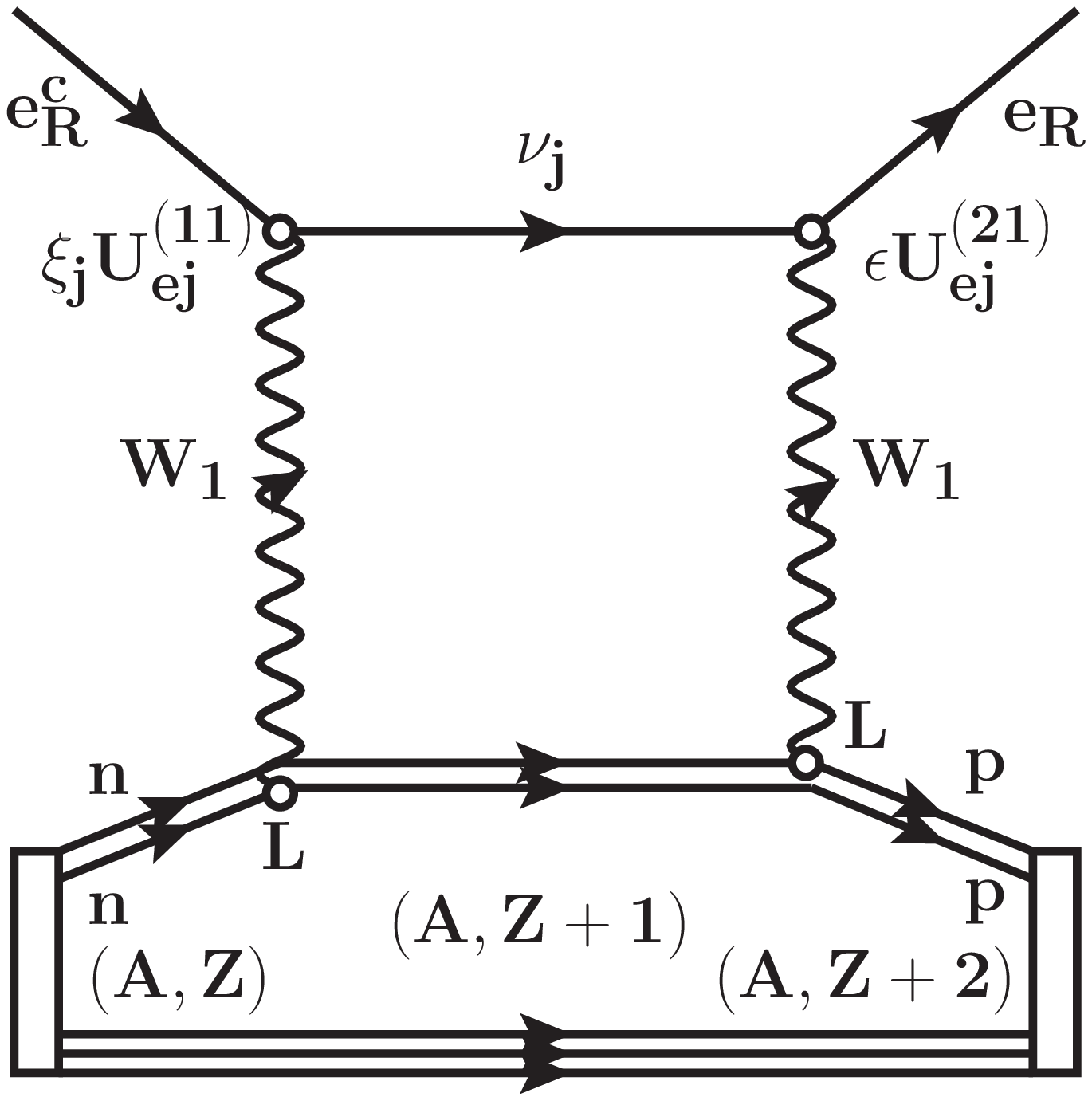}
}
\subfloat[]
{
\includegraphics[scale=0.36]{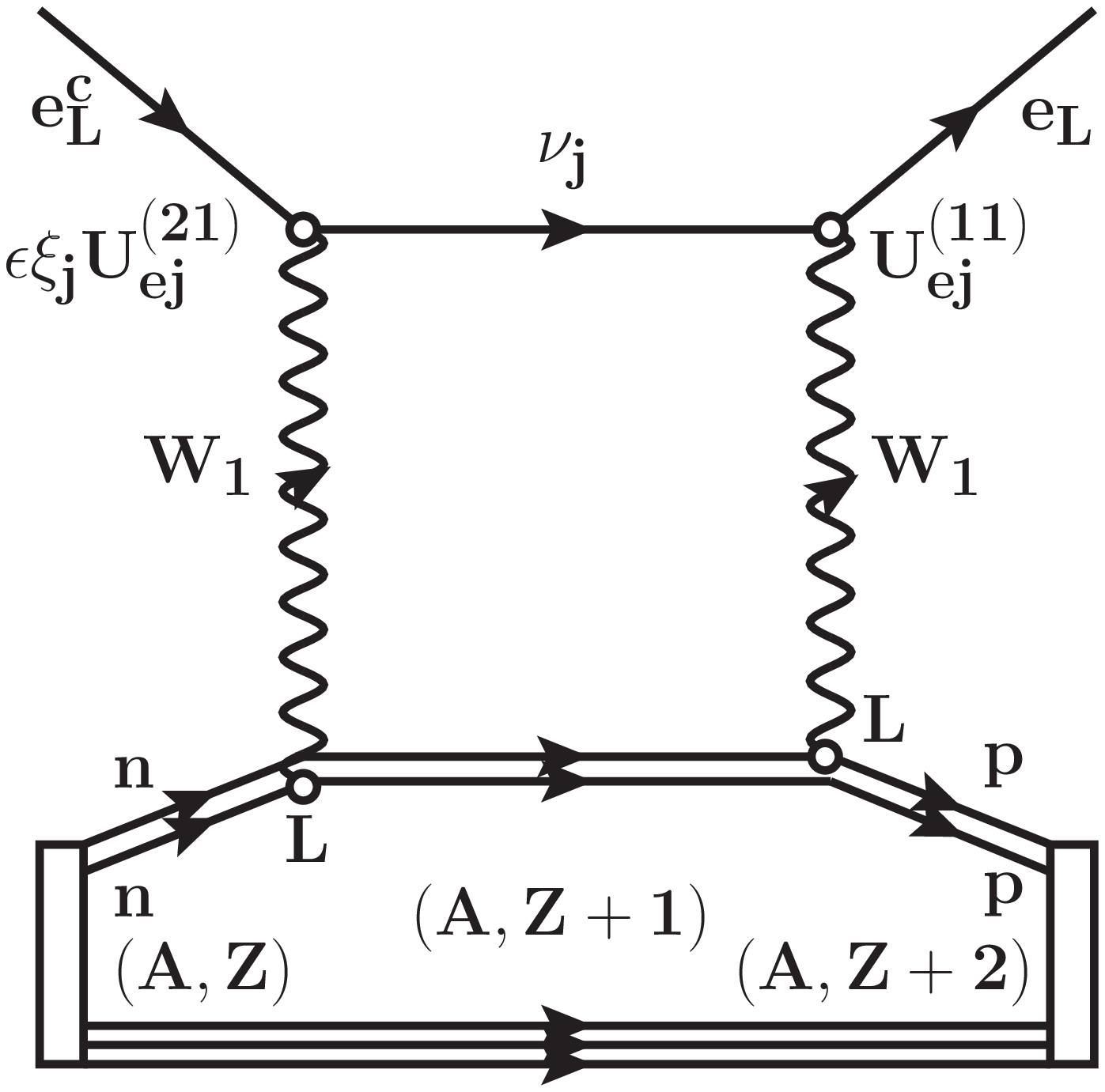}
}
 \caption{ 
	The Feynman diagrams at the nucleon level when the leptonic currents are
	of opposite chirality leading to  the dimensionless lepton number violating
	parameters 
	$\langle\lambda\rangle$ ( panels (a) and (b) of the figure) and $\langle\eta\rangle$ (panels (c) and (d) of the figure)
        of $0\nu\beta\beta$-decay. Note that in part (a) the process proceeds
        via the right handed vector boson, while in part (b) through the mixing
        of the left and right handed bosons.}
	\label{Fig. 2}
\end{center}
\end{figure}

	 As we have already mentioned in the presence of right handed currents one
	can have interference between the leptonic currents of opposite chirality
	(see Fig. \ref{Fig. 2}). The elementary amplitude is now proportional to the
         4-momentum transfer. We thus have a space component and a time component
         in the relevant amplitude.	This leads to different kinematical functions and to two new lepton number violating
	parameters \cite {Ver86} $\langle\lambda\rangle$ and $\langle\eta\rangle$ defined by
	\beq
	\langle\eta\rangle~ = ~ \epsilon \,\eta_{RL} ,\quad \langle\lambda\rangle = \kappa \,\eta_{RL},\quad \eta_{RL} =
	 \sum^{3}_1~ (U^{(21)}_{ek}U^{(11)}_{ek}) ~\xi_k.
	\label{eq:1.7}   
	\eeq
	 The parameters $\langle\lambda\rangle$ and $\langle\eta\rangle$ are small not only due to the smallness
	of the parameters $\kappa$ and $\epsilon$ but, in addition, due to 
         the smallness of $U^{(21)}$.  As we have already mentioned the
         $\langle\lambda\rangle$ can also have a different origin (see section \ref{subsubseclight}). 

	All the above contributions, even though the relevant amplitudes are not
        explicitly dependent 
	on the neutrino mass,  vanish in the limit in which the neutrino is a Dirac 
	particle.

The  above expression for the lifetime is now modified to yield \cite{BilPet87,PSVF96}:
\begin{eqnarray}
[T_{1/2}^{0\nu}]^{-1} = G^{0\nu}_{01} |M^{0\nu}_{GT}|^2 
\left\{ |X_{L}|^2 + |X_{R}|^2 - 
{\tilde C}_1^\prime X_{L} X_{R}+...
\right. ~~~~~~~~~~~   \nonumber \\ 
+ {\tilde C}_2 |\langle\lambda\rangle| X_{L} cos \psi_1 
 + {\tilde C}_3 |\langle\eta\rangle| X_{L} cos \psi_2 
+ {\tilde C}_4 |\langle\lambda\rangle|^2 
\nonumber \\
+ {\tilde C}_5 |\langle\eta\rangle|^2 + {\tilde C}_6 
|\langle\lambda\rangle||\langle\eta\rangle| cos (\psi_1 -\psi_2) + Re ({\tilde C}_2
\langle\lambda\rangle X_{R}  
\nonumber\\ 
\left. + {\tilde C}_3 \langle\eta\rangle  X_{R}) \right\},
\label{eq:69}\nonumber\\
\end{eqnarray}    
where $X_{L}$ and $X_{R}$ 
are defined in Eq.\ (\ref{eq:34}),
$\psi_1$ and $\psi_2$ are the relative phases between $X_{L}$ and
$\lambda$ and $X_{L}$ and $\eta$ respectively. The coefficient ${\tilde C}_1^\prime$, representing the mixing between the left and the right handed currents is kinematically suppressed\cite{FMPSV11}. 
The ellipses \{...\} indicate
contributions arising from other particles, e.g., intermediate SUSY particles 
or unusual particles which are predicted by superstring models
or exotic Higgs scalars etc (see below section \ref{sec:nonumec}).

 Many nuclear matrix elements appear in this case, but they are fairly well
known and they are not going to be reviewed here in detail (see e.g. \cite{Ver86,HS84}
and \cite {DTK85,Tom91,SC98,FS98}). For the reader's convenience 
we are only going to briefly indicate them in our notation \cite {PSV96}) 
the additional nuclear matrix elements, not encountered in the mass mechanism.
 These are: 
$\chi_{F\omega}$, $\chi_{GT\omega}$, 
$\chi_{R}$, $\chi_{1^\pm}$, $\chi_{2^\pm}$
$\chi^{\prime}_{F}$, $\chi^{\prime}_{GT}$, 
$\chi^{\prime}_{T}$,  $\chi^{\prime}_{P}$
where
\begin{equation}
\chi_{F\omega} =  (\frac {g_{V}}
{g_{A}})^2 \frac{M_{F\omega}}{M_{GT}},  
\label{eq:38}   
\end{equation}
\begin{equation}
\chi_{GT\omega} =  
\frac{M_{GT\omega}}{M_{GT}},  
\label{eq:39}   
\end{equation}
\begin{equation}
\chi_{R} =  \frac{M_{R}}{M_{GT}} 
\label{eq:40}   
\end{equation}
and
\begin{equation}
\chi_{1^\pm} = \pm 3\chi^\prime_{F} 
+ \chi^\prime_{GT} -6\chi^\prime_{T},
\label{eq:41}   
\end{equation}
\begin{equation}
\chi_{2^\pm} = \pm\chi_{F\omega} 
+ \chi_{GT\omega} - \frac {1} {9}
 \chi_{1^\pm} 
\label{eq:42}   
\end{equation}
where\cite{PSV96} ($\chi^\prime_{F} = 
M^\prime_{F} /M_{GT}$ etc for the space part) and
 ($\chi_{F\omega} = 
M_{F\omega} /M_{GT}$ etc for the time component).
In the limit in which the average energy denominator 
can be neglected \cite{PSV96},
we obtain 
\begin{equation}
\chi_{F} = \chi^\prime_{F} = \chi_{F\omega}, 
\label{eq:43}   
\end{equation}
\begin{equation}
\chi_{GT} = \chi^\prime_{GT} = 
\chi_{GT\omega} = 1.
\label{eq:44}   
\end{equation}

The quantities $G^{0\nu}_{01}$ have been tabulated 
\cite{DK93,Ver83,DTK85,PSV96,JDV02},
 see also \cite{SC98}
for a recent review.
 The coefficients 
 $\tilde{C}^\prime_1, \tilde{C}_{i}, 
i= 2-6$ are combinations
of kinematical functions and the nuclear matrix elements have been previously  discussed  \cite{JDV02}.
The coefficients  ${\tilde C}_{i}, i = 2,...,6$ with
and without p-n pairing can be found in the literature \cite{PSV96}.
For a more conventional formulation, restricted, however, in the light
neutrino sector, see \cite{SC98}.

        It is worth mentioning 
	that in the case of the $\eta$, in addition to the usual Fermi
        Gamow-Teller and tensor terms, we have additional contributions coming
	from the nucleon recoil term ($\chi_R$) and the kinematically favored
        spin antisymmetric term ($\chi_P$). Due to these two effects  
        the limit extracted for $\eta$ is much smaller than that for $\lambda$
	\cite {PSV96}.
        Effective operators of a similar structure also appear in the
	context of R-parity violating interactions when a neutrino appears in the
	intermediate states (see below).

        There seem to be significant changes in the nuclear matrix elements, when
        the $p-n$ pairing is incorporated. This point needs special care and 
        further exploration is necessary.
        It has only been examined in some
        exactly soluble models, e.g. SO(8), or better approximation schemes \cite{PITTEL97,ENGEL97},
        but only in connection with the $2\nu\beta\beta$-decay, or shell model
         calculations but for systems, which do not double beta decay \cite{PINEDO}.

       Returning back to the question of the availability of nuclear matrix elements
       relevant for neutrinoless double beta decay, we refer once again   
       to existing excellent recent reviews \cite{SC98,FS98,JDV02}. These reviews 
       also provide a more detailed description of the nuclear models employed.

	\subsection{Another neutrino mass independent mechanism (majoron emission)}
\label{sec:majoron}
\begin{figure}[!t]
\begin{center}
\includegraphics[scale=0.4]{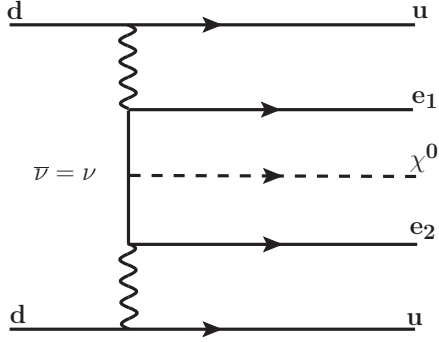}
 \caption{ 
	The Feynman diagrams at quark level leading to  majoron emission
	in the $0\nu\beta\beta$-decay instead of the more well known mass term.
        Here $\chi ^0$ stands for the majoron, not to be confused
        with the neutralino, which we will encounter later in connection
        with supersymmetry.
        } 
	\label{Fig.3}
\end{center}
\end{figure}

\begin{figure}[!t]
\begin{center}
\includegraphics [width=0.4\textwidth]{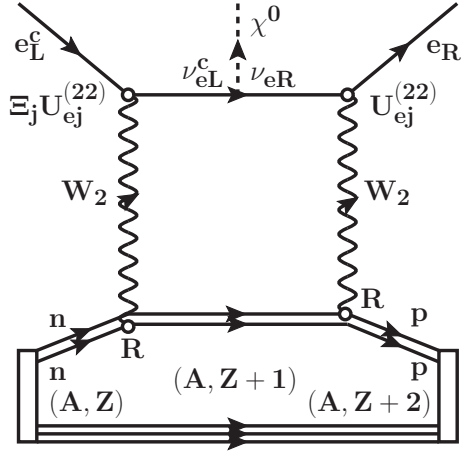}
 \caption{ 
        The same process as in Fig \ref{Fig.3}, but written at the nucleon
        level in the
        case of the isosinglet majoron, which couples to the right handed 
        neutrinos.
        } 
	\label{majoron}
\end{center}
\end{figure}

	It is well known that in some  theories lepton number is associated with
	a global, not a local, symmetry. When such theories are broken spontaneously,
	one encounters  physical Nambu-Goldstone bosons, called majorons. These bosons
	only couple to the neutrinos. So in any model which gives rise  to mass term for
        the light neutrino
        (mass insertion in the neutrino propagator), one may naturally have
         a competing majoron-neutrino-antineutrino coupling.
	Such a mechanism is shown at the quark level in Fig. \ref{Fig.3}.
        The majoron, which couples to the left handed
	neutrinos, comes from the neutral member of the isotriplet. Such a
	multiplet, however, cannot easily be be accommodated theoretically.
        So this type of majoron is not present in the usual models.
        On the other hand there is a majoron $\chi ^0$, the imaginary part of an
	isosinglet scalar, which
	couples to the right handed neutrino with a coupling $g^0_{ij}$.  This gives rise to the mechanism shown in
         Fig. \ref{majoron} at the nucleon level.
        The right handed neutrino, however,
        has a small component of light neutrinos (see Eq. (\ref{Eq:rightnu})).
	\begin{equation}
	{\mathcal{L}}_{\nu \nu \chi^0} = \sum_{i<j} g_{ij} 
	\left[\bar{\nu}_{iL} \gamma_5 \nu_{j L} \right ] \chi^0,
	\label{maj1}   
	\end{equation}
	with
	\begin{equation}
	g_{ij} = \sum_{i<j} U^{(21)}_{\alpha i} U^{(21)}_{\beta j} 
		   g^0_{\alpha \beta} \xi_i,
	\label{maj2}   
	\end{equation}
	  with $g^0_{\alpha \beta}$ the coupling of the majoron to the corresponding 
	neutrino flavors.
	The expression for the half-life takes the form
	\begin{equation}
	[T_{1/2}^{0\nu}]^{-1} = G^{\chi}_{01} |\langle g\rangle M^{0\nu}_{\nu}|^2,  
	\label{maj3}   
	\end{equation}
	with $\langle g\rangle = \sum_{i<j} U^{(21)}_{ei}U^{(21)}_{ej} g_{ij}$.
	 Notice that, even if $g_{ij}$ takes natural values, the
	coupling $g_{ij}$ is very small due to the smallness of the mixing matrix
	 $U^{(21)}$. Thus the effective coupling$\langle g\rangle$ is very small.
	So, even though we do not
	suffer in this case from the suppression due to the smallness of the mass
	of the neutrino, the majoron emission mechanism is perhaps unobservable.
        There exist, however, exotic models, which, in principle,  may allow majoron
         emission with a large coupling, like  the bulk majoron \cite {MOHA00} and others
        \cite {TOMAS01,MONTE00}, which 
	we are not going to pursue this theoretically  any further. In any case it is
        straightforward to extract the limits on the effective coupling $\langle g\rangle$, since the
        nuclear matrix elements are the same as in the light neutrino mass
        mechanism. Note, however, that the spectrum of the summed energy of the two electrons is continuous and the kinematical function
        is  different.

\section[Mechanisms without intermediate neutrinos]{Mechanisms in $0\nu \beta \beta$-decay not involving intermediate neutrinos}
\label{sec:nonumec}

\subsection{The direct decay of doubly charged particles to leptons}
\begin{figure}[!t]
\begin{center}
\subfloat[]
{
\includegraphics[scale=0.36]{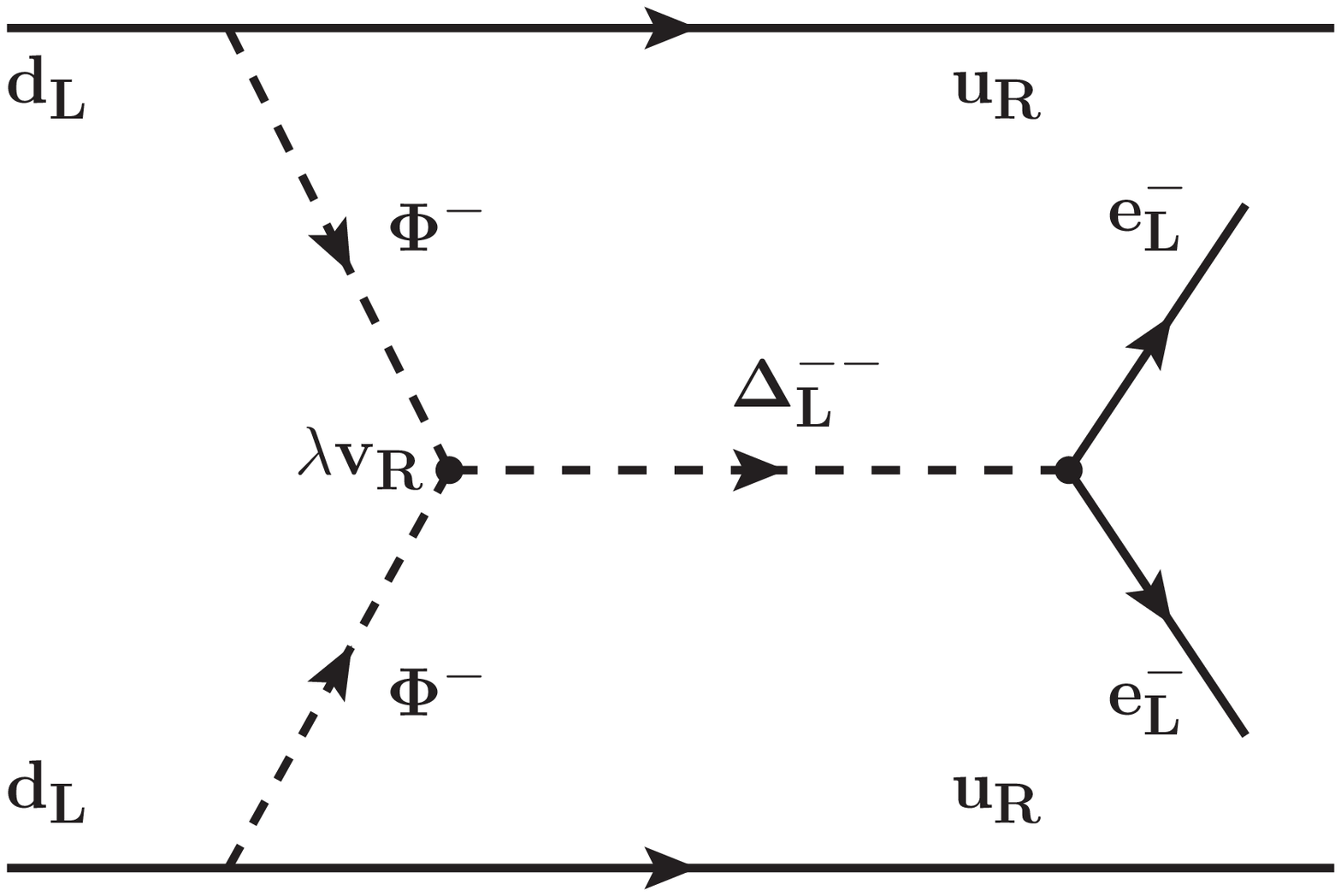}
}
\subfloat[]
{
\includegraphics[scale=0.9]{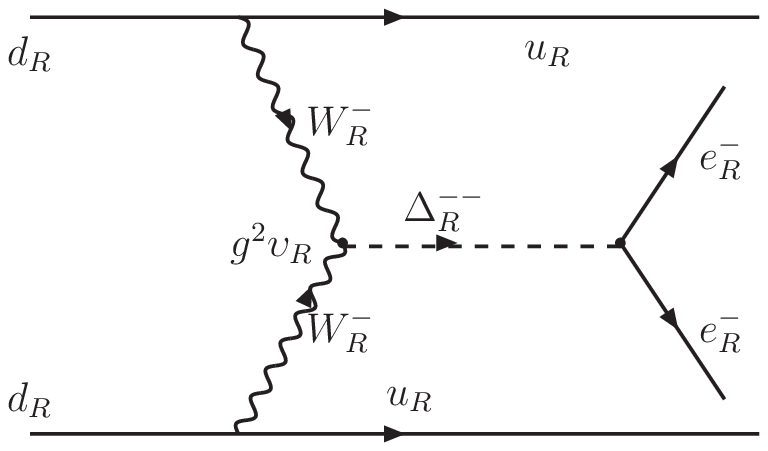}
}
 \caption{ We show the direct decay of the isotriplet into two leptons (a). The coupling to the quarks is achieved via the charged Higgs isodoublet in models where it survives the Higgs mechanism (e.g. SUSY). In (b) we show the direct decay into two leptons of an isosinglet doubly charged Higgs, present in left-right symmetric models. The coupling to the quarks is now achieved via the right handed gauge bosons.}
 \label{Fig:tripletDeltaDir}
\end{center}
\end{figure}

Such a candidate is the isotriplet scalar, which can generate Majorana neutrino mass as  see-saw mechanism II. The doubly charged component can directly decay into two leptons (see Fig. \ref{Fig:tripletDeltaDir}).The coupling to the quarks is achieved via the charged Higgs isodoublet in models where it survives the Higgs mechanism (e.g. SUSY) or through gauge bosons (Figs \ref{Fig:tripletDeltaDir}a and \ref{Fig:tripletDeltaDir}b respectively). For consistency between the two we slightly change the notation here implying that the cubic coupling $\mu_{\Delta}$  originates from a dimensionless quartic coupling $\lambda$ of the Higgs particles of the figure with an isosinqlet scalar acquiring a vacuum expectation value $v_R$.

The lepton number violating parameter associated with Fig. \ref{Fig:tripletDeltaDir} is given by:
\beq
\eta_{\Delta H}=\frac{2 \sqrt{2}\lambda\upsilon_R  m_u^2 m_p g_{ee}}{m_H^4 m^2_{\Delta_L}  G_F},\quad
\eta_{\Delta W}=\frac{\upsilon_R  m_W^2 m_p g_{ee}}{m_{W_R}^4 m^2_{\Delta_R} 4 G_F}
\eeq
Taking $g_{ee}=1$, a natural value, and $\lambda \upsilon_R\approx 1$ TeV, $m_H=m_{\Delta}=100$ GeV  and $m_u=10$ MeV,  one finds \cite{Ver86} $\eta_{\Delta H}\approx 3\times 10^{-8}$. Similarly taking $\upsilon_R=m_{W_R}=m_{\Delta_R}=m_W/\sqrt{\kappa}=10 m_W$, we find $\eta_{\Delta W}\approx 10^{-6}$.\\ Note that the term $\eta_{\Delta H}$ adds to $X_L$ of Eq. (\ref{eq:1.4}), while $\eta_{\Delta W}$ adds to the $X_R$.
		\subsection{The R-parity violating contribution to $0\nu \beta \beta$ decay.}
	 In SUSY theories R-parity is defined as
	\beq
	R = (-1)^{3 B + L + 2 s},
	\label{eq:2.1}   
	\eeq
	with $B=$baryon, $L=$lepton numbers and s the spin. It is +1 for ordinary
	particles and -1 for their superpartners.  R-parity violation has recently
	been seriously considered in SUSY models.
	It allows additional terms in the superpotential given by
	\beq
	W= \epsilon_iL^a_iH_{2a}+\lambda_{ijk} L^a_i L^b_j E^c_k \epsilon_{ab} +
		      \lambda^{\prime}_{ijk} L^a_i U^b_j D^c_k \epsilon_{ab} +
		      \lambda^{\prime\prime}_{ijk} U^c_i U^c_j D^c_k,
	\label{eq:2.2}   
	\eeq
	where a summation over the flavor indices $i,j,k$ and the isospin indices
	a,b is understood ( $\lambda_{ijk}$ is antisymmetric in the indices $i$
        and $j$).
	The last term has no bearing in our discussion, but we will assume that it 
	vanishes due to some discreet symmetry to avoid too fast proton decay. The
        first term is a lepton number  violating bilinear and, since it cannot be rotated
        away, it can lead to neutrinoless double beta decay.
	The $\lambda$'s are dimensionless couplings not predicted by the theory.
	The couplings are assumed to be given in the basis in which the charged
	fermions are diagonal. 
	In the above notation L,Q are isodoublet and $E^c,D^c$ isosinglet chiral
	superfields, i.e they represent both the fermion and the scalar components.
       
        The above R-parity violating superpotential can lead to Majorana neutrino
        masses without the need of introducing the right-handed neutrino and invoking
        the see-saw
        mechanism \cite {HIRSCH01,HVFC00}. One then can have contributions to
        neutrinoless
        double beta decay in the usual way via intermediate massive neutrinos as discussed above. 

	\subsubsection{The contribution arising from the bilinears in the superpotential.}
\begin{figure}[!t]
\begin{center}
\includegraphics [width=0.4\textwidth]{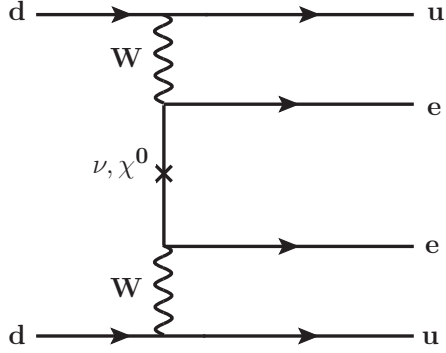}
	\caption[]{
	The R-parity violating contribution to $0\nu \beta \beta$ decay mediated
        by neutralinos arising from the
        bilinear terms in the superpotential. For comparison we give the neutrino
        mediated process of Fig. \ref{Fig:nuLcontr} expressed at the quark level.}
	\label{Fvs. 1}
\end{center}
\end{figure}

	\hspace{2.5cm}
        The first term in the superpotential, Eq. (\ref{eq:2.2}), can cause mixing between the neutrino and the neutralinos as soon as the s-neutrino develops a vacuum expectation value. As a result it can directly lead to neutrinoless
        double beta decay \cite{FKS98b,HIRSCH01,BabMoh95}
        via the effective W-charged lepton-neutralino interaction:
        \beq
        \mathcal{L}=-\frac{g}{\sqrt{2}}\kappa_n~W^-_{\mu}\bar{e}_L\gamma^{\mu}
        \chi^0_{nL}. 
        \label {eq:2.2a}
        \eeq
        where $\kappa_n$ is a dimensionless quantity, associated with each of the four
        neutralinos, which arises due to neutrino neutralino mixing \cite{FKS98b}.
        This term then 
        gives rise to a diagram analogous to that of Fig. \ref{Fig:nuLcontr} with the
        intermediate particle now being the neutralino, which is heavy with mass $M_{\chi^0_n}$ and leads
        to a short ranged operator. For the reader's convenience this 
        is shown in Fig. \ref {Fvs. 1} at the quark level.
        One thus obtains an analogous lepton number violating parameter:
        \beq
        \eta^L_N\rightarrow\eta^L_{\chi^0}=\sum_n~\kappa_n^2\frac{m_p}
        {M_{\chi^0_n}}.
        \label {eq:2.2b}
        \eeq
	\subsubsection{The contribution arising from the cubic terms in the superpotential.}
\begin{figure}[!t]
\begin{center}
\includegraphics [width=0.8\textwidth]{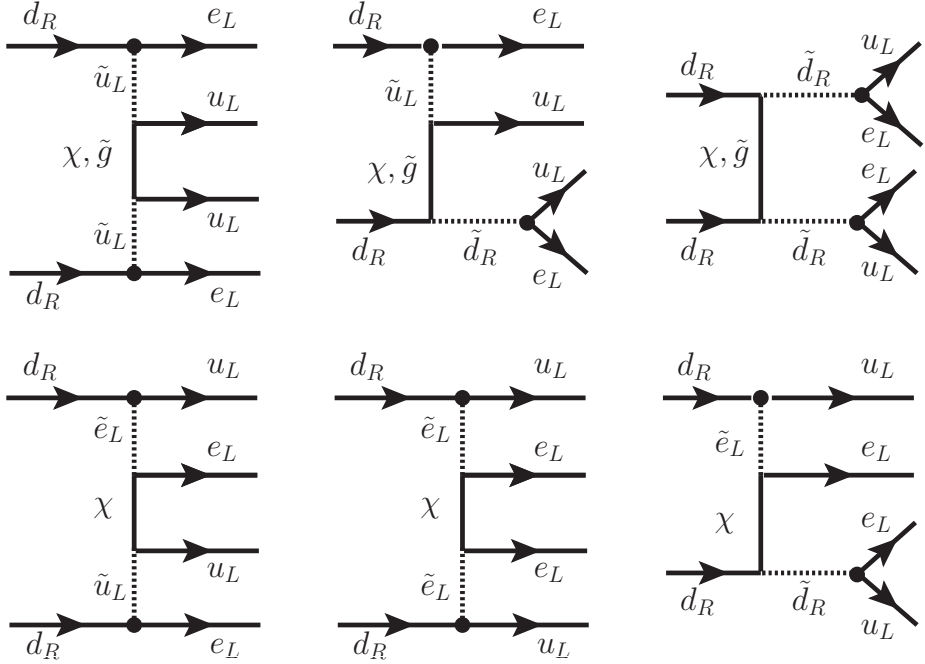}
	\caption[]{
	The R-parity violating contribution to $0\nu \beta \beta$ decay mediated by
	sfermions and neutralinos (gluinos) arising from the cubic terms in the
        superpotential. The dots indicate the lepton violating R-parity interactions.}
	\label{Fig.4}
\end{center}
\end{figure}

	 It has also been recognized quite sometime ago that the second and third terms
        (cubic terms) in the
	superpotential could lead to neutrinoless double beta decay \cite {FKSS97,Moh86,Ver87}.
	Typical diagrams at the quark
	level are shown in Fig. \ref{Fig.4}. Note that as intermediate states, in
	addition to the sleptons and squarks, one must consider the neutralinos,
	4 states which are linear combinations of the gauginos and higgsinos, and
	the colored gluinos (supersymmetric partners of the gluons). 
	 Whenever the process is mediated by gluons a Fierz transformation is
	needed to lead to a colorless combination. The same thing is necessary
	whenever the fermion line connects a quark to a lepton. As a result
	one gets at the quark level not only scalar (S) and pseudoscalar (P) couplings,
	but tensor (T) couplings as well. This must be contrasted to the V and A 
	structure of the traditional mechanisms. One, therefore, must face the problem of
	how to transform these operators from the quark to the nucleon level.

	\subsubsection{The lepton number violating parameters from the cubic  terms without intermediate neutrinos.}

	As we have mentioned the effective lepton number violating parameter
        arising from the bilinear
        terms in the superpotential is analogous to that arising from the heavy  intermediate
       neutrinos and, thus, it will not be further discussed. We will concentrate on the 
       cubic terms in the superpotential \cite {FKSS97,FS98,HKK96,PHK-KK99,FGKS07}. Then 
        the effective lepton number violating parameter
       arising from  these terms,  assuming that the pion exchange mode
	dominates, as the authors of Refs. \cite {FKSS97,FS98} find, can be
        written as
	\beq
	\eta_{SUSY} = (\lambda^{\prime}_{111})^2 \frac{3}{8}(\chi_{PS}~ 
         \eta_{PS}+ \eta_T),
	\label{eq:2.3}   
	\eeq
	with $\eta_{PS}(\eta_T)$ associated with the scalar and pseudoscalar (tensor)
	quark couplings given by
	\beq
	\eta_{PS} = \eta_{\tilde{\chi},\tilde{e}}+ \eta_{\tilde{\chi},
        \tilde{q}}+ \eta_{\tilde{\chi},\tilde{f}}+
		       \tilde{\eta}_{\tilde{g}}+ 7 \eta^{\prime}_{\tilde{g}},
	\label{eq:2.4}   
	\eeq
	\beq
	\eta_{T} =  \eta_{\tilde{\chi},\tilde{q}}- \eta_{\tilde{\chi},\tilde{f}}+
		      \tilde{\eta}_{\tilde{g}} - \eta^{\prime}_{\tilde{g}}.
	\label{eq:2.5}   
	\eeq
	 These authors  find $\chi_{PS}=(5/3)$, but, as we shall see below, it depends,
        in general, on ratios of nuclear matrix elements.
	For the diagram of Fig. \ref{Fig.4}a one finds
	\beq
	\eta_{\tilde{\chi},\tilde{e}} = \frac{2\pi \alpha}{(G_Fm_W^2)^2} 
		     (\kappa_{\tilde{e}})^2 \langle\frac{m_p}{m_{\tilde{\chi}}}
		     \rangle_{\tilde{e}\tilde{e}}.        
	\label{eq:2.6}   
	\eeq
	For the diagram of Fig. \ref{Fig.4}b one finds
	\beq
	 \tilde{\eta}_{\tilde{\chi},\tilde{q}} = \frac{\pi \alpha}{2(G_Fm_W^2)^2} 
		     [(\kappa_{\tilde{d}})^2 \langle\frac{m_p}{m_{\tilde{\chi}}}
		     \rangle_{\tilde{d}\tilde{d}} +        
		     (\kappa_{\tilde{u}})^2 \langle\frac{m_p}{m_{\tilde{\chi}}}
		     \rangle_{\tilde{u}\tilde{u}}],       
	\label{eq:2.7}   
	\eeq
	\beq
	 \tilde{\eta}_{\tilde{g}} = \frac{\pi}{6} \alpha _s \frac{1}{(G_F m_W^2)^2} 
	   [(\kappa_{\tilde{d}})^2 +(\kappa_{\tilde{u}})^2] \frac{m_p}{m_{\tilde{g}}}.
	\label{eq:2.8}   
	\eeq
	For the diagram of Fig. \ref{Fig.4}c one finds
	\beq
	 \tilde{\eta}_{\tilde{\chi},\tilde{f}} =  \frac{\pi \alpha}{2(G_Fm_W^2)^2} 
	  [\kappa_{\tilde{e}} \kappa_{\tilde{d}} \langle\frac{m_p}{m_{\tilde{\chi}}}
		     \rangle_{\tilde{e}\tilde{d}}  +      
	  \kappa_{\tilde{e}} \kappa_{\tilde{u}} \langle\frac{m_p}{m_{\tilde{\chi}}}
		     \rangle_{\tilde{e}\tilde{u}}  +      
	  \kappa_{\tilde{d}} \kappa_{\tilde{u}} \langle\frac{m_p}{m_{\tilde{\chi}}}
		     \rangle_{\tilde{d}\tilde{u}}],      
	\label{eq:2.9}   
	\eeq 
	\beq
	 \tilde{\eta}_{\tilde{g}^{\prime}} = \frac{\pi}{12} \alpha _s 
	\frac{1}{(G_F m_W^2)^2} 
	   \kappa_{\tilde{d}}\kappa_{\tilde{u}} \frac{m_p}{m_{\tilde{g}}},
	\label{eq:2.10}   
	\eeq 
	where
	\beq 
	  \kappa_X = (\frac{m_W}{m_X})^2 \, , \, X=\tilde{e}_L,\tilde{u}_L~~~~~,~~~~~
	  \kappa_{\tilde{d}} = (\frac{m_W}{m_{\tilde{d}_R}})^2 .
	\label{eq:2.12}   
	\eeq 
	\beq 
	\langle\frac{m_p}{m_{\tilde{\chi}}} \rangle_{\tilde{f}\tilde{f}^{\prime}} =  
			      ~\sum_{i=1}^4~ \epsilon_{\tilde{\chi}_{i},\tilde{f}} 
	\epsilon_{\tilde{\chi}_{i},
	\tilde{f}^{\prime}} 
	 \frac{m_p}{m_{\tilde{\chi}_{i}}},
	\label{eq:2.13}   
	\eeq
	where $\epsilon_{\tilde{\chi}_{i},\tilde{f}}$ and 
	$\epsilon_{\tilde{\chi}_{i},\tilde{f}^{\prime}}$ are the couplings of the 
	$i^{th}$ neutralino to the relevant fermion and sfermion. These are calculable
	(see, e.g., Ref. \cite{Ver86}). Thus ignoring the small Yukawa couplings coming
	via the Higgsinos and taking into account only the gauge couplings, we find
	\beq 
	\epsilon_{\tilde{\chi}_{i},\tilde{e}} = \frac{Z_{2i}+tan \theta _W Z_{1i}}
						      {sin \theta _W}, 
	\label{eq:2.14}   
	\eeq 

	\beq 
	\epsilon_{\tilde{\chi}_{i},\tilde{u}} = \frac{Z_{2i}+(tan \theta_W ~/3) Z_{1i}}
						      {sin \theta _W}, 
	\label{eq:2.15}   
	\eeq 
	\beq 
	\epsilon_{\tilde{\chi}_{i},\tilde{d}} = -\frac{Z_{1i}}
						      {3 cos \theta _W}, 
	\label{eq:2.16}   
	\eeq 
	where $Z_{1i},Z_{2i}$ are the coefficients in the expansion of the $\tilde{B},
	\tilde{W}_{3}$ in terms of the neutralino mass eigenstates. Note that in this
	convention some of the masses $m_{\tilde{\chi}_{i}}$  may be negative.
	
	We should mention here that, if the gluino exchange is dominant, the lepton number 
violating parameter $\eta_{\mbox{\tiny{SUSY}}}$ simplifies and becomes $\eta_{\lambda^{'}}$ with $\eta_{\lambda^{'}}$  
given in section \ref{sec:TranOper}.

		\subsubsection{The case of light intermediate neutrinos}
\label{subsubseclight}
\begin{figure}[!t]
\begin{center}
\begin{tabular}{cc}
\includegraphics[width=0.32\textwidth]{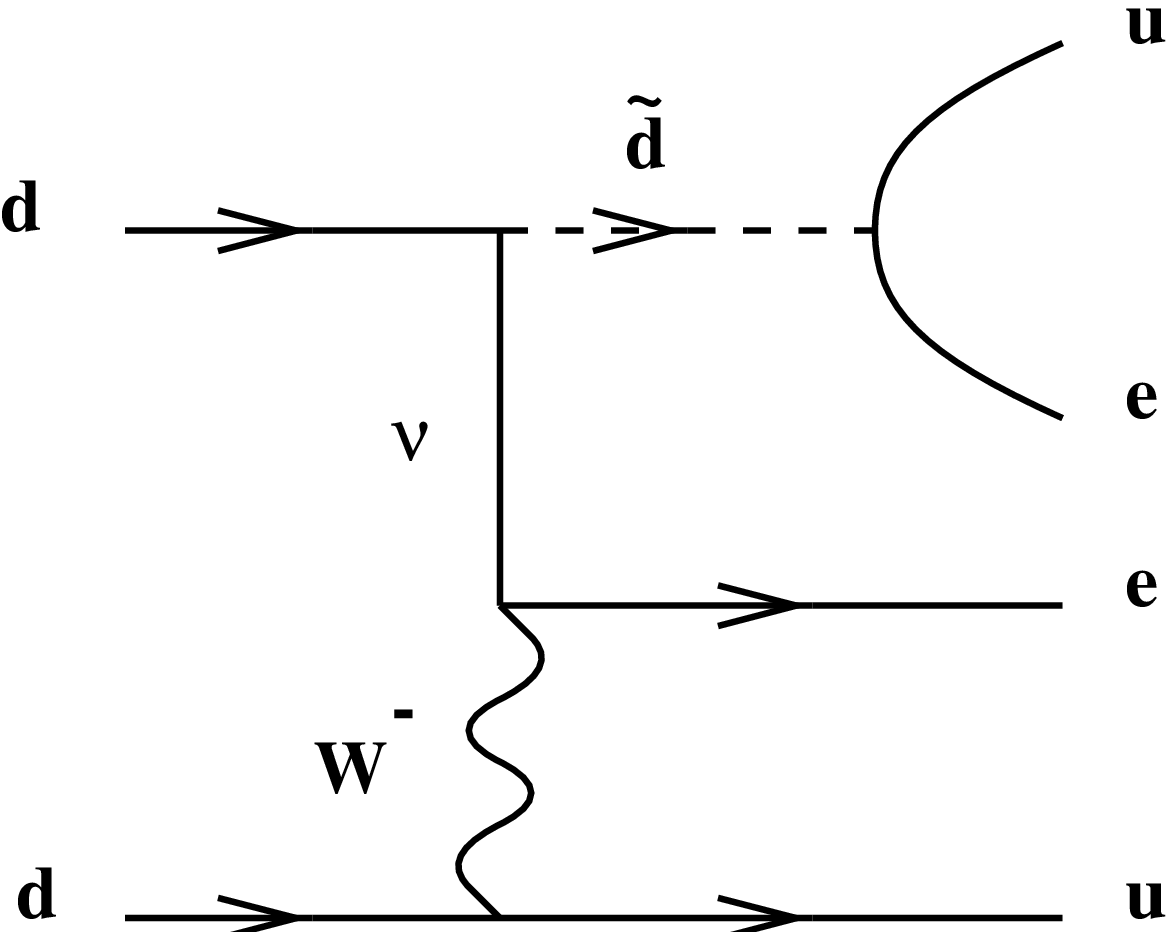}\, & \,
\includegraphics[width=0.32\textwidth]{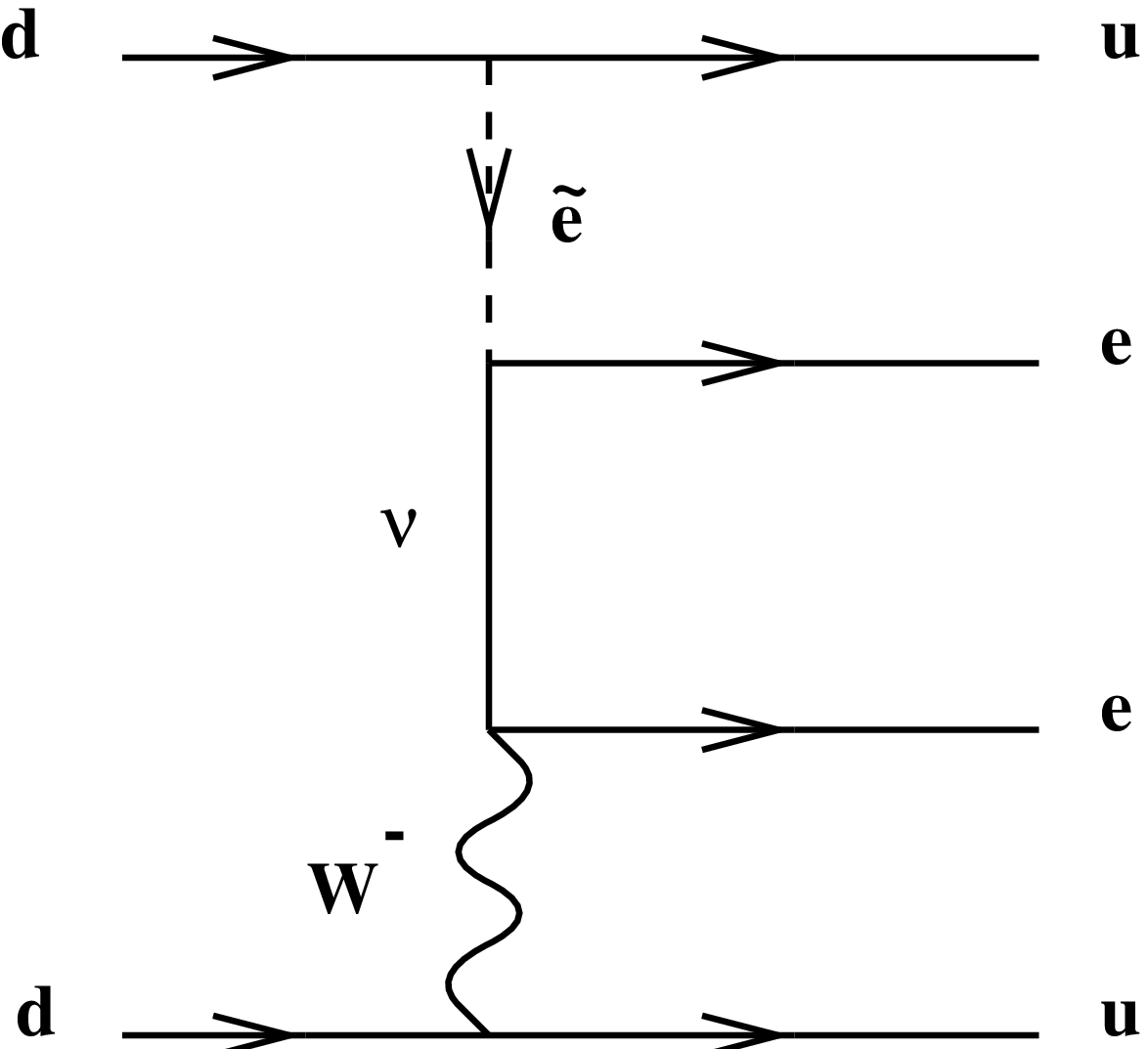}
\end{tabular}
	\caption[]{
	The light neutrino mediated $0\nu\beta\beta$-decay in R-parity violating
	SUSY models. In addition to the usual gauge vertex one has a scalar vertex
	mediated by sleptons  or down-type squarks. The lepton number violation
        proceeds via the mixing between the isodoublet and isosinglet
        sfermions. This is not  indicated  by an $\times$ on the scalar line,
        as it is customary,
        since we use the sfermion mass
        eigenstates.
        } 
	\label{Fig. 7}
\end{center}
\end{figure}

	 It is also possible to have light neutrino mediated $0\nu ~\beta \beta$
	decay originating from R-parity violating interactions. In this case one has
	the usual $\beta$ decay vertex of the $V-A$ type in one vertex and the
	 sfermion mediated vertex, of the $S-P$ type, in the other end
	 \cite{HIRSCH01,BabMoh95,HKK96} (see Fig. \ref{Fig. 7}).
	 The lepton number violation is achieved via
	the mixing of isodoublet and isosinglet sfermions. The simplest diagram, which
	 involves intermediate sleptons, can arise from the following interactions:
	\begin{equation}
	{\mathcal{L}}_{L Q D^c} \rightarrow \lambda^{'}_{111} \sum_{k} V^{*L} _{ek} 
	\bar{u}_{L} d_R \tilde{\ell}^{*}_k
	\label{4.2.1}   
	\end{equation}
	for the d-quark vertex and
	\begin{equation}
	{\mathcal{L}}_{L L E^c} \rightarrow  \lambda_{111} \sum_{j,k} V^R _{ek}U_{ej} 
	\bar{\nu}_{jL} e^c_R \tilde{\ell}_k
	\label{4.2.2}   
	\end{equation}
	for the neutrino vertex.
	In the above expression  $V^L(V^R)$ are the mixing matrices which express the 
	doublet (singlet) selectrons in terms of the mass eigenstates. 
	 U is the usual neutrino mixing matrix. 
	The effective s-lepton propagator is
	\begin{equation}
	P=\sum_{k=1}^3 \frac{V^L_{ek}~V^R_{ek}}{M^2_k}\approx \sum_{k=1}^2 V^L_{ek}V^R_{ek}
	\frac{\delta M^2_k}{M^4_e},\quad \delta M^2_k=M_k^2-M_3^2.
	\label{4.2.3a}   
	\end{equation}
	The last equation follows from the orthogonality condition on the mixing matrices and the fact that the splitting $\delta M^2_k$ is small compared
	to the average s-lepton mass $M^2_e$. If, further, the mixing between 
	generations can be ignored we get
	\begin{equation}
	P=\frac{\sin{2 \theta_{\tilde{e}}}}{2} \frac{\Delta M^2_e}
	{M^4_{\tilde{e}}}  
	\label{4.2.3b}   
	\end{equation}
	Combining the above results with the usual $V-A$ coupling one gets:
	\begin{equation}
	{\mathcal{M}} = (\frac{G_F}{\sqrt{2}})^2 \Lambda_{\tilde {e}} \bar{u}_{L} d_R 
	\bar{e}_{L} \frac{ k_{\alpha}\gamma^{\alpha}}{k^2}\gamma^{\lambda}e^c_R 
	\bar{u}_L\gamma_{\lambda}d_L,
	\label{4.2.3}   
	\end{equation}
	with
	\begin{equation}
	\Lambda_{\tilde{e}} =  \frac{\sqrt{2}}{G_F} \lambda^{'}_{111} \lambda_{111} 
	\sum_{j} U^2_{ej}
	\frac{\sin{2 \theta_{\tilde{e}}}}{2} \frac{\Delta M^2_e}
	{M^4_{\tilde{e}}}  
	\label{4.2.4}   
	\end{equation}

	In the case of squark exchange  of Fig. \ref{Fig. 7} the above expressions become
\begin{equation}
{\mathcal{L}}_{L Q D^c} \rightarrow \lambda^{'}_{111} \sum_{k} V^{*L} _{dk} 
\bar{u}_{L} e^c_R \tilde{d}^*_k
\label{4.2.5}   
\end{equation}
for the u-quark vertex and 
\begin{equation}
{\mathcal{L}}_{L Q D^c} \rightarrow  \lambda^{'}_{111} \sum_{k} V^{R} _{dk} 
U_{ej} \bar{\nu}_{jL} d_R \tilde{d}_k
\label{4.2.6}   
\end{equation}
for the neutrino vertex. Combining them we get
\begin{equation}
{\mathcal{M}} = ( \frac{G_F}{\sqrt{2}})^2 \Lambda_{\tilde{d}} \bar{u}_{L} e^c_R 
\bar{e}_{L} \gamma^{\lambda}\frac{ k_{\alpha}\gamma^{\alpha}}{k^2}d_R, 
\bar{u}_L\gamma_{\lambda}d_L
\label{4.2.7}   
\end{equation}
with
\begin{equation}
\Lambda_{\tilde{d}} =  \frac{\sqrt{2}}{G_F} (\lambda^{'}_{111})^2
\sum_{j} (U_{ej})^2
\frac{\sin{2 \theta _{\tilde{d}}}}{2} \frac{\Delta M^2_{\tilde{d}}}
{M^4_{\tilde{d}}} 
\label{4.2.8}  
\end{equation}
in a rather obvious notation.

In the case $\tilde{e}$ and $\tilde{d}$ contributions, in the context of
 perturbation theory, one can simplify the above expressions
using explicitly the coupling between the singlet and the doublet sfermions of
the lower charge. In this case 
\begin{equation}
\frac{\sin{2 \theta _{\tilde{x}}}}{2} \Delta M^2_{\tilde{x}}=(\mu +A \tan{\beta})m_x~
~~x=e,d. 
\label{4.2.8b}  
\end{equation}
Before proceeding farther we have to perform a Fierz transformation:
\barr
\bar{u}_{L} e^c_R 
\bar{e}_{L} \gamma^{\lambda} k_{\alpha} \gamma^{\alpha} d_R & =& 
-\frac{1}{2}[\bar{u}_{L} d_R \bar{e}_{L} e^c_R k_{\lambda} +
\bar{u}_{L} d_R \bar{e}_{L} i\sigma_{\lambda \nu} k^{\nu}e^c_R\\
\nonumber
 & + & 
\bar{u}_{L} i\sigma_{\lambda \nu} k^{\nu} d_R \bar{e}_{L} e^c_R ]
-\frac{1}{8}\bar{u}_{L} i\sigma_{\alpha \beta} d_R 
\bar{e}_{L} i\sigma^{\alpha \beta} e^c_R  k_{\lambda}.
\label{4.2.9}   
\earr
We must now go to the nucleon level and perform a Fourier transform to the
coordinate space. For $0^+\rightarrow 0^+$ transitions the space component 
yields:
\barr
{\mathcal{M}} &=&  (\frac{G_F}{\sqrt{2}})^2(f_A)^2~\left (-\lambda~ [M^{'}_{T}
+M^{'}_{GT}+ r_F~M^{'}_{F}] \bar{e}\gamma_0 {\bf {q.\gamma}}(1+\gamma_5)e^c \right )\nonumber\\
&-&\frac{\Lambda_d}{2}\left (\tilde{M}_{GT}+(1-2\frac{\Lambda_e}{\Lambda_d})\frac{1}{f_A^2}\tilde{M}_F \right )q_0\bar{e}(1+\gamma_5)e^c
\label{4.2.10}  
\label{Eq:SROSUSY} 
\earr
with 
\begin{equation}
\lambda=\Lambda_{\tilde{d}}/96~~,~~ r_F=\frac{3}{4f_A^2}
(-2 \frac{ \Lambda_{\tilde{e}}}{ \Lambda_{\tilde{d}}}+1).
\label{4.2.11}   
\end{equation}
The parameter $\lambda$ as well as the quantities $M^{'}$ and $\tilde{M}$
have the same meaning as in the mass independent contribution in the
 conventional approach (see section \ref{sec:etalam}). Note, however, that in the present
mechanism there is no term analogous to the $\eta$ of section  \ref{sec:etalam}. We should
stress that this novel mechanism can lead to transitions $J^+$, $J\neq0$.
 So, contrary to conventional wisdom,
from the observation of such transitions one cannot definitely infer the
existence of right handed currents.


	\section[Handling the  short range transition operators]{Handling the  short range transition operators}
	\label{sec:srop}
	We have seen that there exist many mechanisms contributing  to neutrinoless double beta decay, involving the exchange of only  heavy particles. These result to short range transition operators.
	\subsection{The mode involving only nucleons}
	If the nucleons are treated as point like particles, then the effective transition operator essentially behaves like a $\delta$ function in the inter nucleon distance.  Thus their contribution vanishes, due to the presence of a nuclear hard core. This can be cured, if the nucleons can be treated as extended objects. This can be done by introducing into the nucleon current a nucleon form factor \cite{Ver81}, e.g. like a dipole shape with a characteristic mass $m_A\approx 850$ MeV. This can also be accomplished, if one utilizes a quark model for the nucleon \cite{VerFaeTok10}. 
	\begin{itemize}
	\item V-A theories\\
	In this case the approach has become pretty standard, i.e. the spin isospin  structure is similar to that for the light neutrino mass term except for the radial part \cite{Ver86}, which now becomes:
	$$
	 \frac{m_A^2}{m_e m_p}F_N(x_A)\frac{R_0}{r_{k\ell}},\quad x_A=r_{k\ell}\, m_A,\quad F_N(x)=\frac{x}{48}\left (x^2+3x+3 \right )e^{-x}
	$$
	and we are not going to discuss them further. We only mention that it can also proceed via the 2-pion mode with the nuclear operator associated with $\alpha_{2 \pi}=0.1$ (see next section)
	\item $S$, $PS$ theories\\
	In the scalar case we find that the operator is spin independent and has the same radial part as in the previous case.\\
	 In the case of the pseudo scalar part we find that the operator becomes
	 \barr
	 \Omega_{PS} = \frac{m_A^2}{m_e m_p}\frac{1}{3}\left (\frac{m_A}{2 m_p/3}\right )^2\sum_{k\ne\ell}\tau_+(k)\tau_+ 
	 (\ell)  \times~~~~~~~\nonumber\\
	 \left(  \sigma_k .\sigma_{\ell} \frac{2}{x_A}\frac{dF_{N}(x_A)}{d x_A}+T(\hat{x}_A,\sigma_k,\sigma_{\ell})\left ( -\frac{dF_{N}(x_A)}{x_Ad x_A}+\frac{dF^2_{N}(x_A)}{d x^2_A}\right )\right )\nonumber\\
	 \earr
	 where $T(\hat{x}_A,\sigma_k,\sigma_{\ell}) $ is the tensor component.\\
	 This process can also proceed via the 2-pion mode (see next section). In this case one can use the lepton number violating parameter $\eta_{\Delta H}$ with nuclear matrix element as given in the next section with $\alpha_{2 \pi}=0.20$.
	 
	\end{itemize}
	\subsection{The pion mode in R-parity induced $0\nu \beta \beta$ decay.}
\label{sec:pionmode}

\begin{figure}[!t]
\begin{center}
\includegraphics [width=0.95\textwidth]{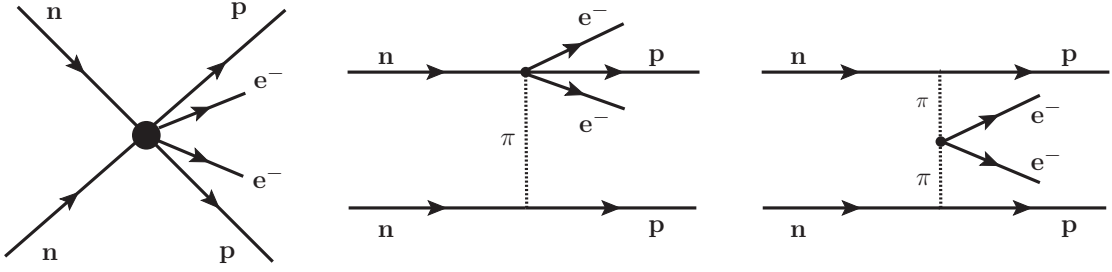}
	\caption[]{
	The pion mediated $0\nu\beta\beta$-decay
	as a contact interaction (a).
	 It arises as an  $1\pi$, (b) (an analogous $\pi$-vertex in the other
	 nucleon is understood),
	and $2\pi$ exchange contributions (c). 
	The lepton number violation occurs either in one nucleon
	(b) or in the pions (c).} 
	\label{Fig:pionmode}
\end{center}
\end{figure}

	 Even though the pion model, (\ref{eq:1}),
         may be important in other cases when the intermediate
         particles are heavy, giving rise to short range operators, in this section we will
         elaborate a bit further on its application in the extraction of the R-parity
        violating parameters associated with the processes discussed above.
The nuclear matrix elements can now be calculated using the effective transition operators
	\beq
	ME_k~ = (\frac{m_A}{m_p})^2~ \alpha_{k\pi}~\frac{m_p}{m_e}~
		[ M^{k\pi}_{GT}+M^{k\pi}_T].
	\label{eq:6.6}   
	\eeq
	Where the two above matrix elements are the usual GT and T matrix elements
	with the additional radial dependence given by
	\beq
	F^{1\pi}_{GT} ~ = e^{-x}~~~  ,~~~
        F^{1\pi}_T ~ = (3+3x+x^2)~\frac{e^{-x}}{x},
	\label{eq:6.7}   
	\eeq
	\beq
	F^{2\pi}_{GT} ~ = (x-2)e^{-x}~~~  ,~~~ F^{2\pi}_T ~ = (1+x)~e^{-x},
	\label{eq:6.8}   
	\eeq
	In the case of the pion exchange mechanism, in particular for $1\pi$ exchange, it is important to include the nucleon form factors \cite{FGKS07,adler}. We are not going to elaborate further on this point. The complete expressions for the transition operators are given in section \ref{sec:TranOper}.

	 We will, instead, concentrate on the coefficients $\alpha_{2\pi}$ and
$\alpha_{1\pi}$. We will begin by considering the elementary particle treatment \cite{FKS98}.
 
\begin{itemize}
\item The coupling coefficients $\alpha_{2\pi}$.
 One finds
  \beq
  \alpha_{2 \pi}=\frac{1}{6 f^2_A}g^2_r h^2_{\pi}\left( \frac{m_{\pi}}{m_p}\right)^4
  \eeq
  Obtained under the factorization approximation\cite{FKS98}:
  \begin{eqnarray}
 < \pi^+|J_P J_P|\pi^- > &=& \frac{5}{3}< \pi^+|J_P|0><0|j_P|\pi^- >, \nonumber\\
 <0|J_P|\pi^-> &=& m^2_{\pi}h_{\pi}
 \label{Eq:EP2pi}
  \end{eqnarray}
  with the parameter $h_{\pi}$ given by
  \beq
  h_{\pi}= 0.668 \sqrt{2}\, i \frac{m_{\pi}}{m_d+m_u}
  \eeq
  Thus using the current quark masses these authors \cite{FKS98} find $\alpha_{2 \pi}=0.20$.
\item The coupling coefficients $\alpha_{1\pi}$\\
One finds
\beq
\alpha_{1\pi}=-F_P \frac{1}{36 f^2_A}g_r h_{\pi}\left( \frac{m_{\pi}}{m_p}\right)^4
  \eeq
 The needed parameters were obtained using the factorization approximation  in  the case of $1-{\pi}$ mode
\begin{eqnarray}
<p|J_P J_P|n~\pi > &=& \frac{5}{3}<p|J_P|n><0|J_P|\pi^->, \nonumber\\
<p|J_P|n> &=& F_P\approx 4.41
 \label{Eq:EP1pi}
\end{eqnarray}
The matrix element $<0|J_P|\pi^->$ was given above (see Eq. (\ref{Eq:EP2pi})). Thus these authors \cite{FKS98}
 find:
\beq
\alpha_{1\pi}=-4.4\times 10^{-2}
\label{Eq:alpha1pi}
\eeq
\end{itemize}

\begin{figure}[!t]
\centering
\begin{tabular}{cc}
\includegraphics[scale=0.3]{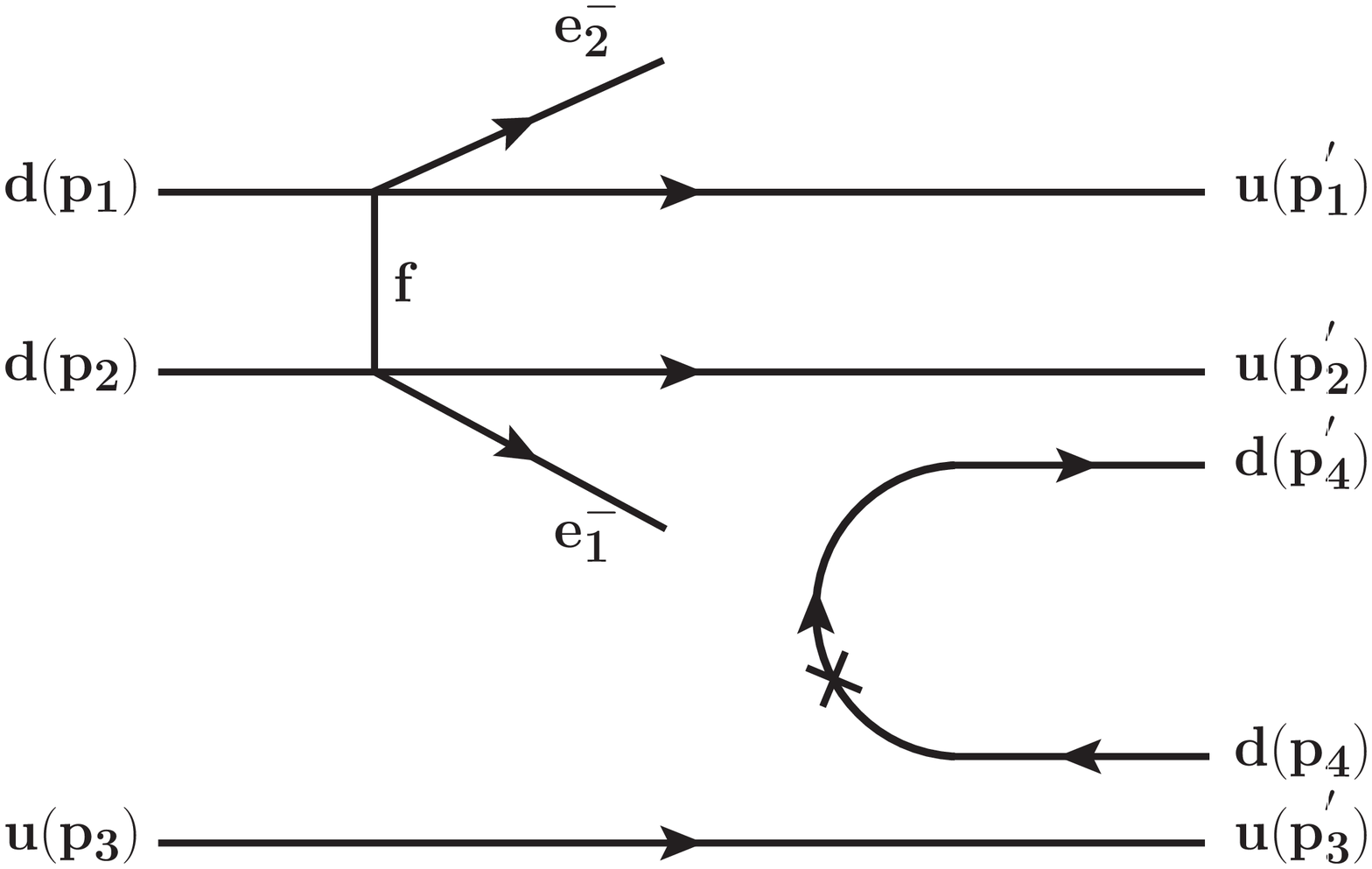} & 
\includegraphics[scale=0.3]{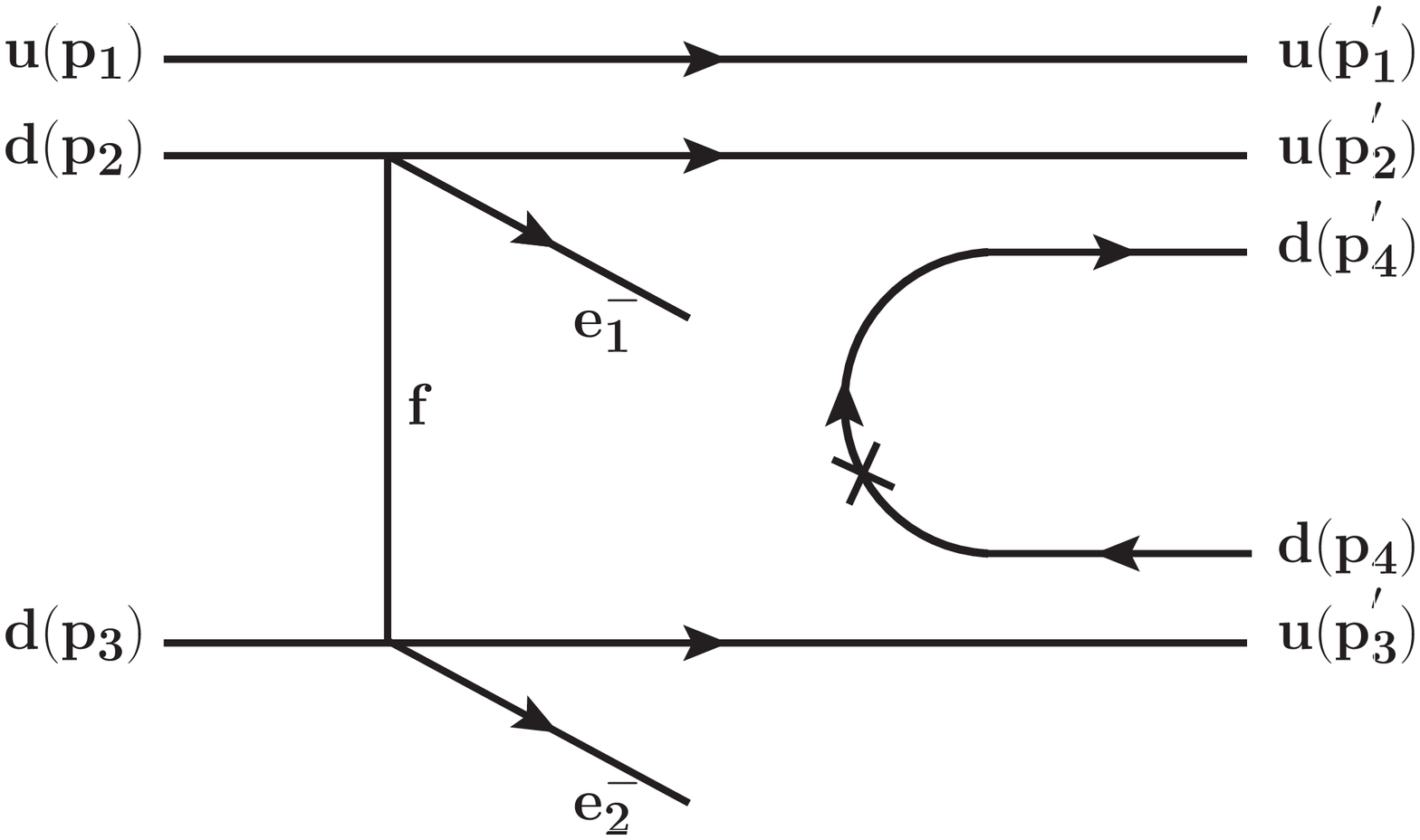} 
\end{tabular}
\begin{tabular}{c}
\includegraphics[scale=0.3]{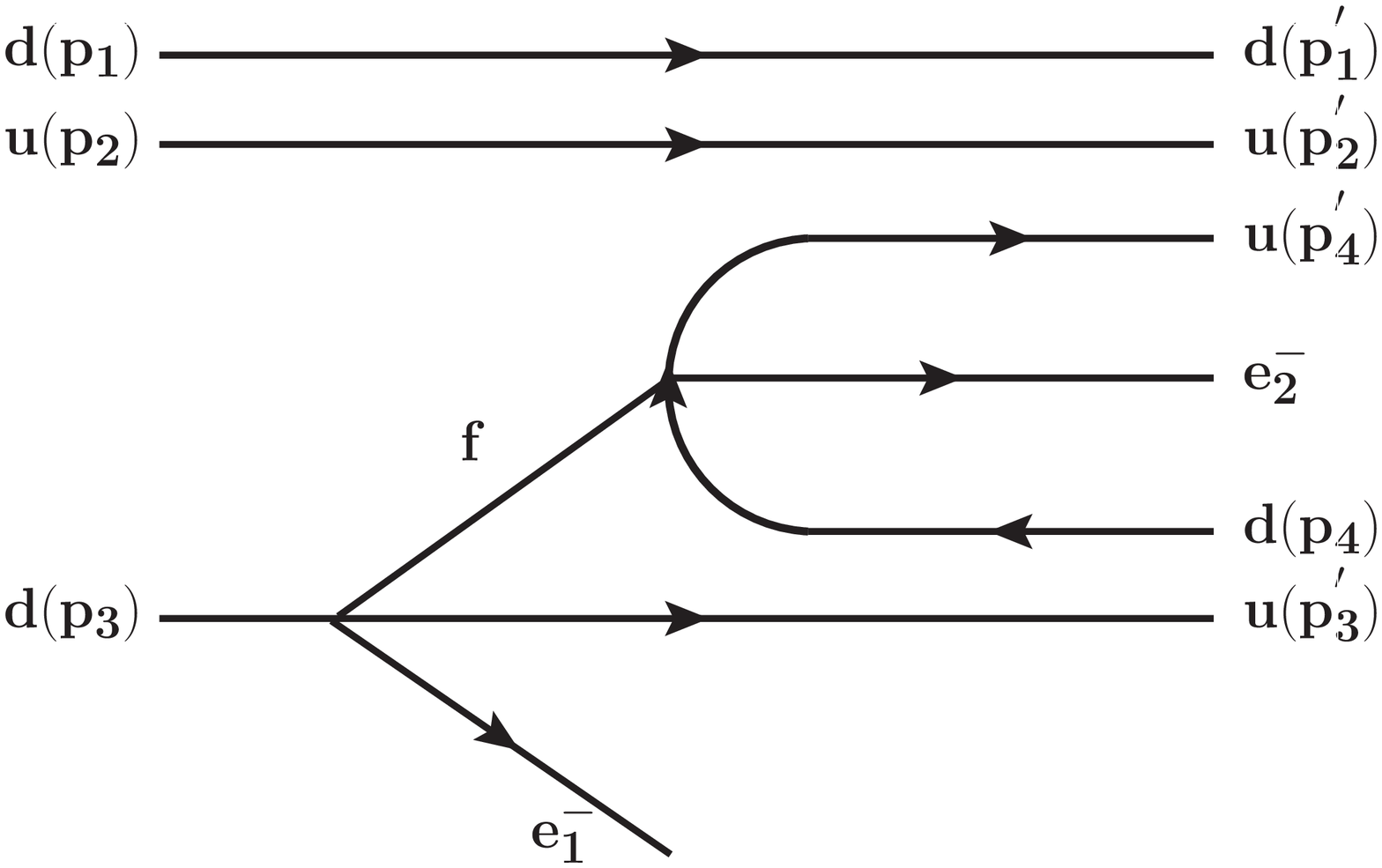}  
\end{tabular}
 \caption{  The pion mediated $0\nu\beta\beta$-decay in the so-called  $1\pi$ mode. In   diagram (a) the quarks of the pion are spectators
, i.e. the  heavy intermediate heavy fermion f is exchanged between the other two quarks. In (b) one of the interacting quarks is in the pion. In (c) the $q\bar{q}$  pair is produced by the weak interaction, while in (a) and (b) the $q\bar{q}$ is produced by the strong interaction out of the vacuum in the context of a multigluon exchange (a la 3$p_0$ mode) indicated by  $\times $. $f$ stands for an intermediate neutral fermion (heavy Majorana neutrino, gluino or neutralino).}
    \label{Fig:onepion}
\end{figure}

In order to provide a check of the approximations involved in the above treatment and to explore the new possibilities appearing in the microscopic treatment, e.g. the role played by the non local terms or the three possibilities entering in  Fig. \ref{Fig:onepion}, which are not distinguished in the standard treatment, we will consider the quark structure of the pion and the nucleon \cite{VerFaeTok10}.
	Thus  we will
 evaluate the relevant amplitude by making a non relativistic expansion of the hadronic current employing a constituent quark mass equal to 1/3 of the nucleon mass. Furthermore for the pion and the nucleon  internal relative quark  wave functions we will employ harmonic oscillator
	wave functions, adjusting the size parameters to fit related experiments \cite{VerFaeTok10}. Then we will  compare this amplitude to that obtained by more standard  techniques \cite{FKS98a}.

        a) {\it The $1\pi$ mode}. 

	 Let us begin with the second process of Eq. (\ref{eq:1}) (see diagrams
	(b) of Fig.  \ref{Fig:pionmode}), which is further analyzed at the quark level in
        Fig. \ref{Fig:onepion}.
	In this case it is clear that the amplitude must be of the 
        PS type only. The tensor contribution cannot lead to a pseudoscalar
        coupling at the nucleon level. Such a coupling is needed to be combined
        with the usual pion nucleon coupling in the other vertex
	to get the relevant operator for a $0^+~ \rightarrow~0^+$ decay.

%
	Let us begin with diagram \ref{Fig:onepion} (a), which is simply the decay
	of the pion into two leptons with a simultaneous 
	change of a neutron to a proton by the relevant nucleon current.
        Then one finds that, if the non local terms, which lead to new type operators not studied up to now, are ignored, the "direct" diagram makes no contribution.
%

	The "exchange" contribution, Fig. \ref{Fig:onepion} (b), in which the produced up quark of the meson is not
	produced from the "vacuum" but  comes from the initial nucleon, is a bit 
	more complicated. The result is:
\beq
c_{1\pi}=1.37~ f^{con}_{1\pi}(x),\quad \alpha_{1\pi}=0.071~f^{con}_{1\pi}(x)\approx 0.071,
\label{Eq:con}
\eeq
which is in size almost a factor of 2 larger  than that obtained in elementary particle treatment \cite{FKS98} (see Eq.(\ref{Eq:alpha1pi})) . Note, however, that our results  depend on the pion size parameter.
	 Finally diagram \ref{Fig:onepion} (c), in which the $q\bar{q}$ is produced by the weak interaction, leads to the expression:
	 \beq
c_{1\pi}=\frac{20 \sqrt{2} \sqrt[4]{\pi
   }  \sqrt{m_{\pi }}}{3
   g_r \sqrt{b_N^3}
   m_N^2} f^A_{1\pi}(x)=3.4\times 10^{-2}~~f^{A}_{1\pi}(x) \mbox{ (constituent masses)}
\eeq
with
\beq
f^{A}_{1\pi}(x)=x^{3/2},\quad x=\frac{b_N}{b_{\pi}}
\eeq
The corresponding coefficient that must multiply the nuclear matrix element for $x=2$ is
\beq
\alpha_{1\pi}=c_{1\pi} \frac{f^2_{\pi NN}}{f^2_A}=5.0\times 10^{-3}
\eeq

        b). {\it The $2\pi$ mode}. 

\begin{figure}[!t]
\begin{center}
\includegraphics[scale=0.46]{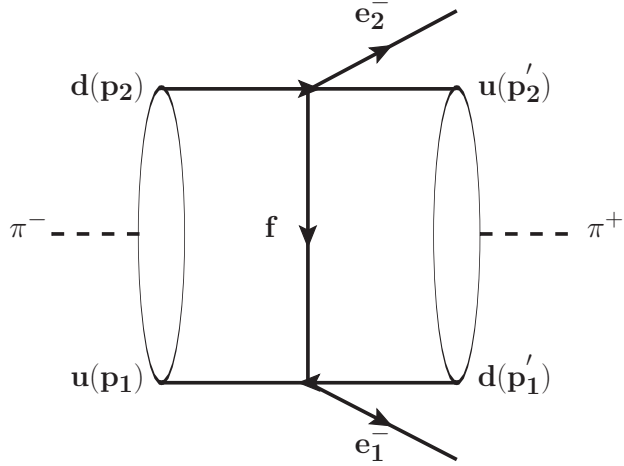}
 \caption{The $0\nu\beta\beta$-decay of pions in flight (2$\pi$ mode of Fig. \ref{Fig:pionmode}) illustrated at the quark level.
 $f$ stands for a effective exchange of a heavy Majorana fermion ( heavy neutrino or, as in
 R-parity violating supersymmetry, a neutralino, gluino etc). The arcs around the pion line merely indicate that the pion is a bound state of two quarks.}
    \label{Fig:2pions}
   \end{center}
  \end{figure}

	 The first process of Eq. \ref{eq:1} is described by diagram (c) of
	 Fig. \ref{Fig:pionmode} and is further illustrated
	 in Fig. \ref{Fig:2pions} at the quark level \cite{VerFaeTok10}. In Fig. \ref{Fig:2pions} $f$ stands for an intermediate fermion, heavy
	 Majorana neutrino, neutralino or gluino.

	In the case of $V-A$ theories we find \cite{VerFaeTok10}
	 \beq
  \alpha_{2 \pi}=\frac{2}{3 g^2_A}  f^2_{\pi N N} c_{2\pi},\quad 
c_{2 \pi}=\frac{1}{ \sqrt{2 \pi}} \frac{16}{b^3_{\pi} m_p^2
m_{\pi}}
\label{alpha2pi}
  \eeq
  The actual value critically depends on the harmonic oscillator size parameter. For $b_{\pi}=0.5$ fm we get $\alpha_{2\pi}\approx 0.1$.
  
 In supersymmetry one encounters scalar, pseudoscalar and tensor interactions. For the pseudoscalar interaction the situation is a bit more complicated \cite{VerFaeTok10}. One finds 
 \beq
 c_{2 \pi}=\frac{1}{ \sqrt{2 \pi}} \frac{4}{b^3_{\pi} m_p^2
m_{\pi}} \frac{b_N^2}{b^2_{\pi}}\left ( \frac{1}{4}( \kappa^2_d+\kappa^2_u )\right ),\quad\kappa_d=\frac{1}{2 m_d b_N}~,~\kappa_u=\frac{1}{2 m_u b_N}
 \eeq 
 Taking $m_u=m_d=(1/3)m_p$, $b_N=1$fm and $b_{\pi}=0.5$fm we find that $\alpha_{2 \pi}$ is much smaller than the value obtained in the elementary particle treatment \cite{FKS98a} , i.e. we find
 $\alpha_{2 \pi}=-0.05$. A larger value can be obtained, if one uses the current quark masses. For $m_u\approx 2 m_d$=10 MeV one finds $\alpha_{2 \pi}=-1.3$, but then the validity of non relativistic expansion may be questionable.
%

\section[Experimental aspects of double beta decays]{Experimental aspects of double beta decays}
\label{sect:exp}
\subsection{{ Progress of DBD experiments}} 

\subsubsection{Experimental aspects of neutrinoless double beta decays}

Neutrinoless double beta decays are concerned with fundamental properties of neutrinos 
and weak interactions, which bear some signatures of the high energy scale, and  are 
of great interest from the view point of particle physics and cosmology, as it has 
been described in previous sections. On the other hand, DBD processes are nuclear 
rare-decays in the low energy scale, which are studied experimentally by low-energy 
and low-background nuclear spectroscopy, as given in review articles \cite{AEE08,HS84,DTK85,eji05}.     

Since double beta decays are low-energy second-order weak processes, the decay rates of 2$\nu\beta\beta$ within the SM are of the order of 10$^{-20}$/y, and the rates of 
0$\nu\beta\beta$-decay beyond the SM are even much smaller than  
2$\nu\beta\beta$-decay rates, depending on the assumed light neutrino mass
in the case of the  Majorana neutrino mediated  process. Then the expected 0$\nu\beta\beta$ 
half-lives are of the order of $T^{0\nu}_{1/2} \approx $ 10$^{27}$ y and 10$^{29}$ y in the case of the 
IH (inverted hierarchy) mass of 30 meV and the NH (normal hierarchy) 
mass of 3 meV, respectively.  
   
For experimental studies of such rare decays, large detectors with ton-scale DBD isotopes are needed  
to get 0$\nu\beta\beta$-decay signals in case of the IH neutrino mass. Here, the signals are very rare 
and occur  as low as  $E_{\beta\beta}\approx $ 2 - 3 MeV. Background (BG) signal rates, however, are huge 
in the energy region of $E_B \le$ 3 MeV.  Thus it is crucial to build ultra low BG detectors to find 
the rare and small 0$\nu\beta\beta$-decay signals among huge BGs in the low energy region.

In spite of this, double beta decays have several unique features that make it realistic to search 
for the low-energy ultra-rare 0$\nu\beta\beta$ signals among huge BGs. 

i. Since $\beta\beta$ half-lives of the order of $T_{1/2} \approx $ 10$^{19}$ - 10$^{21}$y are 
10 orders of magnitude longer than the age of the earth, the $\beta \beta $ isotopes are 
available as almost stable isotopes on the earth, and it is possible to get ton-scale 
(10$^{28}$) $\beta\beta$ isotopes in order to observe some rare 0$\nu\beta\beta$-decay 
events with $T^{0\nu}_{1/2} \approx 10^{27}$ y.

ii. There are even-even nuclei, where the double beta decays are allowed due to the 
pairing interaction, but single $\beta$-decays are energetically forbidden. 
Using such DBD nuclei, one can be free from huge single $\beta$ BGs, which would be 
larger than $\beta\beta$ rates by factors around $10^{30}$.   

iii. The $0\nu\beta\beta$-decay process with a virtual Majorana neutrino exchange between 
two nucleons in a nucleus is greatly enhanced because the nucleons are close to each other 
in the nucleus. Then it is feasible to access the small neutrino masses of the orders of 
$\delta m_{\mbox{\tiny{SUN}}}$ (solar $\nu$)  - $\delta m_{\mbox{\tiny{ATM}}}$ (atmospheric $\nu $). 

iv. High energy-resolution and/or correlation studies of $\beta\beta$ rays select  
low-energy rare $\beta\beta$ signals from huge BG events. There are several DBD nuclei 
to be studied to confirm ultra-rare 0$\nu\beta\beta$-decay events and possible DBD processes.\\ 

Then $\beta\beta$ experiments studies neutrino properties in nuclei, which are around 10$^{-15}$ m 
in diameter. Thus the $\beta\beta$ nuclei are regarded as excellent femto (10$^{-15}$) laboratories 
with a large filtering power to reject single $\beta$ and other RI BG signals and a large 
enlargement factor to enhance the 0$\nu\beta\beta$-decay signal of the neutrino physics interest. 
In the nuclear femto laboratory, two nucleons collide with each other.  The luminosity is around 
$L$ = 10$^{48}$ cm$^{-2}$s$^{-1}$ in a single DBD isotope. Then the summed luminosity for one ton 
(10$^{28}$) isotopes is around 10$^{76}$ cm$^{-2}$s$^{-1}$ . The cross section for exchange of the 
Majorana neutrinos with IH mass of 40 meV is of the order of 10$^{-83}$ cm$^2$. Then the event rate 
is around 2 - 3 per year. The huge luminosity of the DBD femto collider enables one to search 
for the utra-rare $0\nu \beta \beta $ event and the very small neutrino mass  \cite{eji10}.  

\subsubsection{Progresses of DBD experiments.}

Progresses of DBD experiments are well described in review papers \cite{AEE08,HS84,DTK85,eji05,ell02} 
and refs. therein. Here brief remarks on the progresses are given. 
      
Early experimental studies of $\beta\beta$-decays were made by geochemical methods by measuring 
the number of the $\beta\beta$-decay products. They are inclusive measurements of both the 
2$\nu\beta\beta$ and 0$\nu\beta\beta$-decay rates to the ground and excited states. In most realistic cases, 
the geochemistry methods give the 2$\nu\beta\beta$-decay rate to the ground state because 0$\nu\beta\beta$-decay 
rates are much reduced due to the small neutrino mass and decays to excited states are disfavored due 
to the small phase space volume.

Advanced  mass spectroscopy was used to measure the number of $\beta\beta$ product isotopes accumulated for 
the long geological period of around 10$^{6-7}$ years in old ores, and to evaluate the long half-lives of 
the orders of 10$^{20}$y. Extensive studies have been made on such rare-gass isotopes as $^{82}$Kr, 
$^{128}$Xe, and $^{130}$Xe, which are $\beta\beta$-decay products of $^{82}$Se $^{128}$Te, and $^{130}$Te, 
respectively, as described in the review articles. Among them three groups have obtained half lives 
of around 10$^{21}$ for $^{130}$Te \cite{tak66,kir67,kir67b,sir72}. 

Direct counter experiments of $\beta\beta$ decays are exclusive measurements to identify 0$\nu\beta\beta$-decay 
signals. They have been made by coincidence counter measurements of two $\beta$ rays or by two $\beta$ 
measurements with tracking chambers, as described in review articles.

High-sensitivity counter experiments were made by using detectors as $\beta\beta$ sources. Der Mateosian 
and Goldharber have got a limits on the $^{48}$Ca 0$\nu\beta\beta$-decay rate by using  large CaF$_2$ 
scintillators \cite{der66}. Stringent limits on the $^{76}$Ge 0$\nu\beta\beta$-decay rate were obtained 
with high energy-resolution Ge semiconductor detectors by Fiorini et al \cite{fio67,fio73}. 
Coincidence measurements of two $\beta$ tracks with spark and streamer chambers were made to get limits 
on the $^{48}$Ca and $^{82}$Se 0$\nu \beta \beta $ rays \cite{sha67,bar67,bar70,cle75}.

In 1980's, the Ge experiments for $^{76}$Ge 0$\beta\beta$-decays are improved much by the Milano group 
\cite{bel82} and Avignone et al \cite{avi83}. Ejiri et al used the ELEGANT III with a Ge detector 
surrounded by Nal scintillators to get  limits on the  $^{76}$Ge 0$\nu\beta\beta$-decays to the ground 
and excited states \cite{eji84,eji86,eji87}. 
 
The first measurement of the 2$\nu\beta\beta$ rays by the direct counting method was made on $^{82}$Se 
by Elliott, Hahn and Moe \cite{ell87}. The observed rate agrees with the $\beta\beta$ rate measured 
previously by the geochemical method \cite{kir69,kir69b}. The first measurement of the 2$\nu\beta\beta$-decay 
rate by the direct counting method alone on $^{100}$Mo, where no geochemical measurement was made beforehand, 
was carried out by the ELEGANT group (Ejiri et al.) \cite{eji91,eji91b}. 

So far, high-sensitivity counter experiments with the mass sensitivities of the orders of eV and sub eV 
have been made on several $\beta\beta$-decay nuclei, and half lives of 2$\nu\beta\beta$-decay rates 
on many nuclei have been measured by direct counting methods, as reported in the review papers \cite{AEE08,eji05,ell02}. 
Recent experiments are discussed in \ref{sec:PresStat}. 

\subsection{{ Methods and detectors for DBD experiments}}

\subsubsection{Methods for DBD experiments.}

Double beta decays proceed normally through the 2$\nu\beta\beta$-decay process within the SM. Transition rates 
of the 0$\nu\beta\beta$-decay processes beyond SM are much rarer than those of the 2$\nu\beta\beta$-decay 
process in most cases. It is thus necessary to separate experimentally the 0$\nu\beta\beta$-decay 
processes from the 2$\nu\beta\beta$-decay process.  


Geochemical methods counts the number of the decay product isotopes in ores of DBD nuclei 
for geological time of $T_{1/2} \approx 10^{6-7}$ years, which are mostly due to the 
2$\nu\beta\beta$-decay process and 0$\nu\beta\beta$-decay processes are not separated 
from the 2$\nu\beta\beta$-decay process. 

Direct counting methods have been extensively used for measuring various DBD processes. 
The 0$\nu\beta\beta$-decay processes are identified from the sum energy spectrum of 
$E_{\beta\beta}=E(\beta _1) + E(\beta _2)$ for two $\beta$ rays, as shown in Fig. 2 
in Ref. \cite{eji05}. They show a sharp peak characteristic of the 2-body kinematics 
at $E_{\beta\beta} = Q_{\beta \beta }$, while 2$\nu\beta\beta$ shows a continuum 
spectrum characteristic of the 4-body kinematics. Neutrinoless DBD followed by the 
Majoron (see section \ref{sec:majoron}), which is the Goldstone boson associated with 
spontaneous breaking of $B-L$ symmetry, shows the spectrum characteristic of the 3-body kinematics.

The 0$\nu\beta\beta$-decay processes due to the left handed weak current and the 
right-handed one (RHC) are experimentally identified by measuring the energy and 
angular correlations of the two $\beta$ rays, as shown in Fig.4 in ref.\cite{eji05}.


The left handed weak current
0$\nu\beta\beta$ process includes several modes such as the light $\nu $ exchange, 
the heavy $\nu$ exchange, the SUSY particle exchange, and others, as discussed in the previous sections.  
Relative contributions of these modes to the 0$\nu\beta\beta$-decay rate may be investigated 
by observing several DBD isotopes as well as those for the ground and excited states, 
provided that the matrix elements are properly evaluated. Then experimental studies 
of several DBD isotopes are necessary.

\subsubsection{Sensitivity of DBD experiment}

DBD event rates are so low that DBD experiments are necessarily be carried out by using 
high-sensitivity detectors at low-background underground laboratories.

In case of the 0$\nu\beta\beta$-decay process with the light Majorana neutrino exchange, 
the transition rate $\Gamma^{0\nu}$ per year(y) per 1 ton (t) of DBD isotopes is expressed 
in terms of the nuclear sensitivity $S_n$  and the effective mass of the light Majorana neutrinos 
of $|\langle m_\nu\rangle|$ as 
\begin{equation}
\Gamma^{0\nu} = |\langle m_\nu\rangle|^2 S_n.
\end{equation}
The nuclear sensitivity is written as 
\begin{equation}
S_n=(78~\rm{meV})^{-2} |M^{0\nu}_\nu|^2 ~G^{0\nu }(0.01A)^{-1}, 
\label{defsn}
\end{equation}
where $M^{0\nu}$ is the nuclear matrix element, $A$ is the mass number, $G^{0\nu }$ is the phase space factor in units 
of 10$^{-14}$/y.

The mass sensitivity $|\langle m_\nu\rangle|$ is defined as the minimum mass to be measured 
by the 0$\nu\beta\beta$-decay experiment. It is expressed in terms of the detector sensitivity $D$ as follows:
\begin{equation}
|\langle m_m\rangle | = S_{n}^{-1/2} D^{-1/2}, ~~~D = (\epsilon NT)(\delta )^{-1},   
\end{equation}     
where $\epsilon $ is the 0$\nu\beta\beta$ peak efficiency, $N$ is the number of the DBD isotopes 
in unit of ton, $T$ is the  run time in unit of year and $\delta$ is the minimum 
counts required for the peak identification with 90 $\%$ CL (confidence level). 
It is given as $\delta \approx $1.6 + 1.7 ($BNT)^{1/2}$  with $B$ being the 
BG rate /t/y, and  $\delta \approx $2.3 for $BNT~\ge$ 2 and $\le$ 2, respectively.  

The nuclear sensitivity $S_n$ is proportional to the phase space factor $G^{0\nu}$ and 
$|M^{0\nu}_\nu|^2$. Thus DBD nuclei with large $G^{0\nu}$ and large $M^{0\nu}_\nu$ are selected 
for the high nuclear sensitivity. DBD detectors with a large efficiency $\epsilon$, a large amount 
($N$) of isotopes, and a small BG rate ($B$) are used for the high detector sensitivity.

\subsubsection{DBD detectors}

Neutrinoless $\beta ^-\beta ^-$ decays of $(A,Z) \rightarrow (A,Z+2)$ are studied by measuring 
two $\beta^-$ rays, while neutrinoless $\beta^+\beta^+$ decays of $(A,Z) \rightarrow (A,Z-2)$ 
proceed through $\beta^+\beta^+$, $\beta^+$ EC, and EC EC $\gamma$, 
where EC (electron capture) is detected by measuring the X ray, and $\beta^+$ 
by measuring the $\beta^+$ and the two 511 keV annihilation $\gamma$ rays.

High-sensitivity DBD experiments require DBD isotopes with high nuclear sensitivity $S_n$, 
i.e. the large $Q_{\beta\beta}$ and the large phase space factor $G^{0\nu }$, as given by 
Eq. (\ref{defsn}). 

There are several such DBD isotopes of $\beta^-\beta^-$ decays. However, no $\beta^+ \beta^+$ 
nuclei with large $S_n$ are available, although BG rates for the $\beta^+\beta^+$ and
$\beta^+$ EC are quite small by measuring both $\beta^+$ and the annihilation $\gamma$ 
rays and/or the X ray. Accordingly most of high-sensitivity DBD experiments are concentrated 
on the 0$\nu \beta^-\beta^-$ decays with the large $Q_{\beta \beta} $ and $G^{0\nu}$. Thus 
hereafter, we discuss mainly $\beta^-\beta^-$ decays. The large $Q_{\beta \beta} \approx$3 
MeV helps reduce BG rates since most of BGs from natural RIs are below 3 MeV.

The 0$\nu\beta^-\beta^-$  experiments with the IH (30 meV) mass sensitivity are carried out 
by using low BG ($B \approx $1/t y) detectors and ton-scale ($N \approx $0.5 - 1) DBD isotopes 
with the large nuclear matrix element of $|M^{0\nu}_\nu |\approx $3 (see Table \ref{tab.nme})
and the large phase space factor of $G^{0\nu} \approx 5$ in unit of 10$^{-14}$/y  for 
a long ($T \approx $ 2 - 4 year) run time. On the other hand one needs DBD isotopes 
of around $N \approx 50-100$ ton and ultra-low BG detectors with $B \approx 0.01$/t/y to reach  
the NH (2 - 4 meV) mass sensitivity. 

Possible DBD isotopes to be used for high-sensitivity 0$\nu\beta^-\beta^-$ experiments 
are given in Table \ref{tab:ej1}. Among them, $^{82}$Se, $^{100}$Mo, $^{116}$Cd, $^{130}$Te, $^{136}$Xe 
have the large $Q_{\beta \beta} \approx $ 3 MeV and the large $G^{0\nu }\approx $ 5 in unit of 10$^{-14}$/y, 
as shown in Table \ref{tab:ej1}. $^{130}$Te has the largest abundance ratio of 34.5$\%$, 
while others have the abundance ratios of around 10$\%$, and are enriched to 85-90$\%$ by means of 
centrifugal separation. 

$^{76}$Ge is a special case with smaller $Q_{\beta \beta}\approx$ 2 MeV and the smaller  $G^{0\nu }\approx$ 0.71, 
but high sensitivity experiments are possible by using low-BG $^{76}$Ge detectors with high energy-resolution 
\cite{fio67,fio73}. $^{150}$Nd has a very large phase space factor of $G^{0\nu }\approx $ 23.2 in units 
of  10$^{-14}$/y, but the natural abundance ratio is only 5.6$\%$, and the enrichment is hard. 
The matrix element may be reduced because of the large difference in nuclear shape between the initial 
and final nuclei.  

DBD experiments are carried out by using either calorimetric detectors or spectroscopic detectors. 
Calorimetric detectors are made partly of DBD isotopes, and thus the detection efficiency is as 
large as $\epsilon \approx$ 0.6-0.9. Cryogenic detectors with Te \cite{fio84}, semiconductor 
detectors with $^{76}$Ge and $^{116}$Cd, and scintillation detectors with $^{116}$Cd and $^{136}$Xe 
are used as calorimetric detectors.

Cryogenic bolometers of ZnS, CaMoO$_4$, ZnMoO$_4$ and others are shown to be used as high 
energy-resolution DBD detectors. BG rates of these detectors are reduced by measuring 
both thermal signals as well as scintillation signals \cite{ale92,ale98,gir10}.

Spectroscopic tracking detectors with DBD isotopes outside the detectors are used to measure 
individual $\beta$ rays from the isotopes. Then the low BG measurements are possible even 
though the efficiency is low and the energy resolution is modest. $^{82}$Se and $^{100}$Mo 
are studied by spectroscopic detectors with PL (plastic scintillator) arrays. Spectroscopic 
detectors measure $\beta - \beta$ energy and angular correlations, which are used to confirm 
and identify the 0$\nu\beta\beta$-decay process.   

So far, high-sensitivity DBD experiments have been made mainly on the ground state 0$^+$ 
transition because of the large phase space factor. Transitions to the excited 0$^+$ states 
are experimentally measured in coincidence with $\gamma$ rays. Then BG rates are much reduced, 
even though the phase space factors are smaller by almost one order of magnitude. Measurements 
of both the ground and excited transitions are interesting to confirm the $0\nu\beta\beta-$decay
signal and to study the 0$\nu \beta \beta $ mechanism. The excited 0$^+$ states
in $^{82}$Se, $^{100}$Mo, $^{136}$Xe, and $^{150}$Nd are located in low excitation region. 
Some excited states are different in shape from the ground state. Thus  0$\nu\beta\beta$-decay 
measurements for both the ground and excited states are very interesting.

\subsection{{ Present status and future projects of DBD experiments}}
\label{sec:PresStat}

\subsubsection{Neutrinoless double beta decays}

Experimental studies of $0\nu\beta\beta$-decay  have been carried out on several 
$\beta\beta$ nuclei \cite{eji05,AEE08,eji10}. Some of them by counter measurments are listed in Table \ref{tab:ej1}. $^{128}$Te was studied 
by a geochemical method, and the hallife and mass limits are 7.7 $\pm$ 10$^{24}$ y and 1.1 - 1.5 eV \cite{ber93}.
Here the effective mass given in the 6th column shows a range of the values evaluated 
from the experimental half-life by using various matrix elements. In fact the mass 
depends much on the matrix element as discussed in section \ref{sec:0nuME}.
Calorimetric detectors have been used for isotopes such as $^{48}$Ca, $^{76}$Ge, $^{116}$Cd 
and $^{130}$Te. Among them, $^{76}$Ge experiments (Heidelberg-Moscow, IGEX) with Ge semiconductors 
\cite{bau99,kla01,kla01b,aal99,aal02} and the $^{130}$Te experiment(CUORICINO) with 41 kg TeO$_2$ 
cryogenic bolometers \cite{arn08,arn11} have good energy-resolution and give stringent 
limits on the absolute value of effective Majorana neutrino mass 
of the order of 0.3 - 0.5 eV. Currently, the most stringent limit on $|\langle m_\nu\rangle |$
comes from the lower limit on the $T^{0\nu}_{1/2}(^{136}Xe)$ measured in KamLAND-Zen 
experiment \cite{kamlandzen} (see Table \ref{tab:ej1}). 

Spectroscopic detectors (ELEGANT V, NEMO III) have been used for $^{82}$Se, $^{100}$Mo, 
$^{116}$Cd and other isotopes with large $Q_{\beta \beta }$ values \cite{tre11,eji01,arn04,arn05}. 
They are $\beta$-ray tracking detectors with $\beta\beta$ sources separated from detectors. 
NEMO III provides stringent limits on the 0$\nu\beta\beta$ half-lives for $^{82}$Se, $^{100}$Mo,  
and other isotopes \cite{tre11,arn04,arn05}. 

\begin{table}[!t]
\caption{Limits on neutrinoless double beta  decays $T^{0\nu -exp}_{1/2}$ (claim for evidence
is denoted with upper index c).  
$Q_{\beta \beta }$ : $Q$ value for the 0$^+\rightarrow 0^+$ ground state transition. 
$G^{0\nu}$: kinematical (phase space volume) factor ($g_A=1.25$ and R = 1.2 fm $A^{1/3}$).
$\langle m_{\nu }\rangle$: The upper limit on the effective Majorana neutrino mass,
deduced from $T^{0\nu -exp}_{1/2}$ by assuming the ISM \cite{lssm} ($g^{eff}_A=1.25$, UCOM src),
the EDF \cite{edf} ($g^{eff}_A=1.25$, UCOM src), the (R)QRPA  
($1.00 \le g^{eff}_A \le 1.25$, the modern self-consistent treatment of src),
and  the IBM-2 \cite{IBM09} ($1.00 \le g^{eff}_A \le 1.25$, Miller-Spencer src),
nuclear matrix elements (see section \ref{sec:0nuME}). src means short-range correlations.} 
\label{tab:ej1}
\begin{center}
\begin{tabular}{lccccccc} \hline
isotope & $A$ & $Q_{\beta \beta }$ & $G^{0\nu}$ & 
$T^{0\nu -exp}_{1/2}$ &  NME &  $|\left< m_{\nu }\right> |$ eV  & Future\\ 
& $[\%]$  & [MeV] & [$10^{-14}$ y] & [$10^{24}$ y] & & [eV]  & experiments \\ \hline
$^{48}$Ca  & 0.19 & 4.276 &  7.15 & $0.014^a$  & ISM & 19.1  &CANDLES  \\
           &     &       &       &     & EDF & 7.0  &    \\
$^{76}$Ge  & 7.8  & 2.039 &  0.71 & $19^b$     & ISM, EDF &  0.51, 0.31 & GERDA  \\
           &     &       &       &     & (R)QRPA & (0.20,0.32)  &    \\
           &     &       &       &     & EDF     & (0.26,0.35)  &    \\
          & 7.8  & 2.039 &  0.71 & $22^c$     & ISM, EDF & 0.47, 0.29 & -  \\
           &     &       &       &     & (R)QRPA & (0.18,0.30)  &    \\
           &     &       &       &     & EDF     & (0.24,0.32)  &    \\
          & 7.8  & 2.039 &  0.71 & $16^d$     & ISM, EDF  &  0.55, 0.34 & MAJORANA \\
           &     &       &       &     & (R)QRPA & (0.22,0.35)  &    \\
           &     &       &       &     & EDF     & (0.28,0.38)  &    \\
$^{82}$Se  & 9.2  & 2.992 & 3.11 & $0.36^e$  & ISM, EDF & 1.88, 1.17   & SuperNEMO  \\
           &     &       &       &     & (R)QRPA & (0.76,1.28)  &  MOON   \\
           &     &       &       &     & EDF     & (1.12,1.49)  &    \\
$^{100}$Mo & 9.6  & 3.034 & 5.03 & $1.0^f$   & EDF &  0.46  &   MOON  \\
           &     &       &       &     & (R)QRPA & (0.38,0.73)  & AMoRE  \\
           &     &       &       &     & EDF     & (0.62,1.06)  &    \\
$^{116}$Cd & 7.5  & 2.804 & 5.44 & $0.17^g$  & EDF  &   1.15  & COBRA \\
           &     &       &       &     & (R)QRPA & (1.20,2.16)  & CdWO$_4$\\ 
$^{130}$Te & 34.5 & 2.529 & 4.89 & $3.0^h$   & ISM, EDF & 0.52, 0.27 & CUORE \\
           &     &       &       &     & (R)QRPA & (0.25,0.43)  &   \\
           &     &       &       &     & EDF     & (0.33,0.46)  &    \\
$^{136}$Xe & 8.9  & 2.467 & 5.13 & $5.7^i$   & ISM, EDF  & 0.44, 0.23  & EXO, NEXT \\
           &     &       &       &     & (R)QRPA & (0.17,0.30)  &  KamLAND-Zen \\
$^{150}$Nd & 5.6  & 3.368 & 23.2 & $0.018^j$ & EDF  &  4.68 & SuperNEMO \\
           &     &       &       &     & (R)QRPA & (2.13,2.88)  &  SNO+ DCBA \\
\hline
\end{tabular}
\end{center}
a:\cite{oga04}, b:\cite{bau99,kla01,kla01b}, c:\cite{evidence2}, d:\cite{aal99,aal02}, 
e:\cite{tre11,arn04,arn05}, f:\cite{tre11,arn04}, g:\cite{dan03}, h:\cite{arn08,arn11,cre11}, 
i:\cite{kamlandzen}, j:\cite{tre11,fla08}.
\medskip 
\end{table} 



Recently, a claim for the $0\nu\beta\beta$-decay peak, corresponding to the
effective Majorana neutrino mass  of 0.32 eV, was made by a part of the Heidelberg-Moscow 
collaboration \cite{evidence1,evidence2}. The result depends on the off-line analysis method. 
It should be checked by future GERDA/MAJORANA experiments with lower BG rates.   

Neutrino-mass sensitivities of the CUORICINO and  NEMO III detectors are limited to be around 
300 - 500 meV because of the limited $\beta\beta$ isotopes of 11 - 7 kg and the large BG rates. 
Thus future experiments with higher mass sensitivity are necessary to prove/disprove the 
Heidelberg-Moscow claim and to search for the Majorana neutrino in the lower $\nu $ mass regions.

The neutrino oscillation studies have given a strong impact to high-sensitivity studies 
of $\beta\beta$ experiments since the effective mass suggested is of the order of 
$\surd {\delta m^2 }\sim $ 2 meV - 50 meV, which next-generation $\beta\beta$ detectors can 
access if the $\nu $'s are Majorana particles. Future experiments with higher mass sensitivities 
are in progress (see below).

\subsubsection{Two neutrino double beta decays}
 
The 2$\nu\beta\beta$-decay is a process fully consistent with the SM of electroweak interaction. 
The inverse half-life of 2$\nu\beta\beta$-decay is free of unknown parameters on the particle physics. 
It can be expressed as a product of an accurately known phase-space factor 
$G^{2\nu}(E_0,Z)$, which includes fourth power of $g_A$, and the double Gamow Teller transition matrix element 
$M^{2\nu}(A,Z)$, which is a quantity of the second order in the perturbation
theory:
\begin{equation}
\left(T^{2\nu}_{1/2}\right)^{-1} =  G^{2\nu}(E_0,Z) |M^{2\nu}(A,Z)|^2.
\end{equation}

Here $M^{2\nu}$ includes nuclear effects due to the nuclear residual interactions, the nuclear medium and the nuclear quenching. It is obtained from the experimental decay rate. Experimental studies of $2\nu\beta\beta$-decay half-lives have been made on some nuclei by geochemical 
methods, and several nuclei by direct counting methods \cite{recbb}. 
The $2\nu\beta\beta$-decay has been observed so far in 12 nuclides ($^{48}$Ca, $^{76}$Ge $^{82}$Se, $^{96}$Zr, 
$^{100}$Mo, $^{116}$Cd, $^{128}$Te, $^{130}$Te, $^{136}$Xe, $^{150}$Nd, $^{130}$Ba and $^{238}$U) and in two excited states
\cite{recbb}. 
Recent NEMO III experiments provide high-statistic spectroscopic studies of the 2$\nu\beta\beta$-decay rates 
\cite{tre11, fla08}. Spectroscopic measurements of two $\beta$ rays are useful to reduce BG rates and energy 
correlations of two $\beta$ rays are used to identify the 2$\nu \beta \beta $ mechanism.

The measurement of 2$\nu\beta\beta$-decay rates gives us information about the product of the
squared effective axial-vector coupling constant and  2$\nu\beta\beta$-decay 
matrix elements. They are presented in Table \ref{tab.2nbb}. The 2$\nu\beta\beta$-decay matrix elements 
are sensitive to nuclear-spin isospin correlations.
The observed values for $M^{2\nu}$ are used to investigate the nuclear structure and the nuclear 
interactions associated with the 0$\nu\beta\beta$-decays. 

\begin{table}[!t]  
  \begin{center}  
    \caption{The $2\nu\beta\beta$ matrix elements $|M^{2\nu}|$ 
deduced from the measured half-life $T^{2\nu}_{1/2}$ by counter experiments \cite{recbb} \cite{kamlandzen}.
$g_A = 1.269$ is assumed.}   
\label{tab.2nbb}  
\renewcommand{\arraystretch}{1.2}  
\begin{tabular}{lcccc}\hline\hline 
Nucleus &  $T^{2\nu}_{1/2}$ years & & $|M^{2\nu}|$ (MeV)$^{-1}$   \\ \hline
${}^{48}$Ca  &  $4.4^{+0.6}_{-0.5}\times 10^{19}$  &  &  $0.046^{+0.003}_{-0.003}$   \\
${}^{76}$Ge  &  $1.5^{+0.1}_{-0.1}\times 10^{21}$  &  &  $0.137^{+0.005}_{-0.004}$ \\
${}^{82}$Se  &  $9.2^{+0.7}_{-0.7}\times 10^{19}$  &  &  $0.095^{+0.004}_{-0.003}$ \\
${}^{96}$Zr   &  $2.3^{+0.2}_{-0.2}\times 10^{19}$ &   &  $0.091^{+0.004}_{-0.004}$  \\
${}^{100}$Mo   &  $7.1^{+0.4}_{-0.4}\times 10^{18}$ &   &  $0.234^{+0.007}_{-0.006}$ \\
${}^{100}$Mo$^*$ &  $5.9^{+0.8}_{-0.6}\times 10^{20}$  &  &  $0.189^{+0.010}_{-0.012}$ \\
${}^{116}$Cd   &  $2.8^{+0.2}_{-0.2}\times 10^{19}$  &   &  $0.128^{+0.005}_{-0.004}$ \\
${}^{128}$Te   &  $1.9^{+0.4}_{-0.4}\times 10^{24}$  &   & $0.047^{+0.007}_{-0.003}$  \\
${}^{130}$Te   &  $6.8^{+1.2}_{-1.1}\times 10^{20}$   &   & $0.034^{+0.003}_{-0.003}$ \\
${}^{136}$Xe   &  $2.38^{+0.14}_{-0.14}\times 10^{21}$   &   & $0.018^{+0.003}_{-0.001}$ \\
${}^{150}$Nd   &  $8.2^{+0.9}_{-0.9}\times 10^{18}$   &   &  $0.061^{+0.004}_{-0.003}$ \\
${}^{150}$Nd$^*$ &  $1.33^{+0.45}_{-0.26}\times 10^{20}$ &   & $0.045^{+0.005}_{-0.006}$ \\
\hline\hline
\end{tabular}\\
 \end{center}  
\end{table}

\subsubsection{High sensitivity experiments}

In the case of the inverted mass hierarchy, $\beta\beta$ detectors with the IH mass sensitivity of 
$\langle m_{m } \rangle \approx $ 20 - 50 meV can be used to study the $0\nu\beta\beta$-decay, 
while in the case of the normal hierarchy one needs higher sensitivity detectors with 
$\langle m_{m } \rangle \approx $ 2 - 4 meV. Several groups are now working for next-generation 
$\beta\beta$ experiments with the IH mass sensitivities of 20$\sim $50 meV, as discussed in the reviews 
\cite{eji05,AEE08,ablimit}.
 	
Experimental proposals for future $\beta\beta$ experiments  have been made on several $\beta\beta$ 
isotopes, and some of them are listed in Table \ref{tab:ej1}. They are mostly $\beta ^-\beta ^-$ experiments because 
of large kinematical (phase space) factors. DBD experiments with different isotopes and different
methods are indispensable to confirm  and identify the 0$\nu\beta\beta$-decay event and the 
0$\nu\beta\beta$-decay mechanism. Some of them are briefly described below.


$^{76}$Ge experiments with low-BG high resolution $^{76}$Ge detectors are of special interest for proving 
or disproving the Heidelberg-Moscow claim of the large 0$\nu\beta\beta$-decay peak \cite{evidence2}, and for further 
high-sensitivity ton-scale experiments.

{\it GERDA}:
It aims at high energy-resolution studies of $^{76}$Ge 0$\nu\beta\beta$-decays
by using high-purity $^{76}$Ge detectors to check the Heidelberg-Moscow claim and the possible 
Majorana neutrinos in the QD region at LNGS (Gran Sasso). GERDA uses 18 kg $^{76}$Ge detectors 
in Phase I, and add 20 kg detectors in Phase II. The Ge detectors are immersed into high purity 
liquid Ar in order to avoid BG contributions from cryostats \cite{ger08}. GERDA is now running.  

{\it MAJORANA}:
The MAJORANA demonstrator uses 40 kg Ge detectors at the Sanford underground lab. to test/investigate 
the half life of the Heidelberg-Moscow claim and the QD mass regions  and to prove the feasibility for a future 
ton-scale IH (10$^{27}$y, 20-40 meV) experiment. The Ge detectors are PPC (P-type Point Contact) 
detectors with excellent PSA(Pulse Shape Analysis). They are cooled by using ultra-pure electro-formed 
Cu cryostat \cite{ell09}. The BG goal is 4/t y, which scales to 1/t y for the 1-ton experiment. 
The enriched detectors will be on-line in 2013 - 2014. 

These detectors can also be used to study DM and neutrino scattering in the low energy region. 
The GERDA and the Majorana collaboration will be merged for one ton-scale future experiment 
by selecting the best techniques  developed and tested by GERDA and MAJORANA. Recent developments 
are given in the report \cite{hen11}.

There are other experimental plans for QD masses. Among them, CANDLES is for $^{48}$Ca $\beta\beta$-decays 
with an array of CaF$_2$ crystals \cite{oga09}, which is based on the ELEGANT VI experiment with CaF$_2$. 
The $Q_{\beta \beta }$ is large, but the natural abundance of $^{48}$Ca is only 0.2$\%$. Thus the efficient 
isotope enrichment is crucial. 

Several groups are working for future experiments with the IH mass sensitivities of around 
$\langle m_\nu\rangle\approx $ 20 - 50 meV. DBD isotopes required are those with the large 
$Q_{\beta\beta}$ = 2.5 - 3 MeV and $G^{0\nu}$ = 3 - 5 10$^{-14}$/y to get large nuclear 
sensitivities of the order of $(S_N)^{-1/2}$=15 - 20 meV. The experiments use large-scale 
low-BG detectors with ton-scale enriched isotopes and $B \approx $1/t y.   

{\it MOON (Molybdenum Observatory Of Neutrinos)}:
This is an extension of ELEGANT V \cite{eji01}. It is a hybrid $\beta\beta$ and solar $\nu$ 
experiment with $^{100}$Mo to study the  Majorana $\nu$ masses with the QD - IH mass sensitivities 
of 100 - 20 meV and the low energy  solar $\nu$s \cite{eji00a,eji07a,nak07,eji08,geh09}. 
Double beta decays to both the ground and the 1.132 MeV excited 0$^+$ 
states are studied to confirm the 0$\nu \beta \beta $ events and to study the 0$\nu\beta\beta$-decay mechanism. 
MOON can be used for supernova neutrinos as well \cite{eji01c}.

Detectors under considerations are : A) the super-module of PL plate and fiber scintillators 
\cite{eji07a,nak07,eji08} and B) the cryogenic bolometer of ZnMoO$_4$.  The PL scintillator module 
(A) is used for spectroscopic study of two $\beta$-ray energy and angular correlations to identify 
the 0$\nu\beta\beta$-decay process. Here one module consists of a plate (PL) scintillator for the $\beta$ 
energy and two sets of X-Y fiber scintillator planes for the vertex identification, between which 
a thin $^{100}$Mo film is interleaved. The energy resolution is $\sigma\approx $2.2$\%$ at $E=Q_{\beta\beta}$ 
to reduce the $2\nu\beta\beta$-decay tail in the $0\nu\beta\beta$-decay window. The half life (mass) 
sensitivity is $3 \times10^{26}$ y (45 meV) with 480 kg $^{100}$Mo for 5 years. Proto-type detectors were 
built to show the energy resolution as required \cite{eji07a,nak07}.
 
The ZnMoO$_4$ bolometer(B), which is under discussion with the Milano-Rome group is 
for calorimetric study of the sum of two $\beta$-ray energy. The high energy-resolution 
($\Delta E \approx $ 5 keV), the particle identification by the scintillation and/or 
the pulse-shape analysis and the high efficiency($\epsilon\approx $0.8) make 
the high-sensitivity study possible. Thus it is good to start with the detector B, 
and proceed to A to confirm the 0$\nu\beta\beta$-decay process by two $\beta $-ray measurement.     
The half-life (mass) sensitivities are $4 \times 10^{25}$ y (120 meV) for 3 y run with 12 kg $^{100}$Mo 
and $7 \times 10^{26}$ y (30 meV) for 5 y run with 220 kg $^{100}$Mo. 
 
{\it SuperNEMO}:
The goal of SuperNEMO is to reach a sensitivity of 10$^{26}$ y, which corresponds to the IH mass of 
40-110 meV \cite{cha11}. The detector consists of huge tracking chambers and scintillation detectors 
with 100 kg of $\beta \beta $ isotopes of $^{150}$Nd or $^{82}$Se  to search for the $\nu $ mass 
below 0.1 eV. The efficiency is 30$\%$ and the resolution is 4$\%$ in FWHM. The BG impurities are 
deduced to be $^{208}$Tl $\le $ 2 and $^{214}$Bi $\le $ 10 in unit of $\mu $ Bq/kg
It plans to use 20 modules, each module with 5 kg $\beta \beta $ isotopes. The first module 
is a demonstrator with 7 kg of $^{82}$Se. The mass sensitivity of the demonstrator with 15 kg y 
is 210-570 meV, while that of the full detector array with 500 kg y is 53-145 meV, depending 
on the nuclear matrix element. It is based on NEMOIII, and thus it is crucial to improve the energy 
resolution and the efficiency. 

{\it AMoRE Advanced Molybdenum based Rare process Experiment}:
Large volume CaMoO$_4$ crystals with enriched material have been developed to study the $0\nu\beta\beta$-decays 
of $^{100}$Mo and to search for cold dark matter \cite{kim11}. Pilot experiments of 1 kg with scintillation 
technique and Cs(I) active veto are in preparation. In order to improve the energy resolution, 
cryogenic CaMoO$_4$ detectors are being developed. To avoid BGs from the 2$\nu\beta\beta$-decays of 
$^{48}$Ca, depletion of Ca in $^{48}$Ca $\le 0.001 \%$ is made by using ALSIS (Advanced Laser 
Stable Isotope Separation). Additional light sensor and time constant of phonon signal are effective 
to select signals. The goal of AMoRE is to study 0$\nu \beta \beta $ decays of $^{100}$Mo in 
the region of IH mass of 50 meV (3 10$^{26}$ y ) by using 100 kg CaMoO$_4$ cryogenic detectors.  

{\it COBRA}:
It uses a large amount of high energy-resolution CdZnTe semiconductors at room temperature \cite{zub11}. 
The modular design makes coincidence measurements possible to reduce BG rates. The crystal includes several 
$\beta\beta$ isotopes to be studied. The collaboration now tests 64 CZT 1 cm$^3$ detectors at LNGS. 
The goal is to study the Majorana neutrino in the IH mass region by measuring the 0$\nu \beta \beta $ 
from $^{116}$Cd with $Q_{\beta\beta}$ =  2.809 MeV. The detector is composed by 64 K crystals with 
0.42 ton Cd isotopes enriched in $^{116}$Cd. Reduction of BGs from RI impurities inside and 
around detectors are important. Pixelisation(Semiconductor tracker) can be a major step forward.
 
{\it CUORE (Cryogenic Underground Observatory for Rare Event)}:
This is an expansion of CUORICINO. It is a high energy resolution bolometer array to measure the  
$^{130}$Te $0\nu \beta \beta $ decays with $Q = 2.529$ MeV at LNGS \cite{cre11}. It uses natural TeO$_2$ 
crystals with the natural $^{130}$Te abundance of 34\,\%. It consists of 988 TeO$_2$ crystals with 
the net $^{130}$Te mass of 203\,kg. The detector array is under construction since 2005. 

The experiment emphases 20 times more massive than CUORETINO, better energy resolution of 5 keV, 
high granularity, and thus low BG rates. The CUORICINO BG rate, which is around 0.16/keV/kg/y 
(4 $10^3$/5keV/y/ton of $^{130}$Te), is expected to be reduced to 0.02 - 0.01/keV/kg/y in CUORE. 
Then, in cases of the BG rates of $B$ = 0.01 - 0.001/keV/kg/y (2.5 10$^2$ - 2.5 $10^1$/5keV/y/ton of $^{130}$Te),
the half-life and the $\nu $ mass sensitivities are expected to be around 2.1 - 6.5 10$^{26}$ y and 50 - 25 meV, which depend on the nuclear matrix element. The first CUORE tower is CUORE-0.

{\it EXO (Enriched Xenon Observatory)}:
The $\beta\beta$ experiment of $^{136}$Xe with $Q = 2.467$\,MeV \cite{gra08} is made by using 
the laser tagging technique to select  the residual nuclei of $^{136}$Ba to suppresses all kinds 
of RI BGs. The energy-resolution of around $\sigma\sim 2 $ \% is achieved by measuring 
both the ionization and scintillation signals. The 1\,ton enriched Xe detector
with the energy-resolution of $\sigma = 1.6$\,\% gives the $\nu$-mass sensitivity of 50-70 meV 
for a 5\,y run. The 10\,ton Xe detector with the improved energy resolution of $\sigma = 1$\,\% 
will give the sensitivity\footnote{In a recent    paper the EXO collaboration\cite{EXO2012}
  reported that no signal has appeared  in  a search for neutrinoless double-beta decay of $^{136}$Xe with an exposure of  32.5 kg-yr and a background of $ \approx 1.5 \times 10^{-3}$ kg$^{-1}$keV$^{-1}$y$^{-1}$ in the $\pm\sigma$ region of interest. This implies a lower limit on the half life, $T_{1/2}^{0\nu\beta\beta}\ge 1.6\times 10^{25}$ y  (90$\%$ CL), corresponding to an effective Majorana mass of less than 140-380 meV, depending on the nuclear matrix element.} of 11-15 meV. The 200 kg $^{136}$Xe liquid detector is used at 
WIPP to study the 2$\nu\beta\beta$-decay and the quasi-degenerate $\nu$-mass, as the first 
step without the Ba tagging. EXO observed the 2$\nu \beta \beta $ half life of $2.1 \times 10^{21}$ \cite{EXO11}. 
The key point of this experiment is the tagging efficiency of the $^{136}$Ba nuclei.
 
{\it KamLAND-Zen (Kamioka Large Anti Neutrino Detector Zenon)}:  
It studies the $^{136}$Xe DBD by means of the KamLAND detector with the 1 kton liquid scintillator 
at Kamioka \cite{efr11}.  A mini balloon with 3.2 m in diameter is set at the center for the 
$^{136}$Xe-loaded liquid scintillator. It includes $^{136}$Xe isotopes around 400 kg. The energy-resolution 
and the vertex-resolution are 6.8 $\%$/$\surd E$ and 12.5 cm/$\surd E$.  The collaboration neasured the $2\nu \beta \beta $ half-life \cite{kamlandzen}, and hopes to reach 
the sensitivity of around 50 meV.  

{\it NEXT (Neutrino Experiment with a Xe TPC)}:
A Xe TPC with 100-150 kg enriched $^{136}$Xe is used at LSC \cite{NEXT12}. It is a low BG and good E-resolution TPC with separate readout planes for tracking and energy. The  NEXT-100 sensitivity for 5 y run is about 5.9 10$^{25}$ y (better than 100 meV).

{\it Borexino with 2 ton $^{136}$Xe}: The sensitivity is around 100 meV \cite{rag09}.
 
{\it DCBA (Drift Chamber Beta-ray Observatory)}:
It uses a tracking chamber in a magnetic field to study $^{150}$Nd $\beta \beta $ decays \cite{ish09}. 
The $\beta $ energy is obtained by the $\beta $-ray trajectories. DCBA-T3 is now under construction. 
The good energy-resolution and efficient enrichment of $^{150}$Nd isotopes are necessary. 

{\it SNO+ (Sudbury Neutrino Observatory +)}: It uses the 1 k ton scintillation detector with 0.1 
$\%$ Nd isotopes to study QD-IH $\nu $ masses by using natural (5.6 $\%$) and enriched (50$\%$) 
$^{150}$Nd isotopes \cite{sno09}. The mass sensitivities are 100 meV with the natural 
(5.6 \% $^{150}$Nd 56 kg) and 40 meV with enriched isotopes (50 \% enriched $^{150}$Nd 500 kg). 
The collaboration is trying to find a realistic way of the Nd isotope separation, which is of 
great interest to study the IH neutrino mass. It aims at the scintillator filling at the beginning of 2013.

{\it DBD experiments for NH mass}: 
Higher sensitivity DBD experiments for NH masses of $\langle m_{\nu }\rangle$ = 2 - 4 meV require a 
large amount of high nuclear-sensitivity ($S_n \approx $15 meV) DBD isotopes of the order of $N\approx 10-50$ 
tons, and extremely low BG detectors with $B \le $0.1-0.01/t y. DBD isotopes to be studied are $^{82}$Se, 
$^{100}$Mo, and $^{136}$Xe. High energy-resolution cryogenic detectors such as ZnSe, ZnMoO$_4$ with 
pulse shape analyzes and/or scintillation signals \cite{ale92,ale98} may be used to search for the NH masses 
by 0$\nu\beta\beta$-decay experiments of $^{82}$Se and $^{100}$Mo.

\subsection{{Experimental studies of DBD matrix elements}}
\label{sebsec:ExpNME}
\subsubsection{Experimental probes for DBD matrix elements}\

Nuclear matrix elements ($M^{0\nu}_\nu$) for 0$\nu\beta\beta$-decay are crucial for extracting 
the effective Majorana $\nu$ mass and other parameters, relevant to particle physics models 
beyond the SM, from 0$\nu\beta\beta$ experiments, while nuclear matrix elements ($M^{2\nu}$) 
for 2$\nu\beta\beta$-decay can be derived experimentally from the observed  2$\nu\beta\beta$-decay 
half-lives. Extensive calculations of $M^{2\nu}_{GT}$ and $M^{0\nu}_\nu$ have been made in terms 
of QRPA, RQRPA, shell model, and so on, as given elsewhere in the theoretical sections. 

Most $\beta\beta$ strengths are located in $\beta\beta$ (double Gamow-Teller) giant resonances, i.e.,  
in the high-excitation region \cite{eji00}. Thus the $\beta\beta$ matrix elements get very small 
in comparison with single particle estimates and are sensitive to nuclear structures, nuclear 
spin-isospin correlations, nuclear deformations, nuclear interactions, nuclear medium effects 
on the weak coupling constant $g_A$ and others. The theoretical evaluations for them are hard.   
Experimental studies of  nuclear structures and nuclear interactions, which are relevant to 
$2\nu\beta\beta$-decay and  $0\nu\beta\beta$-decay matrix elements, are very interesting to 
get reliable evaluations for them \cite{eji10,eji05,eji00,eji06}. 

\begin{figure}[!t]
\begin{center}
\includegraphics[width=0.6\textwidth, angle=-90]{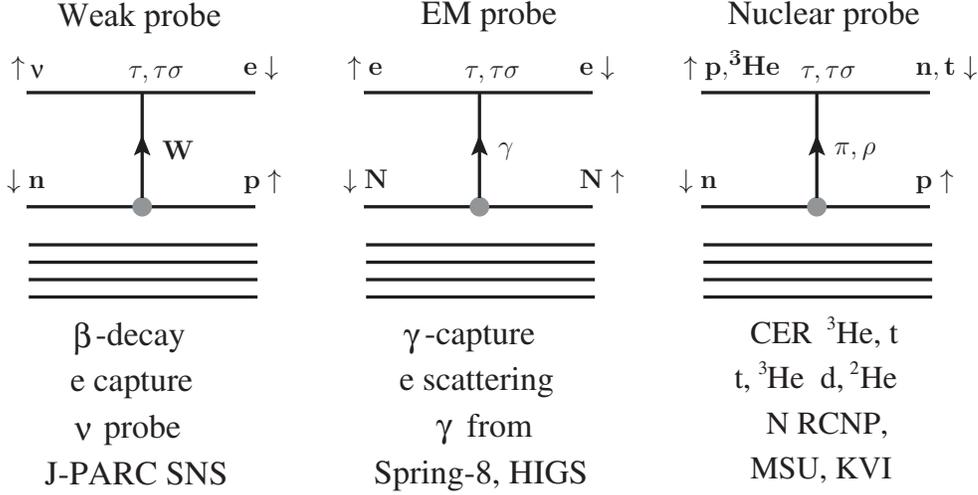}
\vspace{-0.5cm}
\caption{Nuclear spin isospin responses for weak interactions. 
They are studied by $\nu$ probes via weak interactions,  
EM ($\gamma $) probes via EM interactions and by nuclear probes via 
strong interactions \cite{eji10,eji05,Ejiri00}}
\label{fig:6.5}
\end{center}
\end{figure}

Nuclear matrix elements of $M^{2\nu_{GT}}$ and $M^{0\nu }_\nu$ are expressed in terms of 
the successive single $\beta $ processes through intermediate $\left|J^{\pi }\right>$ states. 
Among the intermediate states, low-lying single particle-hole states play dominant roles 
\cite{eji00,eji96}. The single $\beta $ matrix elements are given by spin-isospin responses for 
$Q_{TSLJ} = \tau ^{\pm}[i^Lr^LY_L \times \sigma ^S]$. They are studied experimentally using hadron, 
photon, and lepton probes  as shown in Fig. \ref{fig:6.5}. 

{\it  Lepton probes}
 
Nuclear weak responses for low-lying intermediate states are obtained from single 
$\beta $ decay rates and EC rates. However, they are limited to $\beta^{\pm}$ decays 
from the ground state in the intermediate nucleus. The decays are mostly allowed Gamow-Teller (GT) 
decays with $\tau ^{\pm} \sigma $, and are first forbidden decays with $\tau ^{\pm} irY_1 $ 
and $\tau ^{\pm}[ir^LY_L \times \sigma ^S]$ in some nuclei. 

 Muon capture reactions of ($\mu ^-,\nu_{\mu}$) are used to get the $\beta^+$
 strengths in the intermediate nucleus  \cite{eji06,suh06}. Excitation energies 
and angular momenta involved in this reaction are $E$= 0 - 50 MeV and 0$^{\pm}$, 1$^{\pm}$ and 2$^{\pm}$. 
The capture cross section is quite large, and most muons are stopped and captured 
in the case of medium heavy nuclei. Then the $\tau ^+$ weak strength distribution 
in the intermediate nucleus is derived by measuring the decaying neutrons and $\gamma $-rays 
from excited states and radioactive isotopes produced by the muon capture. 
 
One direct way to get the weak responses is to use $\nu$ beams. Since $\nu$ nuclear cross sections 
are as small as $\sigma = 10^{-40 \sim 42}$ cm$^2$, one needs high-flux $\nu $ beams 
and large detectors \cite{eji10,eji05,Ejiri00}. Low energy $\nu$ beams with $E \le 100$\,MeV 
can be obtained from pion decays. Intense pions are produced by nuclear interaction with 
GeV protons. Weak decays of stopped $\pi ^+$ give low energy neutrinos as
\begin{equation}
\pi ^+ \rightarrow \mu ^+ +\nu _{\mu }, ~~~~\mu ^+ \rightarrow \nu _e + \bar{\nu }_{\mu } + e^+ ,
\end{equation}
where $\nu _{\mu } $ and $\nu _e, \bar{\nu }_{\mu }$ are well separated by the decay time.  
Intense 1 GeV protons from SNS at ORNL provides intense neutrinos around 10$^{15}$ 
per second \cite{avi00} and the J-PARC booster synchrotron with 3 GeV protons 
produces neutrinos around 3$\times $10$^{14}$ per second \cite{eji03a}. 
 
{\it Photon probes}

\begin{figure}[!t]
\begin{center}
\includegraphics [width=0.72\textwidth]{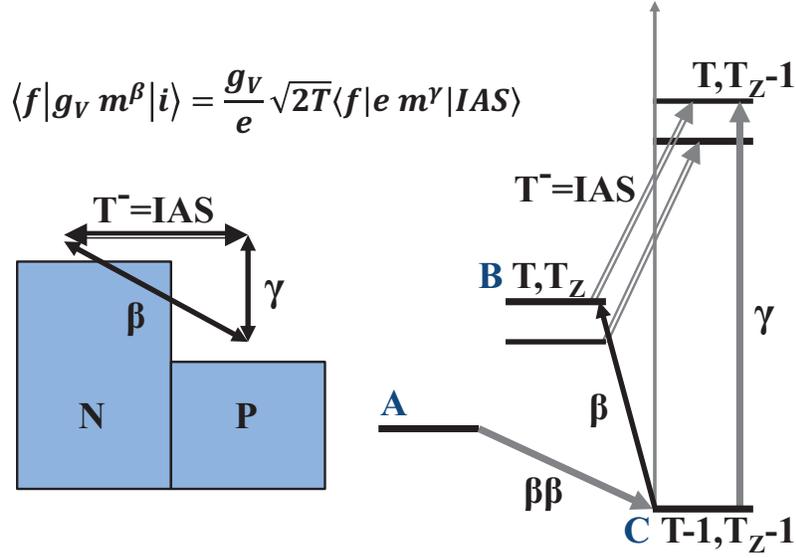}
\caption{Level and transition schemes of $\beta\beta$, single $\beta^+$, and $\gamma$ via IAS. 
T$^-$ is the isospin lowering operator.}
\label{fig:6.7}
\end{center}
\end{figure}

Weak responses for $\beta^+$ decays are studies by using photo-nuclear ($\gamma $,X) reactions 
through isobaric analogue states (IAS) as shown for the first time by (p,$\gamma $) reactions 
\cite{eji68,eji68b}. The $\beta $ and $\gamma $ matrix elements are related as 
\begin{equation}
 <f|g_V~m^{\beta }|i>\approx \frac{g_V}{e}~(2T)^{1/2}<f|e~m^{\gamma }|IAS> ,
\end{equation}
where $|IAS> = (2T)^{-1/2}T^-|i>$, $T$ is the isospin of the parent nucleus and $m^{\beta }$ 
and $m^{\gamma }$ are analogous $\beta$ and $\gamma$ transition operators. Thus one can 
obtain the $\beta$ matrix element for  $|i> \rightarrow |f> $ by observing the analogous 
$\gamma $ absorption $|f> \rightarrow |IAS>$ through the IAS of $|i>$, where $|f>$ and 
$|i>$ are the final state and the intermediate state in the $\beta\beta$-decay, 
as shown in Fig. \ref{fig:6.7}. These photo-nuclear reactions through IAS are used to 
get the $\beta ^+ $ matrix elements to excited states in the intermediate nucleus.

In medium heavy nuclei, IAS is located on the E1 giant resonance (GR) 
at the high excitation region. Accordingly IAS shows up as a sharp 
isobaric analogue resonance (IAR), and the photo-nuclear reaction 
includes IAR, GR, and the interference term as given by \cite{eji68A}
\begin{equation}
\frac{d\sigma (\gamma ,N)}{d\Omega } = k [ |A(I)_{J'}|^2 + \Sigma _J 
|A(G)_J|^2 + 2Re ( A(I)_{J'}~ A(G)_{J'}~ e^{i\phi})], 
\end{equation}
where  $A(I)_{J'},  A(G)_{J'}$ and $\phi $ are the IAR and GR amplitudes 
and the relative phase at IAR. Then one can get the phase of 
the matrix element from the interference. 
 
Laser electron photons, which are obtained from laser photons scattered 
off GeV electrons, are used for the photonuclear reaction. The polarization 
of the photon can be used to study E1 and M1 matrix elements separately \cite{eji06}.  
HI$\gamma $S (High Intensity $\gamma $-ray Source), New SUBARU and other 
electron synchrotrons with $E_e$=1 - 3 GeV provide laser electron photons used 
for the photo-nuclear reaction.  

{\it Nucleon and nuclear probes}
  
Nuclear charge exchange reactions with nuclear (hadron) probes are used 
to study nuclear spin-isospin responses. Charge exchange spin-nonflip 
reactions are used for vector weak responses, while charge exchange 
spin-flip reactions for axial-vector weak responses \cite{eji05,eji00}. 
Extensive studies of charge exchange reactions have been made to get GT(1$^+$) 
responses with $\tau ^{\pm} \sigma $. The reactions studied are (p,n) (n,p), 
(d,$^2$He), ($^3$He,t) (t,$^3$He), and ($^7$Li,$^7$Be) at IUCF, KVI, MSU, RCNP, 
Triumf and others. Some charged particle experiments are in \cite{aki97,aki08, doh08,zeg02,rak04}. Medium-energy 
projectiles with $E_i/A $= 0.1 - 0.3 GeV are used for studying $\tau\sigma$ 
responses because of the relatively large spin-isospin interaction ($V_{\tau \sigma }$) 
and the small distortion interaction ($V_0$) at the medium energy. 

The charge exchange reaction with the medium-energy projectile is mainly 
due to the central isospin and spin-isospin interactions. Then the cross section with the 
transferred momentum ($q$) and energy ($\omega $) is given as 
\begin{equation}
\sigma _{\alpha }(q,\omega ) = K(E_i,\omega)\exp(-\frac{1}{3}q^2\left< r^2\right> ) N^D_{\alpha }|J_{\alpha }|^2B(\alpha ), 
\end{equation}
where $K(E_i,\omega), N^D_{\alpha }, J_{\alpha }$, and $B(\alpha )$ are the kinematical factor, 
the nuclear distortion factor, the volume integral of the spin-isospin interaction, and 
the nuclear spin-isospin response, respectively. $\alpha$ denotes the isospin and spin channel; 
$\alpha $=F for isospin Fermi and $\alpha $= GT for spin-isospin GT. 

The cross section at 0 deg for $q \approx 0$ and $\omega \approx 0$ 
 is corrected for the kinematical and normalization factors, and is expressed  as
\begin{equation}
\frac{d\sigma_{\alpha }(0^{o})}{d\Omega}\frac{1}{K(E_i,0)N^D_{\alpha }} = 
 |J_{\alpha }|^2B(\alpha ), ~~~~ \alpha = \rm{F,~GT}.
 \end{equation}
 The proportionality of the cross section at the forward angle of $\theta \sim $0 deg to 
the $\tau\sigma$ response $B(\alpha )$ has been studied for charge exchange reactions 
with medium-energy light projectiles. In fact, the proportionality is good for spin-flip 
reactions with $B(GT) \ge 0.1 $, but some deviation from the proportionality is found in 
the reactions with smaller $B(GT)$ due to contributions from tensor/non-central interactions. 
 
Charge exchange ($^3$He,t) reactions (ChER) relevant to the $\beta\beta$-decays were studied at 
RCNP by using the 420 MeV $^3$He beam and the high energy-resolution beam line and spectrometer 
system \cite{aki97,aki08,Raker05,Grewe07,Grewe08,Yako09,Frekers10,Puppe11,Guess11}. 
The beam stability and the beam line system have been improved to give 
the fantastic energy-resolution of $\Delta E/E \approx 5 \times 10^{-5}$ and $\Delta E \approx $ 25 keV, 
and ChERs have been studied for several $\beta\beta$ nuclei.
 
The GT strengths are mostly located at the GR resonances in the high excitation region, and accordingly, 
the strengths for the low-lying states are small. In cases of  $^{96}$Zr, $^{100}$Mo, and $^{116}$Cd, 
where valence neutrons and valence protons are in the different major shells, there is only one strong 
GT state (ground state) of (g7/2)$_n$(g9/2)$_p$. Thus the ($^3$He, t) reaction shows a 
strong GT strength only at the ground state. The high resolution $GT^\pm$ strength measurement can give 
significant insight into the details of the nuclear structure and can help to 
determine the $0\nu\beta\beta$-decay NME less nuclear model dependent.

Nucleon transfer reactions were studied to get the occupation and vacancy probabilities 
of nucleons involved in double beta decays \cite{sch08}. They are used to evaluate $\beta\beta$ 
matrix elements.

\subsubsection{DBD matrix elements via low lying intermediate states}
In this subsestion, we briefly discuss experimental GT strengths for low-lying states and $2\nu \beta \beta $ matrix elements\cite{Ejiri00,eji05}. The matrix element $M^{2\nu}$ is given approximately by a sum of products of two single GT matrix elements, $M^{(+)}_{GT}(m)$ from the direction of $\beta^+$ and $M^{(-)}_{GT}(m)$ from $\beta^-$. It is expressed as $M^{2\nu} \approx M^{2\nu}_{GT}$, and
\begin{eqnarray}
M^{2\nu}_{GT} &=&  \sum_m 
\frac{M^{(+)}_{GT}(m) M^{(-)}_{GT}(m)}
{ Q_{\beta\beta}/2 + m_e + E_x(1^+_m) - E_0},
\label{2nbbme}
\end{eqnarray}
Here, $(E_x(1^+_m) - E_0)$ is the energy difference between the 
$m^{th}$ intermediate $1^+$ state and the initial ground state. 
$Q_{\beta\beta}$ is the Q-value of the $\beta\beta$-decay. 

The single GT matrix elements can be derived 
from charge exchange ($^3$He, t), (d,$^2$He) and other reactions and $\beta ^{\pm} $/EC-decay rates. Measurements of $GT^\pm$ strengths have been performed for A = 100, 116 \cite{aki97,aki08}, 48 \cite{rak04,Grewe07}, 76 \cite{Grewe08}, A = 96 \cite{doh08,Frekers10}, 116 \cite{Raker05}, A=136 \cite{Puppe11} and 150 \cite{Guess11}.  For $A = 128$ and $130$ the data are expected to be 
available soon. 

The coherent sum of the weighted products of the measured $M^{+}_{GT}(m)$ and $M^{(-)}_{GT}(m)$ are in accord with the 
observed $M^{2\nu}_{GT}$\cite{aki97,Frekers10}. Actually it has been shown that the $2\nu \beta \beta $ proceeds through low-lying (Fermi Surface) states, and not much through the GT giant resonance\cite{Ejiri00,eji96}, in agreements with observations by charge exchange reactions. The  measured $2\nu \beta \beta$ half-lives and the matrix elements\cite{recbb} are given in Table \ref{tab.2nbb}. 

SSD (Sigle state dominance) hypothesis \cite{aba84} suggests that $2\nu\beta\beta$-decay transitions, 
where the first $1^+$ state of the intermediate nucleus is the ground state
(e.g., $^{100}$Mo, $^{116}$Cd, $^{128}$Te) are governed by the transition through that $J^\pi = 1^+$ state. The SSD hypothesis looks fine in case of nuclei like $^{100}$Mo, where there is only one $J^\pi = 1^+$ state.  

Recently the nuclear matrix elements $M^{2\nu}_{GT}$ are shown to be expressed 
in terms of single $\beta$ matrix elements via Fermi-surface quasi particle states 
(FSQP) \cite{eji96,eji09,eji12}. 

\begin{eqnarray}
M^{2\nu}_{GT} &=&  \sum_k M(k) (\Delta (k))^{-1},~~M(k) = M^{(+)}_{GT}(k) M^{(-)}_{GT}(k),
\end{eqnarray}
where the sum is over the FSQP (low-lying) states ($k$) in the intermediate nucleus, $\Delta (k)$ is the energy denominator and the matrix elements of $M^{(+)}_{GT}(k)$ and $ M^{(-)}_{GT}(k)$ are the experimental single GT matrix elements deduced from charge exchange reactions, $\beta $ decay rates and EC rates. Thus it includes no adjustable parameters. The possible deviation of $g_A$ from 1.26 is embedded in the observed matrix elements of $M^{2\nu}_{GT}$ and similarly in $M^{(+)}_{GT}(k), M^{(-)}_{GT}(k)$. 

In the quasi-particle representation, the experimental matrix elements are given by
\begin{equation}
M^{(-)}_{GT}(k) = k^{eff_i}_A~  m(j_k,J_k) P_i(k),~~~
M^{(+)}_{GT}(k) = k^{eff_f}_A~  m(j_k,J_k) P_f(k),                                 
\label{gaeff}
\end{equation}
where $m(j_k J_nk)$ is the single particle matrix element for the GT transition between single 
particle $j_k$ and $J_k$ states with $j_k$ and $J_k$ being the neutron and proton spins, and 
$P_i(k) = U_p(J_k) V_n(j_k)$ and  $P_f(k) = V_p(J_k) U_n(j_k)$ are the pairing reduction 
factors for transitions from the initial ground state to k-th intermediate state and from this state
to final state, respectively. 
The effective coupling constants, $k^{eff_i}_A$  and $k^{eff_f}_A$, in unit of $g_A$, represent the nuclear core 
effects such as the spin-isospin correlations, the short-range correlations, and others, in addition to the quenching effect, 
while the nuclear surface (shell) effects are given by 
the pairing factors of $P_i(k)$ and $P_f(k)$. 
In fact values for $k^{eff_i}_A$  and $k^{eff_f}_A$ do not depend much on individual states, and thus one can evaluate the single $\beta $ matrix elements, if not available experimentally, by using the $k^{eff_i}_A$  and $k^{eff_f}_A$ for other states in neighboring nuclei and the caluculated values for $m(j_k,J_k) P_{i,f}(k)$.

\begin{figure}[!t]
\begin{center}
\includegraphics[width=0.9\textwidth]{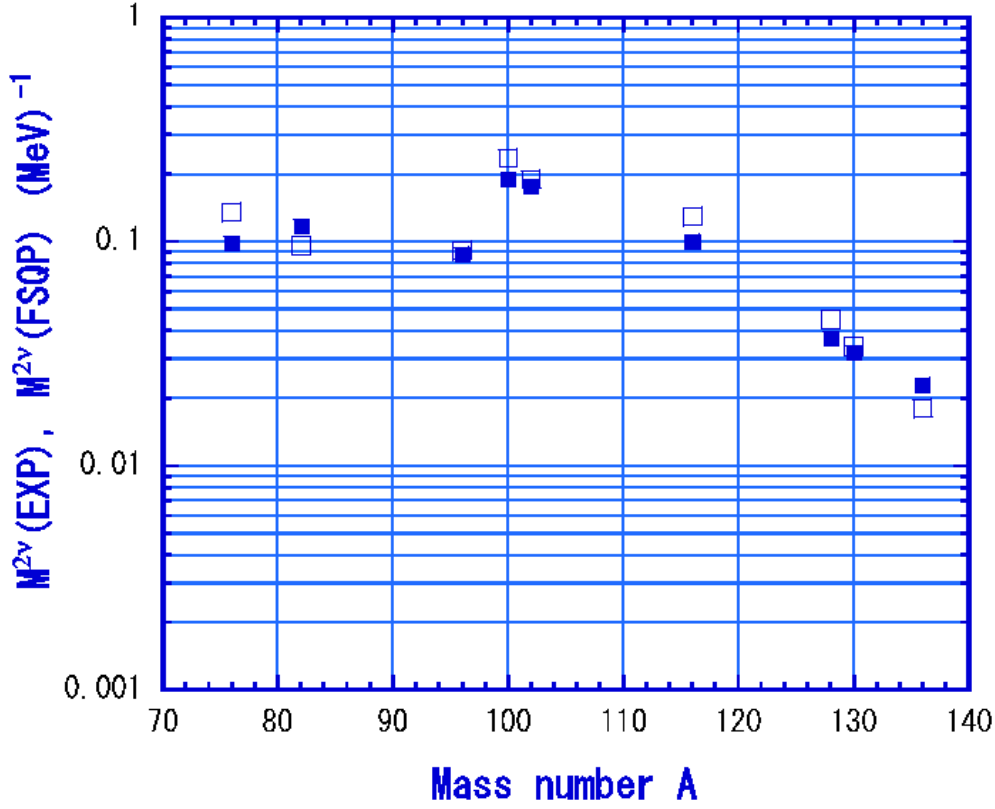}
\caption{ Nuclear  matrix elements for 2$\nu \beta \beta $ decays . Open squares: observed values ($M^{2\nu }$) , closed squares: FSQP values ($M=\Sigma M(k)/\Delta (k)$). The matrix elements for the excited 0$^+$ state in $^{100}$Ru are plotted at $A$=102 
\cite {eji09,eji12}.} 
\label{fig:6.8}
\end{center}
\end{figure}
As shown in Fig. \ref{fig:6.8},
the observed 2$\nu\beta\beta$-decay matrix elements are well reproduced by the sum of the matrix elements through the low-lying FSQP 1$^+$ states in intermediate nucleus. 

In general there are several low-lying 1$^+$ states, and thus 2$\nu\beta\beta$-decay proceeds  
not only through the lowest state but also other FSQP states. The product of the matrix elements is $M^{(+)}_{GT}(k)$ $M^{(-)}_{GT}(k) = k^{eff_i}_A k^{eff_f}_A (m(j_k,J_k))^2 P_i(k)P_f(k)$, which is positive. Thus contributions from the FSQP 
states are not cancelled with each other, but are constructive. 

The $0\nu\beta\beta$ transition operator is a two-body operator. It is shown that the matrix element is approximately given by the sum of the matrix elements for successive processes via intermediate 
states \cite{eji05, Ejiri00}. Then the matrix elements for the successive $\beta$ transitions via 
low-lying intermediate states are derived from the single $\beta$ matrix elements as 
in case of the $2\nu\beta\beta$ process.   

Since the $0\nu\beta\beta$-decay process is a virtual neutrino exchange between 2 nucleons 
in the nucleus, intermediate states with higher spins are involved. They are  studied 
by measuring angular distributions of charge exchange reactions to intermediate 1$^{\pm}, 2^{\pm}$, states and photo-excitations to IAS of the intermediate 1$^{\pm}$ states \cite{eji06}. Charge exchange reactions and photo-excitations of IAS are under progress by Muenster, MSU, NC, RCNP, and other groups 
to study nuclear structures relevant to $2\nu\beta\beta$ and $0\nu\beta\beta$ processes.

\subsection{{Two-neutrino double beta decay and bosonic neutrino}}

Neutrinos may possibly violate the spin-statistics theorem, and hence 
obey Bose statistics or mixed statistics despite having spin half. A violation 
of the spin-statistics relation for neutrinos would lead to a number 
of observable effects in cosmology and astrophysics. In particular, bosonic 
neutrinos might compose all or a part of the cold cosmological dark matter 
(through bosonic condensate of neutrinos) and simultaneously provide some 
hot dark matter. A change of neutrino statistics would have 
an impact on the evolution of supernovae and on the spectra of supernova neutrinos. 
The idea of bosonic neutrinos has been proposed independently 
in Ref.~\cite{dosm}, where cosmological and astrophysical 
consequences of this hypothesis have been studied.  

If neutrinos obey at least partly the Bose-Einstein statistics
the Pauli exclusion principle (PEP) is violated for neutrinos.
The parameter $\sin^2 \chi$ can be introduce to characterize the 
bosonic (symmetric) fraction of the neutrino wave function \cite{bosonic}.
A smooth change of $\sin^2\chi$ from 0 to  1 
transforms fermionic neutrinos into bosonic ones.  
The assumption of violation of the PEP 
leads to a number of fundamental problems which include   
loss of a positive definiteness of energy, violation of 
the CPT invariance, and possibly, of the  Lorentz invariance as well 
as of the unitarity of S-matrix.  
No satisfactory  and consistent mechanism of the Pauli 
exclusion principle violation  has been proposed so far.

\begin{figure}[tb]
\begin{center}
\includegraphics[scale=0.5]{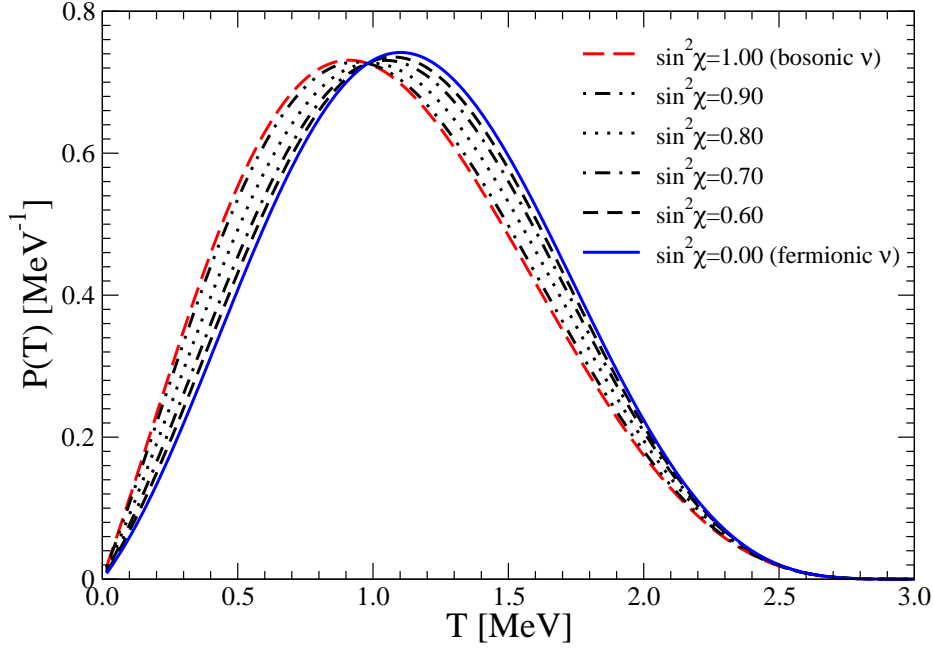}
\caption{(Color online) The differential decay rates normalized to the total decay rate 
vs. the sum of the kinetic energy of outgoing electrons $T$ for 
$2\nu\beta\beta$-decay of $^{100}{\rm Mo}$ 
to the ground state of final nucleus.
The results are presented for different values of 
the squared admixture $\sin^2\chi$ of the bosonic component.
The spectra have been calculated in the single state dominance
approximation \cite{SDS01,DKSS05}.
} 
\label{mosum}
\end{center}
\end{figure}

The lepton number conserving $2\nu\beta\beta$-decay)
can be used to study the statistical properties of neutrinos \cite{bosonic}. 
In the presence of bosonic neutrinos  two contributions to the amplitude of the decay 
from diagram with permuted neutrino momenta have relative plus sign instead of minus
in the Fermi-Dirac case. 

The PEP violation strongly changes the rates of the $2\nu\beta\beta$-decays 
and modifies the energy and angular distributions of the emitted electrons. 
The effect of bosonic neutrinos is different for transitions 
to $0^+$ ground states and $2^+$ excited states. In Fig.~\ref{mosum} 
the energy spectra of two electrons for different values of the bosonic-fraction 
$\sin^2 \chi$ is presented for the $2\nu\beta\beta$-decay of $^{100}$Mo 
to ground state of final nucleus. 
With increase of  $\sin^2 \chi$ the spectra shift to smaller energies. 
We note that  substantial shift occurs only when 
$\sin^2 \chi$ is close to 1.0 Pure bosonic neutrinos are excluded by the present data. 
\cite{bosonic}.
In the case of partly  bosonic (or mixed-statistics) neutrinos 
the analysis of the existing  data allows to put the conservative 
upper bound $\sin^2 \chi < 0.6$ \cite{bosonic}.


\section[Effective transition operators]{Effective transition operators}
\label{sec:TranOper}

The subject of interest are the lepton number violating (LNV)
parameters associated with the exchange of light and heavy Majorana 
neutrinos and with R-parity breaking SUSY mechanisms.

By assuming the dominance of a single mechanism determined by the
LNV parameter $\eta_{\kappa}$   the inverse value 
of the $0\nu\beta\beta$-decay half-life for a given isotope $(A,Z)$ 
can be written as
\begin{eqnarray}
\frac{1}{T^{0\nu}_{1/2}} &=& |\eta_{\kappa}|^2~|{M'}^{0\nu}_{\kappa}|^2~
G^{0\nu}(E_0,Z). 
\end{eqnarray}
Here, $G^{0\nu}(E_0,Z)$ and ${M'}^{0\nu}_\kappa$ are, respectively,
the known phase-space factor ($E_0$ is the energy release)
and the nuclear matrix element, which depends on the nuclear 
structure of the particular isotopes $(A,Z)$, $(A,Z+1)$ and $(A,Z+2)$ under study. 

The phase space factor $G^{0\nu}(E_0,Z)$ includes fourth
power of unquenched axial-vector coupling constant $g_A$ and the inverse
square of the nuclear radius $R^{-2}$, compensated by the
factor $R$ in ${M'}^{0\nu}_\kappa$. The assumed value of the 
nuclear radius is $R = r_0 A^{1/3}$ with $r_0 = 1.1~fm$
or $r_0 = 1.2~fm$ in different publications. The implicit 
radius and  $g_A$ dependencies in $G^{0\nu}(E_0,Z)$ and nuclear
matrix element and the problem of the correct use of them
were discussed in \cite{grabm10}. 

The nuclear matrix element ${M'}^{0\nu}$ is defined as 
\begin{equation}
{M'}^{0\nu}_\kappa =  \left(\frac{g^{eff}_A}{g_A}\right)^2 {M}^{0\nu}_\kappa.
\label{nmep}
\end{equation}
Here, $g^{eff}_A$ is the quenched axial-vector coupling constant.
This definition of ${M'}^{0\nu}_\kappa$ \cite{qrpa2,qrpa3} allows
to display the effects of uncertainties in $g^{eff}_A$ and to use
the same phase factor $G^{0\nu}(E_0,Z)$ when calculating 
the $0\nu\beta\beta$-decay rate.

Before we proceed with the discussion of the nuclear matrix elements 
we will summarize the various types of transition operators entering 
the neutrinoless double beta decay process. We recall that in the various 
particle models the lightest particle exchanged between the two nucleons 
participating in this process is either much lighter than the electron 
or much heavier than the proton. It is thus possible to separate the 
particle physics parameters from those of nuclear physics. Furthermore 
the nature of this exchanged particle dictates the form of the transition 
operators. The LNV parameters of interest 
together with corresponding nuclear matrix elements
are presented briefly below.

\subsection{Transition operators resulting from light neutrino exchange}

In the case of light-neutrino mass mechanism of the $0\nu\beta\beta$-decay we have
\begin{eqnarray}
\eta_{\nu} = \left|\frac{\langle m_\nu\rangle}{m_e}\right|^2.
\end{eqnarray}

The nuclear matrix element associated with the light Majorana-neutrino
exchange ${M}^{0\nu}_\nu$  
consists of the Fermi (F), Gamow-Teller (GT) and tensor (T) parts as
\begin{eqnarray}
{M}^{0\nu}_\nu &=&  - \frac{M^{0\nu}_{F}}{(g^{eff}_A)^2} + M^{0\nu}_{GT} - M^{0\nu}_T \nonumber\\
&=& \langle 0^+_i|\sum_{kl} \tau^+_k \tau^+_l
[ -\frac{H_F(r_{kl})}{(g^{eff}_A)^2} + H_{GT}(r_{kl}) \sigma_{kl}
- H_T(r_{kl}) S_{kl}]
|0^+_f\rangle .\nonumber\\
\end{eqnarray}
Here
\beq
S_{kl} = 3({\vec{ \sigma}}_k\cdot \hat{{\bf r}}_{kl})
       ({\vec{\sigma}}_l \cdot \hat{{\bf r}}_{kl})
      - \sigma_{kl},\quad
\sigma_{kl}={\vec{ \sigma}}_k\cdot {\vec{ \sigma}}_l.
\label{Eq:tensor}
\eeq
The radial parts of the exchange potentials are
\begin{eqnarray}
H_{F,GT,T}(r_{kl}) &=& \frac{2}{\pi} R
\int_0^\infty \frac{j_{0,0,2}(q r_{kl}) h_{F,GT,T}(q^2) q}{q + \overline{E}}
dq.
\end{eqnarray}
where $R$ is the nuclear radius and $\overline{E}$ is the average energy of the virtual intermediate states
used in the closure approximation. The closure approximation  is adopted in 
the calculation of  the NMEs relevant for $0\nu\beta\beta$-decay  with the exception
of the QRPA.  The functions $h_{F,GT,T}(q^2)$ are given by \cite{sim08}
\begin{eqnarray}
h_F(q^2) &=& f^2_{V}(q^2), \nonumber\\
h_{GT}(q^2) &=& \frac{2}{3} f^2_V(q^2) \frac{(\mu_p-\mu_n)^2}{(g^{eff}_A )^2} \frac{q^2}{4 m^2_p} + 
\nonumber\\
&& f^2_A(q^2)
\left(1 - \frac{2}{3}\frac{q^2}{q^2+m_\pi^2 } + \frac{1}{3} \frac{q^4}{(q^2+ m^2_\pi )^2}\right),
\nonumber\\
h_{T}(q^2) &=& \frac{1}{3} f^2_V(q^2) \frac{(\mu_p-\mu_n)^2}{(g^{eff}_A )^2} \frac{q^2}{4 m^2_p} + 
\nonumber\\
&&\frac{1}{3} f^2_A(q^2)\left(2 \frac{{q}^{2}}{(q^2 + m^2_\pi)}-\frac{{q}^{4}}{(q^2 + m_\pi^2)^2}\right).
\label{a14}   
\end{eqnarray}
For the normalized to unity vector and axial-vector form factors the usual dipole approximation is adopted:
$f_V({q}^{2}) = 1/{(1+{q}^{2}/{M_V^2})^2}$, $f_A({q}^{2}) = 1/{(1+{q}^{2}/{M_A^2})^2}$. 
$M_V$ = 850 MeV, and $M_A$ = 1086 MeV.  $g_A$ = 1.254 is assumed and the difference in 
magnetic moments of proton and neutron is $(\mu_p-\mu_n)= 4.71$. 

The above definition of the ${M}^{0\nu}_\nu$ includes contribution of the higher order terms of the 
nucleon current and  for the induced pseudoscalar the Goldberger-Treiman PCAC 
relation, $g_P({q}^{2}) = {2 m_p g_A({q}^{2})}/({{q}^{2} + m^2_\pi})$ was employed \cite{SPVF}.

\subsection{Transition operators resulting the  heavy neutrino exchange mechanism}

We assume that the neutrino mass spectrum include heavy Majorana
states $N_k$ with masses $M_k$ much larger than the typical energy
scale of the $0\nu\beta\beta$-decay. These
heavy states can mediate this process as the previous 
light neutrino exchange  mechanism. The difference is that the 
neutrino propagators in this case can be contracted to points 
and, therefore, the corresponding effective transition operators are 
local unlike in the light neutrino exchange mechanism with long 
range internucleon interactions. 
The corresponding LNV parameters are $\eta_N^L$ and $\eta_N^R$.

Separating the
Fermi (F), Gamow-Teller (GT) and the tensor (T) contributions we
write down for the NME 
\begin{eqnarray}
{\mathcal M}^{0\nu}_{_N} &=& - \frac{M_{F(N)}}{g^2_A}
+ M_{GT(N)} - M_{T(N)} \nonumber \\
&=&
\langle 0^+_i|\sum_{kl} \tau^+_k \tau^+_l
[ \frac{H^{(N)}_F(r_{kl})}{g^2_A} + H^{(N)}_{GT}(r_{kl}) \sigma_{kl} 
- H^{(N)}_T(r_{kl}) S_{kl}]
|0^+_f\rangle ,\nonumber\\
\end{eqnarray}
where $ S_{kl}$ and $\sigma_{kl}$ are given in Eq. (\ref{Eq:tensor}).
The radial parts of the exchange potentials are
\begin{eqnarray}
H^{(N)}_{F,GT,T}(r_{kl}) &=& \frac{2}{\pi} \frac{R}{m_p m_e}
\int_0^\infty j_{0,0,2}(q r_{kl}) h_{F,GT}(q^2) q^2 dq.
\end{eqnarray}

\subsection{Transition operators resulting from the R-parity breaking SUSY mechanism}

Assuming the dominance of gluino exchange, we obtain for the LNV parameter
the following simplified expression
\begin{equation}
\eta_{\lambda'} =
\frac{\pi \alpha_s}{6}
\frac{\lambda^{'2}_{111}}{G_F^2 m_{\tilde d_R}^4}
\frac{m_p}{m_{\tilde g}}\left[
1 + \left(\frac{m_{\tilde d_R}}{m_{\tilde u_L}}\right)^2\right]^2.
\label{eta-N}
\end{equation}
Here, $G_F$ is the Fermi constant,
$\alpha_s = g^2_3/(4\pi )$ is $\rm SU(3)_c$ gauge coupling constant.
$m_{{\tilde u}_L}$, $m_{{\tilde d}_R}$ and $m_{\tilde g}$ are masses 
of the u-squark, d-squark and gluino, respectively.  

We should mention again that for this type of interactions 
the pion-exchange mechanism ($1\pi$ and $2\pi$) discussed in subsection 
\ref{sec:pionmode} dominates over the conventional 
two-nucleon  mechanism. Thus, denoting the $0\nu\beta\beta$-decay 
NME associated with gluino and
neutralino exchange as  ${\mathcal M}^{0\nu}_{\lambda'}$,
we have  \cite{WKS99,FKS98a}
\begin{equation}
{\mathcal M}^{0\nu}_{\lambda'} = c^{1\pi} \left(M_{GT}^{1\pi} - M_{T}^{1\pi} \right)
+ c^{2\pi}\left(M_{GT}^{2\pi} - M_{T}^{2\pi}  \right)
\end{equation}
with 
\begin{eqnarray}
c^{1\pi} = -\frac{2}{9} \frac{\sqrt{2} f_\pi m_\pi^4}
{m_p^3 m_e (m_u + m_d)} \frac{g_s F_P}{g_A^2},~~~
c^{2\pi} = \frac{1}{18} \frac{ f^2_\pi m_\pi^4}
{m_p^3 m_e (m_u + m_d)^2} \frac{g_s^2}{g_A^2}.
\end{eqnarray}
Here, $g_S$  and $F_P$ stand for the standard pion-nucleon
coupling constant ($g_s=13.4$) and the nucleon pseudoscalar 
constant (we take the bag model value $F_P \approx 4.41$ from
Ref. \cite{adler}), respectively. $f_\pi = 0.668 m_\pi$ and $m_\pi$
is the mass of pion. $m_u$ and $m_d$ denote current quark masses. 
The partial nuclear matrix elements of 
the $R_p \hspace{-1em}/\;\:$  SUSY
mechanism for the $0\nu\beta\beta$ process are:
\begin{eqnarray}
{M}_{GT}^{k\pi} &=&
\langle 0^+_f|~\sum_{k\neq l} ~\tau_k^+ \tau_l^+ ~
~H_{GT}^{k\pi}(r_{kl})
~{\bf{\sigma}}_i\cdot{\bf{\sigma}}_j
~| 0^+_i \rangle , \nonumber \\
{M}_{T}^{k\pi} &=&
\langle 0^+_f|~\sum_{k\neq l} ~\tau_k^+ \tau_l^+ ~
~H_{T}^{k\pi}(r_{kl})
~{S}_{kl}
~| 0^+_i \rangle 
\end{eqnarray}
with 
\begin{eqnarray}
H^{1\pi }_{GT, T}(r_{kl}) &=& -\frac{2}{\pi} R
\int_0^\infty j_{0,2}(q r_{kl}) 
\frac{q^4/m^4_{\pi}}{1 + q^2/m^2_{\pi}}
f^2_{A}(q^2) dq, \nonumber\\
H^{2\pi }_{GT, T}(r_{kl}) &=& -\frac{4}{\pi} R
\int_{0, 2}^\infty j_{0,2}(q r_{kl}) 
\frac{q^4/m^4_{\pi}}{(1 + q^2/m^2_{\pi})^2}
f^2_{A}(q^2) dq.
\end{eqnarray}
The two-nucleon exchange  potentials are expressed in momentum space 
as the momentum dependence of normalized to unity nucleon formfactors ($f_A(q^2)$) 
is taken into account.

\subsection{Transition operators resulting from the squark-neutrino mechanism}

In the case of squark-neutrino mechanism \cite{FGKS07} 
due to the chiral structure
of the R-parity breaking (\rp) SUSY interactions, the amplitude of \dbd-decay does not vanish 
in the limit of zero neutrino mass unlike the ordinary 
Majorana neutrino exchange mechanism proportional to the light neutrino mass.
Instead, the squark-neutrino mechanism is roughly proportional to 
the momentum of the virtual neutrino which is of the order of 
the Fermi momentum of the nucleons inside of nucleus $p_F\approx 100$MeV. 
This is a manifestation of the fact that the LNV necessary for \dbd-decay 
is supplied by the \rp SUSY interactions instead of the Majorana neutrino 
mass term and therefore this mechanism is not suppressed by the small 
neutrino mass. The corresponding SUSY LNV parameter is defined as
\begin{eqnarray}\label{eta}
\eta_{\tilde q} &=& \sum_{k} \frac{\lambda'_{11k}\lambda'_{1k1}}{2
\sqrt{2} G_F}
\sin{2\theta^{d}_{(k)} }\left( \frac{1}{m^2_{\tilde d_1 (k)}} -
\frac{1}{m^2_{\tilde d_2 (k)}}\right).\nonumber\\
\end{eqnarray}
Here we use the notation $d_{(k)} = d, s, b$.
This LNV parameter vanishes in the absence of $\tilde q_L-\tilde q_R$ - mixing 
when $\theta^{d}=0$.

At the hadron level we assume dominance of the pion-exchange mode. Then, 
the nuclear matrix element associated with squark-neutrino mechanism 
can be written as a sum of GT ad tensor contributions \cite{FGKS07} 
\begin{eqnarray}
{\mathcal M}^{0\nu}_{\tilde q} = M_{GT({\tilde q})} 
- M_{T({\tilde q})}.
\end{eqnarray}
The exchange potentials are given by
\begin{eqnarray}
H^{({\tilde q})}_{GT, T}(r_{kl}) &=& \frac{2}{\pi} R
\int_0^\infty 
\frac{j_{0,2}(q r_{kl}) h^{\tilde q}(q^2) q^2}{q ( q + \overline{E})
} dq
\end{eqnarray}
with 
\begin{eqnarray}
\label{pi N-GT}
h^{\tilde q}({q}^{2} ) = -\frac{1}{6}f^2_{A}({q}^{2}) 
\frac{m^4_\pi}{m_e (m_u + m_d)}\frac{{q}^{2}}{({q}^{2}+m^2_\pi)^2}.
\end{eqnarray}

\section[Nuclear matrix elements]{$0\nu\beta\beta$-decay nuclear matrix elements}
\label{sec:0nuME}

Interpreting existing results as a measurement of $|\langle m_{\nu}\rangle|$ 
and planning new experiments depends crucially on the knowledge 
of the corresponding nuclear matrix elements (NMEs) that govern 
the decay rate. The NMEs for $0\nu\beta\beta$-decay must be evaluated 
using tools of nuclear structure theory. There are no 
observables that could be directly linked to the magnitude of 
$0\nu\beta\beta$-decay nuclear matrix elements and, thus, could be used 
to determine them in an essentially model independent way. A reliable
calculation of NMEs will be of help in predicting
which are the most favorable nuclides to be employed for $0\nu\beta\beta$-decay searches. 

The evaluation of the nuclear matrix elements can be separated in two steps.
\begin{itemize}
\item The evaluation of the transition matrix elements between the two interacting particles (two body ME).\\ 
Each particle is assumed to occupy a set of single particle states, determined by the assumed model. The spin as well the orbital structure of the operator as it has been discussed in sections \ref{sec:numec} and \ref{sec:nonumec}. The operators discussed in section \ref{sec:numec} are long ranged, except when the intermediate neutrino is heavy leading to  short range transition operators. In section \ref{sec:nonumec}, except for the case of Eq. (\ref{Eq:SROSUSY}), all operators are short ranged. The way of dealing with the short range operators has been discussed in section \ref{sec:srop}. Taking all these into account the  effective transition operators have been constructed in section \ref{sec:TranOper}.
\item The second step involves the  construction of the many body wave functions.\\ 
One needs the wave functions of the initial and final nuclear systems. If closure is not employed, as in the case of QRPA, one also needs the wave functions of all the virtual (intermediate states) allowed by the assumed nuclear model. Some many body features arising from the nuclear medium are: i) the renormalization effects on the $g_A$  coupling and the modification of the nucleon currents, which have already been discussed in section  \ref{sec:TranOper} and ii) the short range correlations, which will be discussed below. The main techniques of the construction of the many body wave functions will be reviewed in this section. We remind the reader that in some cases information about the nuclear ME can be extracted from experiments (see \ref{sebsec:ExpNME}). 
\end{itemize}

The calculation  of the $0\nu\beta\beta$-decay NMEs is a difficult problem because ground 
and many excited states (if closure approximation is not adopted)
of open-shell nuclei with complicated nuclear 
structure have to be considered. In the last few years the reliability of the calculations 
has greatly improved. Five different many-body approximate methods have been applied for the
calculation of the $0\nu\beta\beta$-decay NME:\\
\begin{enumerate}
\item The Interacting Shell Model (ISM) \cite{lssm,ism1,ism2,horoi07,horoi10}. \\
The ISM allows to consider only a limited number
of orbits close to the Fermi level, but all possible correlations within the space are included.
Proton-proton, neutron-neutron and proton-neutron (isovector and isoscalar) pairing
correlations in the valence space are treated exactly. Proton and neutron numbers are
conserved and angular momentum conservation is preserved.
Multiple correlations are  properly described in the laboratory frame. 
The effective interactions are constructed starting from 
monopole corrected G matrices,
which are further adjusted to describe sets of experimental energy levels.
The Strasbourg-Madrid codes can deal with problems involving
basis of $10^{11}$ Slater determinants, using relatively
modest computational resources. A good spectroscopy for parent and daughter nuclei is achieved.
Due to the significant progress in shell-model  configuration mixing
approaches, there are now calculations performed with these
methods for several nuclei.
\item   Quasiparticle Random Phase Approximation (QRPA) \cite{VZ86,CAT87}. \\
The QRPA has the advantage
of large valence space but is not able to comprise all the possible configurations.
Usually, single particle states in the Wood-Saxon potential are considered. One is able
to include to each orbit in the QRPA model space also the spin-orbit partner, which
guarantees that the Ikeda sum rule is fulfilled. This is crucial to describe correctly
the Gamow-Teller strength. The proton-proton and neutron-neutron pairings are considered.
They are treated in the BCS approximation. Thus, proton and neutron numbers are not exactly
conserved. The many-body correlations are treated at the RPA level within the quasiboson approximation.
Two-body G-matrix elements, derived from realistic one-boson exchange potentials within the
Brueckner theory, are used for the determination of nuclear wave functions.
\item  Interacting Boson Model (IBM) \cite{IBM09}.\\ In the IBM the low lying states of the nucleus are
modeled in terms of bosons. The bosons are in either L=0 (s boson) or L=2 (d boson) states.
Thus, one is restricted to $0^+$ and $2^+$ neutron pairs transferring into two protons.
The bosons interact through one- and two-body forces giving rise to bosonic wave functions.
\item The Projected Hartree-Fock-Bogoliubov Method (PHFB) \cite{phfb}.\\ In the PHFB wave functions of
good particle number and angular momentum are obtained by projection on the axially
symmetric intrinsic HFB states. In applications to the calculation of the
$0\nu\beta\beta$-decay NMEs the nuclear Hamiltonian was restricted only to quadrupole
interaction. The PHFB is restricted in its scope. With a real Bogoliubov transformation
without parity mixing one can describe only neutron pairs with even angular momentum
and positive parity.
\item The Energy Density Functional Method (EDF) \cite{edf}.\\ The EDF is considered to be an improvement
with respect to the PHFB. The density functional methods based on the Gogny functional
are taken into account.  The particle number and angular momentum projection for mother
and daughter nuclei is performed and  configuration mixing within the generating coordinate
method is included. A large single particle basis (11 major oscillator shells)
is considered. Results are obtained for all nuclei of experimental interest.
\end{enumerate}

The differences among the listed methods of NME calculations for 
the  $0\nu\beta\beta$-decay are  due to the following reasons:\\ 
(i) The mean field is used in different ways. As a result, 
single particle occupancies of individual orbits of various
methods differ significantly from each other \cite{ocup09}. \\
(ii) The residual interactions are of various origin and
renormalized in different ways.\\
(iii) Various sizes of the model space are taken into account.\\ 
(iv) Different many-body approximations are used in the
diagonalization of the nuclear Hamiltonian.\\
Each of the applied methods has some advantages and drawbacks, whose effect in the
values of the NME can be sometimes explored. The advantage of the ISM calculations
is their full treatment of the nuclear correlations, which tends to diminish
the NMEs. On the contrary, the QRPA, the EDF, and the IBM underestimate the multipole 
correlations in different ways and tend to overestimate the NMEs. 
The drawback of the ISM the limited number of orbits in the valence space
and as a consequence the violation of Ikeda sum rule and underestimation of the 
NMEs. 

In Table \ref{tab.nme}, recent results of the different methods are summarized. 
The presented numbers have been obtained with 
the unquenched value of the axial coupling constant ($g_A^{eff}=g_A=1.25$)\footnote{
A modern value of the axial-vector coupling  constant is $g_A =1.269$. We note 
that in the referred calculations of the $0\nu\beta\beta$-decay NMEs
the previously accepted value $g_A^{eff}=g_A=1.25$ was assumed.}, 
Miller-Spencer Jastrow short-range
correlations \cite{miler} (the EDF values are multiplied by 0.80 in order to account 
for the difference between the unitary correlation operator method
(UCOM) and the Jastrow approach \protect\cite{sim08}), 
the same nucleon dipole form-factors, higher order corrections to 
the nucleon current and the nuclear radius $R = r_0 A^{1/3}$, with
$r_0 = 1.2$ fm (the QRPA values \cite{qrpa2,qrpa3} 
for $r_0 =1.1$ fm are rescaled with the factor $1.2/1.1$). Thus, 
the discrepancies among the results of different 
approaches are solely related to the approximations on which a given 
nuclear many-body method is based. 

\begin{table*}[!t]
\caption{The NME of the $0\nu\beta\beta$-decay $M^{0\nu}_\nu$ calculated in the framework of
different approaches:
interacting shell model (ISM) \protect\cite{lssm,horoi10}, quasiparticle random
phase approximation (QRPA) \protect\cite{fang10,qrpa3,kor07,korte3,suho11}, 
projected Hartree-Fock Bogoliubov approach (PHFB, PQQ2 parametrization) ) 
\protect\cite{phfb}, energy density functional method (EDF) \cite{edf} and
interacting boson model (IBM) \protect\cite{IBM09}. QRPA(TBC) and QRPA(J)
denote QRPA results of Tuebingen-Bratislava-Caltech (TBC) and Jyvaskyla (J) groups,
respectively. The Miller-Spencer Jastrow two-nucleon short-range correlations 
are taken into account.  
The EDF results are multiplied by 0.80 in order to account for the difference
between UCOM and Jastrow \protect\cite{sim08}.
$g_A^{eff}=g_A = 1.25$ and $R = 1.2 fm A^{1/3}$ are assumed. }
\label{tab.nme}
\begin{tabular}{lcccccc}\hline\hline
Transition  & \multicolumn{6}{c}{$|M^{0\nu}_\nu|$} \\ \cline{2-7} 
            & ISM & QRPA (TBC) & QRPA (J)  & IBM-2 & PHFB  & EDF  \\  
            & \cite{lssm,horoi10} & \cite{qrpa3,fang10} &  
\cite{kor07,korte3,suho11} &  \cite{IBM09} 
            &  \cite{phfb} &  \cite{edf}  \\  \hline
${^{48}}$Ca  $\rightarrow {^{48}}$Ti  & 0.61, 0.57 &        &      &      &       &  1.91 \\
${^{76}}$Ge  $\rightarrow {^{72}}$Se  & 2.30       &  4.92  & 4.72 & 5.47 &       &  3.70 \\
${^{82}}$Se  $\rightarrow {^{82}}$Kr  & 2.18       &  4.39  & 2.77 & 4.41 &       &  3.39 \\
${^{96}}$Zr  $\rightarrow {^{96}}$Mo &            &  1.22  & 2.45 &      &  2.78 &  4.54 \\
${^{100}}$Mo $\rightarrow {^{100}}$Ru &            &  3.64  & 2.91 & 3.73 &  6.55 &  4.08 \\
${^{110}}$Pd $\rightarrow {^{110}}$Cd &            &        & 3.86 &      &       &       \\
${^{116}}$Cd $\rightarrow {^{116}}$Sn &            &  2.99  & 3.17 &      &       &  3.80 \\
${^{124}}$Sn $\rightarrow {^{124}}$Te & 2.10       &        & 2.65 &      &       &  3.87 \\
${^{128}}$Te $\rightarrow {^{128}}$Xe & 2.34       & 3.97   & 3.56 & 4.52 &  3.89 &  3.30 \\
${^{130}}$Te $\rightarrow {^{130}}$Xe & 2.12       & 3.56   & 3.28 & 4.06 &  4.36 &  4.12 \\
${^{136}}$Xe $\rightarrow {^{136}}$Ba & 1.76       & 2.30   & 2.54 &      &       &  3.38 \\
${^{150}}$Nd $\rightarrow {^{150}}$Sm &            & 3.16   &      & 2.32 &  3.16 &  1.37 \\
\hline\hline
\end{tabular}
\end{table*}

From Table \ref{tab.nme} we can make the following conclusions:
\begin{enumerate}
  \item The ISM values of NMEs, with the exception of the NME for the double magic
  nucleus $^{48}\mathrm{Ca}$, practically do not depend on the nucleus.
  They are significantly smaller, by about a factor 2-3, when compared with NMEs 
  of other approaches.
  \item The largest values of NME  are obtained 
  in the IBM (${^{76}}$Ge and ${^{128}}$Te), PHFB (${^{100}}$Mo, 
   ${^{130}}$Te and ${^{150}}$Nd), QRPA (${^{150}}$Nd) and EDF 
  (${^{48}}$Ca, ${^{96}}$Zr, ${^{116}}$Cd, ${^{124}}$Sn  and ${^{136}}$Xe) approaches.  
  \item  NMEs obtained by the QRPA(TBC) and IBM methods are in a good 
  agreement (with the exception of ${^{150}}$Nd). 
  \item
  In the case of  $^{130}\mathrm{Te}$ all discussed methods, with the exception of the ISM,
  give practically the same result.
  \item The disagreement between IBM-2 and ISM is particularly
   troublesome, because IBM-2 is a truncation of the shell-model
   space to the S and D pair space and, in the limit of spherical
   nuclei, IBM-2 and ISM should produce the same results.
  \item The disagreement between the QRPA(TBC) and QRPA(J) results is not large but
  it needs to be clarified.
\end{enumerate}

\begin{figure}[!t]
\begin{center}
  \includegraphics[height=.33\textheight]{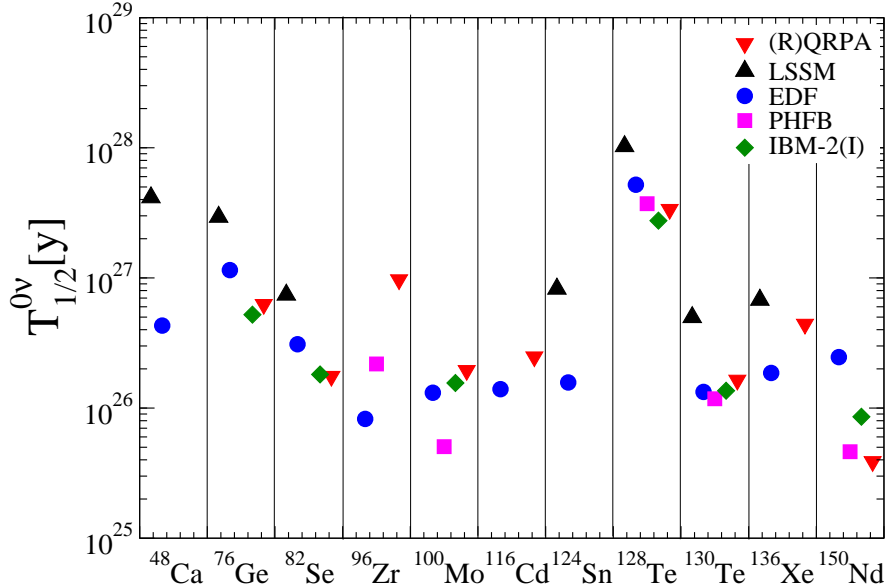}
  \caption{(Color online.) The $0\nu\beta\beta$-decay half-lives 
of nuclei of experimental interest for $|\langle m_{\nu}\rangle|$ = 50 meV 
and NMEs of different approaches. The Miller-Spencer Jastrow 
two-nucleon short-range correlations are considered. The axial-vector
coupling constant $g_A$ is assumed to be 1.25.}
\label{fig.t12}
\end{center}
\end{figure}

Comparing $0\nu\beta\beta$-decay nuclear matrix elements calculated by different methods 
gives some insight in the advantages or disadvantages of different candidate nuclei. However, 
matrix elements are not quite the only relevant quantities (see section \ref{sect:exp} 
for the nuclear sensitivity factor). Experimentally, half-lives are measured 
or constrained, and the effective Majorana neutrino mass $\langle m_{\nu}\rangle$ is the ultimate goal. 
For $|\langle m_{\nu}\rangle|$ equal to 50 meV the calculated half-lives for double $\beta$-decaying nuclei 
of interest are presented in Fig. \ref{fig.t12}. We see that the spread of half-lives for given isotope 
is up to the factor of 4-5.  

It is worth to noticing that due to the theoretical efforts made over
the last years the disagreement among different NMEs is now much less
severe than it was about a decade before. Nevertheless the present-day situation
with the calculation of $0\nu\beta\beta$-decay NMEs can not be considered
as completely satisfactory. Further progress is required and  it is believed
that the situation will be improved with time.  Accurate determination of the NMEs, and 
a realistic estimate of their uncertainty, is of great importance. Nuclear matrix elements 
need to be evaluated with uncertainty of less than 30\% to establish the neutrino mass spectrum 
and CP violating phases of the neutrino mixing 

\subsection{Uncertainties in calculated NMEs}

The improvement of the calculation of the $0\nu\beta\beta$-decay NMEs is a very 
important and challenging problem. The uncertainty associated with the calculation 
of the $0\nu\beta\beta$-decay NMEs can be diminished by suitably chosen nuclear probes. 
Complementary experimental information from related processes 
like charge-exchange and particle transfer reactions,
muon capture and charged current (anti)neutrino-nucleus reactions is very relevant. 
A direct confrontation of nuclear structure models with data from these processes 
improve quality of nuclear structure models (see section \ref{sect:exp}). The constrained parameter space 
of nuclear models is a promising way to reduce uncertainty in the calculated 
$0\nu\beta\beta$-decay NMEs.

A steady progress in nuclear structure approaches is gradually leading to a 
better understanding and to a reduction of the differences among their results. 
However, even in the most refined approaches, the estimates of $M^{0\nu}_\nu$ 
remain affected by various uncertainties, whose reduction is of great importance.

{\it i) QRPA calculation of NMEs.}\\
Due to its simplicity the QRPA is a  popular technique
to calculate the $0\nu\beta\beta$-decay NMEs. 
One of the most important factors of the QRPA calculation of the
$0\nu\beta\beta$-decay NMEs  is the way the particle-particle 
strength of the nuclear Hamiltonian $g_{pp}$  is fixed. The 
Tuebingen-Bratislava-Caltech (TBC) group has  
shown that by adjusting $g_{pp}$ to the $2\nu\beta\beta$-decay rates 
the uncertainty associated with variations in QRPA calculations of 
the $0\nu\beta\beta$-decay NMEs can be significantly eliminated 
\cite{qrpa2,qrpa3,sim08,qrpa1}. In particular, the results obtained in this way are essentially 
independent of the size of the basis, the form of different realistic 
nucleon-nucleon potentials, or on whether QRPA or renormalized QRPA 
(take into account Pauli exclusion principle) is used. This new way of
fixing parameter space was criticized by the Jyvaskyla group in series 
of papers maintaining the role of single $\beta$-transitions. It was 
claimed that careful study of single $\beta$ and $2\nu\beta\beta$ 
observables points to serious shortcomings of adopted procedure
\cite{crit_plb05,crit_npa05}. These objections were refuted in 
\cite{qrpa2,qrpa3}.
In the recent publications \cite{kor07,korte3,suho11} also
the Jyvaskyla group adopted the procedure of fixing of $g_{pp}$ 
proposed by the TBC group \cite{qrpa1}.

Usually, two variants of the QRPA are considered. The standard QRPA,
which is based on the quasiboson approximation, and the renormalized
QRPA (RQRPA) \cite{TS95,src09,qrpa2}, which takes into account 
the Pauli exclusion principle. Further improvement is achieved within 
the self-consistent QRPA (SRQRPA) \cite{ocup09,delion97,krmpotic98} 
by conserving the mean particle number in the correlated 
ground state. The restoration of Pauli exclusion principle 
and of particle number conservation lead to a reduction of the 
$0\nu\beta\beta$-decay NMEs \cite{src09,qrpa2,qrpa3,ocup09}.

There is some controversy about the importance of the tensor $M_T$ contribution
to $M^{0\nu}_\nu$. According to the ISM \cite{lssm} the tensor term is small, 
a fact  understood  by the small model space adopted. The Jyvaskyla group performing the calculation
within the QRPA claims that $M_T$ is negligible \cite{kor07,korte3}. $M_T$ was neglected
in the PHFB \cite{phfb} and EDF \cite{edf} calculations of the $0\nu\beta\beta$-decay NMEs. 
Contrary, results of the IBM-2 \cite{IBM09} and the QRPA(TBC) \cite{SPVF} calculations
show that $M_T$ can not be neglected and its absolute value can be up to 10\% of $M^{0\nu}_\nu$. 

{\it ii) The closure approximation}\\
The $0\nu\beta\beta$-decay matrix elements are usually calculated using the closure 
approximation for intermediate nuclear states. Within this approximation energies 
of intermediate states $(E_n - E_i)$ are replaced by an average value
$\overline{E} \approx 10$ MeV, and the sum over intermediate states 
is taken by closure, $\sum_n |n><n| =$ 1. This simplifies the numerical 
calculation drastically. The calculations with exact treatment of the energies 
of the intermediate nucleus were achieved within the QRPA-like methods
\cite{src09,qrpa2,qrpa3,sim08,qrpa1}. The effect of the closure approximation 
was studied in detail in \cite{muto94}. It was found that the differences 
in nuclear matrix elements are within 10\%. The the dependence of the 
NMEs on the average energy of the intermediate states $\overline{E}$ was 
studied within the nuclear shell model. By varying $\overline{E}$ from 
2.5 to 12.5 MeV the variation in the NME was obtained to be less than 5\% 
\cite{horoi10}.

{\it iii) The two-nucleon short range correlations and finite nucleon size.}\\
The physics of finite nucleon size (FNS) and two-nucleon short-range
correlations (SRC) is different. Both reduce magnitude of the 
$0\nu\beta\beta$-decay NME by competing each with other. The importance
each of them depends on the type of SRC and involved form-factor
parameters.

The FNS is taken into account via momentum dependence 
of the nucleon form-factors. For the vector, weak-magnetism  (axial-vector) 
the usual dipole approximation with cut-off parameter $M_V =$ 850 MeV
($M_A =$ 1 086 MeV), which comes come from electron scattering experiments 
(neutrino charged-current scattering experiments), is considered.
The form-factors suppress high-momentum exchange. We note that 
in the limit of point-like nucleon ($M_{V,A}\rightarrow \infty$) 
the weak magnetism contribution to the $0\nu\beta\beta$-decay
would be divergent. 

The SRCs are included via the correlation function f(r),
that modifies the relative two-nucleon wave functions 
at short distances:
\begin{equation}
\Psi_{nl}(r) \rightarrow  [1 + f(r)] \Psi_{nl}(r), 
\end{equation}
where f(r) can be parametrized as \cite{src09}
\begin{equation}
f (r) = -c e^{-a r^2 } (1 - b r^2 ).
\end{equation}
Previously, Miller-Spencer Jastrow SRC
($a=1.1~fm^{-2}$, $b = 0.68~fm^{-2}$, $c = 1.0$) have been added 
into the involved two-body transition matrix elements, 
changing two neutrons into  two protons, to achieve healing 
of the correlated wave functions. A suppression of 
$M^{0\nu}_\nu$ by 20\% was found \cite{IBM09,src09,lssm,horoi10}. 
However, recent work has questioned this prescription
\cite{src09,kor07,korte3}.
   
The two-nucleon short range correlations were studied within
the coupled clusters method (CCM) in \cite{src09}. The
Jastrow function fit for T=1 channel  
reported for Argonne V18 and Bonn-CD NN interactions set of parameters 
($a=1.59~fm^{-2}$, $b = 1.45~fm^{-2}$, $c = 0.92$) and
($a=1.52~fm^{-2}$, $b = 1.88~fm^{-2}$, $c = 0.46$), respectively
\cite{src09}. These correlation functions were confirmed by exploiting 
a construction of an effective shell model operator for 
$0\nu\beta\beta$-decay of $^{82}$Se \cite{engel_src}. 
The notable differences between the results calculated with
Miller-Spencer Jastrow and CCM SRC are about of $20\%$-$30\%$ \cite{src09,horoi10}. 
The previous results with Miller-Spencer treatment of SRC 
certainly overestimates the quenching due to short-range correlations.
Of course, the results obtained with the CCM SRC are preferable. We note
that in the case of Bonn-CD CCM SRC the $0\nu\beta\beta$-decay NMEs are
slightly increased \cite{src09}.

The two-nucleon short range correlations were treated also
through the Unitary-Correlation Operator Method
(UCOM) \cite{kor07,korte3,korte1}, which has the advantages of wave-function
overlap preservation and a range of successful applications \cite{roth05}. 
The drawback of this approach, when applied to the $0\nu\beta\beta$-decay,
is that it violates some general properties of the Fermi and Gamow-Teller 
matrix elements \cite{src09}.

Recently, a question of many-body short-range correlations 
in the evaluation of the $0\nu\beta\beta$-decay NMEs 
was addressed within a simple model \cite{engel_m11}. The existing 
calculations include long-range many-body correlations in 
model dependent (ISM, QRPA, PHFB, EDF, IBM) nuclear
wave functions but allow only two particles to be 
correlated at short distances. There are some indications
that it is not sufficient.

{\it iv) The effect of deformation.}\\
The nuclei undergoing double beta decay, which are  of experimental interest, 
are spherical or weakly deformed nuclei with exception of $^{150}Nd$,
which is strongly deformed. 
It was found in \cite{deform04} that deformation introduces a
mechanism of suppression of the $2\nu\beta\beta$-decay matrix element 
which gets stronger when deformations of the initial and final nuclei
differ from each other \cite{deform04,alvar04}.
A similar dependence of the suppression
of both $M^{2\nu}$ and $M^{0\nu}_\nu$ matrix elements on the difference
in deformations has been found in the PHFB \cite{phfb05,cha08} 
and the ISM \cite{lssm}. The NMEs have a well-defined maximum
when the deformations of parent and daughter nuclei
are similar, and they are quite suppressed when the
difference in the deformations is large. The ISM results 
suggest that a large mismatch of deformation
can reduce the  matrix elements by factors as large
as $2­3$. Within the IBM-2  the
effects of the deformation are introduced through the bosonic 
neutron-proton quadrupole interaction. For weakly deformed  
nuclei the effect is a reduction by about 20\%. 

The QRPA calculation of the $0\nu\beta\beta$-decay NMEs requires a 
construction of all states of the intermediate nucleus, even 
if closure approximation is considered. The results
were obtained in spherical limit, which  is a significant
simplication. Recently, the proton-neutron deformed QRPA with 
a realistic residual NN interaction was developed 
\cite{fang10,saleh09,saleh10,fang11}. This approach was applied
in the case of  $^{76}$Ge, $^{150}$Nd and $^{160}$Gd and lead to the  conclusion that
 the $0\nu\beta\beta$-decay of $^{150}$Nd, to be measured 
soon by the SNO+ collaboration,  provides one of the best probes 
of the Majorana neutrino mass \cite{fang10,fang11}. 

{\it v) The occupancies of individual orbits.}\\
The occupancies of valence neutron and proton orbits determined 
experimentally represent important constraints for nuclear models 
used in the evaluation of the $0\nu\beta\beta$-decay NME.
For the $^{76}$Ge and  ${^{76}}$Se they have been
extracted by accurate measurements of one nucleon
adding and removing transfer reactions by J. Schiffer and collaborators 
\cite{schiffer08,schiffer09}. The main motivation
to study these nuclei was the fact that they are the initial and final
states of $0\nu\beta\beta$-decay  transitions. These
measurements offer a possibility to compare 
these experimental results with the theoretical occupations and, 
if necessary, detect which modifications would be required 
in the mean field or the effective interaction
in order to obtain improved agreement with the experiment. 

In a theoretical study \cite{ocup09} measured proton and
neutron occupancies were used as a guideline for a modification 
of the effective mean field energies, which resulted in a better 
description of these quantities. The calculation of the 
$0\nu\beta\beta$-decay NME for $^{76}Ge$ performed with 
an adjusted Woods-Saxon mean field combined 
with the self-consistent RQRPA (SRQRPA) method \cite{ocup09}, 
which conserves the mean particle number in correlated ground state, 
led to a reduction of $M^{0\nu}_\nu$ by 20\%-30\% when
compared to the previous QRPA values.

In the ISM the variation of the nuclear matrix element 
(NME) for $0\nu\beta\beta$-decay of $^{76}$Ge was studied 
after the wave functions were constrained to reproduce 
the experimental occupancies of the two nuclei involved in the transition. 
It was found \cite{menen09} that in the ISM description the
value of the NME is enhanced about 15\% compared to previous calculations.
This diminishes the discrepancies between the ISM and the QRPA approaches.

The role of occupancies of the single-particle orbitals in the 
standard QRPA calculation of $M^{0\nu}_\nu$ to ground and   
$0^+_1$ states of final nucleus were studied also in \cite{suhoc08,suhoc11}. 
Unlike the treatment of \cite{ocup09}, whereby occupancies in respect to the correlated
SRQRPA ground state were considered, the occupancies were evaluated 
at the level of uncorrelated BCS ground state. The basic features of the 
ground and excited state decays were found to be quite different.

{\it vi)  The axial-vector coupling constant $g_A$.}\\
It is well known that the calculated strengths of Gamow­Teller 
$\beta$-decay transitions to individual final states  are 
significantly larger than the experimental ones. That effect is
known as the axial-vector current matrix elements quenching. 
To account for this, it is customary to quench the calculated 
GT matrix elements up to 70\%. Formally, this is accomplished by replacing 
the true value of the coupling constant $g_A = 1.269$ 
(the previously $g_A=1.254$ was considered) by a quenched value 
$g^{eff}_A =1.0$. The origin of the quenching is not completely
known. This effect is assigned to the $\Delta$-isobar admixture 
in the nuclear wave function or to the shift of the GT
strength to higher excitation energies due to the short-range 
tensor correlations. It is not clarified yet whether similar 
phenomenon exists for other multipoles, besides $J = 1^+$. 

Quenching is very important for the double beta decay because 
$g^{eff}_A$ appears to the fourth power in the decay rate. If it occurs 
also for the $0\nu\beta\beta$-decay, it could significantly reduce 
the $0\nu\beta\beta$-decay half-life by as much
as a factor of 2-3. The axial-vector coupling constant $g^{eff}_A$ 
or in other words, the treatment of quenching, is also a source of 
differences in the calculated $0\nu\beta\beta$-decay NMEs. 
${M'}^{0\nu}$  is a function of squared 
$g^{eff}_A$, which appears by vector and weak-magnetism terms
following the definition of Eq. (\ref{nmep}).

In \cite{lisi08} three independent lifetime data ($2\nu\beta\beta$-decay,
EC, $\beta$-decay) were accurately reproduced in the QRPA by means of 
two free parameters $(g_{pp}, g^{eff}_A)$, resulting in an overconstrained parameter space. 
The general trend in favor of $g^{eff}_A < 1$ was confirmed.
This novel possibility to reconcile QRPA results with experimental data, 
which deserves further discussions and tests, warrants a reconsideration 
of the quenching problem from a new perspective. 

As it was manifested above nuclear NMEs for $0\nu\beta\beta$-decay are affected by 
relatively large theoretical uncertainties. Within the QRPA approach, 
it was shown that, within a given set of nuclei, the correlations
among NME errors are as important as their size \cite{lisi09}. This 
represents a first attempt to quantify the covariance matrix of the NMEs, 
and to understand its effects in the comparison of current and prospective
$0\nu\beta\beta$-decay results for two or more nuclei. 
It would be useful if other theoretical groups in the $0\nu\beta\beta$ 
field could present ``statistical samples'' of NME calculations  as well, 
in order to provide independent estimates of (co)variances for their
NME estimates. A covariance analysis like the one proposed \cite{lisi09}
represents a useful tool to estimate correctly current or prospective sensitivities
to effective Majorana neutrino mass $\langle m_{\nu}\rangle$.

Currently, the uncertainty in calculated $0\nu\beta\beta$-decay NMEs 
can be estimated up to factor of 2 or 3
depending on the  considered isotope, mostly due to differences 
between the ISM results  and the results of other approaches 
(QRPA, PHFB, EDF, IBM) and also due to unknown value of $g^{eff}_A$.

\subsection{Anatomy of NMEs}

The anatomy of the $0\nu\beta\beta$-decay NME was performed in
\cite{lssm,sim08}. $M^{0\nu}_\nu$ was decomposed on the angular 
momenta and parities ${\mathcal J}^\pi$ of the pairs of neutrons 
that are transformed into protons with the same ${\mathcal J}^\pi$.
It was found that the final value of $M^{0\nu}_\nu$ reflects two 
competing forces: the like particle pairing interaction that 
leads to the smearing of Fermi levels and the residual 
neutron-proton interaction that, through ground state correlations, 
admixes "broken-pair" (higher-seniority) states. The function 
$C^{0\nu}(r)$ that describes the dependence of the $M^{0\nu}_\nu$
on internucleon distances r,
\begin{equation}
M^{0\nu}_\nu = \int_0^\infty C^{0\nu}(r) dr,
\end{equation}
was subject of interest. It was shown that the above competition implies 
that only  internucleon distances $r < 2-3$ fm contribute to $M^{0\nu}_\nu$ 
\cite{sim08}. The maximum value of $C^{0\nu}(r)$ occurs around r = 1 fm,
which means that almost the complete value of $M^{0\nu}_\nu$ comes from 
contribution of decaying nucleons which are close to each other. This 
distance correspond to a neutrino momentum of $q\approx 200$ MeV, twice
larger value as was expected before. This finding, which explains a small 
spread of results  for different nuclei, was confirmed also by the ISM 
\cite{lssm}  and a similar 
behavior for $C^{0\nu}(r)$ was obtained also within the PHFB \cite{phfb}.
The QRPA and ISM functions $C^{0\nu}(r)$ differ only by a scaling factor, 
which is expected to be related with the ratio of the average number of pairs 
in both calculations.

The largest component of  $M^{0\nu}_\nu$ is the GT part. We have
\begin{equation}
M^{0\nu}_\nu = M^{0\nu}_{GT} \left( 1 + \chi_F + \chi_T \right),
\end{equation}
where $\chi_F$ and $\chi_{GT}$ are matrix element ratios that are smaller 
than unity and, presumably, less dependent on the details of the applied
nuclear model. In \cite{0n2n11} it was shown that $M^{0\nu}_{GT}$ is related 
to the closure $2\nu\beta\beta$-decay  NME $M^{2\nu}_{cl}$. That 
relation is revealed when these matrix elements are expressed as functions of 
the relative distance between the pair of neutrons that are transformed into 
a pair of protons. We have 
\begin{eqnarray}
C^{0\nu}_{GT}(r) =  H_{GT}(r,\overline{E}) C^{2\nu}_{cl}(r), 
\label{0n2n}
\end{eqnarray}
where $H(r,\overline{E})$ is the neutrino exchange potential in nucleus
and  $C^{2\nu}_{cl}(r)$ is defined as 
\begin{eqnarray}
M^{2\nu}_{cl} &=& \int_0^\infty C^{2\nu}_{cl}(r) dr.
\end{eqnarray}
While the matrix element $M^{2\nu}_{cl}$ get contributions only from $1^+$
intermediate states, the function $C^{2\nu}_{cl}(r)$ gets contributions from
all intermediate multipoles. 

The Eq. (\ref{0n2n}) represents the basic relation 
between the $0\nu\beta\beta$- and $2\nu\beta\beta$-decay modes. 
An analysis of this relation allowed to explained the contrasting 
behavior of $M^{0\nu}_{GT}$ and $M^{2\nu}_{cl}$ when A and Z is changed, 
namely that $M^{0\nu}_{GT}$  changes slowly and smoothly unlike 
$M^{2\nu}_{cl}$, which has pronounced shell effects \cite{0n2n11}.

In \cite{vad_mf}  a connection of the Fermi $0\nu\beta\beta$-decay 
NME $M^{0\nu}_{F}$ with an energy-weighted double Fermi transition
matrix element was presented.  It is argued that $M^{0\nu}_{F}$
can be reconstructed, if the  
isospin-forbidden  Fermi transition between the ground state of the 
final nucleus and the isobaric analog state in the intermediate
nucleus can be measured, e.g. by means of (n,p) charge-exchange reactions.  
By knowing $M^{0\nu}_{F}$ one can evaluate $M^{0\nu}$ by assuming 
an approximate relation $M^{0\nu}_{F}/M^{0\nu}_{GT}\approx -2.5$, 
which follows from the QRPA calculations \cite{src09}.

\section[Distinguishing the various mechanisms]{Distinguishing the $0\nu\beta\beta$-decay mechanisms}
\label{sec:DistMec}

Many extensions of the SM generate Majorana neutrino masses 
and offer a plethora of $0\nu\beta\beta$-decay mechanisms.
Among these we should mention the exchange of heavy neutrinos, 
the exchange of SUSY superpartners with R-parity violation,
leptoquarks, right-handed W bosons, or Kaluza-Klein
excitations, among others, which have been discussed in
the previous sections or can be found in the  literature\cite{RODEJ11}.

An unambiguous  detection of  $0\nu\beta\beta$-decay will prove that
the total lepton number is broken in nature and neutrinos are 
Majorana particles. However, after neutrino oscillations have established that the neutrinos are massive, as we have already mentioned,  the observation of $0\nu\beta\beta$-decay is expected to play a crucial role in determining the neutrino mass scale. This prospect  generates the questions: 
What is the mechanism that triggers the decay? 
What happens if several mechanisms are active for the decay? 

\subsection{Dominance of a single mechanism}

Usually, the $0\nu\beta\beta$-decay is discussed by assuming that 
one mechanism at a time dominates. Then the half-life in a given 
nucleus $i \equiv (A,Z)$ can be written as
\begin{eqnarray}
\left(T^{0\nu}_{1/2}(i)\right)^{-1} ~= 
~|\eta_{\kappa}|^2~|{M'}^{0\nu}_{\kappa}(i)|^{2}~G^{0\nu}_{\kappa}(i). 
\label{lnv}
\end{eqnarray}
Here, $\eta_{\kappa}$, ${M'}^{0\nu}_{\kappa}$, $G^{0\nu}_{\kappa}(A,Z)$ 
are the LNV parameter ($\kappa$ denotes a given mechanism of the 
$0\nu\beta\beta$-decay), associated NME and kinematical factor, respectively.
The calculation of ${M'}^{0\nu}_{\kappa} = (g^{eff}_A/g_A)^2 {M}^{0\nu}_{\kappa}$ 
(${M}^{0\nu}_{\kappa}$ in some cases depends also on $g^{eff}_{A}$) allows to deduce
constraint on $\eta_{\kappa}$ from the measured lower bound on the 
$0\nu\beta\beta$-decay half-life.  The definition of ${M'}^{0\nu}_{\kappa}$ 
in (\ref{lnv}) allows to display the effects of uncertainties in
 $g^{eff}_A$ and to use  the same phase factor $G^{0\nu}_{\kappa}$ when calculating 
the $0\nu\beta\beta$-decay rate \cite{qrpa2,qrpa3}.

In connection with the neutrino oscillations  much attention 
is attracted  to the light neutrino mass mechanism of the 
$0\nu\beta\beta$-decay ($\eta_\nu = \langle m_\nu\rangle/m_e$) (see section \ref{sec:extrnumass}). 
Small neutrino masses and neutrino mixing are commonly considered 
as a signature of physics beyond the SM. Several beyond the SM 
mechanisms of neutrino-mass generation were proposed. The most viable and plausible
mechanism is the famous see-saw mechanism which is based
on the assumption that the total lepton number L is violated 
at a scale much larger than the electroweak scale. 

The $0\nu\beta\beta$-decay is ruled by the light Majorana 
 neutrino-mass mechanism in the case of the standard see-saw 
mechanism of neutrino-mass generation, 
which is based on the assumption that the lepton number is violated at a large 
($10^{15}$ GeV) scale. In \cite{bilpot11} it was shown that if 
$0\nu\beta\beta$-decay  will be observed in future experiments sensitive 
to the effective Majorana mass in the inverted mass hierarchy region, 
then a comparison of the derived ranges with measured half-lives
will allow us to probe the standard see-saw mechanism (see section  \ref{sec:numass}), assuming that 
future cosmological data will establish the sum of the neutrino masses 
to be about 0.2 eV .

A primary purpose of type I see-saw, see section  \ref{sec:numass}, which is the simplest extension of the SM, 
is to account for light neutrino masses in a renormalizable gauge model. 
Only heavy sterile  neutrino states are added to the spectrum of the 
$SU(3)_C \times SU(2)_L \times U(1)_Y$ theory. These heavy states might 
lead to measurable effects also for the $0\nu\beta\beta$-decay.
The possible contribution of sterile neutrino dominated Majorana mass 
eigenstate $\nu_h$ with mass $m_h$ to the $0\nu\beta\beta$-decay was 
examined in \cite{benes05}.
From the most stringent lower bound on the $0\nu\beta\beta$-decay 
half-life of $^{76}$Ge upper limits on the neutrino mixing matrix element 
$|U_{eh}|^2$ in wide region of values of $m_h$ (below and above TeV scale)
were derived. It was assumed that the value of $|\langle m_\nu\rangle|$ 
is significantly smaller than the current limit on this quantity 
($|\langle m_\nu\rangle| \ll 0.2-0.3$ eV). 

Recently, the $0\nu\beta\beta$-decay associated with the exchange of 
virtual sterile neutrinos, that mix with ordinary neutrinos and are heavier 
than 200 MeV \footnote{The case of light sterile neutrinos has 
already been discussed in section  \ref{sec:extrnumass}.}, was revisited \cite{sterile11}. The question of having 
a dominant heavy sterile neutrino contribution in $0\nu\beta\beta$-decay 
was explored in detail. Due to the improved result of the NMEs \cite{lisi11, FMPSV11}, 
the bounds on active-sterile mixing coming from $0\nu\beta\beta$-decay 
has become one order of magnitude stronger. The possibility that the sterile 
neutrino contribution become dominant over the light neutrino contribution
was addressed for the two flavor and the three flavor scenarios \cite{sterile11}. 
The dominant sterile neutrino contribution in $0\nu\beta\beta$ process provide 
a way to overcome the conflict between cosmology and the claim for evidence of 
the $0\nu\beta\beta$-decay by Klapdor and collaborators \cite{evidence1,evidence2}.
  
There is a possibility that the total lepton number is violated at TeV scale 
\cite{ibarra11,vissa11,AKP09a,AKP09b}, which is accessible at the Large Hadron Collider. 
The Large Hadron Collider can determine the right-handed neutrino masses and mixings. In \cite{vissa11}
it was manifested that the discovery of left-right (LR) symmetry at the Large Hadron Collider would 
provide a strong motivation for $0\nu\beta\beta$ searches. By 
exploiting the LR model with type-II see-saw (see section  \ref{sec:numass}) it was shown that 
the exchange of heavy neutrinos may dominate the $0\nu\beta\beta$-decay rate
 depending on the mass of right-handed charged gauge boson 
and the mixing of right-handed neutrinos \cite{vissa11} (see Eq. (\ref{eq:1.6})). A complementary 
study of lepton-flavor violating processes (e.g., $\mu \rightarrow e \gamma$),
which can provide constraints on masses of right-handed neutrinos and 
doubly charged scalars is of great importance  \cite{vissa11,egamma}.  

The LR symmetric models \cite{pasa74,mopa75,lrsm2} are popular models of particle physics 
due to restoration of parity at high energy scale and because they can naturally account 
for the smallness of neutrino masses. They allow not only the light and heavy neutrino 
mass mechanisms of the $0\nu\beta\beta$-decay  but also those associated with 
effective neutrino mass independent  parameters $\langle\eta\rangle$ and $\langle\lambda\rangle$, see section \ref{sec:etalam}. As it was showed
already before there is an exchange of light neutrinos between two $\beta$-decaying
nucleons in nucleus in this case. 

The three terms $\langle m_\nu\rangle$, $\langle\eta\rangle$ and $\langle\lambda\rangle$ in the $0\nu\beta\beta$-decay
rate  show different  characteristics in the angular correlations and energy spectrum 
\cite{DTK85}. By knowing the single electron energy spectrum and the angular correlation of the
two electrons with sufficient accuracy, one could distinguish between decays due
to coupling to the left handed and right-handed hadronic currents \cite{DTK85}. This 
possibility was studied in the context of $^{76}$Ge and SuperNEMO (Isotopes under 
consideration for SuperNEMO are $^{82}$Se, $^{150}$Nd and $^{48}$Ca \cite{deppisch11,ayrg11}) detectors. 
In \cite{klapmech1,klapmech2,klapmech3,klapmech4} the expected pulse shapes to be observed 
for the $0\nu\beta\beta$-decay events in a big $^{76}$Ge detector have been calculated 
starting from their Monte Carlo calculated time history and spatial energy distribution. 
The conclusion was  that with the spatial resolution of a large size Ge detector 
for the majority of $0\nu\beta\beta$ events it is not possible to
differentiate between the contributions of $\langle m_\nu\rangle$ and the right-handed weak 
current parameters $\langle\eta\rangle$ and $\langle\lambda\rangle$. Contrary, the SuperNEMO experiment \cite{deppisch11,ayrg11} 
has a unique potential to measure the decay electron's angular and energy distributions 
and thus to disentangle these possible mechanisms for
$0\nu\beta\beta$ decay \cite{deppisch11,deppisch10}. We note that the planned experiment 
SuperNEMO, which allows the measurement of $0\nu\beta\beta$-decay in several isotopes
to both the ground and excited states, is able to track the trajectories of the emitted 
electrons and determine their individual
energies. We should mention here that a measurement to both states can also be useful 
in reducing the background \cite{DuLindZub11}.  Other planned experiments that will be able to measure the energy and angular 
distributions are EXO \cite{ack11}, MOON \cite{mooncol} and COBRA \cite{cobracol}. 

There is a motivation to consider the $0\nu\beta\beta$-decay rate in a general framework,
parameterizing the new physics contributions in terms  of all effective low-energy currents 
allowed by Lorentz-invariance \cite{PHK-KK99,PHK-KK01}, e.g.  within the effective field theory 
\cite{PR-MV03}. This approach allows to separate the nuclear physics part of 
the $0\nu\beta\beta$-decay  from the underlying particle physics model, and derive limits 
on arbitrary lepton number violating  theories. A general Lorentz invariant effective Lagragian 
for leptonic and hadronic charged weak currents was used to perform a comparative analysis 
of various $0\nu\beta\beta$-decay long-range mechanisms in \cite{borisov07,ali10}. 
It was shown that by measuring of angular correlations of emitted electrons in the 
$0\nu\beta\beta$-decay together with the ability of observing these
decays in several nuclei, would help significantly in identifying
the dominant mechanism underlying this process.

There is a class of $0\nu\beta\beta$-decay mechanisms, which one cannot distinguish 
from each other kinematically. The light ($\eta_\nu$) and heavy ($\eta^L_N$, $\eta^R_N$) 
Majorana neutrino mass, the trilinear R-parity breaking mechanisms - both  
the short-range mechanism ($\eta_{\lambda'}$) with the exchange of heavy superpartners 
(gluino and squarks and/or  neutralinos and selectron) \cite{FKS98,Moh86,Ver87,FKSW97}
and the long-range mechanism ($\eta_{\tilde q}$) involving both the exchange of heavy 
squarks and light neutrino \cite{FGKS07}  (called squark-neutrino mechanism), 
constitute such a group. A discussed possibility to distinguish between these
mechanisms is a comparison of results for $0\nu\beta\beta$-decay in two or more 
isotopes \cite{fsdistin,bilratio,depaes07}.

Under the assumption of a dominance of a single mechanism of the $0\nu\beta\beta$-decay
the LNV  parameter $\eta_{\kappa}$ drops out in the ratio of experimentally determined 
half-lives for two different isotopes. This ratio depend on the mechanism of
the $0\nu\beta\beta$-decay due nuclear matrix elements and kinematical factors,
but is free of LNV parameter. Thus it can be compared with the theoretical prediction
for different mechanisms. In addition, it is assumed that in ratio of nuclear matrix 
elements theoretical uncertainties are reduced due to cancellations of systematic effects.
Relative deviations of half-life ratios for various new physics contributions, which
were normalized to the half-life of $^{76}$Ge and compared to the
ratio in the light neutrino mass mechanism, were studied in \cite{depaes07}. 
It was found that the change in ratios of half-lives varies from 60\% for supersymmetric models 
up to a factor of 5­20 for extra-dimensional and LR-symmetric mechanisms.
It is concluded that complementary measurements in different
isotopes would be strongly encouraged \cite{depaes07,gehman07}. 

Another possibility  to distinguish between the various $0\nu\beta\beta$-decay mechanisms is 
a study of the branching ratios of $0\nu\beta\beta$-decays to excited $0^+$  \cite{fsdistin,simk1p} 
and $2^+$ \cite{Tom91,tomo00} states and a comparative study of the $0\nu\beta\beta$-decay 
and neutrinoless electron capture with emission of positron ($0\nu EC\beta^+$) \cite{HMOK94}.
Unfortunately, the search for the $0\nu EC\beta^+$-decay is complicated due 
to small rates and the experimental challenge to observe the produced x rays 
or Auger electrons, and most double beta experiments of the next generation are not
sensitive to electron tracks or transitions to excited states.

\subsection{Nuclear matrix elements of exotic mechanisms}

Recently, the most interest was paid to the calculation of NMEs associated with the
light neutrino mass mechanism and ground state to ground state transition.
Less progress was achieved in the calculation of NMES of exotic 
$0\nu\beta\beta$-decay mechanisms.

Experimental studies of transitions to an excited $0^+_1$ and $2^+_1$  
final states \footnote{As we have already mentioned, transitions to  non zero angular momentum final states can only occur via the leptonic  $j_L,j_R$ interference term associated with the $\langle\lambda\rangle$ and $\langle\eta\rangle$ parameters. In this section we will refer to it as right handed current contribution.}   allow us to reduce the background by gamma-electron coincidences. 
Drawbacks are lower Q values and  suppressed nuclear matrix elements. 
The theoretical studies of the corresponding nuclear transitions were performed within the 
ISM \cite{lssm},  Hartree­Fock­Bogoliubov \cite{tomo00} and QRPA 
\cite{suhoc11,simk1p,suhex00a,suhex00b} approaches. In the ISM 
the $0\nu\beta\beta$-decays of ${^{48}}$Ca, ${^{76}}$Ge, ${^{82}}$Se, ${^{124}}$Sn, ${^{130}}$Te,  
and ${^{136}}$Xe  to $0^+_1$ excited final state were found at least 25 times 
more suppressed with respect to the ground state to  ground state transition
in the case of light neutrino mass mechanism. A similar conclusion was found 
for the $0\nu\beta\beta$-decays of $^{76}$Ge, $^{82}$Se, $^{100}$Mo and  $^{136}$Xe
to the excited collective $0^+$ state suggesting a suppression 10-100 larger than 
that of the transition to ground state \cite{suhoc11,simk1p,suhex00a}. In addition
to light neutrino mass also right-handed current \cite{suhex00a,suhex00b} and 
R-parity breaking mechanisms \cite{simk1p} were considered.

Quite the opposite is claimed  in a different study \cite{suhex00b}, namely it was found that the transition rate of the $0\nu\beta\beta$-decay  
of $^{76}$Zr to first excited $0^+$ state is favored by the enhanced transition matrix elements 
attributed to the monopole-vibrational structure of this state.

\begin{table}[!t]
  \begin{center}
\caption{\label{nmedifm} 
Nuclear matrix elements ${M'}^{0\nu}_\nu$ (light Majoran neutrino mass mechanism),
${M'}^{0\nu}_N$ (heavy Majorana neutrino mass mechanism), ${M'}^{0\nu}_{\lambda'}$ 
(trilinear R-parity breaking SUSY mechanism)  and 
${M'}^{0\nu}_{\tilde q}$ (squark mixing mechanism)
for the $0\nu\beta\beta$-decays of $^{76}$Ge, $^{82}$Se, $^{100}$Mo,
$^{130}$Te  and $^{136}$Xe  within the Selfconsistent
Renormalized Quasiparticle Random Phase Approximation (SRQRPA). 
R = 1.1 fm  $A^{1/3}$ is assumed.
}
\begin{tabular}{lccccccccc}
\hline\hline
 Nucleus & NN pot.&  $g^{eff}_A$ & $|{M'}^{0\nu}_\nu|$ &  & $|{M'}^{0\nu}_N|$ & &
 $|{M'}^{0\nu}_{\lambda'}|$ & & $|{M'}^{0\nu}_{\tilde q}|$ \\ 
 &  & & & & & & & &  \\ \hline
$^{76}$Ge &  Argonne  & 1.25 & 5.44 & & 265 & & 700 & & 718 \\
                          &          & 1.00 & 4.39 & & 196 & & 461 & & 476 \\
                          & CD-Bonn  & 1.25 & 5.82 & & 412 & & 596 & & 728 \\
                          &          & 1.00 & 4.69 & & 317 & & 393 & & 483 \\
$^{82}$Se  & Argonne  & 1.25 & 5.29 & & 263 & & 698 & & 710 \\
                          &          & 1.00 & 4.18 & & 193 & & 455 & & 465 \\
                          & CD-Bonn  & 1.25 & 5.66 & & 408 & & 594 & & 720 \\
                          &          & 1.00 & 4.48 & & 312 & & 388 & & 472 \\
$^{100}$Mo  &  Argonne & 1.25 & 4.79 & & 260 & & 690 & & 683 \\
                          &          & 1.00 & 3.91 & & 192 & & 450 & & 449 \\
                           & CD-Bonn  & 1.25 & 5.15 & & 404 & & 589 & & 691 \\
                          &          & 1.00 & 4.20 & & 311 & & 384 & & 455 \\
$^{130}$Te &  Argonne & 1.25 & 4.18 & & 240 & & 626 & & 620  \\
                           &           & 1.00 & 3.34 & & 177 & & 406 & & 403 \\
                           & CD-Bonn  & 1.25 & 4.70 & & 385 & & 540 & & 641 \\
                          &           & 1.00 & 3.74 & & 294 & & 350 & & 416 \\
$^{136}$Xe &   Argonne & 1.25 & 2.75 & & 160 & & 428 & & 418  \\
                          &           & 1.00 & 2.19 & & 117 & & 277 & & 271 \\
                           & CD-Bonn  & 1.25 & 3.36 & & 172 & & 460 & & 459 \\
                          &           & 1.00 & 2.61 & & 125 & & 297 & & 297 \\
\hline\hline
\end{tabular}
  \end{center}
\end{table}

The $0\nu\beta\beta$-decay of $^{76}$Ge and $^{100}$Mo  to $2^+_1$ final state was investigated
for light neutrino mass and right-handed current mechanisms by taking into account recoil 
corrections to the nuclear currents in \cite{tomo00}.
The initial $0^+$ and final $2^+_1$ nuclear states were described
in terms of the Hartree­-Fock-­Bogoliubov type wave functions, which were 
obtained by a variation after particle-number and angular-momentum projection 
\cite{tomo86}.  By the numerical calculation of relevant NMEs, 
it was found that the relative sensitivities of $0^+\rightarrow 2^+$ decays 
to $\langle m_\nu\rangle$ and $\langle\eta\rangle$ are comparable to those of $0^+ \rightarrow 0^+$ decays. 
At the same time it was noted that  the $0^+ \rightarrow 2^+_1$ decay 
is relatively more sensitive to $\langle\lambda\rangle$. We should remind the reader that the observation of $0^+\rightarrow 2^+$ transition does not establish the presence of right handed currents. For a more complete analysis one should consider not only the right handed currents, but supersymmetric contribution as well (see the comment at the end of section \ref{sec:nonumec})

The right-handed current mechanisms are associated with many different NMEs.
Within the  nuclear shell model they were evaluated just for $0\nu\beta\beta$-decay
of $^{48}Ca$ to the final ground state \cite{RCN95}. The VAMPIR approach 
was exploited to calculate them in the case of $^{76}$Ge \cite{tomo86}. 
Many calculations of NMEs related to right-handed current mechanisms were performed
within the QRPA and for all nuclei of experimental interest  
\cite{suhex00a,suhex00b,MBK89,khadki91,PSVF96}. However, they do not 
include recent improvements concerning the fixing of parameter space
of nuclear Hamiltonian \cite{qrpa2,qrpa3,sim08,qrpa1} 
and concerning the description of two-nucleon short-range  correlations \cite{src09}.

There is a revived interest to heavy neutrino mass ($\eta^{L,R}_N$)
and R-parity breaking supersymmetric ($\eta_{\lambda'}$, $\eta_{\tilde q}$)
mechanisms of the $0\nu\beta\beta$-decay. The NMEs governing these mechanisms were calculated
only within the QRPA \cite{WKS99,SPVF} with exception of the PHFB calculation for the heavy neutrino 
mass mechanism \cite{ratheavy}, which, however, neglects the role of induced hadron currents. 

Recently, nuclear matrix elements ${M'}^{0\nu}_\nu$ (light neutrino mass mechanism),
${M'}^{0\nu}_N$ (heavy neutrino mass mechanism), ${M'}^{0\nu}_{\lambda'}$ 
(trilinear R-parity breaking SUSY mechanism)  and 
${M'}^{0\nu}_{\tilde q}$ (squark-neutrino mechanism) were calculated for the $0\nu\beta\beta$-decay
of $^{76}$Ge, $^{82}$Se, $^{100}$Mo, $^{130}$Te and $^{136}$Xe  within  the 
SRQRPA \cite{FMPSV11,lisi11}. Unlike in previous calculations the particle-particle strength was 
adjusted to the $2\nu\beta\beta$-decay half-life and the two-nucleon short-range correlations  
derived from same potential as residual interactions, namely from the CD-Bonn and 
Argonne potentials \cite{src09}, were considered. These refinements affect mainly heavy neutrino
mass NMEs, which became significantly larger. In the case of NMEs related to 
LNV parameter $\eta_{SUSY}$ the finite nucleon size effect was taken into account, which 
plays an important role in the case of one-pion exchange. For large model space and quenched and unquenched
value of weak coupling constant NMEs of these four  mechanisms are presented in Table \ref{nmedifm}. 
We note that a large model space is important to describe reliably especially tensor matrix element
contribution to the full matrix element.

\begin{table}[t]
\begin{center}
\caption{Upper bounds on the lepton number violating parameters $\langle m_\nu\rangle$, $\eta^{L,R}_N$,
$\eta_{\lambda'}$ and $\eta_{\tilde q}$ deduced from current lower bounds on the half-life ($T^{0\nu-exp}_{1/2}$) 
of $0\nu\beta\beta$-decay for $^{76}$Ge \cite{bau99}, $^{82}$Se, $^{100}$Mo \cite{tre11}, 
$^{130}$Te \cite{te130} and $^{136}$Xe \cite{kamlandzen}.
Limits on \rp SUSY coupling $\lambda'_{111}$ and on the products of the trilinear \rp-couplings
$\lambda^{\prime}_{11k}\lambda^{\prime}_{1k1}$ (k=1,2,3) for $\Lambda_{SUSY}=100$ GeV are given 
by assuming that the gluino and squark masses and the trilinear soft SUSY breaking parameters 
are approximately equal to a common SUSY breaking scale $\Lambda_{SUSY}=100$ GeV.  Nuclear matrix 
elements calculated within the Selfconsistent Renormalized Quasiparticle Random Phase Approximation  
(CD-Bonn potential, $g_A = 1.25$, see Table \ref{nmedifm}) are considered.
}
\label{table.limits}
\renewcommand{\arraystretch}{1.6}
\begin{tabular}{lccccccccccc}\hline\hline
nucl. & $T^{0\nu-exp}_{1/2}$ & $|\langle m_\nu\rangle |$ & $|\eta_N^{L,R}|$ &  & $|\eta_{\lambda'}^{}|$ & $\lambda'_{111}$ &   
& $|\eta^{}_{\tilde q}|$ & $\lambda^{'}_{111}\lambda^{'}_{111}$ & $\lambda^{'}_{112}\lambda^{'}_{121}$ & $\lambda^{'}_{113}\lambda^{'}_{131}$\\
 & [years] & [eV] & $\times 10^{9}$ & & $\times 10^{9}$ & $\times 10^{4}$ & & $\times 10^{9}$ & $\times 10^{6}$ & $\times 10^{7}$ & $\times 10^{8}$ \\
\hline
$^{76}$Ge  & $1.9~10^{25}$ & 0.23 & 6.2 & & 4.3 & 1.2 & & 3.5
           & 6.3 & 3.3 & 1.4 \\
$^{82}$Se  & $3.2~10^{23}$ & 0.85 & 23. & & 16. & 2.3 & & 13.
           & 24. & 12. & 5.1 \\
$^{100}$Mo & $1.0~10^{24}$ & 0.41 & 10. & & 7.1 & 1.5 & & 6.0
           & 11. & 5.7 & 2.4 \\
$^{130}$Te & $3.0~10^{24}$ & 0.27 & 6.4 & & 4.5 & 1.2 & & 3.8
           & 6.9 & 3.6 & 1.5 \\
$^{136}$Xe & $5.7~10^{24}$ & 0.26 & 10. & & 3.7 & 1.1 & & 3.8
           & 6.8 & 3.5 & 1.5 \\
\hline\hline
\end{tabular}
\end{center}
\end{table}

The lepton number violating parameters $\langle m_\nu\rangle$, $\eta^{L,R}_N$,
$\eta_{\lambda'}$ and $\eta_{\tilde q}$ deduced from current lower bounds on the half-life ($T^{0\nu-exp}_{1/2}$) 
of $0\nu\beta\beta$-decay for $^{76}$Ge \cite{bau99}, $^{82}$Se, $^{100}$Mo \cite{tre11}, 
$^{130}$Te \cite{te130} and $^{136}$Xe \cite{kamlandzen} are shown in Table \ref{table.limits}. 
The SRQRPA NMEs of \ref{nmedifm}, in particular those evaluated with CD-Bonn potential 
and $g_A=1.25$, were considered. We see that upper limits on $|\langle m_\nu\rangle|$ 
and $|\eta_{\tilde q}|$ from 
CUORICINO ($^{130}$Te) \cite{te130} and KamLAND-Zen ($^{136}$Xe) \cite{kamlandzen} experiments
are already comparable with those from the Heidelberg-Moscow ($^{76}$Ge) experiment \cite{bau99}.
The running KamLAND-Zen experiment is even slightly more sensitive to the $0\nu\beta\beta$-decay signal
as already finished Heidelberg-Moscow experiment in the case of the gluino exchange mechanism.

The $\eta_{\lambda'}$ and $\eta_{\tilde q}$ parameters are related with the 
with the \rp-coupling $\lambda'_{111}$ and products of 
the trilinear \rp-couplings $\lambda^{'}_{11k}\lambda^{'}_{1k1}$ (k=1,2,3), respectively.
The current limits on them, presented in Table \ref{table.limits}, have been derived under 
the conventional simplifying assumptions. We assumed all the squark masses 
and the trilinear soft SUSY breaking parameters $A_d$ to be approximately 
equal to a common SUSY breaking scale $\Lambda_{SUSY}$. Thus we approximately 
have \cite{FGKS07} 
\begin{eqnarray}\label{lim}
\lambda^{\prime}_{11k} \lambda^{\prime}_{1k1} \leq \epsilon_k
\frac{1}{\sqrt{T^{0\nu - exp}_{1/2} G^{01}}}
\frac{1}{|{M'}^{0\nu}_{\tilde q}|}
\left(\frac{\Lambda_{SUSY}}{100 \mbox{GeV}}\right)^3
\end{eqnarray}
with $\epsilon_k=(1.8\times 10^3; 94.2; 3.9)$ calculated for the 
current quark masses $m_d = 9$ MeV, $m_s=175$ MeV and $m_b=4.2$~GeV.
In the case of gluino and neutralino \rp SUSY mechanisms of the
$0\nu\beta\beta$-decay we obtain
\begin{eqnarray}
\lambda^{\prime}_{111} &\leq& 1.8~\frac{1}{\sqrt{T^{0\nu - exp}_{1/2} G^{01}}}
\frac{1}{|{M'}^{0\nu}_{\lambda'}|}
\left(\frac{m_{\tilde q}}{100 \mbox{GeV}}\right)^2
\left(\frac{m_{\tilde g}}{100 \mbox{GeV}}\right)^{1/2}\nonumber\\
\lambda^{\prime}_{111} &\leq& 12.5~\frac{1}{\sqrt{T^{0\nu - exp}_{1/2} G^{01}}}
\frac{1}{|{M'}^{0\nu}_{\lambda'}|}
\left(\frac{m_{\tilde e}}{100 \mbox{GeV}}\right)^2
\left(\frac{m_{\chi}}{100 \mbox{GeV}}\right)^{1/2}
\end{eqnarray}
with $m_{\tilde q} \simeq m_{\tilde g} \simeq m_{\tilde e} \simeq m_{\chi} = \Lambda_{SUSY}$.
$m_{\tilde q}$, $m_{\tilde g}$, $m_{\tilde e}$ and  $m_{\chi}$ are masses of squark, gluino,
selectron and neutralino, respectively. This approximation is well motivated 
by the constraints from the flavor changing neutral currents.  

It goes without saying that the calculated NMEs of light neutrino mass and exotic mechanisms of 
the $0\nu\beta\beta$-decay depend on the assumption about the nuclear model.
In order to improve their reliability and reliability of the upper limits on the 
$0\nu\beta\beta$-decay lepton number violating parameters, further investigations are necessary. 

\subsection{Two or more competing mechanisms}

\begin{figure}[!t]
\begin{center}
\includegraphics[width=.72\textwidth,angle=0]{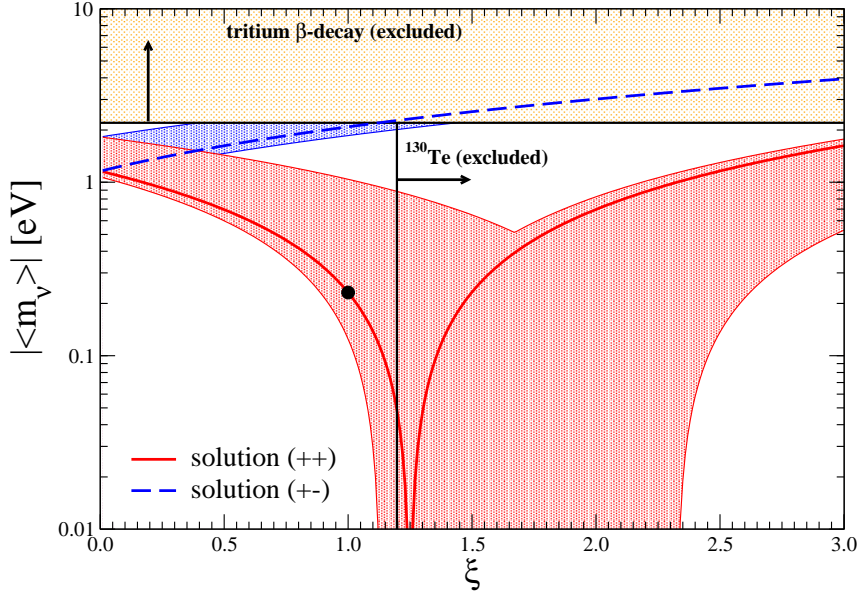}
\end{center}
\caption{(Color online) The effective Majorana mass of neutrinos in the case
of two active mechanisms of the $0\nu\beta\beta$-decay, 
namely light and heavy neutrino exchange mechanisms, as function
of parameter $\xi$ (see Eq. (\ref{xi})).  Here 
$T^{0\nu}_{1/2}(^{76}$\mbox{Ge}$) = 2.23^{+0.44}_{-0.31}\times 10^{25}~\mbox{y}$
\protect\cite{evidence1,evidence2} is assumed. Solutions were obtained 
for equal and opposite signs (++, +-) on the left hand side of Eqs. (\ref{2eqs}).
The bold point indicates the value of $|\langle m_\nu\rangle|$, if the
light neutrino exchange is the only active mechanism.
The dashed regions showed the uncertainty of the obtained predictions
for $|\langle m_\nu\rangle|$ if a $3\sigma$ experimental error of the measured half-lives 
is considered. We also see that a value of  $|\langle m_\nu\rangle| > 2.2 ~eV$ is excluded
due to Mainz tritium $\beta$-decay experiment \protect\cite{otten}.
} 
\label{comptm1}
\end{figure}

There is a general consensus that a measurement of the $0\nu\beta\beta$-decay 
in one isotope does not allow us to determine the underlying physics mechanism. 
Complementary measurements in different isotopes is very important especially
for the case there are competing mechanisms of the $0\nu\beta\beta$-decay.

In the case of coexisting mechanisms with identical phase space factors,
the Eq. (\ref{lnv}) is generalized as
\begin{eqnarray}
\left(T^{0\nu}_{1/2}(i)\right)^{-1} ~= 
~G^{0\nu}(i)~\sum_{\kappa} ~|\eta_{\kappa} ~{M}^{0\nu}_{\kappa}(i)|^{2}.
\label{lnvmore}
\end{eqnarray}
Here, $g^{eff}_A = g_A$ is assumed.
The parameters $\eta_{\kappa}$ may take either sign leading 
to constructive or destructive interference in the decay amplitude,
if CP conservation is assumed. In general case of CP violation
they include complex phases. By exploiting the fact that the associated 
nuclear matrix elements are target dependent, given definite experimental 
results on a sufficient number of targets, in principle one can determine 
or sufficiently constrain all LNV parameters including the light neutrino
mass term.

In \cite{lisi11} up to four coexisting mechanisms for the $0\nu\beta\beta$-decay, 
mediated by light Majorana neutrino exchange ($\eta_\nu$), heavy Majorana neutrino exchange 
($\eta^L_N$), R-parity breaking supersymmetry ($\eta_{\lambda'}$)), and
squark-neutrino ($\eta_{\tilde q}$) were considered. Both, constructive or destructive 
interference in the decay amplitude and the $0\nu\beta\beta$-decay in four different
candidate nuclei ($^{76}$Ge, $^{82}$Se, $^{100}$Mo, $^{130}$Te) with NMEs given 
in Table  \ref{nmedifm} were assumed. It was found that 
unfortunately, current NME uncertainties appear to prevent a robust determination of the
relative contribution of each mechanism to the decay amplitude, even assuming accurate 
measurements of  $0\nu\beta\beta$-decay lifetimes. 

Another important feature in analysis of two or more competing mechanisms 
was pointed out in \cite{SVF10}. For example, in the case of two active 
mechanisms represented by the LNV parameters $\eta_\nu = \langle m_\nu\rangle/m_e$ and $\eta^L_N$,
assuming the measurement of the $0\nu\beta\beta$-lifetime of two isotopes
($i = {^{76}}$Ge, ${^{130}}$Te) and CP conservation, one obtains 
four sets (phases $++, ~+-,~-+,~--$) of two linear equations:
\begin{eqnarray}
\frac{\pm 1}
{\sqrt{T^{0\nu}_{1/2}(i)~G^{0\nu}(i)}} = 
\frac{\langle m_\nu\rangle}{m_e} {M}^{0\nu}_{\nu}(i)  ~+~\eta^L_{N} {M}^{0\nu}_{N}(i).
\label{2eqs} 
\end{eqnarray}
It was found that this improved analysis leads to completely different
results compared to those of one mechanism at a time. 
By making additional assumption that the $0\nu\beta\beta$-decay of ${^{76}}$Ge 
was measured with half-life given in \cite{evidence1,evidence2} the 
two different solutions for $|\langle m_\nu\rangle|$ are plotted as function 
of $\xi$, where
\begin{equation}
\xi = \frac{|M^{0\nu}_\nu (^{130}\mbox{Te})|\sqrt{T^{0\nu}_{1/2}(^{130}\mbox{Te})~G^{0\nu}(^{130}\mbox{Te})}}
{|M^{0\nu}_\nu (^{76}\mbox{Ge})|\sqrt{T^{0\nu}_{1/2}(^{76}\mbox{Ge})~G^{0\nu}(^{76}\mbox{Ge})}},
\label{xi}
\end{equation}
in Fig. \ref{comptm1}. The parameter $\xi$ represents the unknown half-life of the 
$0\nu\beta\beta$-decay of $^{130}$Te. We note that for 
$\xi = 1$ the solution for active only light neutrino mass mechanism 
is reproduced and that $\xi = 0$
means non-observation of the $0\nu\beta\beta$-decay for a considered
isotope. By glancing the Fig. \ref{comptm1} the obtained results allows 
to conclude:\\
i) One of the solutions leads to small values of $|\langle m_\nu\rangle|$, when all mechanisms 
add up coherently. This is compatible also with inverted ($m_i < 50~ meV$) 
or normal ($m_i \approx$ few meV) hierarchy of neutrino masses.\\ 
ii) The second solution allows quite large values of $|\langle m_\nu\rangle|$, even larger than 
1 eV. It can be excluded by cosmology and tritium $\beta$-decay. However, 
if the claim for evidence will be ruled out by running (GERDA \cite{gerda,jochum10,schonert10}, 
EXO \cite{ack11}) and future experiments \cite{AEE08,Gomez12} the values of two solutions will become smaller and  perhaps it
will  not anymore be possible to exclude this solution. \\
iii) There is possibility that the non-observation of the $0\nu\beta\beta$-decay for 
some isotopes could be in agreement with a value of $|\langle m_\nu\rangle|$ in the sub-eV region.\\ 
iv) The obtained results are sensitive to the accuracy of measured
half-lives and to uncertainties in calculated nuclear matrix elements.

Other possibilities of getting information about the different LNV parameters
in the case of competing $0\nu\beta\beta$-decay mechanisms were discussed in
\cite{FMPSV11}. First, two competitive mechanisms, 
namely   light left handed Majorana neutrino exchange and heavy 
right-handed Majorana neutrino exchange, were considered. As the 
interference term is negligible the $0\nu\beta\beta$-decay half-life 
for a given isotope is written as
\begin{eqnarray}
\left(T^{0\nu}_{1/2}(i) G^{0\nu}(i) \right)^{-1} ~= ~
|\eta_{\nu} ~{M}^{0\nu}_{\nu}(i)|^{2} + |\eta^R_{N} ~{M}^{0\nu}_{N}(i)|^{2},
\end{eqnarray}
where the index i denotes the isotope. As we have mentioned  in subsection \ref{sec:etalam} the interference between the left and right handed currents is small. Given a pair of nuclei, 
solutions for $|\eta_\nu|^2$ and $|\eta^R_N|^2$ can be found by solving a
system of two linear equations. From the ``positivity'' conditions 
($|\eta_\nu|^2 > 0$ and $|\eta^R_N|^2 > 0$)  it follows that 
the ratio of  half-lives is within the range \cite{FMPSV11}
\begin{eqnarray}
\frac{ G^{0\nu}(i) |{M}^{0\nu}_{N}(i)|^{2}} 
{ G^{0\nu}(j) |{M}^{0\nu}_{N}(j)|^{2}} 
  \le \frac{T^{0\nu}_{1/2}(j)}{T^{0\nu}_{1/2}(i)} \le 
\frac{ G^{0\nu}(i) |{M}^{0\nu}_{\nu}(i)|^{2}} 
{ G^{0\nu}(j) |{M}^{0\nu}_{\nu}(j)|^{2}}. 
\end{eqnarray}
Surprisingly, the physical solutions are possible only 
if the ratio of the half-lives, in particular of three considered isotopes
${^{76}}$Ge, ${^{100}}$Mo and ${^{130}}$Te, takes values in very narrow 
intervals \cite{FMPSV11}.

The $0\nu\beta\beta$-decay can be triggered also by two
competitive mechanisms whose interference contribution
to the decay rates is non-negligible. As an example
the light Majorana neutrino mass  and gluino exchange   
mechanisms were considered in \cite{FMPSV11}. We have
\begin{eqnarray}
\left(T^{0\nu}_{1/2}(i) G^{0\nu}(i) \right)^{-1} ~&=& ~
|\eta_{\nu} ~{M}^{0\nu}_{\nu}(i)|^{2} + |\eta_{\lambda'} ~{M}^{0\nu}_{\lambda'}(i)|^{2}
\nonumber\\
&&+ 2 \cos{\alpha}|\eta_{\nu}| |\eta_{\lambda'}||{M}^{0\nu}_{\nu}(i)|  
|{M}^{0\nu}_{\lambda'}(i)|. 
\label{interfm}
\end{eqnarray}
Here, $\alpha$ is the relative phase of $\eta_{\nu}$ and $\eta_{\lambda'}$.
From (\ref{interfm}) it is possible to extract 
the values of $|\eta_{\nu}|^2$ and $|\eta_{\lambda'}|^2$ and $\cos{\alpha}$
setting up a system of three equations with these three unknowns using as input
the data on the half-lives of three different nuclei. Results of Ref. \cite{FMPSV11}
show that by using of prospective upper bounds on the absolute scale of neutrino
masses stringent constraints on some of new physics mechanisms, which interfere
destructively with light neutrino mass mechanism, can be found or even 
these scenarios can be excluded.

\section[Resonant neutrinoless double electron capture]{Resonant neutrinoless double electron capture}
\label{sec:twoecap}

As it has already been mentioned in section \ref{overview}, the resonant $0\nu$ECEC,
was considered by Winter \cite{WINTER} already in 1955 as a process that would 
demonstrate the Majorana nature of neutrinos and the violation of the total lepton number. 
The two asterisks denote the possibility of leaving the system in an excited nuclear and/or
atomic state. The energy excess given by the Q-value of the initial atom is carried away
by emission of x rays (or Auger electrons) as the daughter atom has two electron holes
and by emission of a single or few photons due to de-excitation of final nucleus. 

The possibility of a resonant enhancement of the $0\nu$ECEC in case of a mass 
degeneracy between the initial and final atoms was pointed out  by Bernab\'eu, De Rujula, 
and Jarlskog as well as by Vergados about 30 years ago \cite{Ver83,BeRuJar83}. 
The half-life of the process was estimated
by considering non-relativistic atomic wave functions at nuclear origin, simplified
evaluation of corresponding NME and assuming that the degeneracy parameter 
$\Delta = M_{A,Z} - M_{A,Z - 2}^{**}$,
being the difference of masses of the initial and final excited atoms 
with masses $M_{A,Z}$ and $M_{A,Z - 2}^{**}$, varies from zero to 10 keV (representing
the accuracy of atomic mass measurement at that time). The range of 
$\Delta$ induced uncertainty of about 5 orders in magnitude 
in calculated $0\nu ECEC$ half-life. 
A list of promising isotopes based on the degeneracy requirement associated with
arbitrary nuclear excitation was presented. The $^{112}Sn \rightarrow ^{112}Cd$ 
resonant $0\nu$ECEC transition  was identified as a good case.  

In 2004 Sujkowski and Wycech \cite{SujWy04} and Lukaszuk et al. \cite{lukas} analyzed the 
resonant $0\nu$ECEC  process for  nuclear $0^+ \to 0^+$ transitions accompanied 
by emission of a single photon. By assuming $|\langle m_\nu\rangle| = 1$ eV and 1 $\sigma$
error in the atomic mass determination the resonant $0\nu$ECEC rates 
of six selected isotopes were calculated by considering the perturbation theory approach. 
The lowest $0\nu$ECEC half-life was found for $^{152}$Gd. 

The main limitation in identifying promising isotopes for experimental search of 
$0\nu$ECEC has been poor experimental accuracy of measurement of $Q$-values which 
until recently were known with uncertainties of 1 - 10 keV only \cite{AUDI03}. 
The resonance enhancement can increase the probability of capture by many orders 
of magnitude. Therefore, accurate mass difference measurements are of great importance
in order to narrow down the possibilities. 
Progress in precision measurement of atomic masses with Penning traps 
\cite{BLAUM06,penning1,BNW10} has revived the interest in the old idea 
on the resonance $0\nu$ECEC capture. Recently, the accuracy of $Q$-values at around 
100 eV was achieved \cite{DEC11,redshaw07,redshaw09,SCIE09,RAKH09,kolhinen,mount10,elis2,EliNov,elis4,elis5,droese11},  
which has already allowed to exclude some of isotopes from the list of the most promising 
candidates (e.g., $^{112}$Sn and $^{164}$Er) for searching the $0\nu$ECEC. 

Recently, a significant progress has been achieved also in theoretical description 
of the resonant $0\nu$ECEC \cite{SIMKO11,SimKriv09,verg11}. A new theoretical framework
for the calculation of resonant $0\nu$ECEC transitions, namely the oscillation of 
stable and quasi-stationary atoms
due to weak interaction with violation of the total lepton number and parity,
was proposed in  \cite{SIMKO11,SimKriv09}. The $0\nu$ECEC transition rate near 
the resonance is of Breit-Wigner form, 
\begin{equation} 
\Gamma^{0\nu\mathrm{ECEC}}_{a b} (J^\pi) = \frac{\left| V_{a b}(J^\pi) \right|^2} 
{\Delta^2  
+ \frac{1}{4} \Gamma^2_{ab}} \Gamma_{ab}, 
\label{e:5} 
\end{equation} 
where $J^\pi$ denotes angular momentum and parity of final nucleus. 
The degeneracy parameter can be expressed as
$\Delta = Q - B_{ab} -E_\gamma$.  $Q$ stands for 
a difference between the initial and final atomic masses in ground states and
$E_\gamma$ is an excitation energy of the daughter nucleus. 
$B_{ a b} = E_{a} + E_{b} + E_C $ is the energy of two electron holes, 
whose quantum numbers $(n, j, l)$ are denoted by indices $a$ and $b$
and $E_C$ is the interaction energy of the two holes.
The binding energies of single electron holes $ E_{a}$ are 
known  with accuracy with few eV \cite{LARKINS}. The 
width of the excited final atom with the electron holes is given by
\begin{equation}
\Gamma_{ab} = \Gamma_{a} + \Gamma_{b} + \Gamma^*.
\end{equation}
Here, $\Gamma_{a, b}$ is one-hole atomic width 
and $\Gamma^*$ is the de-excitation width of daughter nucleus, which can be neglected.
Numerical values of $\Gamma_{ab}$ are about up to few tens eV \cite{CAMP}.

For light neutrino mass mechanism and favorable cases of a capture of $s_{1/2}$ and 
$p_{1/2}$ electrons the explicit form of lepton number violating amplitude 
associated with nuclear transitions $0^+ \rightarrow J^\pi = 0^{\pm 1}, 1^{\pm 1}$
is given in \cite{SIMKO11}. By factorizing the electron shell structure and nuclear 
matrix element one get
\begin{eqnarray}
V_{a b} (J^{\pi}) = 
\frac{1}{4 \pi}~ G^2_{\beta}m_e \eta_{\nu}  \frac{g^2_A}{R} 
<F_{a b}> M^{0\nu ECEC}(J^\pi).
\label{potential}
\end{eqnarray}
Here, $<F_{a b}>$ is a combination of averaged upper and lower bispinor 
components of the atomic electron wave functions \cite{SIMKO11}
and  $M^{0\nu ECEC}(J^\pi)$ is the nuclear matrix element. We note that 
by neglecting the lower bispinor components $M^{0\nu ECEC}(0^+)$ takes 
the form of the $0\nu\beta\beta$-decay NME for ground state to
ground state transition after replacing isospin operators $\tau^-$ 
by $\tau^+$. 

There is a straightforward generalization of the LNV potential $V_{a b} (0^+)$ 
in (\ref{potential}) for the heavy neutrino exchange,
the trilinear R-parity breaking with gluino and neutralino exchange and 
squark-neutrino mechanisms. It is achieved by replacements 
$\eta_\nu = \langle m_\nu\rangle/m_e$ with $\eta_{\kappa}$ ($\kappa = N, \lambda', {\tilde q}$)
and $M^{0\nu ECEC}_\nu (0^+)$ with $M^{0\nu ECEC}_\kappa (0^+)$. The $0\nu$ECEC leading to final states different than $0^+$, possible only in the presence 
 weak right-handed currents due to the leptonic of $j_L-j_R$ interference, has been discussed in \cite{verg11}.

New important theoretical findings with respect of the $0\nu$ECEC were 
achieved in \cite{SIMKO11}. They are as follows:
i) Effects associated with
the relativistic structure of the electron shells reduce the
$0\nu$ECEC half-lives by almost one order of magnitude.
ii) The capture of electrons from the $n p_{1/2}$ states is only 
moderately suppressed in comparison with the capture from the 
$n s_{1/2}$ states unlike in the non-relativistic theory.
iii) For light neutrino mass mechanism  selection rules appear 
to require that nuclear transitions with a change in
the nuclear spin $J \ge 2$ are strongly suppressed. We note
that, if right-handed currents are considered,
selection rules are modified allowing also $J\ne 0^+$ \cite{verg11}. 
iv) New transitions due to the violation of parity in 
the $0\nu$ECEC process were proposed. For example,
nuclear transitions $0^+ \rightarrow   0^\pm, 1^\pm$ 
are compatible with a mixed capture of s- and p-wave electrons.
v) The interaction energy of the two holes $E_C$ has to be taken 
into account by evaluating a mass degeneracy of initial and final
atoms.
vi) Based on the most recent atomic and nuclear data 
and by assuming $M^{0\nu ECEC}(J^\pi) = 6$ the 
$0\nu$ECEC half-lives were evaluated and the complete list of 
the most perspective isotopes for further experimental study 
was provided. Some isotopes such as $^{156}$Dy have several 
closely-lying  resonance levels. A more accurate measurement of 
Q-value of $^{156}$Dy by Heidelberg group confirmed the existence 
of multiple-resonance phenomenon for this isotope  \cite{EliNov}.
vii) In the unitary limit some $0\nu$ECEC half-lives 
were predicted to be significantly
below the $0\nu\beta\beta$-decay half-lives for the same value
of $\langle m_\nu\rangle$.  A probability of finding resonant transition 
with low $0\nu$ECEC half-life was evaluated. vii) The process of the
resonant neutrinoless double electron production ($0\nu$EPEP), i.e. neutrinoless double beta decay to two bound electrons, namely
\begin{equation}
(A,Z) \rightarrow (A,Z+2)^{**} + e^-_b + e^-_b,
\end{equation}
was proposed and analyzed. This process was found to be unlikely 
as it requires that a Q-value is extremely fine tuned to a nuclear
excitation. The two electrons must be placed into any of the
upper most non-occupied electron shells of the final atom leaving
only restricted possibility to match to a resonance condition.

\begin{table*}[!t] 
\centering 
\caption{A comparison of the neutrinoless double beta decay  and the resonant neutrinoless double electron
capture for light neutrino mass mechanism. The lepton number violating amplitude $V_{a b}(J^\pi)$ is given 
in Eq. (\ref{potential}).
}\label{table.ecec} 
\centering 
\renewcommand{\arraystretch}{1.1}  
\begin{tabular}{lcc} 
\hline\hline  
           & $0\nu\beta\beta$-decay &  $0\nu ECEC$ \\ \hline
definition & $(A,Z) \rightarrow (A,Z+2) + e^- + e^-$ & $(A,Z) + e^-_b + e^-_b \rightarrow (A,Z-2)^{**}$ \\
formalism  & perturbation field theory & oscillation of atoms \cite{SimKriv09,SIMKO11} \\

half-life  & $\frac{1}{T^{0\nu}_{1/2}} = \left|\frac{\langle m_\nu\rangle}{m_e}\right|^2 G^{0\nu} |M^{0\nu}(J^\pi)|^2$ & 
$\frac{\ln{2}}{T^{0\nu\mathrm{ECEC}}_{1/2}} = \frac{\left| V_{a b}(J^\pi) \right|^2} 
{(M_{A,Z} - M_{A,Z - 2}^{**})^2 + \frac{1}{4} \Gamma^2_{ab}} \Gamma_{ab}$ \\
nucl. trans. & $0^+ \rightarrow 0^+, 2^+$ & $0^+ \rightarrow 0^+, 0^-, 1^+, 1^-$ \\
fav. at. syst.   & large Q-value (3-4 MeV)  & mass difference \\
                    &                          & of few tens of eV \\
                    & $^{48}$Ca, $^{76}$Ge, $^{76}$Se, $^{100}$Mo,  &  unknown yet ($^{106}$Cd, $^{124}$Xe, \\ 
                    & $^{116}$Cd, $^{130}$Te, $^{136}$Xe, $^{150}$Nd, & $^{152}$Gd, $^{156}$Dy, $^{168}$Yb,... \cite{SIMKO11})\\ 
uncert. in $T^{0\nu}_{1/2}$ & factor $\sim 4-9 $ due & many orders in magn. \\
                               & to calc. of NME        & up to measured mass diff.\\
                               &                     & and due to NMEs \\
exp. sign. & peak at end of sum   &  x rays or Auger el.\\
           & of two el. energy spectra    & plus nucl. de-excitation\\
$T^{0\nu - exp}_{1/2}$ & $> 10^{24}$-$10^{25}$ y & $> 10^{19}$-$10^{20}$ y \\ 
exp. act.  & const. of (0.1-1 ton) exp. &  small exper. yet   \\
               & with sensitivity to inverted  &      \\
               & hierarchy of neutrino masses  &      \\
background & $2\nu\beta\beta$-decay & $2\nu ECEC$ is strongly \\
           & upon resolution of exp.          & suppressed \\
\hline 
\hline 
\end{tabular} 
\end{table*} 

A detailed calculation of the $0\nu$ECEC of $^{152}\mathrm{Gd}$, $^{164}\mathrm{Er}$ and $^{180}\mathrm{W}$  
associated with the ground-state to ground-state nuclear transitions was performed in \cite{DEC11,SKF11,FBE11}.
Improved measurements of Q-value for these transitions with accuracy of about 100 eV  \cite{elis4,droese11,elis5,SKF11}
were considered. The nuclear matrix elements of
${^{152}\mathrm{Gd}}\rightarrow {^{152}\mathrm{Sm}}$, ${^{164}\mathrm{Er}}\rightarrow {^{164}\mathrm{Dy}}$ and  
${^{180}\mathrm{W}} \rightarrow {^{180}\mathrm{Hf}}$ transitions were calculated within spherical and 
deformed QRPA \cite{SKF11,FBE11}. The obtained results excludes $ ^{164} $Er and $ ^{180} $W from the list 
of prospective candidates to search for the $0\nu$ECEC. The  $0\nu$ECEC half-life of $^{152}$Gd is 2-3 orders 
of magnitude longer than the half-life of $0\nu\beta\beta$ decay of $^{76}$Ge corresponding to the same value of 
$\langle m_\nu\rangle$ and is the smallest known half-life among known $0\nu$ECEC transitions at present.

The transition of $^{106}$Cd to an excited state of $^{106}$Pd with  
the nuclear excitation energy of 2717.59 keV was calculated in Ref. \cite{suho2011}
by making assumption that this is the $0^+$ state. However, it was noted in \cite{SIMKO11}
that, as long as this level $\gamma$-decays by 100\% into the $3^+$ state at 1557.68 keV, 
this possibility is excluded. 

There is also an increased experimental activity in the field of the resonant $0\nu$ECEC
\cite{BELLI11,barab1,barab2,belli09,rukh11,FPS11,BELLI11b}. The resonant $0\nu$ECEC
has some  important advantages with respect to experimental signatures and background 
conditions.  The de-excitation of the final excited nucleus proceeds in most cases through 
a cascade of easy to detect  rays. A two- or even higher-fold coincidence setup can cut 
down any background rate right from the beginning,  thereby requiring significantly less 
active or passive shielding \cite{SIMKO11}.  A clear detection of these $\gamma$ rays 
would already signal the resonant $0\nu$ ECEC without any doubt, as there are no 
background processes feeding those particular nuclear levels. It is worth  noting
that lepton number conserving  ECEC with emission of two neutrinos, 
\begin{equation}
(A,Z) + e^-_b + e^-_b \rightarrow (A,Z-2)^{**} + \nu_e + \nu_e,
\end{equation}
is strongly suppressed due to almost vanishing phase space \cite{Ver83,BeRuJar83,SIMKO11}.
The ground state to ground state resonant $0\nu$ECEC transitions can be 
detected by monitoring the x rays or Auger electrons emitted from excited electron 
shell of the atom. This can be achieved, e.g.,  by calorimetric measurements. 

Till now, the most stringent limit on the resonant $0\nu$ECEC were established 
for $^{74}$Se \cite{FPS11}, $^{106}$Cd \cite{rukh11} and $^{112}$Sn \cite{barab2}.
The ground state of $^{74}$Se is almost degenerate to the second excited
state at 1204 keV in the daughter nucleus $^{74}$Ge, which is a $2^+$ state \cite{frekdeg}.
The $2\gamma$-ray cascade has been searched for by using the low-radioactivity detector 
setup at the Comenius University in Bratislava and 3 kg of natural selenium.
A lower limit for the half-life of $T^{0\nu ECEC}_{1/2} \ge 4.3\times 10^{19}$ y 
was determined \cite{FPS11}, which is slightly larger than the value reported in
\cite{barab1}. The resonant $0\nu$ECEC transition to the $0^+_3$ excited state
in $^{112}$Cd (1871.0 keV) has been investigated in an experiment performed 
with natural tin in the Modane Underground Laboratory. A lower bound on half-life 
of $0.92\times 10^{20}$ y was established. It is worth  noticing that 
a new mass measurement \cite{RAKH09} has excluded a complete mass degeneracy for a $^{112}$Sn 
decay and has therefore disfavored significant resonant enhancement of the $0\nu$ECEC
mode for this transition. Within the TGV experiment in Modane \cite{rukh11} 
an interest 
has also arisen in   the $0\nu$ECEC resonant decay mode of $^{106}$Cd (KL-capture) 
to the excited 2741 keV state of $^{106}$Pd. The spin value of this final state 
was unknown  and it was assumed to be $J =(1, 2)^+$. After measurements had begun 
a new value for the spin of the 2741 keV level in $^{106}$Pd of $J = 4^+$ was adopted,
but , following recent theoretical analysis  \cite{SIMKO11}, this channel is now disfavored.
Nevertheless the most stringent limit on the $0\nu$ECEC half-life of 
of $1.1\times 10^{20}$ y was reported \cite{rukh11}.

A comparison of the $0\nu$ECEC with the $0\nu\beta\beta$-decay is presented in 
Table \ref{table.ecec}. It is maintained that these two lepton number violating
processes are quite different and at different levels of both theoretical and experimental
investigation. Precise measurements of Q-values between the initial and final
atomic states, additional spectroscopic information on the excited nuclear states 
(energy, spin and parity) and reliable calculation of corresponding NMEs are highly 
required to improve predictions of half-lives of the resonant $0\nu$ECEC.  It is 
expected that the accuracy of 10 eV in the measurement of atomic masses will be 
achievable in the near future. The electron binding energy depends on the local 
physical and chemical environment. An interesting question is whether it possible and, if so, how 
to manage the atomic structure in such a way as to  implement 
the degeneracy of the atoms and create conditions for the resonant
enhancement, as discussed in a recent work \cite{SIMKO11}.

\section[Concluding remarks]{Concluding remarks}

In this review we discussed in some detail the lepton number violating neutrinoless double beta decay and other similar transitions, involving various nuclear isotopes for which ordinary beta decay and e-capture are forbidden or highly suppressed. Both theoretical and experimental aspects were considered. 

 We have seen that this is a process with long and interesting history with important implications for physics and cosmology, but its observation is still elusive. It is an exotic process, which requires physics beyond the SM. At present a complete theory is missing and, thus, to motivate and guide the experiments we examined a number of  reasonable viable models, beyond the SM, in particular in connection with the neutrino mass matrix and mixing (see sections \ref{sec:numass} and \ref{sec:numixing}). Such models predict that lepton number violation, and consequently neutrinoless double beta decay, must occur at some level, implying that the neutrinos are Majorana particles. These models, however, cannot provide a precise determination of the parameters involved, such as the absolute scale of the neutrino mass. So they must be extracted from the experiments, if and when reliable accurate results become available (see section \ref{sec:extrnumass} for the neutrino mass).  The observed values may, then, be used to differentiate between such models and, hopefully, lead to the ultimate theory.
 
In order to achieve this goal first such processes must be definitely observed. Then the obtained results must be analyzed by considering the various mechanisms implied by the above models, see sections \ref{sec:numec} and \ref{sec:nonumec} for mechanisms involving intermediate neutrinos and other particles respectively. This, however, can only be done, if the corresponding nuclear matrix elements are evaluated with high precision,  accuracy and reliability. We have  seen that this is a formidable task, since the nuclei that can undergo double beta decay have rather complicated structure. 

The evaluation of the nuclear matrix elements involves two steps. In the first step the effective transition operators for each mechanism (see section \ref{sec:TranOper}). Special attention must be paid to the proper treatment of these operators at short distances (short range correlations, nucleon current corrections, inclusion of hadrons other than nucleons etc). The second step consists of selecting the proper nuclear model for constructing the wave functions involved in the evaluation of the nuclear matrix elements. Practically all models available in the nuclear theory artillery have been employed. The most prominent are the large basis shell model, the various refinements of the quasi particle random phase approximation (QRPA) and the interacting boson model (IBM). The essential features of these models and the numerical values of the obtained nuclear matrix elements have been summarized in section \ref{sec:0nuME}. We have seen that great progress has been made in this direction in recent years and it is encouraging that the  nuclear ME obtained with these vastly different nuclear models tend to converge. 

We have discussed in section \ref{sect:exp} the ongoing, planned and future experiments. We have witnessed great progress in  tackling the various background problems, improving the energy resolution and preparing large  masses of the needed isotopes. It is, thus, expected that half lives of the order of $10^{26}$y can be achieved and, consequently, a sensitivity of a few tens of meV for the average neutrino mass can be reached. This may be sufficient to differentiate between the normal and inverted hierarchy scenarios (see section \ref{sec:extrnumass}). Furthermore we have seen that various nuclear charge changing nuclear reactions can be employed in an effort to experimentally extract useful information or provide checks for the nuclear matrix elements.

 It is clear that the observation of neutrinoless double beta decay will be a great triumph for physics and experimental physics in particular. It will demonstrate that the neutrinos are Majorana particles and there exist lepton number violating interactions in the universe. This, however, will not be the end of the story. The data should be analyzed in such a way to determine the mechanism responsible for this process and, in particular, to extract the most important parameter, which is scale of the neutrino mass. Great progress has been in this direction has recently been made as briefly exposed in section \ref{sec:DistMec}. In order to unambiguously accomplish this goal, however, the accuracy of the nuclear matrix elements must be further improved.
 
 Finally recent developments, towards  the accurate determination of atomic masses as well as the evaluation of the inner shell atomic wave functions and energies, have stimulated interest in experiments involving the resonant neutrinoless double e-capture, see section \ref{sec:twoecap}. This new process,  if observed, especially in case it leads to negative parity final nuclear states, will greatly facilitate the analysis of determining the dominant mechanism involved in neutrinoless double beta decay.
 
 It is clear that theoretical attempts in the determination of the nuclear matrix elements and experimental efforts towards achieving the observation of neutrinoless  double beta decays, involving as many as possible nuclear isotopes and utilizing all available techniques, should be encouraged and supported.
\section*{Aknowledgements}
The work of one of the authors (JDV) was supported in part by UNILHC PITN-GA-2009-237920 and the DIBOSON Thalis project.
F.\v S acknowledges the support by the VEGA Grant agency  under the contract
No.~1/0876/12. The authors express their sincere thanks to Rastislav Dvornicky and
  Rastislav Hodak for the preparation of some of the figures.
  
		

\section*{Bibliography}
\begin{thebibliography}{100}

\bibitem{GOEPMAY}
M.~Goeppert-Mayer.
\newblock {\em Phys. Rev.}, 48:512, 1035.

\bibitem{emajorana}
E.~Majorana.
\newblock {\em Nuovo Cim.}, 14:171, 1937.

\bibitem{racah}
G.~Racah.
\newblock {\em Nuovo Cim.}, 14:322, 1937.

\bibitem{Fur39}
W.~Furry.
\newblock {\em Phys. Rev.}, 56:1184, 1939.

\bibitem{Pr52}
H.~Primakoff.
\newblock {\em Phys. Rev.}, 85:888, 1952.

\bibitem{barabhis}
A.~Barabash.
\newblock {\em Phys. Atom. Nucl.}, 74:603, 2011.

\bibitem{davisno}
Jr. R.~Davis.
\newblock {\em Phys. Rev.}, 97:766, 1955.

\bibitem{fire1}
E.~Fireman.
\newblock {\em Phys. Rev.}, 75:323, 1949.

\bibitem{fire2}
E.~Fireman.
\newblock {\em phys. Rev.}, 86:451, 1952.

\bibitem{ing}
M.~G. Ingram and J.~H. Reynolds.
\newblock {\em Phys. Rev.}, 78:822, 1950.

\bibitem{kir67}
T.~Kirsten, W.~Gentner, and O.A. Schaeffer.
\newblock {\em Z. Physik}, 202:273, 1967.

\bibitem{kir67b}
T.~Kirsten, W.~Gentner, and O.~Miller.
\newblock {\em Z. Naturf. A}, 22:1783, 1967.

\bibitem{tak66}
N.~Takaota and K.~Ogata.
\newblock {\em Z. Naturf. A}, 21:84, 1966.

\bibitem{sir72}
B.~Srinvasan, O.K. E.C.~Alexander, and Manuel.
\newblock {\em J. Inorg. Nucl. Chem.}, 34:2381, 1972.

\bibitem{G58}
M.~Goldhaber, L.~Grodzins, and A.W. Sunyar.
\newblock {\em Phys. Rev.}, 109:1015, 1958.

\bibitem{DTNOT}
M.~Doi, T.~Kotani, N.~Nishiura, K.~Okuda, and E.~Takasugi.
\newblock {\em Phys. Lett B.}, 103:219, 1981.

\bibitem{pasa74}
J.~C. Pati and A.~Salam.
\newblock {\em Phys. Rev. D}, 10:275, 1974.

\bibitem{mopa75}
R.~N. Mohapatra and J.~C. Pati.
\newblock {\em Phys. Rev. D}, 11:2558, 1975.

\bibitem{mose75}
R.~N. Mohapatra and G.~Senjanovi\'c.
\newblock {\em Phys. Rev. D}, 12:1502, 1975.

\bibitem{fri}
H.~Fritzsch and R.~Minkowski.
\newblock {\em Phys. Rep.}, 73:67, 1981.

\bibitem{kumo93}
R.~Kuchimanchi and R.~N. Mohapatra.
\newblock {\em Phys. Rev. D}, 48:4352, 1993.

\bibitem{aula98}
A.~Rasin C.~S.~Aulakh, A.~Melfo and G.~Senjanovic.
\newblock {\em Phys. Rev. D}, 58:115007, 1998.

\bibitem{dumo99}
B.~Dutta and R.~N. Mohapatra.
\newblock {\em Phys. Rev. D}, 59:015018, 1999.

\bibitem{SVa82}
J.~Schechter and J.~W.~F. Valle.
\newblock {\em Phys. Rev. D}, 25:2951, 1982.

\bibitem{klko96}
H.V. Klapdor-Kleingrothaus M.~Hirsch and S.~Kovalenko.
\newblock {\em Phys. Lett. B}, 372:181, 1996.

\bibitem{FKSS97}
A.~Faessler, S.~Kovalenko, F.~{\v S}imkovic, and J.~Schwieger.
\newblock {\em Phys. Rev. Lett.}, 78:183, 1997.

\bibitem{FKS98}
A.~Faessler, S.~Kovalenko, and F.~{\v S}imkovic.
\newblock {\em Phys. Rev. D}, 58:055004, 1998.

\bibitem{WKS99}
A.~Wodecki, W.~A. Kami{\' n}ski, and F.~{\v S}imkovic.
\newblock {\em Phys. Rev. D}, 60:115007, 1999.

\bibitem{zdes02}
V.~I. Tretyak and Yu.~G. Zdesenko.
\newblock {\em At. Dat. Nucl. Dat. Tabl.}, 80:83, 2002.

\bibitem{eji05}
H.~Ejiri.
\newblock {\em J. Phys. Soc. Jap.}, 74:2101, 2005.

\bibitem{AEE08}
F.T. Avignone, S.R. Elliott, and J.Engel.
\newblock {\em Rev. Mod. Phys.}, 80:481, 2008.

\bibitem{ell87}
S.~R. Elliott, A.~A. Hahn~A A, and M.~Moe.
\newblock {\em Phys. Rev. Lett.}, 59:2020, 1987.

\bibitem{SUPERKAMIOKANDE}
Y. Fukuda {\it et al}, The Super-Kamiokande Collaboration, {\it Phys. Rev.
  Lett.} {\bf 86}, (2001) 5651; {\it ibid} {\bf 81} (1998) 1562 $\&$ 1158; {\it
  ibid} {\bf 82} (1999) 1810 ;{\it ibid} {\bf 85} (2000) 3999.

\bibitem{SOLAROSC}
Q.R. Ahmad {\it et al}, The SNO Collaboration, {\it Phys. Rev. Lett.} {\bf 89}
  (2002) 011302; {\it ibid} {\bf 89} (2002) 011301; {\it ibid} {\bf 87} (2001)
  071301;\\ K. Lande {\it et al}, Homestake Collaboration, {\it Astrophys, J}
  {\bf 496}, (1998) 505;\\ W. Hampel {\it et al}, The Gallex Collaboration,
  {\it Phys. Lett. B} {\bf 447}, (1999) 127;\\ J.N. Abdurashitov {\it al}, Sage
  Collaboration, {\it Phys. Rev. C} {\bf 80} (1999) 056801;\\ G.L Fogli {\it et
  al}, {\it Phys. Rev. D} {\bf 66} (2002) 053010.

\bibitem{KAMLAND}
K.~Eguchi et~al. (The KamLAND~Collaboration).
\newblock {\em Phys. Rev. Lett.}, 90:021802, 2003.

\bibitem{bau99}
L.~Baudis et~al.
\newblock {\em Phys. Rev. Lett.}, 83:41, 1999.

\bibitem{nemoiii05}
R.~Arnold and others (NEMO3~Collaboration).
\newblock {\em Phys. Rev. Lett.}, 95:182302, 2005.

\bibitem{tre11}
V.I. Tretyak and The~NEMOIII collaboration.
\newblock {\em AIP Conf. Proc.}, 1417:125, 2011.

\bibitem{te130}
C.~Arnaboldi et~al. (CUORE~Collaboration).
\newblock {\em Phys. Lett. B}, 584:260, 2004.

\bibitem{kamlandzen}
A.~Gando et~al.
\newblock {\em Phys. Rev. C}, 85:045504, 2012.

\bibitem{evidence1}
H.L.~Harney H.~V. Klapdor-Kleingrothaus, A.~Dietz and I.~V. Krivosheina.
\newblock {\em Mod. Phys. lett. A}, 16:2409, 2001.

\bibitem{evidence2}
H.~V. Klapdor-Kleingrothaus and I.V. Krivosheina.
\newblock {\em Mod. Phys. Lett. A}, 21:1547, 2006.

\bibitem{gerda}
I. Abt {\it et al.} (GERDA Collaboration), arXiv:0404039[hep-ex].

\bibitem{Ver86}
J.D. Vergados.
\newblock {\em Phys. Rep.}, 133:1, 1986.

\bibitem{BilPet87}
S.~M. Bilenky and S.P. Petcov.
\newblock {\em Rev. Mod. Phys.}, 59:671, 1987.

\bibitem{KING04}
S.F. King.
\newblock {\em Rep. Prog. Phys.}, 67:107, 2004.

\bibitem{MohSmyr06}
R.N. Mohapatra et~al.
\newblock {\em Ann. Rev. Nucl. Part. Sci.}, 56:569, 2006.

\bibitem{MohEtal07}
R.N. Mohapatra et~al.
\newblock {\em Rep. Prog. Phys.}, 70:1757, 2007.

\bibitem{eji10}
H.~Ejiri.
\newblock {\em Progress Particle Nuclear Physics}, 64:249, 2010.

\bibitem{WINTER}
R.~Winter.
\newblock {\em Phys. Rev.}, 100:142, 1955.

\bibitem{Ver83}
J.D. Vergados.
\newblock {\em Nuc. Phys. B}, 218:109, 1983.

\bibitem{BeRuJar83}
J.~Bernabeu, A~de~Rujula, and C.~Jarlskog.
\newblock {\em Phys. Rev. C}, 223:15, 1983.

\bibitem{DK93}
M.~Doi and T.~Kotani.
\newblock {\em Prog. Theor. Phys.}, 89:139, 1993.

\bibitem{SIMKO11}
M.I. Krivoruchenko, F.~{\v S}imkovic, D.~Frekers, and A.~Faessler.
\newblock {\em Nucl. Phys. A}, 859:140, 2011.

\bibitem{DEC11}
S.~Eliseev et~al.
\newblock {\em Phys. Rev. Lett.}, 106:052504, 2011.

\bibitem{BLAUM06}
K.~Blaum.
\newblock {\em Phys. Rep.}, 425:1, 2006.

\bibitem{PR81}
H.~Primakoff and S.P. Rosen.
\newblock {\em Annu. Rev. Nucl. Part. Sci}, 31:145, 1981.

\bibitem{DKFS04}
P.~Domin, S.~Kovalenko, A.~Faessler, and F.~{\v S}imkovic.
\newblock {\em Phys. Rev. C}, 70:065501, 2004.

\bibitem{PSI}
C.~Dohmen et~al. (SINDRUM II~Collaboration).
\newblock {\em Phys. Lett. B}, 317:631, 1993.

\bibitem{MMM94}
J.H. Missimer, R.N. Mohapatra, and N.C. Mukhopadhyay.
\newblock {\em Phys. Rev. D}, 50:2067, 1994.

\bibitem{SFKS02}
F.~{\v S}imkovic, A.~Faessler, S.~Kovalenko, and I.~Schmidt.
\newblock {\em Phys. Rev. D}, 66:033005, 2002.

\bibitem{HS84}
W.C. Haxton and G.S.~Stephenson Jr.
\newblock {\em Prog. Part. Nucl. Phys.}, 12:409, 1984.

\bibitem{DTK85}
M.~Doi, T.~Kotani, and E.~Tagasugi.
\newblock {\em Prog. Theor. Phys. (Supp.)}, 83:1, 1985.

\bibitem{Tom91}
T.~Tomoda.
\newblock {\em Rep. Prog. Phys.}, 54:53, 1991.

\bibitem{SC98}
J.~Suhonen and O.~Civitarese.
\newblock {\em Phys. Rep.}, 300:123, 1998.

\bibitem{FS98}
A.~Faessler and F.~{\v S}imkovic.
\newblock {\em J. Phys. G}, 24:2139, 1998.

\bibitem{Ver02}
J.D. Vergados.
\newblock {\em Phys. Rep.}, 361:1, 2002.

\bibitem{RODEJ11}
W.~Rodejohann.
\newblock {\em Int. J. Mod. Phys. E}, 20:1833, 2011.

\bibitem{SujWy04}
Z.~Sujkowski and S.~Wycech.
\newblock {\em Phys. Rev. C}, 70:052501, 2004.

\bibitem{SimKriv09}
F.~{\v S}imkovic and M.I Krivoruchenko.
\newblock {\em Phys. Part. Nuc.}, 6:298, 2009.

\bibitem{BELLI11}
P.~Belli et~al.
\newblock {\em Eur. Phys. J. A}, 47:91, 2011.

\bibitem{SDSJ97}
J.~Suhonen, P.C. Divari, L.D. Skouras, and I.~D. Johnstone.
\newblock {\em Phys. Rev. C}, 55:714, 1997.

\bibitem{RCN95}
J.~Retamosa, E.~Caurier, and F.~Novacki.
\newblock {\em Phys. Rev. C}, 51:371, 1995.

\bibitem{CNPR96}
E.~Caurier, F.~Novacki, A.~Poves, and J.~Retamosa.
\newblock {\em Phys. Lett.}, 77:1954, 1996.

\bibitem{SSDV92}
J.~Sinatkas, L.D. Skouras, D.~Strottman, and J.D. Vergados.
\newblock {\em J. Phys. G}, 18:1377, 1992.

\bibitem{CPZ90}
E.~Caurier, A.~Poves, and A.P Zucker.
\newblock {\em Phys. Lett B}, 252:13, 1990.

\bibitem{SUPKAM}
Y.~Fukuda and others (SuperKamiokande~Collaboration).
\newblock {\em Phys. Rev. Lett.}, 81:1562, 1998.

\bibitem{AHARMIN}
B.~Aharmin et~al.
\newblock {\em Phys. Rev. C}, 72:055502, 2005.

\bibitem{CHOOZ}
M.~Apollonio and others (CHOOZ~Collaboration).
\newblock {\em Phys. Lett. B}, 446:415, 1999.

\bibitem{ARAKI}
T.~Araki et~al.
\newblock {\em Phys. Rev. Lett.}, 94:081801, 2005.

\bibitem{SCHWETZ}
T.~Schwetz, M.~T\'ortola, and J.W.F. Valle.
\newblock {\em New J. Phys.}, 10:113011, 2008.

\bibitem{DuLindZub11b}
M.~Duerr, M.~Lindner, and A.~Merle.
\newblock {\em JHEP}, 1106:091, 2011.

\bibitem{CDGG-CHMR10}
A.~Cervera, A.~Donini, M.~B. Gavela, J.~J.Gomez Cadenas, P.~Hernandez, O.~Mena,
  and S.~Rigolin.
\newblock {\em Nuc. Phys. B}, 579:17, 2000.

\bibitem{Agar11}
S.K. Agarwalla, arXiv:1110.3681 [hep-ph].

\bibitem{G-GM08}
M.C. Gonzales-Garcia and M.~Maltoni.
\newblock {\em Phys. Rep.}, 460:1, 2008.

\bibitem{Moh98}
R.~N. Mohapatra.
\newblock {\em Nucl. Phys. Proc. Suppl.}, 77:376, 1999.
\newblock hep-ph/9808284.

\bibitem{PSV96}
G.~Pantis, F.~{\v S}imkovic, J.D. Vergados, and A.~Faessler.
\newblock {\em Phys. Rev. C}, 53:695, 1996.

\bibitem{Moh86}
R.~Mohapatra.
\newblock {\em Phys. Rev. D}, 34:3457, 1986.

\bibitem{Ver87}
J.D. Vergados.
\newblock {\em Phys. Lett. B}, 184:55, 1987.

\bibitem{HKK95}
M.~Hirsch, H.V. Klapdor-Kleingrothaus, and S.G Kovalenko.
\newblock {\em Phys. Lett.}, 75:17, 1995.

\bibitem{PAES99}
H.~P{\" a}s, M.~Hirsch, and H.V. Klapdor-Kleingrothaus.
\newblock {\em Phys. Lett. B}, 459:450, 1999.

\bibitem{WKS97}
A.~Wodecki, W.~Kami{\' n}ski, and S.~Pagerka.
\newblock {\em Phys. Lett. B}, 413:342, 1997.

\bibitem{FKS98a}
A.~Faessler, S.~Kovalenko, and F.~{\v S}imkovic.
\newblock {\em Phys. Rev. D}, 58:115004, 1998.

\bibitem{FKS98b}
A.~Faessler, S.~Kovalenko, and F.~{\v S}imkovic.
\newblock {\em Phys. Rev. D}, 58:055004, 1998.

\bibitem{HIRSCH01}
M.~Hirsch and J.W.F. Valle.
\newblock {\em Nucl. Phys. B}, 557:60, 2001.

\bibitem{Ver81}
J.D. Vergados.
\newblock {\em Phys. Rev. C}, 24:640, 2981.

\bibitem{ponte68}
B.~Pontecorvo.
\newblock {\em Phys. Lett. B}, 26:630, 1968.

\bibitem{Ver82}
J.D. Vergados.
\newblock {\em Phys. Rev. D}, 25:914, 1982.

\bibitem{Ver85}
J.D. Vergados.
\newblock {\em Nucl. Phys. B}, 250:618, 1985.

\bibitem{tom85}
T.~Tomoda, A.~Faessler, K.~W. Schmid, and F.~Grummer.
\newblock {\em Phys. Lett. B}, 157:4, 1985.

\bibitem{SPVF}
F.~{\v S}imkovic, G.~Pantis, J.D. Vergados, and A.~Faessler.
\newblock {\em Phys. Rev. C}, 60:055502, 1999.

\bibitem{SEIL92}
F.~{\v S}imkovic nd~G.V.~Efimov, M.A. Ivanov, and V.E. Lyubovitskij.
\newblock {\em Z. Phys. A}, 341:193, 1992.

\bibitem{Ver90}
J.D. Vergados.
\newblock {\em Nuc. Phys. A}, 506:482, 1990.

\bibitem{PV90}
G.~Pantis and J.D. Vergados.
\newblock {\em Phys. Lett B}, 242:1, 1990.

\bibitem{FKPV91}
A.~Faessler, W.A. Kami{\' n}ski, G.~Pantis, and J.D. Vergados.
\newblock {\em Phys. Rev. C}, 43:21, 1991.

\bibitem{SKF90}
S.B.~Khadkikar J.~Suhonen and A.~Faessler.
\newblock {\em Phys. Lett. B}, 237:8, 1990.

\bibitem{Ver76}
J.D. Vergados.
\newblock {\em Phys. Rev. C}, 13:865, 1976.

\bibitem{HSS82}
W.~C. Haxton, G.~S. Stephenson, and D.~Strottman.
\newblock {\em Phys. Rev. D}, 25:2360, 1982.

\bibitem{SV83}
L.~Skouras and J.D. Vergados.
\newblock {\em Phys. Rev. C}, 28:2122, 1983.

\bibitem{ZBR90}
L.~Zhao, B.A. Brown, and W.A. Richter.
\newblock {\em Phys. Rev. C}, 42:1120, 1990.

\bibitem{ZB93}
L~. Zhao and B.A. Brown.
\newblock {\em Phys. Rev. C}, 47:2641, 1993.

\bibitem{Retal96}
R.B. Radha et~al.
\newblock {\em Phys. Rev. Lett.}, 76:2642, 1996.

\bibitem{NSM96}
T.~Sebe H.~Nakada and K.~Muto.
\newblock {\em Nucl. Phys. A}, 607:235, 1996.

\bibitem{KDL97}
D.J.~Dean S.E.~Koonin and K.~Langanke.
\newblock {\em Phys. Rep.}, 278:1, 1997.

\bibitem{VZ86}
P.~Vogel and M.R. Zirnbauer.
\newblock {\em Phys. Rev. Lett}, 57:3148, 1986.

\bibitem{CAT87}
O.~Civitarese, A.~Faessler, and T.~Tomoda.
\newblock {\em Phys. Lett. B}, 194:11, 1987.

\bibitem{MBK88}
K.~Muto and H.V. Klapdor.
\newblock {\em Phys. Lett. B}, 208:53, 1988.

\bibitem{EVJP91}
J.~Engel, P.~Vogel, X.D. Ji, and S.~Pittel.
\newblock {\em Phys. Lett. B}, 225:5, 1989.

\bibitem{RFSK91}
A.~A. Raduta, A.~Faessler, S.~Stoica, and W.~A. Kami{\' n}ski.
\newblock {\em Phys. Lett. B}, 254:7, 1991.

\bibitem{GV92}
A.~Griffiths and P.~Vogel.
\newblock {\em Phys. Rev. C}, 46:181, 1992.

\bibitem{SC93}
J.~Suhonen and O.~Civitarese.
\newblock {\em Phys. Lett. B}, 308:212, 1993.

\bibitem{CS94}
O.~Civitarese and J.~Suhonen.
\newblock {\em Nuc. Phys. A}, 575:251, 1994.

\bibitem{SSVP97}
F.~{\v S}imkovic, J., Schwieger, M.~Veselsk\'y, G.~Pantis, and A.~Faessler.
\newblock {\em Phys. Lett. B}, 393:267, 1997.

\bibitem{SPF98}
F.~{\v S}imkovic, G.~Pantis, and A.~Faessler.
\newblock {\em Prog Part. Phys.}, 40:285, 1998.

\bibitem{cheoun}
M.K. Cheoun, A.~Bobyk, A.~Faessler, F.~{\v S}imkovic, and G.~Teneva.
\newblock {\em Nucl. Phys. A}, 561:74, 1993.

\bibitem{MUT97}
K.~Muto.
\newblock {\em Phys. Lett. B}, 391:243, 1997.

\bibitem{TS95}
J.~Toivanen and J.~Suhonen.
\newblock {\em Phys. Rev. Lett.}, 75:410, 1995.

\bibitem{SSF96}
J.~Schwieger, F.~{\v S}imkovic, and A.~Faessler.
\newblock {\em Nuc. Phys. A}, 600:179, 1996.

\bibitem{phfb}
P.K. Rath, R.~Chandra, K.~Chaturvedi, P.K. Raina, and J.G. Hirsch.
\newblock {\em Phys. Rev. C}, 82:064310, 2010.

\bibitem{IBM09}
J.~Barea and F.~Iachello.
\newblock {\em Phys. Rev. C}, 79:044301, 2009.

\bibitem{edf}
T.R. Rodrigez and G.~Martinez-Pinedo.
\newblock {\em Phys. Rev. Lett.}, 105:252503, 2010.

\bibitem{Ejiri00}
H.~Ejiri.
\newblock {\em Phys. Rep.}, 338:265, 2000.

\bibitem{SMIRNOV04}
A.Yu. Smirnov, arXiv: hep-ph/0411194.

\bibitem{WEINBERG79}
S.~Weinberg.
\newblock {\em Phys. Rev. Lett.}, 43:1566, 1979.

\bibitem{SeeSaw07}
A.~Abada, C.Biggio, F.~Bonnet, M.B. Gavela, and T.~Hambye.
\newblock {\em JHEP}, 0712:061, 2007.
\newblock arXiv:0707.4058 (hep-ph).

\bibitem{MagWet80}
M.~Magg and C.~Wetterich.
\newblock {\em Phys. Lett. B}, 94:61, 1980.

\bibitem{LazShafWet81}
M.~Magg and C.~Wetterich.
\newblock {\em Nuc. Phys. B}, 181:287, 1981.

\bibitem{MohSen81}
R.~N. Mohapatra and G.~Senjanovich.
\newblock {\em Phys. Rev. D}, 23:165, 1981.

\bibitem{Abada08}
A.~Abada, C.Biggio, F.~Bonnet, M.B. Gavela, and T.~Hambye.
\newblock {\em Phys. Rev. D}, 78:033007, 2008.
\newblock arXiv:0803.0481 (hep-ph).

\bibitem{SeeSawIII08}
J.~Chakrabortty, A.~Dighe, S.~Goswami, and S.~Ray.
\newblock {\em Nuc. Phys. B}, 820:116, 2009.
\newblock arXiv:0812.2776(hep-ph).

\bibitem{Pascoli08}
C.~Boehm, Y.~Farzan, T.~Hambye, S.~Palomares-Ruiz, and S.~Pascoli.
\newblock {\em Phys. Rev. D}, 77:043516, 2008.

\bibitem{HVFC00}
O.~Haug, J.D. Vergados, A.~Faessler, and S.~Kovalenko.
\newblock {\em Nucl. Phys. B}, 565:38, 2000.

\bibitem{BFK98}
V.~Bednyakov, A.~Faessler, and S.~Kovalenko.
\newblock {\em Phys. Lett. B}, 442:203, 1998.

\bibitem{GKS04}
M.~G{\' o}{\' z}d{\' z}, W.A. Kami{\' n}ski, and F.~{\v S}imkovic.
\newblock {\em Phys. Rev. D}, 70:095005, 2004.

\bibitem{GKSA06}
M.~G{\' o}{\' z}d{\' z}, W.A. Kami{\' n}ski, F.~{\v S}imkovic, and A.~Faessler.
\newblock {\em Phys. Rev. D}, 74:055007, 2006.

\bibitem{Pascoli06}
S.~Pascoli, S.T. Petcov, and A.~Riotto.
\newblock {\em Nucl. Phys. B}, 774:1, 2007.

\bibitem{AltFer10}
G.~Altarelli and F.~Feruglio.
\newblock {\em Rev. Mod. Phys.}, 82:2701, 2010.

\bibitem{IKOOST10}
H.~Ishimori, T.~Kobayashi, H.~Ohki, H.~Okada, Y.~Shimizu, and M.~Tanimoto.
\newblock {\em Prog. Theor. Phys. Suppl.}, 183:1, 2010.

\bibitem{AntLeoRiz90}
I.~Antoniadis, G.K. Leontaris, and J.~Rizos.
\newblock {\em Phys. Lett. B}, 245:161, 1990.

\bibitem{LeoRiz99}
G.K. Leontaris and J.~Rizos.
\newblock {\em Nucl. Phys. B}, 554:3, 1999.

\bibitem{FarKounRiz90}
A.E. Faraggi, C.~Kounas, and J.~Rizos.
\newblock {\em Phys. Lett. B}, 648:84, 2007.

\bibitem{RamRobRoss93}
P.~Ramond, R.G. Roberts, and G.G. Ross.
\newblock {\em Nucl. Phys. B}, 406:19, 1993.

\bibitem{IbanRich09}
L.E. Iba$\tilde{n}$ez and R.~Richter.
\newblock {\em JHEP}, 0903:090, 2009.

\bibitem{AnasKirLion09}
P.~Anastasopoulos, E.~Kiritsis, and A.~Lionetto.
\newblock {\em JHEP}, 0908:026, 2009.

\bibitem{LEONTARIS09}
G.K. Leontaris.
\newblock {\em J. Mod. Phys. A}, 24:6035, 2009.

\bibitem{LeonVlach10}
G.K. Leontaris and N.~Vlachos.
\newblock {\em JHEP}, 1001:016, 2010.

\bibitem{MohPas05}
R.N. Mohapatra {\it et al}, A White Paper, arXiv:hep-ph/0510213.

\bibitem{Minos}
A.~Habig et~al. (MINOS~Collaboration).
\newblock {\em Mod. Phys. Lett. A}, 25:1219, 2010.

\bibitem{Gando11}
A.~Gando et~al. (KamLAND~Collaboration).
\newblock {\em Phys. Rev. D}, 83:052002, 2011.

\bibitem{dayabay}
F.P. An {\it et al.} (Daya Bay Collaboration), arXiv:1203.1669[hep-ex].

\bibitem{T2K}
K.~Abe et~al. (T2K~Collaboration).
\newblock {\em Phys. Rev. Lett.}, 107:041801, 2011.

\bibitem{dct13}
H.D. Kerret (Double Chooz Collaboration), LowNu11, November 9-12, 2011, Seoul
  National University, Seoul, Korea.

\bibitem{reno12}
J.K. Ahn {\it et al.} (RENO Collaboration), arXiv:1204.0626[hep-ex].

\bibitem{Pascoli05}
S.~Pascoli, S.T. Petcov, and T.~Schwetz.
\newblock {\em Nuc. Phys. B}, 734:24, 2006.

\bibitem{Katrin}
Osipowicz A et al (KATRIN Collaboration) 2001 hep-ex/0109033;\\ Angrik J et al
  (KATRIN Collaboration) 2004 KATRIN Design Report
  http://bibliothek.fzk.de/zb/berichte/FZKA7090.pdf.

\bibitem{otten}
E.W. Otten and C.~Weinheimer.
\newblock {\em Rep. Prog. Phys.}, 71:086201, 2008.

\bibitem{Mare}
E.~Andreotti~(MARE Collaboration).
\newblock {\em Nucl. Instrum. Meth.}, 572:208, 2007.

\bibitem{Abarajan11}
K.~N. Abazajian et~al.
\newblock {\em Astropart. Phys.}, 35:177, 2011.

\bibitem{ThAbdaLah10}
S.~A. Thomas, F.~B. Abdalla, and O.~Lahav.
\newblock {\em Phys.Rev.Let.}, 105:031301, 2010.

\bibitem{bilpot11}
S.M. Bilenky, A.~Faessler, W.~Potzel, and F.~{\v S}imkovic.
\newblock {\em Eur. Phys. J.}, 71:1754, 2011.

\bibitem{Thomas}
S.A. Thomas, F.B. Abdalla, and O.~Lahav.
\newblock {\em Phys. Rev. Lett}, 105:031301, 2010.

\bibitem{Serpico}
P.D. Serpico.
\newblock {\em Phys. Rev. Lett}, 98:171301, 2007.

\bibitem{Abazajian}
K.N.~Abazajian et~al.
\newblock {\em Astropart. Phys.}, 35:177, 2011.

\bibitem{src09}
F.~{\v S}imkovic, A.~Faessler, H.~M{\" u}ther, V.~Rodin, and M.~Stauf.
\newblock {\em Phys. Rev. C}, 79:055501, 2009.

\bibitem{ger08}
S.~Schoenert~(GERDA Collaboration).
\newblock {\em J. Phys. Conf. Ser.}, 203:012014, 2010.

\bibitem{allesa11}
F. Alessandria {\it et al.} (CUORE Collaboration), arXiv:1109.0494 [nucl-ex].

\bibitem{ack11}
N.~Ackerman et~al. (The EXO~Collaboration).
\newblock {\em Phys. Rev. Lett.}, 107:212501, 2011.

\bibitem{Exo}
R.~Gornea~(EXO Collaboration).
\newblock {\em J. Phys. Conf. Ser.}, 259:012039, 2010.

\bibitem{ell09}
S.R. Elliott {\it et al.} MAJORANA Collaboration 2009 {\it Private
  communication, 2009}.

\bibitem{deppisch11}
R.~Arnold et~al. (SuperNEMO~Collaboration).
\newblock {\em Eur. Phys. J}, 70:927, 2010.

\bibitem{snoplus}
Ch. Kraus~(SNO+ Collaboration).
\newblock {\em Prog. Part. Nucl. Phys.}, 64:273, 2010.

\bibitem{ASB07}
A.S. Barabash.
\newblock {\em Phys. Atom. Nucl.}, 70:1191, 2007.

\bibitem{ablimit}
A.S. Barabash.
\newblock {\em AIP Conf. Proc.}, 1417:5, 2011.

\bibitem{fang10}
Dong-Liang Fang, A.~Faessler, V.~Rodin, and F.~{\v S}imkovic.
\newblock {\em Phys. Rev. C}, 82:051301, 2010.

\bibitem{RNA11}
G.~Mention et~al.
\newblock {\em Phys. Rev. D}, 83:073006, 2011.

\bibitem{KMS11}
M.~Maltoni J.~Kopp and T.~Schwetz.
\newblock {\em Phys. Rev. Lett.}, 107:091801, 2011.

\bibitem{BRZ11}
J.~Barry, W.~Rodejohann, and He~Zhang.
\newblock {\em JHEP}, 1107:091, 2011.

\bibitem{Pascoli06b}
S.~Pascoli, S.T. Petcov, and A.~Riotto.
\newblock {\em Phys. Rev. D}, 75:083511, 2007.

\bibitem{Pascoli09}
A.~Atre, T.~Han, S.~Pascoli, and B.~Zhang.
\newblock {\em JHEP}, 0905:030, 2009.

\bibitem{PSVF96}
G.~Pantis, F.~{\v S}imkovic, J.D. Vergados, and A.~Faessler.
\newblock {\em Phys. Rev. C}, 53:695, 1996.

\bibitem{FMPSV11}
A.~Faessler, A.~Meroni, S.~T. Petcov, F.~{\v S}imkovic, and J.~D. Vergados.
\newblock {\em Phys. Rev. D}, 83:113003, 2011.

\bibitem{JDV02}
J.D. Vergados.
\newblock {\em Phys. Rep.}, 361:1, 2002.

\bibitem{PITTEL97}
O.~Civitarese, M.~Reboiro, and P.~Vogel.
\newblock {\em Phys. Rev. C}, 56:1840, 1997.

\bibitem{ENGEL97}
J.~Engel, S.~Pittel, M.~Stoitsov, P.~Vogel, and J.~Dukelsky.
\newblock {\em Phys. Rev. C}, 55:1781, 1997.

\bibitem{PINEDO}
G.~Martinez-Pinedo, K.~Langanke, and P.~Vogel.
\newblock {\em Nucl. Phys. A}, 651:379, 1999.

\bibitem{MOHA00}
R.N. Mohapatra, A.~Perez-Lorenzana, and C.A. de~S.~Pires.
\newblock {\em Phys. Lett. B}, 491:143, 2000.

\bibitem{TOMAS01}
R.~Tomas, H.~P{\" a}s, and J.W.F. Valle.
\newblock {\em Phys. Rev. D}, 64:095005, 2001.

\bibitem{MONTE00}
J.C. Montero, C.A. de~S.~Pires, and V.~Pleitez.
\newblock {\em Phys. Rev. D}, 64:096001, 2001.

\bibitem{BabMoh95}
K.S. Babu and R.N. Mohapatra.
\newblock {\em Phys. Rev. Lett.}, 75:2276, 1995.

\bibitem{HKK96}
M.~Hirsch, H.V. Klapdor-Kleingrothaus, and S.G Kovalenko.
\newblock {\em Phys. Rev. D}, 53:1329, 1996.

\bibitem{PHK-KK99}
H.~P{\" a}s, M.~Hirsch, H.V. Klapdor-Kleingrothaus, and S.G. Kovalenko.
\newblock {\em Phys. Lett. B}, 453:194, 1999.

\bibitem{FGKS07}
A.~Faessler, Th. Gutsche, S.~Kovalenko, and F.~{\v S}imkovic.
\newblock {\em Phys. Rev. D}, 77:113012, 2008.

\bibitem{VerFaeTok10}
J.D. Vergados, A.~Faessler, and H.~Toki.
\newblock {\em Phys. Rev. D}, 81:034018, 2010.

\bibitem{adler}
S.L. Adler et~al.
\newblock {\em Phys. Rev. D}, 11:3309, 1975.

\bibitem{ell02}
S.R. Elliott and P.~Vogel.
\newblock {\em Annu. Rev. Nucl. Part. Sci.}, 52:115, 2002.

\bibitem{der66}
E.~Der Mateosian and M.~Goldharber.
\newblock {\em Phys. Rev.}, 146:810, 1966.

\bibitem{fio67}
E.~Fiorini, A.~Pullia, G.~Bertolini, F.~Cappellani, and G.~Restelli.
\newblock {\em Phys. Lett B.}, 25:602, 1967.

\bibitem{fio73}
E.~Fiorini et~al.
\newblock {\em Nuovo Cimento A}, 13:747, 1973.

\bibitem{sha67}
M.~H. Schapiro, S.~Frankel, S.~Koicki, W.~D. Wales, and G.~T. Wood.
\newblock {\em Phys. Rev.}, 154:1050, 1967.

\bibitem{bar67}
R.~K. Bardin, P.~G. Gollon, J.~D. Ullman, and C.~S. Wu.
\newblock {\em Phys. Lett. B}, 26:112, 1967.

\bibitem{bar70}
R.~K. Bardin, P.~G. Gollon, J.~D. Ullman, and C.~S. Wu.
\newblock {\em Nuc. Phys. A}, 158:337, 1970.

\bibitem{cle75}
B.~T. Cleveland et~al.
\newblock {\em Phys. Rev. Lett.}, 35:757, 1975.

\bibitem{bel82}
E.~Bellotti, E.~Fiorini, C.~Liguori, A.~Pallia, A.~Sarracino, and L.~Zanotti.
\newblock {\em Lett. Nuovo Cimento}, 33:273, 1982.

\bibitem{avi83}
F.~T. Avignone et~al.
\newblock {\em Phys. Rev. Lett.}, 50:721, 1983.

\bibitem{eji84}
H. Ejiri {\it et al.} {\it Proc. Neutrino mass and low energy weak
  interactions, (Teremark)}, 1984 (World Scientific) p. 383.

\bibitem{eji86}
E.~Ejiri et~al.
\newblock {\em Nucl. Phys. A}, 448:271, 1986.

\bibitem{eji87}
E.~Ejiri et~al.
\newblock {\em J. Phys. G; Nucl.Phys.}, 13:839, 1987.

\bibitem{kir69}
T.~Kirsten and H.~W. Muller.
\newblock {\em Earth and Planetary Sci. Lett.}, 6:271, 1969.

\bibitem{kir69b}
T.~Kirsten and H.~W. Muller.
\newblock {\em Earth and Planetary Sci. Lett.}, 7:300, 1969.

\bibitem{eji91}
H.~Ejiri et~al.
\newblock {\em Phys. Lett. B}, 258:17, 1991.

\bibitem{eji91b}
H.~Ejiri et~al.
\newblock {\em J. Phys. G}, 17:S155, 1991.

\bibitem{fio84}
E.~Fiorini and T.~O. Niinikoski.
\newblock {\em Nucl. Instr. Methods}, 224:83, 1984.

\bibitem{ale92}
A.~Alessandrello et~al.
\newblock {\em Nucl. Phys. B (Proc. Suppl.)}, 28:233, 1992.

\bibitem{ale98}
A.~Alessandrello et~al.
\newblock {\em Phys. Lett. B}, 420:109, 1998.

\bibitem{gir10}
L.~Gironi et~al.
\newblock {\em Nucl. Instr. Methods A}, 617:478, 2010.

\bibitem{ber93}
T.~Bernatowicz et~al.
\newblock {\em Phys. Rev. C}, 47:8066, 1993.

\bibitem{kla01}
H.~V. Klapdor-Kleingrothaus et~al.
\newblock {\em Phys. Rev. D}, 63:073005, 2001.

\bibitem{kla01b}
H.~V. Klapdor-Kleingrothaus et~al.
\newblock {\em Phys. Lett. A}, 16:2469, 2001.

\bibitem{aal99}
C.~E.~Aalsethand others.
\newblock {\em Phys. Rev. C}, 59:2108, 1999.

\bibitem{aal02}
C.~E.~Aalsethand others.
\newblock {\em Phys. Rev. D}, 65:092007, 2002.

\bibitem{arn08}
C.~Arnaboldi et~al.
\newblock {\em Phys. Rev. C}, 78:035502, 2008.

\bibitem{arn11}
C.~Arnaboldi et~al.
\newblock {\em Astropart. Phys.}, 34:822, 2011.

\bibitem{eji01}
E.~Ejiri et~al.
\newblock {\em Phys. Rev. C}, 63:065501, 2001.

\bibitem{arn04}
C.~Arnaboldi et~al.
\newblock {\em JETP Lett.}, 80:377, 2004.

\bibitem{arn05}
C.~Arnaboldi et~al.
\newblock {\em Phys. Rev. Lett.}, 95:142501, 2005.

\bibitem{lssm}
J.~Men\'endez, A.~Poves, E.~Caurier, and F.~Nowacki.
\newblock {\em Nucl. Phys. A}, 818:139, 2009.

\bibitem{oga04}
I.~Ogawa et~al.
\newblock {\em Nucl. Phys. A}, 730:215, 2004.

\bibitem{dan03}
F.A. Danevich et~al.
\newblock {\em Phys. Rev. C}, 68:035501, 2003.

\bibitem{cre11}
O. Cremonesi and The CUORE collaboration, {\it Talk given at MEDEX11 conf.,
  Prague, 2011}, http://medex11.utef.cvut.cz/.

\bibitem{fla08}
R. L. Flack, NEMOIII collaboration 2008 {\it Proc. Neutrino08 (Christchurch)},
  2008.

\bibitem{recbb}
A.S. Barabash.
\newblock {\em Phys. Rev. C}, 81:035501, 2010.

\bibitem{hen11}
E.~Aguayo et~al. (The MAJORANA~collaboration).
\newblock {\em AIP Conf. Proc.}, 1417:95, 2011.

\bibitem{oga09}
I. Ogawa I {\it et al.} {\it Private communication}, 2009.

\bibitem{eji00a}
H.~Ejiri et~al.
\newblock {\em Phys. Rev. Lett.}, 85:2917, 2000.

\bibitem{eji07a}
H.~Ejiri.
\newblock {\em Mod. Phys. Lett. A}, 22:1277, 2007.

\bibitem{nak07}
H.~Nakamura et~al.
\newblock {\em J. Phys. Soc. Japan}, 76:114201, 2007.

\bibitem{eji08}
H.~Ejiri et~al.
\newblock {\em Eur. Phys. Jour.}, 162:239, 2008.

\bibitem{geh09}
V.~M. Gehman, P.J. Doe, R.~G.~H. Robertson, D.~J. Will, H.~Ejiri, and
  R.~Hazama.
\newblock {\em Nucl. Instr. Method Phys. Research A}, 622:602, 2010.

\bibitem{eji01c}
H.~Ejiri, J.~Engel, and N.~Kudomi.
\newblock {\em Phys. Lett. B}, 530:27, 2002.

\bibitem{cha11}
A.~Chapon and The~SuperNEMO collaboration.
\newblock {\em AIP Conf. Proc.}, 1417:18, 2011.

\bibitem{kim11}
S.K. Kim, AMoRE Collaboration, {\it Talk given at MEDEX11 conf., Prague, 2011},
  http://medex11.utef.cvut.cz/.

\bibitem{zub11}
K. Zuber and The COBRA collaboration, {\it Talk given at MEDEX11 conf., Prague,
  2011}, http://medex11.utef.cvut.cz/.

\bibitem{gra08}
G. Gratta, {\it et al.}, {\it Talk given at Neutrino 08 conf., New Zealand,
  2008}, http://www2.phys.canterbury.ac.nz/~jaa53/.

\bibitem{EXO2012}
M. Auger et al, EXO Collaboration 2012, arXiv:1205.5608v1 [hep-ex].

\bibitem{EXO11}
N.~Ackermann et~al.
\newblock {\em Phys. Rev. Lett.}, 107:212501, 2011.

\bibitem{efr11}
Y. Efremenko and The KamLAND-Zen Collaboration, {\it Talk given at MEDEX11
  conf., Prague, 2011}, http://medex11.utef.cvut.cz/.

\bibitem{NEXT12}
V. \'{A}lvarez et al (NEXT collaboration), arXiv:1202.0721v2 [physics.ins-det].

\bibitem{rag09}
R. Raghavan, in Proceedings of NDM09, Madison 2009.

\bibitem{ish09}
N. Ishihara {\it et al.} {\it Private communication}, 2009.

\bibitem{sno09}
K. Zuber and The COBRA collaboration, {\it Talk given at MEDEX11 conf., Prague,
  2011}, http://medex11.utef.cvut.cz/.

\bibitem{eji00}
H.~Ejiri.
\newblock {\em Phys. Rep.}, 338:265, 2000.
\newblock and refs. therein.

\bibitem{eji06}
H.~Ejiri.
\newblock {\em Czech. J. Phys.}, 56:459, 2006.

\bibitem{eji96}
H.~Ejiri and H.~Toki.
\newblock {\em J. Phys. Soc. Japan}, 65:7, 1996.

\bibitem{suh06}
J.~Suhonen and M.~Kortelainen.
\newblock {\em Czech J. Phys.}, 56:519, 2006.

\bibitem{avi00}
F. Avignone {\it Workshop Neutr. Nucl. Phys. Stopped $\pi \mu $ Facility (Oak
  Ridge)}, 2000.

\bibitem{eji03a}
H.~Ejiri.
\newblock {\em Nucl. Instr. Meth. Phys. Research}, 503:276, 2003.

\bibitem{eji68}
H.~Ejiri et~al.
\newblock {\em Phys. Rev. Lett.}, 21:373, 1968.

\bibitem{eji68b}
H.~Ejiri et~al.
\newblock {\em Nucl. Phys. A}, 128:388, 1969.

\bibitem{eji68A}
H.~Ejiri and J.~Bondorf.
\newblock {\em Phys. Lett. B}, 28:304, 1968.

\bibitem{aki97}
H.~Akimune et~al.
\newblock {\em Phys. Lett. B}, 394:23, 1997.

\bibitem{aki08}
H.~Akimune et~al.
\newblock {\em Phys. Lett. B}, 665:424, 2008.

\bibitem{doh08}
H.~Dohman et~al.
\newblock {\em Phys. Rev. C}, 78:041602R, 2008.

\bibitem{zeg02}
R.G.T. Zegers et~al.
\newblock {\em Phys. Rev. Lett.}, 99:202501, 2007.

\bibitem{rak04}
S.~Rakers et~al.
\newblock {\em Phys. Rev. C}, 70:054302, 2004.

\bibitem{Raker05}
S.~Rakers et~al.
\newblock {\em Phys. Rev. C}, 71:054313, 2005.

\bibitem{Grewe07}
E.W.~Grewe et~al.
\newblock {\em Phys. Rev. C}, 76:054307, 2007.

\bibitem{Grewe08}
E.W.~Grewe et~al.
\newblock {\em Phys. Rev. C}, 78:044301, 2008.

\bibitem{Yako09}
K.~Yako et~al.
\newblock {\em Phys. Rev. Lett.}, 103:012503, 2009.

\bibitem{Frekers10}
D.~Frekers.
\newblock {\em Prog. Part. Nucl. Phys.}, 64:281, 2010.

\bibitem{Puppe11}
P.~Puppe et~al.
\newblock {\em Phys. Rev. C}, 84:051305, 2011.

\bibitem{Guess11}
C.J.~Guess et~al.
\newblock {\em Phys. Rev. C}, 83:064318, 2011.

\bibitem{sch08}
J.~P. Schiffer et~al.
\newblock {\em Phys. Rev. Lett.}, 100:112501, 2008.

\bibitem{aba84}
J.~Abad, A.~Morales, R.~Nunez-Lagos, and A.~F. Pacheco.
\newblock {\em Nuovo Cim. A}, 75:173, 1983.

\bibitem{eji09}
H.~Ejiri.
\newblock {\em J. Phys. Soc. Japan}, 78:074201, 2009.

\bibitem{eji12}
H.~Ejiri.
\newblock {\em J. Phys. Soc. Japan}, 81:033201, 2012.

\bibitem{dosm}
A.D. Dolgov and A.Yu. Smirnov.
\newblock {\em Phys. Lett. B}, 621:1, 2005.

\bibitem{bosonic}
A.S. Barabash, A.D. Dolgov, F.~{\v S}imkovic R.~Dvornick{\' y}, and A.Yu.
  Smirnov.
\newblock {\em Nucl. Phys. B}, 783:90, 2007.

\bibitem{SDS01}
F.~{\v S}imkovic, P.~Domin, and S.V. Semenov.
\newblock {\em J. Phys. G}, 27:2233, 2001.

\bibitem{DKSS05}
P.~Domin, S.~Kovalenko, F.~{\v S}imkovic, and S.V. Semenov.
\newblock {\em Nucl. Phys. A}, 753:337, 2005.

\bibitem{grabm10}
A.~Smolnikov and P.~Grabmayr.
\newblock {\em Phys. Rev. C}, 81:028502, 2010.

\bibitem{qrpa2}
V.A. Rodin, A.~Faessler, F.~{\v S}imkovic, and P.~Vogel.
\newblock {\em Nucl. Phys. A}, 766:107, 2006.

\bibitem{qrpa3}
V.A. Rodin, A.~Faessler, F.~{\v S}imkovic, and P.~Vogel.
\newblock {\em Nucl. Phys. A}, 793:213, 2007.

\bibitem{sim08}
F.~{\v S}imkovic, A.~Faessler, V.A. Rodin, P.~Vogel, and J.~Engel.
\newblock {\em Phys. Rev. C}, 77:045503, 2008.

\bibitem{ism1}
A.~Poves, E.~Caurier, and F.~Nowacki.
\newblock {\em Eur. Phys. J. A}, 36:195, 2008.

\bibitem{ism2}
E.~Caurier, J.~Menendez, F.~Nowacki, and A.~Poves.
\newblock {\em Phys. Rev.Lett.}, 100:052503, 2008.

\bibitem{horoi07}
M.~Horoi, S.~Stoica, and B.~A. Brown.
\newblock {\em Phys. Rev. C}, 75:034303, 2007.

\bibitem{horoi10}
M.~Horoi and S.~Stoica.
\newblock {\em Phys. Rev. C}, 81:024321, 2010.

\bibitem{ocup09}
F.~{\v S}imkovic, A.~Faessler, and P.~Vogel.
\newblock {\em Phys. Rev. C}, 79:015502, 2009.

\bibitem{miler}
G.A. Miller and J.~E. Spencer.
\newblock {\em Ann. Phys. (NY)}, 100:562, 1976.

\bibitem{kor07}
M.~Kortelainen and J.~Suhonen.
\newblock {\em Phys. Rev. C}, 75:051303(R), 2007.

\bibitem{korte3}
M.~Kortelainen and J.~Suhonen.
\newblock {\em Phys. Rev. C}, 76:024315, 2007.

\bibitem{suho11}
J.~Suhonen.
\newblock {\em Nucl. Phys. A}, 864:63, 2011.

\bibitem{qrpa1}
V.A. Rodin, A.~Faessler, F.~{\v S}imkovic, and P.~Vogel.
\newblock {\em Phys. Rev. C}, 68:044302, 2003.

\bibitem{crit_plb05}
J.~Suhonen.
\newblock {\em Phys. Lett. B}, 607:87, 2005.

\bibitem{crit_npa05}
O.~Civitarese and J.~Suhonen.
\newblock {\em Nucl. Phys. A}, 761:313, 2005.

\bibitem{delion97}
D.S. Delion, J.~Dukelsky, and P.~Schuck.
\newblock {\em Phys. Rev. C}, 55:2340, 1997.

\bibitem{krmpotic98}
F.~Krmpoti{\' c} et~al.
\newblock {\em Nucl. Phys. A}, 637:295, 1998.

\bibitem{muto94}
K.~Muto.
\newblock {\em Nucl. Phys. A}, 577:415c, 1994.

\bibitem{engel_src}
J.~Engel and G.~Hagen.
\newblock {\em Phys. Rev. C}, 79:064317, 2009.

\bibitem{korte1}
M.~Kortelainen, O.~Civitarese, J.~Suhonen, and J.~Toivanen.
\newblock {\em Phys. Lett. B}, 647:128, 2007.

\bibitem{roth05}
R.~Roth, H.~Hergert, P.~Papakonstantinou, T.~Neff, and H.~Feldmeier.
\newblock {\em Phys. Rev. C}, 72:034002, 2005.

\bibitem{engel_m11}
J.~Engel, J.~Carlson, and R.B. Wiringa.
\newblock {\em Phys. Rev. C}, 83:034317, 2011.

\bibitem{deform04}
F.~{\v S}imkovic, L.~Pacearescu, and A.~Faessler.
\newblock {\em Nucl. Phys. A}, 733:321, 2004.

\bibitem{alvar04}
R.~Alvarez-Rodriguez, P.~Sarriguren, E.~Moya de~Guerra, L.~Pacearescu,
  A.~Faessler, and F.~{\v S}imkovic.
\newblock {\em Phys. Rev. C}, 70:064309, 2004.

\bibitem{phfb05}
R.~Chandra, J.~Singh, P.K. Rath, P.K. Raina, and J.~G. Hirsch.
\newblock {\em Eur. Phys. J. A}, 23:223, 2005.

\bibitem{cha08}
R.~Chandra K.~Chaturvedi, P.K. Rath, P.K. Raina, and J.G. Hirsch.
\newblock {\em Phys. Rev. C}, 78:054302, 2008.

\bibitem{saleh09}
M.~Yousuf Saleh, V.~Rodin, A.~Faessler, and F.~{\v S}imkovic.
\newblock {\em Phys. Rev. C}, 79:014314, 2009.

\bibitem{saleh10}
Dong-Liang Fang, A.~Faessler, V.~Rodin, Mohamed Saleh, and F.~{\v S}imkovic.
\newblock {\em Phys. Rev. C}, 81:037303, 2010.

\bibitem{fang11}
V.~Rodin, A.~Faessler, and F.~{\v S}imkovic.
\newblock {\em Phys. Rev. C}, 83:034320, 2011.

\bibitem{schiffer08}
J.P.~Schiffer et~al.
\newblock {\em Phys. Rev. Lett.}, 100:112501, 2008.

\bibitem{schiffer09}
B.P.~Kay et~al.
\newblock {\em Phys. Rev. C}, 79:021301, 2009.

\bibitem{menen09}
J.~Men{\' e}ndez, A.~Poves, E.~Caurier, and F.~Nowacki.
\newblock {\em Phys. Rev. C}, 80:048501, 2009.

\bibitem{suhoc08}
J.~Suhonen and O.~Civitarese.
\newblock {\em Phys. Lett. B}, 668:277, 2008.

\bibitem{suhoc11}
J.~Suhonen.
\newblock {\em Nucl. Phys. A}, 853:36, 2011.

\bibitem{lisi08}
A.~Faessler, G.L. Fogli, V.~Rodin E.~Lisi, A.M. Rotunno, and F.~{\v S}imkovic.
\newblock {\em J. Phys. G}, 35:075104, 2008.

\bibitem{lisi09}
A.~Faessler, G.L. Fogli, E.~Lisi, V.~Rodin, and F.~{\v S}imkovic.
\newblock {\em Phys. Rev. D}, 79:053001, 2009.

\bibitem{0n2n11}
F.~{\v S}imkovic, R.~Hod\'ak, A.~Faessler, and P.~Vogel.
\newblock {\em Phys. Rev. C}, 83:015502, 2011.

\bibitem{vad_mf}
V.~Rodin and A.~Faessler.
\newblock {\em Phys. Rev. C}, 80:041302, 2009.

\bibitem{benes05}
P.~Bene\v{s}, A.~Faessler, S.~Kovalenko, and F.~{\v S}imkovic.
\newblock {\em Phys. Rev. D}, 71:077901, 2005.

\bibitem{sterile11}
M.~Mitra, G.~Senjanovi{\' c}, , and F.~Vissani.
\newblock {\em Nucl. Phys. B}, 856:26, 2012.

\bibitem{lisi11}
A.~Faessler, G.L. Fogli, E.~Lisi, A.M. Rotunno, and F.~{\v S}imkovic.
\newblock {\em Phys. Rev. D}, 83:113015, 2011.

\bibitem{ibarra11}
E.~Molinaro A.~Ibarra and S.T. Petcov.
\newblock {\em Phys. Rev. D}, 84:013005, 2011.

\bibitem{vissa11}
V.~Tello, M.~Nemev\v{s}ek, F.~Nesti, G.~Senjanovi\'{c}, and F.~Vissani.
\newblock {\em Phys. Rev. Lett.}, 106:151801, 2011.

\bibitem{AKP09a}
B.C. Allanach, C.H. Kolm, and H.~P{\" a}s.
\newblock {\em JHEP}, 0910:026, 2009.

\bibitem{AKP09b}
B.C. Allanach, C.H. Kolm, and H.~P{\" a}s.
\newblock {\em Phys. Rev. Let.}, 103:091801, 2009.

\bibitem{egamma}
V.~Cirigliano, A.~Kurylov, M.J. Ramsey-Musolf, and P.~Vogel.
\newblock {\em Phys. Rev. Lett.}, 93:231802, 2004.

\bibitem{lrsm2}
R.N. Mohapatra and J.C. Pati.
\newblock {\em Phys. Rev. D}, 11:566, 1975.

\bibitem{ayrg11}
J.~Argyriades et~al. (SuperNEMO~Collaboration).
\newblock {\em Nucl. Instrum. Meth. A}, 622:120, 2010.

\bibitem{klapmech1}
H.V. Klapdor-Kleingrothaus, I.V. Krivosheina, and I.V. Titkova.
\newblock {\em Phys. Rev. D}, 73:013010, 2006.

\bibitem{klapmech2}
H.V. Klapdor-Kleingrothaus, I.V. Krivosheina, and I.V. Titkova.
\newblock {\em Int. J. Mod. Phys. A}, 21:1159, 2006.

\bibitem{klapmech3}
H.V. Klapdor-Kleingrothaus, I.V. Krivosheina, V.~Mironov, and I.V. Titkova.
\newblock {\em Phys. Lett. B}, 636:235, 2006.

\bibitem{klapmech4}
H.V. Klapdor-Kleingrothaus, I.V. Krivosheina, and I.V. Titkova.
\newblock {\em Mod. Phys. Lett. A}, 21:1257, 2006.

\bibitem{deppisch10}
F.~Deppisch, C.~Jackson, I.~Nasteva, and S.~S{\" o}ldner-Rembold.
\newblock {\em Prog. Part. Nucl. Phys.}, 64:278, 2010.

\bibitem{DuLindZub11}
M.~Duerr, M.~Lindner, and K.~Zuber.
\newblock {\em Phys. Rev. D}, 84:093004, 2011.

\bibitem{mooncol}
H.~Ejiri~(MOON Collaboration).
\newblock {\em Nucl. Phys. A}, 844:10c, 2010.

\bibitem{cobracol}
K.~Zuber.
\newblock {\em Prog. Part. Nucl. Phys.}, 64:267, 2010.

\bibitem{PHK-KK01}
M.~Hirsch H.~P{\" a}s, H.V. Klapdor-Kleingrothaus, and S.G. Kovalenko.
\newblock {\em Phys. Lett. B}, 498:35, 2001.

\bibitem{PR-MV03}
G.~Pr\'{e}zeau, M.~Ramsey-Musolf, and P.~Vogel.
\newblock {\em Phys. Rev. D}, 68:034016, 2003.

\bibitem{borisov07}
A.V.~Borisov A.~Ali and D.V. Zhuridov.
\newblock {\em Phys. Rev. D}, 76:093009, 2007.

\bibitem{ali10}
A.V.~Borisov A.~Ali and D.V. Zhuridov.
\newblock {\em Phys. Atom. Nucl.}, 73:2083, 2010.

\bibitem{FKSW97}
A.~Faessler, S.~Kovalenko, F.~{\v S}imkovic, and J.~Schwieger.
\newblock {\em Phys. Lett. B}, 78:183, 1997.

\bibitem{fsdistin}
F.~{\v S}imkovic and A.~Faessler.
\newblock {\em Prog. Part. Nucl. Phys.}, 48:201, 2002.

\bibitem{bilratio}
S.M. Bilenky and J.A. Grifols.
\newblock {\em Phys. Lett. B}, 550:154, 2002.

\bibitem{depaes07}
F.~Deppisch and H.~P{\" a}s.
\newblock {\em Phys. Rev. Lett.}, 98:232501, 2007.

\bibitem{gehman07}
V.M. Gehman and S.~R. Elliott.
\newblock {\em J. Phys. G}, 34:667, 2007.

\bibitem{simk1p}
F.~{\v S}imkovic, M.~Nowak, W.A. Kami{\' n}ski, A.A. Raduta, and A.~Faessler.
\newblock {\em Phys. Rev. C}, 64:035501, 2001.

\bibitem{tomo00}
T.~Tomoda.
\newblock {\em Phys. Lett. B}, 474:245, 2000.

\bibitem{HMOK94}
M.~Hirsch, K.~Muto, T.~Oda, and H.V. Klapdor-Kleingrothaus.
\newblock {\em Z. Phys. A}, 347:151, 1994.

\bibitem{suhex00a}
J.~Suhonen.
\newblock {\em Phys. Lett. B}, 477:99, 2000.

\bibitem{suhex00b}
J.~Suhonen.
\newblock {\em Phys. Rev. C}, 62:042501, 2000.

\bibitem{tomo86}
T.~Tomoda, A.~Faessler, K.W. Schmid, and F.~Gr{\" u}mmer.
\newblock {\em Nucl. Phys. A}, 452:591, 1986.

\bibitem{MBK89}
K.~Muto, E.~Bender, and H.V. Klapdor.
\newblock {\em Z. Phys. A}, 334:187, 1989.

\bibitem{khadki91}
J.~Suhonen, S.B. Khadkikar, and A.~Faessler.
\newblock {\em Nucl. Phys. A}, 535:509, 1991.

\bibitem{ratheavy}
P.K. Rath, R.~Chandra, P.K. Raina, and K.~Chaturvedi.
\newblock {\em Phys. Rev. C}, 85:014308, 2012.

\bibitem{SVF10}
F.~{\v S}imkovic, J.D. Vergados, and A.~Faessler.
\newblock {\em Phys. Rev. D}, 82:113015, 2010.

\bibitem{jochum10}
J.~Jochum~(GERDA Collaboration).
\newblock {\em Prog. Part. Nucl. Phys.}, 64:261, 2010.

\bibitem{schonert10}
S.~Schonert~(GERDA Collaboration).
\newblock {\em J. Phys. Conf. Ser.}, 203:012014, 2010.

\bibitem{Gomez12}
J.J. Gomez-Cadenas, J.~Martin-Albo, M.~Mezzetto, F.~Monrabal, and M.~Sorel.
\newblock {\em Riv. Nuovo Cim.}, 35:29, 2012.

\bibitem{lukas}
L.~Lukaszuk, Z.~Sujkowski, and S.~Wycech.
\newblock {\em Eur. Phys. J. A}, 27:63, 2006.

\bibitem{AUDI03}
G.~Audi et~al.
\newblock {\em Nucl. Phys. A}, 729:3, 2003.

\bibitem{penning1}
G.~Douysset et~al.
\newblock {\em Phys. Rev. Lett.}, 86:4259, 2001.

\bibitem{BNW10}
K.~Blaum, Yu.N. Novikov, and G.~Werth.
\newblock {\em Contemp. Phys.}, 51:149, 2010.

\bibitem{redshaw07}
M.~Redshaw et~al.
\newblock {\em Phys. Rev. Lett.}, 98:053003, 2007.

\bibitem{redshaw09}
M.~Redshaw et~al.
\newblock {\em Phys. Rev. Lett.}, 102:212502, 2009.

\bibitem{SCIE09}
N.D. Scielzo et~al.
\newblock {\em Phys. Rev. C}, 80:0225501, 2009.

\bibitem{RAKH09}
S.~Rahaman et~al.
\newblock {\em Phys. Rev. Lett.}, 103:042501, 2009.

\bibitem{kolhinen}
V.S. Kolhinen et~al.
\newblock {\em Phys. Lett. B}, 684:17, 2010.

\bibitem{mount10}
B.J. Mount, M.~Redshaw, , and E.~G. Myers.
\newblock {\em Phys. Rev. C}, 81:032501, 2010.

\bibitem{elis2}
S.~Eliseev et~al.
\newblock {\em Phys. Rev. C}, 83:038501, 2011.

\bibitem{EliNov}
S.~Eliseev et~al.
\newblock {\em Phys. Rev. C}, 84:012501, 2011.

\bibitem{elis4}
M.~Goncharov et~al.
\newblock {\em Phys. Rev. C}, 84:028501, 2011.

\bibitem{elis5}
S.~Eliseev et~al.
\newblock {\em Phys. Rev. Lett.}, 107:152501, 2011.

\bibitem{droese11}
C.~Droese et~al.
\newblock {\em Nucl. Phys. A}, 875:1, 2012.

\bibitem{verg11}
J.D. Vergados.
\newblock {\em Phys. Rev. C}, 84:044328, 2011.

\bibitem{LARKINS}
F.B. Larkins.
\newblock {\em At. Data Nucl. Data Tables}, 20:311, 1977.

\bibitem{CAMP}
J.L. Campbell and T.~Papp.
\newblock {\em At. Data Nucl. Data Tables}, 77:1, 2001.

\bibitem{SKF11}
F.~{\v S}imkovic, M.I. Krivoruchenko, and A.~Faessler.
\newblock {\em Prog. Part. Nucl. Phys.}, 66:446, 2011.

\bibitem{FBE11}
Dong-Liang Fang, K.~Blaum, S.~Eliseev, A.~Faessler, M.I. Krivoruchenko,
  V.~Rodin, and F.~{\v S}imkovic.
\newblock {\em Phys. Rev. C}, 85:035503, 2012.

\bibitem{suho2011}
J.~Suhonen.
\newblock {\em Phys. Lett. B}, 701:490, 2011.

\bibitem{barab1}
A.S. Barabash, Ph. Hubert, A.~Nachab, and V.~Umatov.
\newblock {\em Nucl. Phys. A}, 785:371, 2007.

\bibitem{barab2}
A.S. Barabash, Ph. Hubert, A.~Nachab, S.I. Konovalov, I.A. Vanyushin, and V.I.
  Umatov.
\newblock {\em Nucl. Phys. A}, 807:269, 2008.

\bibitem{belli09}
P.~Belli et~al.
\newblock {\em Nucl. Phys. A}, 824:101, 2009.

\bibitem{rukh11}
N.I.~Rukhadze et~al. (TGV~Collaboration).
\newblock {\em Nucl. Phys. A}, 852:197, 2011.

\bibitem{FPS11}
D.~Frekers, P.~Puppe, J.H. Thies, P.~Povinec, F.~{\v S}imkovic, J.~Stani{\v
  c}ek, and I.~S{\' y}kora.
\newblock {\em Nucl. Phys. A}, 860:1, 2011.

\bibitem{BELLI11b}
P.~Belli et~al.
\newblock {\em Nucl. Phys. A}, 859:126, 2011.

\bibitem{frekdeg}
D. Frekers, arXiv:hep-ex/0506002.

\end{thebibliography}
\end{document}